\RequirePackage{fix-cm}
\documentclass[mnsc]{informs4Preprint}
\RequirePackage{tgtermes}
\RequirePackage{newtxtext}
\RequirePackage{newtxmath}
\RequirePackage{endnotes}

\makeatletter
\DeclareRobustCommand{\qed}{%
  \ifmmode 
  \else \leavevmode\unskip\penalty9999 \hbox{}\nobreak\hfill
  \fi
  \quad\hbox{\qedsymbol}}
\newcommand{\qedsymbol}{\openbox}
\renewenvironment{proof}[1][\proofname]{\par
  \normalfont
  \topsep6\p@\@plus6\p@ \trivlist
  \item[\hskip\labelsep\itshape
    #1.]\ignorespaces
}{%
  \qed\endtrivlist
}
\newcommand{\proofname}{Proof}
\makeatother

\OneAndAHalfSpacedXI

\usepackage{algorithm}
\usepackage{algpseudocode}
\usepackage{tikz}

\usepackage{natbib}
 \bibpunct[, ]{(}{)}{,}{a}{}{,}%
 \def\bibfont{\small}%
\EquationsNumberedThrough    

\TheoremsNumberedThrough     
\ECRepeatTheorems  %

\MANUSCRIPTNO{}

\usepackage{pgfplots}
\pgfplotsset{compat=1.18,compat/labels=pre 1.3}
\usepackage{tikz}
\usepackage{subcaption}
\usetikzlibrary{topaths,calc}
\usepackage{soul}
\usepackage{amsmath,amssymb}
\usepackage[dvipsnames]{xcolor}
\usepackage{multicol}
\usepackage{natbib}
\bibpunct{(}{)}{;}{a}{,}{,}
\usepackage[utf8]{inputenc} 
\usepackage[T1]{fontenc}    
\usepackage[colorlinks=true, linkcolor=blue!70!black, citecolor=blue!70!black,urlcolor=black,breaklinks=true, hypertexnames=false]{hyperref}
\usepackage{hyperref}
\usepackage{subcaption}

\usepackage{comment}
\usepackage{url}            
\usepackage{booktabs}       
\usepackage{nicefrac}       
\usepackage{microtype}      

\usepackage{algorithm}
\usepackage{algpseudocode}

\usepackage{bm}
\usepackage{mathtools}
\usepackage{color-edits}
\addauthor{yf}{blue}    
\addauthor{wt}{purple}
\addauthor{tianze}{red}
\addauthor{ms}{Green}
\usepackage{array}
\usepackage{makecell}
\newcolumntype{C}[1]{>{\centering\arraybackslash}p{#1}}

\usepackage{enumitem}

\usetikzlibrary{patterns, decorations.pathreplacing, arrows.meta}

\usetikzlibrary{patterns, decorations.pathreplacing, arrows.meta, shapes.geometric}

\usepackage{multirow}
\usepackage{float}
\usepackage{xspace}
\usepackage{enumitem}
\usepackage{xfrac}
\usepackage{cleveref}

\usepackage{thmtools}
\theoremstyle{plain}

\usepackage{thm-restate}

\makeatletter
\AddToHook{cmd/appendix/before}{\def\cref@section@alias{appendix}\def\cref@subsection@alias{appendix}}
\makeatother

\usepackage{tcolorbox}

\newcommand{\xhdr}[1]{\vspace{6pt}\noindent{\bf {#1.}}}

\newcounter{relctr} 
\everydisplay\expandafter{\the\everydisplay\setcounter{relctr}{0}} 

\AtBeginDocument{} 

\newcommand{\prob}[2][]{\text{Pr}\ifthenelse{\not\equal{}{#1}}{_{#1}}{}\!\left[{\def\givenn{\middle|}#2}\right]}
\newcommand{\expect}[2][]{\mathbb{E}\ifthenelse{\not\equal{}{#1}}{_{#1}}{}\!\left[{\def\givenn{\middle|}#2}\right]}

\newcommand{\indicator}[2][]{\text{\usefont{U}{bbm}{m}{n}1}\ifthenelse{\not\equal{}{#1}}{_{#1}}{}\!\left[{\def\givenn{\middle|}#2}\right]}

\newcommand{\squishlist}{
\begin{list}{{{\small{$\bullet$}}}}
{\setlength{\itemsep}{3pt}      \setlength{\parsep}{1pt}
\setlength{\topsep}{1pt}       \setlength{\partopsep}{0pt}
\setlength{\leftmargin}{1em} \setlength{\labelwidth}{1em}
\setlength{\labelsep}{0.5em} } }
\newcommand{\squishend}{  \end{list}}

\newcommand{\prior}{\lambda}

\newcommand{\stateNum}{m}
\newcommand{\action}{a}

\newcommand{\receiverU}{v}
\newcommand{\state}{i}

\newcommand{\signal}{\sigma}

\newcommand{\posterior}{\mu}

\newcommand{\noiseScale}{\beta}

\newcommand{\scaledNoiseDiff}{\delta}

\newcommand{\dd}{\mathrm{d}}

\newcommand{\OPT}{\textsc{Opt}}

\newcommand{\RRID}[2][]{\textsc{MRRCR}\ifthenelse{\not\equal{}{#1}}{_{#1}}{}\!\left[{\def\givenn{\middle|}#2}\right]}
\newcommand{\MRRP}[2][]{\textsc{MRRP}\ifthenelse{\not\equal{}{#1}}{_{#1}}{}\!\left[{\def\givenn{\middle|}#2}\right]}
\newcommand{\MRRR}[2][]{\textsc{MRRR}\ifthenelse{\not\equal{}{#1}}{_{#1}}{}\!\left[{\def\givenn{\middle|}#2}\right]}
\newcommand{\Payoff}[2][]{\textsc{ Payoff}\ifthenelse{\not\equal{}{#1}}{_{#1}}{}\!\left[{\def\givenn{\middle|}#2}\right]}

\newcommand{\statespace}{\Omega}
\newcommand{\indirectU}{V}
\newcommand{\MPC}{\textsc{MPC}}
\newcommand{\Quant}{W}
\newcommand{\randomstate}{\omega}
\newcommand{\stochastic}{\Delta}
\newcommand{\signalscheme}{\pi}
\newcommand{\parameterizedscheme}{\pi^{(\pooledmass)}}

\newcommand{\signalspace}{\Sigma}

\newcommand{\betaed}{_{\noiseScale}}
\newcommand{\betaone}{_{1}}
\newcommand{\betahat}{_{\hat{\beta}}}

\newcommand{\primed}{^\dagger}
\newcommand{\doubleprimed}{^{\ddagger}}
\newcommand{\fullsetrationalitylevel}{[0,\infty)\cup\{\infty\}}

\newcommand{\noiseDualVar}{\alpha}
\newcommand{\distDualVar}{\eta}

\newcommand{\zerobf}{\mathbf{0}}

\newcommand{\tangentIntersect}{\kappa}

\newcommand{\censorshipsignalInverse}{s\primed}

\newcommand{\reals}{\R}

\newcommand{\cs}{{\small \texttt{threshold-matching}}}
\newcommand{\dfone}{{\small \texttt{tangency}}}
\newcommand{\dftwo}{{\small \texttt{threshold-selection}}}
\newcommand{\dfthree}{{\small \texttt{sign-restrictions}}}
\newcommand{\pf}{{\small \texttt{fraction-bounds}}}

\newcommand{\censorshipprob}{\theta}
\newcommand{\censorshipstate}{\ell}
\newcommand{\censorshipsignal}{s}

\newcommand{\optscheme}{\signalscheme^*}

\newcommand{\optschemeHat}{\hat{\signalscheme}^*}
\newcommand{\censorshipprobHat}{\hat{\theta}}
\newcommand{\censorshipsignalHat}{\hat{s}}
\newcommand{\censorshipstateHat}{\hat{\ell}}
\newcommand{\pooledmassHat}{\hat{q}}

\newcommand{\rationalityClass}{\mathcal{B}}
\newcommand{\robustRatio}{\Gamma}
\newcommand{\instancerobustRatio}[2][]{\Gamma\ifthenelse{\not\equal{}{#1}}{_{#1}}{}\!\left[{\def\givenn{\middle|}#2}\right]}
\newcommand{\optschemebeta}{\signalscheme\doubleprimed}
\newcommand{\optschemeInfty}{\signalscheme\primed}

\newcommand{\censorshipsignalbeta}{s\doubleprimed}
\newcommand{\pooledmassbeta}{q\doubleprimed}

\newcommand{\censorshipsignalInfty}{s^{\dagger}}
\newcommand{\pooledmassInfty}{q\primed}

\newcommand{\noiseScaleUB}{{\overline{\noiseScale}}}

\newcommand{\thresholdfrac}{p}
\newcommand{\thresholdfracHat}{\hat{p}}
\newcommand{\tangentfunc}{f}

\newcommand{\instance}{\mathcal{I}}
\newcommand{\schemespace}{\Pi}
\newcommand{\schemeoptimal}{\pi^*}

\newcommand{\censorshipparameter}{t}
\newcommand{\tailmeanscaled}{u}

\newcommand{\worstcaseratio}{\Gamma^*}
\newcommand{\ratiovalue}{0.8996}
\newcommand{\ratiopercent}{89.96}  
\newcommand{\losspercent}{10.04}   

\newcommand{\schemeoptimalratio}{\pi_{\text{ratio}}^*}
\newcommand{\schemeoptimalpayoff}{\pi_{\text{payoff}}^*}
\newcommand{\schemeoptimalregret}{\pi_{\text{regret}}^*}

\newcommand{\noiseD}{\eta}
\newcommand{\rationalitySubclass}{\mathcal{B}}

\newcommand{\PayoffS}[2][]{U\ifthenelse{\not\equal{}{#1}}{_{#1}}{}\!\left({\def\givenn{\middle|}#2}\right)}
\newcommand{\objectiveF}
[2][]{\robustRatio\ifthenelse{\not\equal{}{#1}}{_{#1}}{}\!\left({\def\givenn{\middle|}#2}\right)}
\newcommand{\objectiveFderivative}
[2][]{\robustRatio^\prime\ifthenelse{\not\equal{}{#1}}{_{#1}}{}\!\left({\def\givenn{\middle|}#2}\right)}
\newcommand{\kernelF}
[2][]{G\ifthenelse{\not\equal{}{#1}}{_{#1}}{}\!\left({\def\givenn{\middle|}#2}\right)}
\newcommand{\generalkernelF}
[2][]{F\ifthenelse{\not\equal{}{#1}}{_{#1}}{}\!\left({\def\givenn{\middle|}#2}\right)}
\newcommand{\generalkernelFderivative}
[2][]{F^\prime\ifthenelse{\not\equal{}{#1}}{_{#1}}{}\!\left({\def\givenn{\middle|}#2}\right)}
\newcommand{\generalkernelFweight}{\alpha}
\newcommand{\kernelFderivative}
[2][]{G^\prime\ifthenelse{\not\equal{}{#1}}{_{#1}}{}\!\left({\def\givenn{\middle|}#2}\right)}
\newcommand{\kernelFsecondderivative}
[2][]{G^{\prime\prime}\ifthenelse{\not\equal{}{#1}}{_{#1}}{}\!\left({\def\givenn{\middle|}#2}\right)}

\newcommand{\rationalityIntclass}{\mathcal{B}_k}
\newcommand{\integer}{k}

\newcommand{\censorshipspace}{\Pi_{C}}
\newcommand{\censorshipspacepositive}{\Pi_C^+}
\newcommand{\ratiointeger}{\gamma_\integer}
\newcommand{\integernew}{j}
\newcommand{\ratiolimit}{\gamma^*}

\newcommand{\negativeratiofn}{f}
\newcommand{\ratio}{\gamma}

\newcommand{\payofffunction}[2][]
{P\ifthenelse{\not\equal{}{#1}}{_{#1}}{}\!\left({\def\givenn{\middle|}#2}\right)}
\newcommand{\payofffunctionderivative}[2][]
{P^{\prime}\ifthenelse{\not\equal{}{#1}}{_{#1}}{}\!\left({\def\givenn{\middle|}#2}\right)}

\newcommand{\instancerobustPayoff}[2][]{P\ifthenelse{\not\equal{}{#1}}{_{#1}}{}\!\left[{\def\givenn{\middle|}#2}\right]}

\newcommand{\PayoffF}[2][]{P\ifthenelse{\not\equal{}{#1}}{_{#1}}{}\!\left[{\def\givenn{\middle|}#2}\right]}
\newcommand{\payoffkernelF}
[2][]{H\ifthenelse{\not\equal{}{#1}}{_{#1}}{}\!\left({\def\givenn{\middle|}#2}\right)}
\newcommand{\payoffinteger}{p_k}
\newcommand{\payoffinfty}{p^*}
\newcommand{\payoff}{p}

\newcommand{\instancerobustRegret}[2][]{L\ifthenelse{\not\equal{}{#1}}{_{#1}}{}\!\left[{\def\givenn{\middle|}#2}\right]}
\newcommand{\RegretF}[2][]{L\ifthenelse{\not\equal{}{#1}}{_{#1}}{}\!\left[{\def\givenn{\middle|}#2}\right]}
\newcommand{\regretkernelF}
[2][]{H\ifthenelse{\not\equal{}{#1}}{_{#1}}{}\!\left({\def\givenn{\middle|}#2}\right)}
\newcommand{\regretconstant}{C_{\eta}}
\newcommand{\noiseDscaled}{\tilde{\eta}}
\newcommand{\regretinteger}{\ell_k}
\newcommand{\regretinfty}{\ell^*}

\newcommand{\coreinstance}{\tilde{\mathcal{I}}}

\newcommand{\thresholdprob}{p}

\newcommand{\negativeratio}{\xi}

\newcommand{\censorshipnegativeprob}[2][]{q\ifthenelse{\not\equal{}{#1}}{_{#1}}{}\!\left({\def\givenn{\middle|}#2}\right)}

\newcommand{\censorshipcontribution}[2][]
{N\ifthenelse{\not\equal{}{#1}}{_{#1}}{}\!\left({\def\givenn{\middle|}#2}\right)}

\newcommand{\optimalcontribution}[2][]
{Q\ifthenelse{\not\equal{}{#1}}{_{#1}}{}\!\left({\def\givenn{\middle|}#2}\right)}

\newcommand{\tailmean}[1][]
{b\ifthenelse{\not\equal{}{#1}}{_{#1}}{}}

\newcommand{\rcdf}[2][]
{F\ifthenelse{\not\equal{}{#1}}{_{#1}}{}\!\left({\def\givenn{\middle|}#2}\right)}

\newcommand{\Upperfunction}[2][]
{\psi\ifthenelse{\not\equal{}{#1}}{_{#1}}{}\!\left({\def\givenn{\middle|}#2}\right)}

\newcommand{\rivalrevealedmeanratio}{\rho^{\textsc{M}}}

\newcommand{\judgefunction}{H}
\newcommand{\judgefunctionf}{J}

\newcommand{\corestate}{\widetilde{m}}
\newcommand{\coreprior}{\widetilde{\lambda}}
\newcommand{\corereceiverU}{\widetilde{\receiverU}}

\newcommand{\Payoffinstance}[3][]{\textsc{ Payoff}\ifthenelse{\not\equal{}{#1}}{_{#1}^{#2}}\!\left[{\def\givenn{\middle|}#3}\right]}

\newcommand{\refinedinstance}{\instance^{\mathrm{sp}}}
\newcommand{\corecandidatesignalscheme}{\tilde\signalscheme\primed}

\newcommand{\corecandidatepooledmass}{\tilde\pooledmass\primed}

\newcommand{\lowerblock}{K^+}
\newcommand{\lowerblockmass}{\Lambda^+}

\newcommand{\pooledmass}{q}

\newcommand{\tailmeanscaledf}{z}

\newcommand{\candidatecensor}{\signalscheme\primed}
\newcommand{\benchmarkcensor}{\signalscheme\doubleprimed}
\newcommand{\candidatecensorstate}{\censorshipstate\primed}
\newcommand{\benchmarkcensorstate}{\censorshipstate\doubleprimed}
\newcommand{\candidatecensorprob}{\censorshipprob\primed}
\newcommand{\benchmarkcensorprob}{\censorshipprob\doubleprimed}
\newcommand{\candidatepoolsignal}{\censorshipsignal\primed}
\newcommand{\benchmarkpoolsignal}{\censorshipsignal\doubleprimed}
\newcommand{\candidatepooledmass}{\pooledmass\primed}
\newcommand{\benchmarkpooledmass}{\pooledmass\doubleprimed}
\newcommand{\candidatepoolpara}{\censorshipparameter\primed}
\newcommand{\benchmarkpoolpara}{\censorshipparameter\doubleprimed}

\newcommand{\candidatetailmean}{\bar{v}^{\textsc{B}}}
\newcommand{\topstates}{\textsc{T}}
\newcommand{\middlestates}{\textsc{M}}
\newcommand{\bottomstates}{\textsc{B}}
\newcommand{\benchmarkcoefficient}{c\doubleprimed}

\newcommand{\variableA}{a}
\newcommand{\variableB}{b}
\newcommand{\variablecapA}{A}
\newcommand{\variablecapB}{B}
\newcommand{\variableZ}{z}

\newcommand{\variableX}{x}

\newcommand{\variableY}{y}
\newcommand{\variableR}{r}
\newcommand{\variableH}{h}

\newcommand{\variableK}{k}

\newcommand{\revealedstates}{D^+}

\newcommand{\splitstate}{k}
\newcommand{\positivestate}{k^+}

\newcommand{\positivefraction}{\alpha^+}

\newcommand{\splitprob}{\alpha}
\newcommand{\lowmean}{L}

\newcommand{\variableZnegativesqrt}{z^{-\frac{1}{2}}}
\newcommand{\variableZpositivesqrt}{z^{\frac{1}{2}}}
\newcommand{\candidatenegativeratio}{\xi_1}
\newcommand{\benchmarknegativeratio}{\xi_2}

\newcommand{\censorshipsignalderivative}{s^\prime}
\newcommand{\derivativegap}{D}
\newcommand{\tangentpoint}{\kappa}
\newcommand{\maxderivative}{M}
\newcommand{\breakpoint}{t}

\newcommand{\priormean}{d}

\newcommand{\rivalpooledmass}{\hat{q}}
\newcommand{\rivalsignalscheme}{\hat{\pi}}
\newcommand{\receiverexpU}{u}
\newcommand{\rivalreceiverexpU}{\hat{u}}

\newcommand{\poly}{\textsc{poly}}

\newcommand{\accuracy}{\epsilon}
\newcommand{\smallnumber}{\eta}

\newcommand{\fptascensorratio}{\hat{\pi}_{\text{ratio}}}
\newcommand{\fptascensorpayoff}{\hat{\pi}_{\text{payoff}}}
\newcommand{\fptascensorregret}{\hat{\pi}_{\text{regret}}}

\newcommand{\gridspace}{\Delta}
\newcommand{\pooledmassgrid}{Q}
\newcommand{\rationalitylevelgrid}{B}

\newcommand{\noiseScaleHat}{\hat{\beta}}
\newcommand{\payofffullyrational}[2][]
{T\ifthenelse{\not\equal{}{#1}}{_{#1}}{}\!\left({\def\givenn{\middle|}#2}\right)}
\newcommand{\payoffcommon}[1][]{R_{#1}}

\newcommand{\instancerobustRatioestimation}[2][]{\hat{\Gamma}\ifthenelse{\not\equal{}{#1}}{_{#1}}{}\!\left[{\def\givenn{\middle|}#2}\right]}
\newcommand{\instancerobustPayoffestimation}[2][]{\hat{P}\ifthenelse{\not\equal{}{#1}}{_{#1}}{}\!\left[{\def\givenn{\middle|}#2}\right]}
\newcommand{\instancerobustRegretestimation}[2][]{\hat{L}\ifthenelse{\not\equal{}{#1}}{_{#1}}{}\!\left[{\def\givenn{\middle|}#2}\right]}

\newcommand{\pooledmassratioestimation}{\hat{q}_{\text{ratio}}}
\newcommand{\pooledmasspayoffestimation}{\hat{q}_{\text{payoff}}}
\newcommand{\pooledmassregretestimation}{\hat{q}_{\text{regret}}}
\newcommand{\pooledmassratio}{q_{\text{ratio}}}
\newcommand{\pooledmasspayoff}{q_{\text{payoff}}}
\newcommand{\pooledmassregret}{q_{\text{regret}}}
\newcommand{\nonpositivemass}{q^{\mathrm{NP}}}

\newcommand{\fullyrationalmass}{q^{\mathrm{FR}}}

\newcommand{\censorshipprogram}{\mathcal{P}_{\mathrm{CENS\text{-}OPT}}}

\newcommand{\optdirectInfy}{\tilde{\pi}}

\newcommand{\cA}{\mathcal{A}}

\newcommand{\R}{\mathbb{R}}

\renewcommand{\FigureCaptionFontStyle}{\rmfamily\mdseries\upshape}
\renewcommand{\FigureNameFontStyle}{\rmfamily\bfseries\upshape}
\makeatletter
\long\def\@makefigurecaption#1#2{%
  \begingroup
  \EGT\FigureCaptionFontStyle
  \leftskip\z@ \rightskip\z@
  \parfillskip\z@ plus 1fil
  \parindent\z@
  \setbox\@tempboxa=\hbox{\FigureNameFontStyle #1\kern16pt}%
  \hangindent\wd\@tempboxa \hangafter\@ne
  \noindent\HD{9}{0}\box\@tempboxa
  \ignorespaces #2\HD{0}{0}\par
  \endgroup}
\makeatother

\begin{document}

\RUNAUTHOR{}

\RUNTITLE{Less Rational, More Transparent: 
Censorship as (Robustly) Optimal Disclosure}

\TITLE{Less Rational, More Transparent: \\
Censorship as (Robustly) Optimal Disclosure}

\ARTICLEAUTHORS{%
\AUTHOR{Yiding Feng}
\AFF{Department of Industrial Engineering and Decision Analytics,
Hong Kong University of Science and Technology, \EMAIL{ydfeng@ust.hk}}

\AUTHOR{Chien-Ju Ho}
\AFF{ Department of Computer Science and Engineering,
Washington University in St.\ Louis, \EMAIL{chienju.ho@wustl.edu}}

\AUTHOR{Wei Tang}
\AFF{ Department of Decisions, Operations and Technology,
Chinese University of Hong Kong, \EMAIL{weitang@cuhk.edu.hk}}

\AUTHOR{Tianze Wei}
\AFF{ Department of Decisions, Operations and Technology,
Chinese University of Hong Kong, \EMAIL{tianzewei@cuhk.edu.hk}}
} 

\ABSTRACT{%
Online platforms often control the information behind user decisions they want to influence, such as whether a driver accepts a trip or a buyer completes a purchase. 
A classical result in information disclosure shows that a {\em censorship} policy is optimal for the platform: the platform pools all sufficiently favorable options under a common message, such as a quality badge, and reveals the rest individually. 
This result, however, presumes fully rational users. 
We study disclosure to boundedly rational users who follow the logit quantal response model and whose rationality level the platform may not know. 

Our results deliver a unifying message: less rational, more transparent. Neither bounded rationality nor uncertainty about it calls for a more complex policy; each changes only how much the platform should pool, and only toward greater transparency. 
When the rationality level is known, censorship remains optimal, and less rational users optimally receive at least as much information as more rational ones. 
When the rationality level is unknown, censorship remains robustly optimal under each of three criteria: the maximin ratio, maximin payoff, and minimax regret, and the robust censorship can be chosen to be at least as transparent as the policy designed for fully rational users. 
This additional transparency weakly benefits both parties and acts as insurance against behavioral uncertainty: rationality-robust censorship secures at least 89.96\% of the payoff attainable with knowledge of the user's rationality level, whereas the policy designed for fully rational users can secure as little as 50\%. 
Taken together, these results suggest that platforms can keep their badge-based disclosure but should narrow its coverage when users may be boundedly rational.\footnote{A preliminary conference version of this work appeared in the Proceedings of the ACM-SIAM Symposium on Discrete Algorithms (SODA'24) \citep{Anon-24}. The present paper subsumes and refocuses the conference version on information disclosure by online platforms and adds the rationality-robust optimality of censorship under three robustness criteria (\Cref{sec:robust_optimal_scheme_unkown_rationality_level}), the tight 89.96\% universal guarantee (\Cref{sec:robust_ratio}), transparency and welfare results (\Cref{apx:full_bounded_rationality}, \Cref{apx:same_pooled_mass_interval,apx:receiver_utility_information_revealing}), computational results (\Cref{sec:computation}), and also significantly rewrites the introduction (\Cref{sec:introduction}) and also \Cref{sec:optimal_scheme_known_rationality_level}.}
}%



\maketitle

\section{Introduction}
\label{sec:introduction}

In many online marketplaces, the party that benefits from a user's decision also controls the information on which that decision is based.
A ride-hailing platform decides how much to reveal about a trip, such as its expected duration, before a driver accepts or declines the request \citep{WAO-21,RZWT-23}.
An e-commerce platform decides whether to display a detailed quality grade for each product or to group products of different quality under a common label before a buyer decides whether to purchase \citep{BPZ-24}.
A freelancing platform decides how much to disclose about a worker's track record before a client decides whether to hire \citep{GJ-21}.
Although these settings differ, they share a common structure: the value of a user's decision depends on information held by the platform, while the platform benefits when the user takes a particular action, such as accepting the trip, completing the purchase, or making the hire.


How the platform discloses this information shapes users' beliefs and, consequently, their decisions.
More transparent disclosure helps users tell suitable options from unsuitable ones, whereas coarser disclosure groups different options under a common message, so that users cannot tell them apart.
By choosing what to reveal and what to conceal, the platform can therefore influence how likely users are to take its preferred action.
The design of disclosure policies is thus an important problem for online platforms and marketplaces, and a growing literature studies this problem through the lens of information design and Bayesian persuasion \citep{KG-11,BM-19}.
In this framework, the platform commits to a disclosure policy, also called a \emph{signaling scheme}, which specifies the set of possible messages (or \emph{signals}) and how a signal is generated, possibly at random, from each realization of the platform's private information.
After observing the signal, the user updates her belief using Bayes' rule and then chooses an action.
Applications in operations and online platforms include recommendation systems that encourage users to explore new products \citep{CH-18,PBS-18}, the disclosure of congestion information in service systems \citep{LI-19}, and the strategic release of information on two-sided platforms \citep{BM-23}.

This literature typically assumes that the user is \emph{fully rational}: after observing a signal, she always chooses the action with the higher expected utility under her posterior belief.
Under this assumption, when the platform seeks to maximize the probability that the user takes its preferred action, the optimal policy has a simple \emph{censorship} form \citep{RSV-17,KMZ-22}.
In the product-quality interpretation, a censorship policy pools all quality levels above a cutoff into a single common message and reveals each remaining quality level individually.
The platform chooses this cutoff so that the pooled message leaves the user just willing to take the preferred action.
As a result, the user takes this action even for some products that she would reject if she were fully informed.
Censorship policies are attractive in practice because they reduce a potentially complicated disclosure problem to the choice of a single cutoff, which makes them easy both to optimize and to implement.
Moreover, the resulting policy resembles the familiar certification practice of grouping all sufficiently good options under a common badge.
Censorship policies and closely related threshold policies also emerge as optimal in a wide range of operational settings, including 
the disclosure of queue information in service systems \citep{LI-19,AIM-23}, 
information disclosure in delegated search \citep{XHW-26}, 
and the disclosure of market-size information on lead-selling platforms \citep{SZLW-26}.
\footnote{Similar structures also arise in matching markets \citep{RS-19}, school grading \citep{OS-10}, clinical trials \citep{K-15}, bank stress tests \citep{GL-18}, and voter persuasion \citep{AC-16b}.}
However, this classical prescription relies critically on full rationality: the user takes the platform's preferred action whenever that action offers even a marginal advantage in expected utility.

Although full rationality is a classic assumption, extensive empirical evidence documents systematic deviations from rational choice across many economic domains \citep{sim-78,KT-13,tha-16,del-09}.
Limited attention, cognitive costs, and idiosyncratic noise can lead a user to forgo an option that is marginally attractive or to accept one that is marginally unattractive.
In this paper, we adopt the logit quantal response model \citep{MP-95}, a canonical model of bounded rationality, to capture this behavior parsimoniously.
Under this model, a user chooses more attractive actions with higher probability, and a single parameter, the rationality level $\noiseScale\in[0, \infty)\cup \{\infty\}$, captures how strongly she responds to differences in expected utility.
The two extreme levels $\noiseScale = 0$ and $\noiseScale = \infty$ correspond to a user who chooses uniformly at random and a fully rational user, respectively, and intermediate levels $\noiseScale\in(0,\infty)$ span the range between them, with a larger $\noiseScale$ corresponding to a more rational user.
The quantal response model originated in behavioral game theory and is now widely used in operations management (see \Cref{subsec:related}).
Beyond its clean structure, 
the model admits a foundation in rational inattention \citep{MM-15}, and shares its logit form with the multinomial logit model of discrete choice \citep{M-72}.

This departure from full rationality matters for information disclosure.
Consider again a censorship policy whose pooled message leaves the user exactly indifferent between taking the platform's preferred action and rejecting it.
Under the standard assumption that ties are broken in the platform's favor, a fully rational user takes this action with certainty, whereas a user who follows quantal response takes it with probability only one half.
A censorship policy designed for a fully rational user can therefore perform much worse than the fully rational benchmark predicts.
Quantal response offers a more flexible model of user behavior, but it leaves open how the platform should adapt its disclosure policy when users respond probabilistically rather than always choosing the action with the higher expected utility.\footnote{Intuitions from the fully rational case need not carry over. In \Cref{exa:full_bounded_rationality} we present an instance in which a boundedly rational user strictly prefers a less informative signaling scheme to a more informative one in the Blackwell sense. This cannot happen for a fully rational user, who by Blackwell's theorem always weakly benefits from more information.}
This leads to our first research question, which concerns the optimal disclosure policy when the user's rationality level is known:
\begin{quote}
    \textbf{(Question 1)}
    \emph{When the user is boundedly rational and her rationality level $\noiseScale$ is known, what is the optimal information disclosure policy?
    In particular, does the censorship structure, which is optimal under full rationality, remain optimal when the user follows quantal response?}
\end{quote}
The characterization of the optimal policy under full rationality does not automatically extend to probabilistic choices.
Under quantal response, the platform must account for how strongly users respond to disclosed information, and the optimal amount of information to pool may depend on their rationality level.

Even if the optimal disclosure policy can be characterized for each known rationality level, implementing it requires the platform to know the user's rationality level $\noiseScale$, which may be difficult to determine or costly to learn in practice.
For example, a platform may observe only a few decisions from each user.
It may also serve users who differ in responsiveness but need to apply a common disclosure policy to all of them, such as a badge displayed to every visitor of a product page, rather than tailor the policy to each user.
In such settings, even an estimate of average responsiveness may not accurately reflect the behavior of any particular user.
We therefore focus on uncertainty about the rationality level alone: the distribution of the platform's private information and users' payoffs from each option can often be estimated from aggregate historical outcomes, whereas a user's rationality level is an individual behavioral trait whose estimation requires that user's own choice data.
These practical considerations motivate our second research question, which concerns the design of disclosure policies under uncertainty about the user's rationality:
\begin{quote}
    \textbf{(Question 2)}
    \emph{When the user's rationality level $\noiseScale$ is unknown, what disclosure policy performs best in the worst case over all possible rationality levels? Does this rationality-robust policy preserve the censorship structure, and how much performance does the platform lose because of this uncertainty?}
\end{quote}
The second question introduces an additional challenge: even if censorship is optimal for each known rationality level, it does not follow that a single censorship policy is optimal under a robust objective.
Different rationality levels may favor different pooling cutoffs, and a policy that performs well for one rationality level may perform poorly for another.
A robust policy must therefore balance its performance across rationality levels.
A natural way to do so is to randomize between policies with different cutoffs, but such a mixture is no longer a censorship policy.
A priori, it is therefore ambiguous whether a single cutoff can strike this balance optimally.
A simple candidate is the optimal censorship policy for a fully rational user, but it is not obvious how well this policy performs when the user is boundedly rational, relative to the payoff the platform could attain if it knew the user's rationality level.

The answers to these questions matter for how platforms design disclosure policies in practice.
In particular, they bear on whether the classical censorship structure remains a sound design when users are boundedly rational, and how much performance a platform may lose when it does not know its users' rationality levels.

\subsection{Our Contributions and Techniques}
\label{subsec:contributions}
We address the two research questions in sequence.
Perhaps surprisingly, we show that the censorship structure remains optimal when the platform knows the user's rationality level $\noiseScale$, and remains robustly optimal when it does not.
Bounded rationality and uncertainty about it may change how much information the platform should pool, and this amount can depend on both the user's rationality level and the platform's robustness criterion.
In both cases, any change is toward greater transparency: less rational users optimally receive weakly more information, and a rationality-robust policy can be chosen to reveal weakly more information than the policy designed for a fully rational user.
However, neither departure from the classical setting calls for a more complicated disclosure policy.
We also establish tight bounds on the performance loss caused by uncertainty about the user's rationality level.
In particular, in every problem instance, a rationality-robust censorship policy guarantees at least 89.96\% of the payoff attainable with knowledge of the user's rationality level, whereas the optimal censorship policy designed for a fully rational user guarantees only 50\%.
These bounds provide a quantitative justification for adopting a common disclosure policy without estimating each user's rationality level.
We explain and discuss our main contributions below.

\xhdr{Censorship is optimal at a known rationality level}
We first study the information design problem when the platform knows the user's rationality level $\noiseScale$.
We show that, for every rationality level, there exists an optimal censorship policy (\Cref{thm:opt_state_independent}).
This result extends the classical optimality of censorship under full rationality to users who follow the logit quantal response model.
In particular, even though users now respond probabilistically to disclosed information, it remains optimal to pool the states (i.e., the possible realizations of the platform's private information) on one side of a threshold into a common message and to reveal the remaining states individually, possibly randomizing at the threshold state.
We further characterize the optimal threshold through a feasibility program derived from a primal-dual analysis of the information design problem.
This characterization also yields a polynomial-time algorithm for computing the optimal policy (\Cref{prop:complexity_SISU_opt}).

An important implication of this characterization is that less rational users, that is, users who respond less sharply to differences in expected utility, should receive weakly more information in the Blackwell sense (\Cref{prop:SISU_threshold_state_monotonicity}).
Intuitively, the classical cutoff makes the pooled message only barely persuasive, leaving the user just willing to take the preferred action, so a boundedly rational user takes this action only about half the time.
By pooling less, the platform makes the pooled message more persuasive, which raises this probability at the cost of sending the message less often.
This comparative-statics result offers a useful managerial insight: a platform serving less responsive users should become more transparent rather than pool information more aggressively.
This additional transparency also benefits the user, who obtains weakly higher expected utility than under the policy designed for a fully rational user.

\xhdr{Rationality-robust optimality of censorship}
We next consider the more challenging setting in which the platform does not know the user's rationality level and must commit to a single disclosure policy.
We study three standard robustness criteria, which capture different attitudes toward this uncertainty: the \emph{maximin ratio} criterion maximizes the worst-case fraction of the payoff that the platform could attain if it knew the user's rationality level; the \emph{maximin payoff} criterion maximizes the worst-case payoff; and the \emph{minimax regret} criterion minimizes the worst-case shortfall from that benchmark payoff.
Our main structural result shows that, under each criterion, some censorship policy is rationality-robust optimal (\Cref{thm:robust_optimal_censorship_three_criteria}).
Thus, censorship remains optimal not only when users depart from full rationality, but also when the platform must design its policy without knowing the extent of that departure. Computationally, our characterization also yields a fully polynomial-time approximation scheme (FPTAS) for computing a rationality-robust optimal censorship policy under each criterion (\Cref{prop:computation_rationality_robust_scheme}).

These results also yield an important managerial insight: accounting for uncertainty about user rationality need not come at the expense of either the platform's performance or user welfare.
Specifically, under each robustness criterion, a rationality-robust optimal censorship policy can be chosen to reveal weakly more information than the optimal policy designed for a fully rational user.
This additional disclosure benefits both parties: relative to the policy designed for a fully rational user, it weakly improves the platform's worst-case performance under the chosen criterion and weakly increases the user's expected utility at every rationality level.
In the product-quality interpretation, a platform that is uncertain about its users' rationality can therefore narrow the coverage of its common badge relative to the fully rational benchmark, while still covering every product that a fully informed user would buy.

\xhdr{Tight universal performance guarantees}
Finally, we quantify the cost of not knowing the user's rationality level, focusing on the maximin ratio criterion.
We first show that the optimal censorship policy designed for a fully rational user guarantees at least 50\% of the payoff attainable with knowledge of the user's rationality level, for every rationality level and every problem instance (\Cref{thm:rationality_robust}\ref{item:baseline_ratio}).
We then show that optimizing the censorship policy specifically for robustness substantially improves this guarantee: in every instance, the rationality-robust censorship policy attains a worst-case ratio of at least $\worstcaseratio\approx\ratiovalue$ (\Cref{thm:rationality_robust}\ref{item:optimal_ratio}).
Both guarantees are tight: a single instance with only two states attains both bounds.
Hence, the worst case arises in the simplest possible setting, and restricting attention to simple instances would not improve either guarantee.
Overall, the worst-case loss from rationality uncertainty falls from as much as 50\% under the policy designed for a fully rational user to at most \losspercent\% under an optimally designed rationality-robust policy.

Our tight guarantees offer two managerial insights for platforms that design disclosure policies under uncertainty about user behavior.
First, the value of personalization is limited.
Tailoring the disclosure policy to each user's rationality level can improve performance, but by at most \losspercent\% of the payoff attainable with perfect knowledge of that level.
Hence, when individual-level behavioral data are scarce or personalization is costly, a platform can retain most of the attainable payoff with a single policy, without estimating each user's rationality level.
Second, ignoring bounded rationality can be costly.
Simply applying the policy optimized for fully rational users can lose up to 50\% of the attainable payoff, even in a setting with only two states.
Thus, while precisely estimating users' rationality levels may not be essential, explicitly accounting for uncertainty about their rationality is crucial for avoiding substantial performance losses.

\xhdr{Key technical ingredient: a pooled-mass representation of censorship}
A unifying technical ingredient in our analysis is to represent censorship policies by their \emph{pooled mass}, that is, the ex ante probability that the common pooled message is sent, rather than by the cutoff state as in conventional formulations \citep[e.g.,][]{AC-16,Z-20,KMZ-22}.
For a fixed prior distribution over the states, the pooled mass uniquely determines the cutoff state and any randomization at the cutoff.
Thus, whereas the conventional representation pairs a discrete cutoff state with a randomization probability, ours indexes every censorship policy by a single scalar $\pooledmass$ in the convex interval $[0,1]$.
This representation provides a common analytical foundation for our three main results.
First, when the rationality level $\noiseScale$ is known, we characterize the optimal censorship policy by exploiting the S-shaped curvature of the quantal response function, which is concave when the user favors the platform's preferred action and convex when she does not, together with the primal-dual structure of the information design problem.
The pooled-mass representation further allows us to compare optimal censorship policies across rationality levels and to establish how the optimal amount of pooling varies with $\noiseScale$.

Second, the pooled-mass representation plays a central role in establishing the rationality-robust optimality of censorship.
As discussed above, this result does not follow from the optimality of censorship at each known rationality level, because different rationality levels may favor different cutoffs.
We therefore formulate the robust problem as a minimax problem in which an adversary chooses the user's rationality level, possibly at random.
Against any fixed distribution over rationality levels, the platform faces a standard persuasion problem whose indirect utility, a weighted average of quantal response functions, remains S-shaped; hence, censorship is a best response by the characterization of \citet{KMZ-22}.
However, this best-response property alone does not establish robust optimality: the set of censorship policies is nonconvex, since a mixture of two censorship policies with different cutoffs is not itself a censorship policy, so standard minimax arguments do not apply directly.
The pooled-mass representation resolves this difficulty by turning the platform's optimization over this nonconvex family into a one-dimensional optimization over the compact and convex interval $[0,1]$.
Crucially, we show that the platform's objective under any such distribution is single-peaked, and hence quasi-concave, in $\pooledmass$.
This property allows us to apply Sion's minimax theorem \citep{S-58} over $[0,1]$, which establishes that a single censorship policy is optimal for the original robust problem.

Third, the pooled-mass representation also underpins our tight universal performance guarantees.
Their proofs develop a state-reduction framework for comparing the payoffs of different censorship policies.
In particular, the pooled masses of any two censorship policies naturally partition the states into three groups: states pooled by both policies, states pooled by only one policy, and states revealed by both.
Exploiting this decomposition and the curvature of the quantal response function, we reduce payoff comparisons involving arbitrarily many states to comparisons in instances with only two states.
This reduction allows us to establish both tight guarantees and to identify a simple worst-case instance.

\subsection{Related Work}
\label{subsec:related}
\xhdr{Quantal response to model bounded rationality}
Departures from full rationality in human decision-making have long been recognized \citep{S-56}, motivating a growing literature that incorporates bounded rationality into operational models (see, e.g., \citealp{GP-08}).
Our work is most closely related to studies that model bounded rationality through logit-based stochastic choice \citep{MP-95}.
In these models, actions with higher expected utilities are chosen with higher probabilities, and a rationality parameter captures how strongly individuals respond to utility differences.

Logit-based models have been widely adopted in operations management.
For example, \citet{S-08} uses a continuous logit rule to model a newsvendor's ordering decisions, while \citet{HAB-13} consider customers' queue-joining decisions under logistic errors in perceived waiting times.
\citet{KCE-11} employ a logit rule to model suppliers' choices of contract terms based on learned profitability scores.
Other applications include pricing contracts \citep{LH-07, HZ-08, TTH-22}, strategic ordering under capacity rationing \citep{CSZ-12}, and service-provider selection \citep{LGL-16}.
Building on this literature, we study information disclosure when users update their beliefs according to Bayes' rule and subsequently make binary decisions following logit quantal response.
Our focus differs from these studies in examining how the optimal information disclosure policy depends on the user's rationality level and, more importantly, how to design robust disclosure policies when this level is unknown.

\xhdr{Information design in operations management}
Our work builds on the rich literature on information design, which originates from the Bayesian persuasion model of \citet{KG-11}; we refer the reader to \citet{BM-19} and \citet{kam-19} for surveys, and to \citet{can-20} for its applications in operations. 
A growing body of work in operations management uses this framework to study how a firm or a platform should share its private information with customers, suppliers, or the public. 
In service systems, the provider signals queue or wait-time information to arriving customers \citep{LI-19,AIM-23}. 
In revenue management, a seller shares information about product availability or earlier purchases to
shape purchase timing \citep{DJR-21,KOW-22}. 
On online platforms, the designer discloses information to induce consumer exploration
\citep{PBS-18} or to steer suppliers' quality, entry, and pricing decisions
\citep{BPZ-24,BM-23,GMMS-23}. 
Further applications include demand disclosure in supply chains \citep{CG-25}, 
information sharing among competing producers \citep{WYSH-25},
machine prediction disclosure \citep{LT-26},
and public warnings about health risks \citep{AVW-20,VGW-21}. 
A recurring finding across these applications is that the optimal policy takes a simple threshold form.
A parallel literature in computer science
studies problems in 
selection games \citep{BMSW-25},
recommendation and pricing \citep{FTX-22,AFT-23}, 
and competition \citep{DTWZ-24}.

Our work also relates to a literature that relaxes the rationality
assumption in information design. \citet{CZ-22} study how the receiver's mistakes in probabilistic inference affect optimal persuasion.
\citet{AIL-24} study persuasion when the receiver's utility may be nonlinear in her belief, and they apply their approach to signaling in queues with uncertainty-averse customers. 
\citet{TPG-23} consider shared medical decision making in which the patient's utility is noisy and the doctor departs from
exact Bayesian updating. 
\citet{TH-21} and \citet{YTNH-23} conduct behavioral
experiments and relax the Bayesian rationality assumption in simple persuasion settings. 

\xhdr{Censorship} 
Censorship, a policy that reveals states on one side of a cutoff and pools those on the other, arises in various operational settings, including inventory disclosure \citep{APR-17, CCW-23, LCMT-26}, product-quality disclosure \citep{G-09, GZ-09}, corporate disclosure \citep{JM-23}, and over-the-counter market benchmarks \citep{DDZ-17}.
For example, online retailers may disclose exact inventory levels when stock is scarce but withhold numerical counts when inventory is sufficient.

The optimality of censorship has also been extensively studied in the Bayesian persuasion literature.
\citet{KMZ-22} characterize its optimality in terms of the quasi-concavity of the sender's marginal utility, while \citet{K-18} and \citet{AC-16} provide sufficient conditions in related settings.
\citet{DM-19} illustrate optimal censorship in an effort-inducement problem.
Building on this literature, we establish that censorship remains optimal when receivers follow logit quantal response, even when their rationality levels are unknown, and derive tight guarantees for its robust performance.

\xhdr{Robust optimization}
Our work also relates to robust optimization and decision-making under uncertainty.
We consider three standard robustness criteria: maximin ratio \citep{ST-85}, maximin payoff \citep{W-45}, and minimax regret \citep{S-51}.
These criteria have been widely applied in operational settings, including 
airline booking \citep{LGBK-08}, 
assortment optimization \citep{RT-12}, 
pricing \citep{CHP-22}, 
and auctions \citep{ABB-25, ABB-26}.

While many studies focus on a single robustness criterion, several recent works examine multiple criteria jointly.
\citet{CHP-22} compare deterministic monopoly pricing under all three criteria and show that policies optimized for one criterion may perform poorly under another.
\citet{ABB-24} study how randomized pricing can perform across all three criteria, while \citet{ABB-26} develop a generalized regret framework that encompasses these criteria in auction design.
Complementing this literature, we establish that the same information policy structure---censorship---remains optimal under all three robustness criteria in rationality-robust information design, although the optimal disclosure threshold may differ across criteria.

\section{Preliminaries}
\label{sec:preliminary}

In this section, we present our model.
We first describe the environment, then the receiver's bounded rationality, which we capture by the logit quantal response model, and finally the class of censorship signaling schemes on which our analysis focuses.

\xhdr{Environment}
We consider the canonical Bayesian persuasion setting with a finite state space $[\stateNum] \triangleq \{1, 2, \dots, \stateNum\}$.
The state $\randomstate$ is drawn from a common-knowledge prior distribution\footnote{Throughout, we identify distributions with their probability mass functions.} $\prior \in \stochastic([\stateNum])$.
We denote by $\prior_i \triangleq \prob{\randomstate = i}$ the prior probability of state $i \in [\stateNum]$.

There are two players, a sender (platform) and a receiver (user).
The receiver chooses an action from the binary set $\cA = \{0, 1\}$.
Given the realized state $i \in [\stateNum]$ and action $\action \in \cA$, the receiver obtains utility $\receiverU_i(\action)$, while the sender obtains utility $\action$, that is, $1$ if the receiver takes action $1$ and $0$ otherwise.\footnote{This follows the standard convention in which the sender's utility depends only on the receiver's action, and is thus state-independent, and the sender strictly prefers action $1$ to action $0$ in every state \citep{KG-11,AC-16b,LI-19,LR-20,ZIX-25}.}
Before the state is realized, the sender commits to a signaling scheme $\signalscheme: [\stateNum] \to \stochastic(\signalspace)$, which maps each state to a probability distribution over an arbitrary signal space $\signalspace$.
Let $\schemespace$ denote the set of all such schemes, and let $\signalscheme_i(\signal)$ denote the probability of sending signal $\signal \in \signalspace$ when the state is $i$.
After observing a signal $\signal$ sent with positive probability, the receiver updates her belief by Bayes' rule to form the posterior belief $\posterior(\signal) \in \stochastic([\stateNum])$, where the posterior probability of state $i$ is
\begin{align*}
    \posterior_i(\signal) \triangleq \frac{\prior_i \signalscheme_i(\signal)}{\sum_{j \in [\stateNum]} \prior_j \signalscheme_j(\signal)}~.
\end{align*}
The sender's objective is to maximize her expected utility, i.e., the probability that the receiver takes action $1$.

\xhdr{Decision-making with quantal response}
We model the receiver as boundedly rational using the logit quantal response framework \citep{MP-95}.
Specifically, instead of deterministically choosing the action that maximizes her expected utility, the receiver chooses each action with probability proportional to the exponential of its expected utility multiplied by a parameter $\noiseScale \geq 0$.

Formally, given a posterior belief $\posterior \in \stochastic([\stateNum])$, let $\receiverU(\action \mid \posterior) \triangleq \sum_{i \in [\stateNum]} \posterior_i \receiverU_i(\action)$ denote the receiver's expected utility from action $\action \in \cA$.
The probability that the receiver takes action $1$ is then
\begin{align*}
    \frac{\exp(\noiseScale \cdot \receiverU(1 \mid \posterior))}{\exp(\noiseScale \cdot \receiverU(1 \mid \posterior)) + \exp(\noiseScale \cdot \receiverU(0 \mid \posterior))} = \frac{1}{1 + \exp(\noiseScale \cdot (\receiverU(0 \mid \posterior) - \receiverU(1 \mid \posterior)))}\,.
\end{align*}
Here, $\noiseScale \in \fullsetrationalitylevel$ is the receiver's \emph{rationality level}, which captures how strongly she responds to differences in expected utility: a larger $\noiseScale$ corresponds to a more responsive receiver.
When $\noiseScale = 0$, the receiver chooses each action uniformly at random regardless of her posterior belief; as $\noiseScale \to \infty$, she chooses the action with the higher expected utility with probability approaching one, and the model approaches the classic Bayesian persuasion setting with a fully rational receiver (we specify the boundary case $\noiseScale = \infty$ below).

Define the \emph{quantal response function} $\Quant\betaed(\variableX) \triangleq \sfrac{1}{(1 + \exp(\noiseScale \variableX))}$.
For every finite $\noiseScale > 0$, the function $\Quant\betaed$ is decreasing, concave on $(-\infty, 0]$, convex on $[0, \infty)$, and satisfies $\Quant\betaed(\variableX) + \Quant\betaed(-\variableX) = 1$; this S-shaped curvature plays a central role in our analysis.
For each state $i \in [\stateNum]$, with a slight abuse of notation, let $\receiverU_i \triangleq \receiverU_i(0) - \receiverU_i(1)$ denote how much the receiver prefers action $0$ to action $1$ in state $i$.
For any posterior belief $\posterior$, the expected utility difference is then $\receiverU(0 \mid \posterior) - \receiverU(1 \mid \posterior) = \sum_{i \in [\stateNum]} \posterior_i \receiverU_i$, so the probability that the receiver takes action $1$ can be written compactly as $\Quant\betaed\bigl(\sum_{i \in [\stateNum]} \posterior_i \receiverU_i\bigr)$.
Without loss of generality, we assume that $\{\receiverU_i\}_{i \in [\stateNum]}$ is strictly increasing in $i$, which can be ensured by relabeling the states and merging states with equal values.\footnote{Since each state $i$ is uniquely identified by $\receiverU_i$, we also refer to $\receiverU_i$ as the \emph{value} of state $i$.}
For the fully rational boundary case $\noiseScale = \infty$, we adopt sender-favorable tie-breaking and define $\Quant_\infty(\variableX) \triangleq \indicator{\variableX \le 0}$.
That is, when the receiver is indifferent between the two actions, she chooses the sender's preferred action\footnote{Note that $\Quant_\infty$ coincides with the pointwise limit of $\Quant\betaed$ as $\noiseScale \to \infty$ everywhere except at $\variableX = 0$, where this limit equals $\sfrac{1}{2}$.
This gap underlies the potential performance loss of signaling schemes designed for a fully rational receiver, discussed in \Cref{sec:introduction}.} $1$.

Because the probability that the receiver takes action $1$ depends on her posterior belief only through the expected utility difference $\sum_{i \in [\stateNum]} \posterior_i \receiverU_i$, any two signals that induce the same difference can be merged without affecting the sender's expected utility.
Hence, it is without loss of generality to restrict attention to signaling schemes in which each signal $\scaledNoiseDiff \in \mathbb{R}$ equals the expected utility difference that it induces, i.e., $\scaledNoiseDiff = \sum_{i \in [\stateNum]} \posterior_i(\scaledNoiseDiff) \receiverU_i$, where $\posterior_i(\scaledNoiseDiff)$ denotes the posterior probability of state $i$ given signal $\scaledNoiseDiff$.
Given a signaling scheme\footnote{Although we allow signaling schemes with a continuous signal space $\signalspace$, in which case $\signalscheme_i(\cdot)$ is a probability density function, all signaling schemes considered in this paper, including the optimal ones, have finite signal spaces. We therefore abuse notation and treat $\signalscheme_i(\cdot)$ as a probability mass function when this is clear from the context.} $\signalscheme$, let $\Payoff[\noiseScale]{\signalscheme}$ denote the sender's expected utility when facing a receiver with rationality level~$\noiseScale$:
\begin{align*}
    \Payoff[\noiseScale]{\signalscheme} \triangleq \sum\nolimits_{i \in [\stateNum]} \prior_i \int_{-\infty}^{\infty} \signalscheme_i(\scaledNoiseDiff) \Quant\betaed(\scaledNoiseDiff) \,\dd\scaledNoiseDiff~.
\end{align*}
When the sender knows the receiver's rationality level $\noiseScale \in \fullsetrationalitylevel$, she chooses a signaling scheme $\signalscheme \in \schemespace$ to maximize her expected utility.
We denote the corresponding optimal expected utility by
\begin{align*}
    \OPT\left(\noiseScale\right) \triangleq \max\nolimits_{\signalscheme \in \schemespace} \Payoff[\noiseScale]{\signalscheme}~,
\end{align*}
and call a signaling scheme attaining this maximum an \emph{optimal signaling scheme} for rationality level $\noiseScale$.
Throughout the paper, a \emph{problem instance} is the tuple $\instance = (\stateNum, \{\prior_i\}_{i\in[\stateNum]}, \{\receiverU_i\}_{i\in[\stateNum]})$.

\xhdr{Censorship signaling schemes}
We now introduce the \emph{censorship signaling scheme}, the main focus of this paper.
A censorship signaling scheme pools the lowest states, that is, those under which the receiver most favors action $1$, into a single signal and fully reveals the remaining states, possibly splitting the threshold state between the two groups.
Since the states are ordered by their values $\{\receiverU_i\}_{i\in[\stateNum]}$, we can parameterize a censorship signaling scheme by its \emph{pooled mass} $\pooledmass\in[0,1]$, namely, the total prior probability of the states pooled into the common signal, or equivalently, the ex ante probability that this signal is sent.
In particular, $\pooledmass = 0$ corresponds to full revelation and $\pooledmass = 1$ to full pooling, which reveals no information.
Once the pooled mass $\pooledmass$ is fixed, the prior distribution uniquely determines the \emph{threshold state}, that is, the highest state that is at least partially pooled, and the \emph{threshold-state fraction}, that is, the fraction of this state that is pooled.
\Cref{def:censorship} formalizes this construction, and we denote by $\censorshipspace$ the set of all censorship signaling schemes.

\begin{definition}
\label{def:censorship}
Fix any pooled mass $\pooledmass\in[0,1]$.
If $\pooledmass=0$, the censorship signaling scheme fully reveals every state, i.e., $\signalscheme_i(\scaledNoiseDiff) = \indicator{\scaledNoiseDiff = \receiverU_i}$ for every $i \in [\stateNum]$.
For $\pooledmass \in(0,1]$, define the threshold state $\censorshipstate(\pooledmass)$, the threshold-state fraction $\censorshipprob(\pooledmass)$, and the pooling signal $\censorshipsignal(\pooledmass)$ as follows:
\begin{align*}
    \censorshipstate(\pooledmass) 
    & \triangleq 
    \min\left\{ i\in[\stateNum] : \sum\nolimits_{j\in[i]} \prior_j \ge \pooledmass \right\}\;; \quad
    \censorshipprob(\pooledmass)
    \triangleq 
    \frac{\pooledmass-\sum\nolimits_{j\in[ \censorshipstate(\pooledmass) -1]}\prior_j}{\prior_{\censorshipstate(\pooledmass)}} \in (0,1]\;;\\
    \censorshipsignal(\pooledmass)
    & \triangleq
    \frac{1}{\pooledmass}\cdot \left({\sum\nolimits_{j\in[ \censorshipstate(\pooledmass)-1]}\prior_j \receiverU_j
    \;+\;
    \censorshipprob(\pooledmass)\prior_{\censorshipstate(\pooledmass)}\receiverU_{\censorshipstate(\pooledmass)}}\right)
    \;.
\end{align*}
The censorship signaling scheme $\signalscheme$ with pooled mass $\pooledmass$ is then given by
\begin{align*}
    i\in [\censorshipstate(\pooledmass)-1]: 
    &\qquad \signalscheme_i(\scaledNoiseDiff)=\indicator{\scaledNoiseDiff=\censorshipsignal(\pooledmass)}\;,\\
    i= \censorshipstate(\pooledmass): 
    & \qquad 
    \signalscheme_i(\scaledNoiseDiff)=
    \censorshipprob(\pooledmass)\cdot \indicator{\scaledNoiseDiff=\censorshipsignal(\pooledmass)}
    +\left(1-\censorshipprob(\pooledmass)\right)\cdot\indicator{\scaledNoiseDiff=\receiverU_{\censorshipstate(\pooledmass)}}\;,\\
    i\in [\censorshipstate(\pooledmass)+1:\stateNum]: 
    & \qquad \signalscheme_i(\scaledNoiseDiff)=\indicator{\scaledNoiseDiff=\receiverU_i}\;.
\end{align*}
Equivalently, the scheme pools all states strictly below the threshold state $\censorshipstate(\pooledmass)$, together with a $\censorshipprob(\pooledmass)$-fraction of the threshold state, into a single pooling signal, and fully reveals all remaining states.
The pooling signal $\censorshipsignal(\pooledmass)$ equals the conditional mean value of the pooled states, consistent with our convention that each signal equals the expected utility difference it induces.
When the pooled mass $\pooledmass$ is clear from the context, we drop the argument and write $\censorshipstate,\censorshipprob,\censorshipsignal$ instead of $\censorshipstate(\pooledmass),\censorshipprob(\pooledmass),\censorshipsignal(\pooledmass)$.
\end{definition}

We call a state $\state$ positive if $\receiverU_{\state} > 0$ and negative if $\receiverU_{\state} < 0$.
Throughout the remainder of the paper, we focus on instances with $\receiverU_1 < 0 < \receiverU_{\stateNum}$, so the receiver strictly prefers action 1 in some state and action 0 in another.
The remaining cases admit a single scheme that is optimal at every rationality level: 
if all state values are non-positive, full pooling is optimal by the concavity of the quantal response function $\Quant\betaed(\cdot)$ on $(-\infty,0]$; if all state values are non-negative, full revelation is optimal by its convexity on $[0,\infty)$.

\section{Censorship is Optimal at a Known Rationality Level}
\label{sec:optimal_scheme_known_rationality_level}

In this section, we study the sender's problem when the receiver's rationality level $\noiseScale$ is fixed and known to the sender.
We show that the sender can restrict attention to censorship signaling schemes without loss of optimality.
By \Cref{def:censorship}, choosing such a scheme amounts to choosing its pooled mass, which in turn determines the threshold state, the threshold-state fraction, and the pooling signal.
\Cref{thm:opt_state_independent} characterizes an optimal pooled mass, and this characterization serves as the foundation for our rationality-robust results in \Cref{sec:robust_optimal_scheme_unkown_rationality_level}.
\begin{restatable}{theorem}{knownrationalityoptimality}
\label{thm:SISU_opt}
\label{thm:opt_state_independent}
Fix any instance $\instance$ and any rationality level $\noiseScale \in [0,\infty)\cup\{\infty\}$, there exists an optimal censorship signaling scheme $\optscheme$ with pooled mass $\pooledmass= \sum\nolimits_{i\in[\censorshipstate-1]}\prior_i
+\censorshipprob\prior_{\censorshipstate}$
where
     the threshold state $\censorshipstate$
     and 
     the threshold-state fraction 
    $\censorshipprob$
    satisfy the following 
    feasibility program~\ref{eq:SISU_opt_condition}:
    \begin{align}
        \label{eq:SISU_opt_condition}
        \tag{$\mathcal{P}_{\texttt{CENS-OPT}}$}
        \begin{split}
    (1 - \censorshipprob)
    (\censorshipsignalInverse - 
    \receiverU_{\censorshipstate}) &= 0
    \qquad\qquad(\text{threshold-matching})
    \\
    \left({\Quant\betaed(\censorshipsignalInverse)  - 
    \Quant\betaed(\censorshipsignal)}\right)
    -
    \Quant\betaed'(\censorshipsignal) ({
    \censorshipsignalInverse
    - \censorshipsignal}) &= 0
    \qquad\qquad (\text{tangency})
    \\
\max\nolimits_{i\in[\stateNum]}\{
    \receiverU_i:\receiverU_i \leq \censorshipsignalInverse\}
    &=
    \receiverU_{\censorshipstate}
    \qquad\,\,\,\,\,\,\,\,\,\,\,(\text{threshold-selection})
    \\
            \censorshipsignal \leq 0,~
            \censorshipsignalInverse &\geq 0
           \qquad\qquad (\text{sign-restrictions})
           \\
           0\leq \censorshipprob&\leq 1
           \qquad\qquad(\text{fraction-bounds})
        \end{split}
    \end{align}
where $\censorshipsignal = 
    \frac{\sum_{i \in [\censorshipstate-1]} \prior_i\receiverU_i + \censorshipprob \prior_{\censorshipstate}\receiverU_{\censorshipstate}}{\sum_{i \in[ \censorshipstate-1]} \prior_i + \censorshipprob \prior_{\censorshipstate}}$ is the pooling signal and $ 
    \censorshipsignalInverse$ is an auxiliary variable.
\end{restatable}
\Cref{thm:opt_state_independent} reduces the sender's optimization over all signaling schemes to a one-dimensional choice of pooled mass: $\max\nolimits_{\signalscheme \in \schemespace} \Payoff[\noiseScale]{\signalscheme} = \max\nolimits_{\pooledmass \in [0,1]} \Payoff[\noiseScale]{\parameterizedscheme}$, where $\parameterizedscheme$ denotes the censorship signaling scheme with pooled mass $\pooledmass$.
This characterization also implies that, for every rationality level $\noiseScale\in\fullsetrationalitylevel$, an optimal signaling scheme can be computed in polynomial time (see \Cref{prop:complexity_SISU_opt}).

For a platform, this result offers a simple implementation principle: a common badge or recommendation pools all sufficiently favorable realizations, while the remaining realizations are disclosed individually.
A platform that already uses such a badge need not redesign its disclosure format when its users are boundedly rational; it only needs to adjust the badge's coverage and, as we show below, to narrow it relative to the fully rational benchmark.
In short, bounded rationality changes the optimal extent of pooling but preserves the censorship structure.

Below, we first explain the constraints in the feasibility program~\ref{eq:SISU_opt_condition} and the trade-off that they capture.
We then examine how the optimal pooled mass varies with the rationality level, and we conclude with a proof overview of \Cref{thm:opt_state_independent}.

\begin{figure}[tb]
    \centering
    \begin{tikzpicture}[scale=0.7, transform shape]
\begin{axis}[
axis line style={->,thick,gray},
axis lines=middle,
xtick style={draw=none},
ytick style={draw=none},
xticklabels=\empty,
yticklabels=\empty,
xmin=-9,xmax=12,ymin=-0.15,ymax=1.3,
width=0.9\textwidth,
height=0.5\textwidth,
samples=50]

\node[anchor=north east, font=\large, inner sep=3pt] at (axis cs:0,0) {0};

\addplot[domain=-9:12, gray!40!white, line width=2.mm] (x, {1/(1+e^(0.7*x))});

\addplot[mark=*,only marks, fill=white] coordinates {(6.34907,0.0116084)} node[above, pos=1]{};
\addplot[gray, thick] coordinates {(6.34907,0.01)
(6.34907,-0.01)};
\addplot[] coordinates {(6.34907,0.)} node[below, pos=1]{\Large$\censorshipsignalInverse$};

\addplot[mark=*,only marks, fill=white] coordinates {(-2.34798,0.838022)} node[above, pos=1]{};
\addplot[gray, thick] coordinates {(-2.34798,0.01)
(-2.34798,-0.01)};
\addplot[] coordinates {(-2.34798,0.)} node[below, pos=1]{\Large$\censorshipsignal$};

\addplot[gray, dotted] coordinates {(-2.34798,0.838022)
(-2.34798,0.)};

\addplot[gray, thick] coordinates {(5,0.01)
(5,-0.01)};
\addplot[] coordinates {(5,-0.035)} node[below, pos=1]{\Large$\receiverU_{\censorshipstate}$};


\addplot[gray, thick] coordinates {(11,0.01)
(11,-0.01)};
\addplot[] coordinates {(11,-0.035)} node[below, pos=1]{\Large$\receiverU_{\censorshipstate+1}$};

\addplot[domain=-6:6.34907, black,line width=0.5mm] (x, {-0.09502 * x + 0.614897});
\end{axis}

\end{tikzpicture}
    \caption{Graphical interpretation of the feasibility program $\censorshipprogram$.
    The gray solid curve is the quantal response function $\Quant\betaed\left(\cdot\right)$.
    The black line is tangent to the curve at the pooling point $\left(\censorshipsignal,\Quant\betaed\left(\censorshipsignal\right)\right)$ and intersects it at the auxiliary point $\left(\censorshipsignalInverse,\Quant\betaed\left(\censorshipsignalInverse\right)\right)$.
    The threshold state $\censorshipstate$ has the largest state value no greater than the auxiliary value $\censorshipsignalInverse$.
    The tangent line, extended to the left of this auxiliary value, together with the quantal response curve to its right, forms a convex upper bound on the quantal response function.
    This upper bound underlies the optimality proof.
    \label{fig:SISU_opt_condition}
    }
\end{figure}
\xhdr{Economic and graphical interpretation}
The optimal pooled mass characterized in \Cref{thm:opt_state_independent} balances the coverage of the pooling signal against its effectiveness in inducing the sender's preferred action.
Because the receiver finds action $1$ more attractive in states with smaller values, expanding the pool adds states that are less favorable to action $1$.
This expansion increases the probability that the pooling signal is sent, but it also raises the pooling signal $\censorshipsignal$ and thus lowers the conditional probability $\Quant\betaed(\censorshipsignal)$ that the receiver takes action $1$.
The feasibility program identifies where these two effects balance.

The curvature of the quantal response function provides intuition for the optimal structure.
Recall that $\Quant\betaed$ is concave on $(-\infty,0]$ and convex on $[0,\infty)$.
By Jensen's inequality, concavity favors pooling states in which the receiver prefers action $1$, whereas convexity favors revealing states in which she prefers action $0$, at least when each group is considered in isolation.
Consistent with this intuition, the optimal censorship signaling scheme can be selected to pool every state with a non-positive value at every rationality level; the rationality level affects only how much probability mass from states with positive values joins the pool.

The constraint \dfone{} identifies a line that is tangent to the quantal response curve at the pooling point $(\censorshipsignal,\Quant\betaed(\censorshipsignal))$ and intersects the curve again at the auxiliary point $(\censorshipsignalInverse,\Quant\betaed(\censorshipsignalInverse))$.
The constraint \dfthree{} places the pooling point in the concave region of the curve and the auxiliary point in its convex region.
For each auxiliary value $\censorshipsignalInverse \geq 0$, the tangency condition determines a unique pooling point $\censorshipsignal \leq 0$, and the resulting mapping from $[0,\infty)$ to $(-\infty,0]$ is a decreasing bijection: the farther the auxiliary point lies to the right, the farther the pooling point lies to the left (see \Cref{fig:SISU_opt_condition}).
The constraint \dftwo{} then selects as the threshold state $\censorshipstate$ the state with the largest value no greater than the auxiliary value $\censorshipsignalInverse$; this state is well defined because $\receiverU_1 < 0 \leq \censorshipsignalInverse$.

The remaining constraints connect this geometric construction to a feasible signaling scheme.
The constraint \pf{} ensures that the threshold-state fraction $\censorshipprob$ is a valid probability.
The constraint \cs{} requires $\censorshipsignalInverse = \receiverU_{\censorshipstate}$ whenever part of the threshold state remains revealed, i.e., whenever $\censorshipprob < 1$.
In that case, the tangency condition becomes $\Quant\betaed(\censorshipsignal) + \Quant\betaed'(\censorshipsignal)(\receiverU_{\censorshipstate} - \censorshipsignal) = \Quant\betaed(\receiverU_{\censorshipstate})$.
The left-hand side is the marginal payoff from moving a small amount of the threshold state into the pool, which accounts for the resulting shift in the pooling signal and hence in the receiver's response to the entire pool; the right-hand side is the payoff from revealing that amount instead, so the sender is indifferent at the margin.
Together, these conditions pin down the optimal pooled mass.

\xhdr{Less rational, more transparent}
Although censorship remains optimal across rationality levels, its optimal pooled mass varies with the receiver's responsiveness.
To see how, consider an instance with a positive prior mean, i.e., $\sum\nolimits_{\state \in [\stateNum]} \prior_{\state}\receiverU_{\state} > 0$, so that a receiver without additional information would prefer action $0$.
For a fully rational receiver, the optimal censorship signaling scheme $\optschemeHat$ selects the pooled mass $\pooledmassHat$ at which the pooling signal equals $\censorshipsignalHat = 0$, leaving the receiver exactly indifferent between the two actions.
Under sender-favorable tie-breaking, this signal induces action $1$ with certainty, whereas a boundedly rational receiver takes action $1$ with probability only one half at the same signal.
The sender therefore has an incentive to pool less mass: doing so lowers the pooling signal below zero and raises this probability above one half, even though the pooling signal is then sent less often.
Indeed, for every finite rationality level $\noiseScale < \infty$, the optimal censorship signaling scheme $\optscheme$ selects a pooled mass $\pooledmass$ whose pooling signal satisfies $\censorshipsignal \leq 0$ (\Cref{thm:opt_state_independent}).
Because the pooling signal increases with the pooled mass, it follows that $\pooledmass \leq \pooledmassHat$; that is, the optimal scheme for a boundedly rational receiver pools weakly less mass than the fully rational benchmark.
(If the prior mean is non-positive, the fully rational benchmark pools all states, i.e., $\pooledmassHat = 1$, so the comparison holds trivially.)
The following proposition generalizes this comparison to any two rationality levels.
\begin{restatable}{proposition}{sisustatemonotone}
\label{prop:SISU_threshold_state_monotonicity} 
Fix any two rationality levels $\noiseScale \leq \hat\noiseScale$.
Let $\optscheme$ and $\optschemeHat$ be the optimal censorship signaling scheme
for a boundedly rational receiver with 
rationality level $\noiseScale$
and $\hat\noiseScale$, respectively.
Then
the pooled mass $\pooledmass$
of censorship scheme $\optscheme$ is weakly smaller than the pooled mass $\pooledmassHat$
of censorship scheme $\optschemeHat$.
\end{restatable}
In other words, the sender optimally reveals weakly more information to a less rational receiver, in the sense of the Blackwell order \citep{bla-53}: a censorship signaling scheme with a larger pooled mass can be obtained from one with a smaller pooled mass by merging additional states into the pooling signal, so it is a garbling of the latter.

The curvature of the quantal response function also provides geometric intuition for this result.
When the receiver is less rational, i.e., $\noiseScale$ is smaller, the quantal response curve is flatter.
As a result, for any given auxiliary point $(\censorshipsignalInverse,\Quant\betaed(\censorshipsignalInverse))$, the tangent line in \Cref{fig:SISU_opt_condition} touches the curve farther to the left, so the tangency condition requires a lower pooling signal $\censorshipsignal$.
Because the pooling signal increases with the pooled mass, a lower pooling signal requires pooling less probability mass from states with positive values, that is, a lower threshold state or a smaller threshold-state fraction.
Hence, the optimal pooled mass decreases and more information is revealed.
The formal proof of \Cref{prop:SISU_threshold_state_monotonicity} is deferred to \Cref{apx:SISU_proof_SISU_threshold_state monotonicity}.

For a platform, \Cref{prop:SISU_threshold_state_monotonicity} indicates how to adapt disclosure to user responsiveness, holding the prior distribution and utility differences fixed.
In the product-quality interpretation, a platform serving less responsive users should narrow the coverage of its common badge by removing the least favorable products from it and disclosing these products individually.
Thus, when users respond less sharply to disclosed information, the platform benefits from greater transparency rather than from more aggressive pooling.
Greater transparency also benefits the user.
Holding the user's rationality level fixed, among censorship signaling schemes that pool every state with a non-positive value and pool no more mass than the fully rational benchmark, a smaller pooled mass weakly increases her expected utility (see \Cref{prop:receiver_utility_information_revealing}).
In particular, a user with any rationality level is weakly better off under the optimal scheme for her rationality level than under the scheme designed for a fully rational user.
The platform's optimal response to less responsive users is therefore aligned with the interests of these users.

\xhdr{Proof overview of \Cref{thm:opt_state_independent}}
We now outline the proof of \Cref{thm:opt_state_independent}; the formal proof is deferred to \Cref{apx:SISU_opt}.
The proof is based on linear programming duality.
We formulate the sender's problem as a linear program whose variables are the probabilities $\signalscheme_i(\scaledNoiseDiff)$ of sending each signal $\scaledNoiseDiff$ in each state $i$; the constraints require that each $\signalscheme_i(\cdot)$ be a probability distribution and that each signal equal the expected utility difference it induces.
We then derive the corresponding dual program (see \Cref{apx:primal_dual_program}).
Given a solution to the feasibility program~\ref{eq:SISU_opt_condition}, we explicitly construct a feasible primal assignment and a feasible dual assignment and show that they satisfy complementary slackness, which implies that both are optimal.

The construction proceeds in two steps.
First, for every tuple $(\censorshipsignal,\censorshipsignalInverse,\censorshipstate,\censorshipprob)$ that satisfies constraints \dfone, \dftwo, and \dfthree, that is, every configuration described in the graphical interpretation above, we construct a feasible dual assignment.
By weak duality, its objective value is an upper bound on the sender's expected utility under every signaling scheme.
This dual assignment corresponds to the convex upper bound on the quantal response function $\Quant\betaed(\cdot)$ illustrated in \Cref{fig:SISU_opt_condition}: the tangent line to the left of the auxiliary point $\censorshipsignalInverse$, together with the curve itself to its right.
Second, to show that the censorship signaling scheme attains this upper bound, we additionally require constraint \pf, which ensures that the corresponding primal assignment is feasible, and constraint \cs, which ensures that complementary slackness holds at the threshold state $\censorshipstate$.
Finally, we show that a tuple satisfying all constraints exists and is unique.



\section{Censorship is Rationality-Robustly Optimal}
\label{sec:robust_optimal_scheme_unkown_rationality_level}

\Cref{sec:optimal_scheme_known_rationality_level} shows that when the receiver's rationality level is known, a censorship maximizes the sender's expected utility (\Cref{thm:opt_state_independent}).
In practice, however, the sender may not know the receiver's rationality level, or may be able to learn it only at significant cost.
Motivated by this concern, this section studies the rationality-robust problem, in which the sender commits to a single signaling scheme without knowing the receiver's rationality level and evaluates it by its worst-case performance across all rationality levels.
We consider three robustness criteria (\Cref{def:worst-case_ratio_payoff_regret}): the \emph{maximin ratio} criterion maximizes the worst-case fraction of the benchmark payoff $\OPT(\noiseScale)$ that the scheme attains; the \emph{maximin payoff} criterion maximizes the worst-case payoff; and the \emph{minimax regret} criterion minimizes the worst-case shortfall from $\OPT(\noiseScale)$.
Perhaps surprisingly, \Cref{thm:robust_optimal_censorship_three_criteria} shows that under each of these criteria, a censorship remains optimal.
This result strengthens \Cref{thm:opt_state_independent}: censorship remains optimal even when the sender does not know the receiver's rationality level.

\begin{definition}
\label{def:worst-case_ratio_payoff_regret}
Fix any instance $\instance$, and recall that $\OPT(\noiseScale)$ denotes the sender's optimal expected utility when she knows the receiver's rationality level $\noiseScale$.
For a signaling scheme $\signalscheme$, define
\begin{itemize}
    \item its \emph{worst-case ratio} $\instancerobustRatio{\signalscheme} \triangleq \inf\nolimits_{\noiseScale\in \fullsetrationalitylevel} \frac{\Payoff[\noiseScale]{\signalscheme}}{\OPT(\noiseScale)}$;
    \item its \emph{worst-case payoff} $\instancerobustPayoff{\signalscheme} \triangleq \inf\nolimits_{\noiseScale \in \fullsetrationalitylevel} \Payoff[\noiseScale]{\signalscheme}$; and
    \item its \emph{worst-case regret} $\instancerobustRegret{\signalscheme} \triangleq \inf\nolimits_{\noiseScale\in \fullsetrationalitylevel} \left(\Payoff[\noiseScale]{\signalscheme} - \OPT(\noiseScale)\right)$, that is, the negative of the largest shortfall from the benchmark.\footnote{The literature typically defines regret as the shortfall $\OPT(\noiseScale) - \Payoff[\noiseScale]{\signalscheme}$ and seeks a scheme that minimizes its supremum (minimax regret). To treat all three criteria within a common $\sup_{\signalscheme}\inf_{\noiseScale}$ framework, we instead define the worst-case regret through the negative shortfall. Since $\instancerobustRegret{\signalscheme} = -\sup_{\noiseScale}\left(\OPT(\noiseScale) - \Payoff[\noiseScale]{\signalscheme}\right)$, maximizing $\instancerobustRegret{\signalscheme}$ is equivalent to the standard minimax regret objective.}
\end{itemize}
\end{definition}
Under each criterion, a signaling scheme is \emph{rationality-robust optimal} if it maximizes the corresponding worst-case objective over all signaling schemes in $\schemespace$; we denote such schemes by $\schemeoptimalratio$, $\schemeoptimalpayoff$, and $\schemeoptimalregret$, respectively.
Their worst-case values $\instancerobustRatio{\schemeoptimalratio}$, $\instancerobustPayoff{\schemeoptimalpayoff}$, and $\instancerobustRegret{\schemeoptimalregret}$ are called the \emph{maximin optimal ratio}, the \emph{maximin optimal payoff}, and the \emph{minimax optimal regret}, respectively.

\begin{restatable}{theorem}{robustoptimalcensorship}
\label{thm:robust_optimal_censorship_three_criteria}
For every instance $\instance$ and each of the three robustness criteria (maximin ratio, maximin payoff, and minimax regret), there exists a censorship that is rationality-robust optimal.
\end{restatable}

In the remainder of this section, we first discuss several implications of \Cref{thm:robust_optimal_censorship_three_criteria} and then prove the theorem for the maximin ratio criterion.
The proofs for the maximin payoff and minimax regret criteria follow a similar structure and are deferred to \Cref{apx:missing_proof_robust_optimal_payoff_regret}.

\begin{figure}
    \centering

    \begin{subfigure}[b]{0.495\textwidth}
        \centering
        \begin{tikzpicture}
            \begin{axis}[
                width=\linewidth, height=6cm,
                xlabel={$\pooledmass$},
                xlabel style={at={(rel axis cs:1,0)}, anchor=south east, yshift=2pt, inner sep=1pt},
                ylabel={Worst-Case Performance}, 
                ylabel style={at={(rel axis cs:0,1)}, anchor=south east, yshift=-8pt, xshift=-10pt, inner sep=2pt},
                xmin=0, xmax=1.1,
                ymin=-1, ymax=1.05, 
                axis x line=bottom, 
                axis y line=left,   
                axis line style={-stealth, thick, gray}, 
                clip=false,           
                ytick={-1, 0}, 
                yticklabel style={font=\tiny, inner sep=1pt},
                xtick={0}, 
                xticklabels={0},
                xticklabel style={
                    rotate=0,           
                    anchor=north,       
                    font=\tiny,            
                    inner sep=1pt,
                }, 
                label style={font=\footnotesize}
            ]

            \draw[black, thick] (axis cs:1.0, -1.02) -- (axis cs:1.0, -0.98);
                \node[anchor=south, font=\tiny, inner sep=2pt, yshift=1pt] at (axis cs:1, -1) {$\fullyrationalmass$};
                
            \draw[black, thick] (axis cs:0.5, -1.02) -- (axis cs:0.5, -0.98);
            \node[anchor=south, font=\tiny, inner sep=2pt, yshift=1pt] at (axis cs:0.5, -1) {$\nonpositivemass$};
            
            \addplot[black, solid, very thick] table [x=q, y=ratio, col sep=comma] {plots/robust_criteria_data_2.csv};
            \addplot[black, densely dashed, very thick] table [x=q, y=payoff, col sep=comma] {plots/robust_criteria_data_2.csv};
            \addplot[black, densely dashdotted, very thick] table [x=q, y=regret, col sep=comma] {plots/robust_criteria_data_2.csv};

            \draw[gray, loosely dotted, thick] (axis cs:0.8994, 0.8994) -- (axis cs:0, 0.8994) node[black,left, font=\tiny] {0.90};
            \draw[gray, loosely dotted, thick] (axis cs:0.8994, 0.8994) -- (axis cs:0.8994, -1) node[black,below left, font=\tiny, inner sep=1pt, xshift=1pt] {0.90};

            \draw[gray, loosely dotted, thick] (axis cs:0.7, 0.5) -- (axis cs:0, 0.5) node[ text=black, left, font=\tiny] {0.50};
            \draw[gray, loosely dotted, thick] (axis cs:0.7, 0.5) -- (axis cs:0.7, -1) node[black,below, font=\tiny, inner sep=1pt] {0.70};

            \draw[gray, loosely dotted, thick] (axis cs:0.9145, -0.0855) -- (axis cs:0, -0.0855) node[black, left, font=\tiny, yshift = -2pt] {-0.09};
            \draw[gray, loosely dotted, thick] (axis cs:0.9145, -0.0855) -- (axis cs:0.9145, -1) node[black, below right, font=\tiny, inner sep=1pt, xshift=-2pt] {0.91};

            \addplot[only marks, mark=star, mark options={scale=1.8, fill=black}] coordinates {(0.8994, 0.8994)};
            \addplot[only marks, mark=triangle*, mark options={scale=1.5, fill=black, draw=black}] coordinates {(0.7, 0.5)};
            \addplot[only marks, mark=square*, mark options={scale=1.0, fill=black, draw=black}] coordinates {(0.9145, -0.0855)};
            
            \end{axis}
        \end{tikzpicture}
        \caption{}
        \label{fig:robust_criteria_a}
    \end{subfigure}
  \hfill
    \begin{subfigure}[b]{0.495\textwidth}
        \centering
        \begin{tikzpicture}
            \begin{axis}[
                width=\linewidth, height=6cm, 
                xlabel={$\pooledmass$},
                xlabel style={at={(rel axis cs:1,0)}, anchor=south east, yshift=2pt, inner sep=2pt},
                ylabel={Worst-Case Performance}, 
                ylabel style={at={(rel axis cs:0,1)}, anchor=south east, yshift=-8pt, xshift=-10pt, inner sep=2pt},
                xmin=0, xmax=1.0,
                ymin=-1, ymax=1.05, 
                axis x line=bottom, 
                axis y line=left,   
                axis line style={-stealth, thick, gray}, 
                clip=false,           
                ytick={-1, 0}, 
                yticklabel style={font=\tiny, inner sep=1pt},
                xtick={0}, 
                xticklabels={0 },
                xticklabel style={
                    rotate=0,           
                    anchor=north,       
                    font=\tiny,            
                    inner sep=1pt,
                }, 
                label style={font=\footnotesize}
            ]
                \draw[black, thick] (axis cs:0.7, -1.02) -- (axis cs:0.7, -0.98);
                \node[anchor=south, font=\tiny, inner sep=2pt, yshift=1pt] at (axis cs:0.7, -1) {$\fullyrationalmass$};

                \draw[black, thick] (axis cs:0.3, -1.02) -- (axis cs:0.3, -0.98);
            \node[anchor=south, font=\tiny, inner sep=2pt, yshift=1pt] at (axis cs:0.3, -1) {$\nonpositivemass$};

            \addplot[black, solid, very thick, restrict x to domain=0:0.7000] table [x=q, y=ratio, col sep=comma] {plots/robust_criteria_data.csv};
            \addplot[black,  densely dashed, very thick, restrict x to domain=0:0.7000] table [x=q, y=payoff, col sep=comma] {plots/robust_criteria_data.csv};
            \addplot[black, densely dashdotted, very thick, restrict x to domain=0:0.7000] table [x=q, y=regret, col sep=comma] {plots/robust_criteria_data.csv};

            \addplot[black, solid, very thick, restrict x to domain=0.7001:1] table [x=q, y=ratio, col sep=comma] {plots/robust_criteria_data.csv};
            \addplot[black,  densely dashed, very thick, restrict x to domain=0.7001:1] table [x=q, y=payoff, col sep=comma] {plots/robust_criteria_data.csv};
            \addplot[black, densely dashdotted, very thick, restrict x to domain=0.7001:1] table [x=q, y=regret, col sep=comma] {plots/robust_criteria_data.csv};

            \draw[gray, loosely dotted, thick] (axis cs:0.6296, 0.8996) -- (axis cs:0, 0.8996) node[black, left, font=\tiny] {0.90};
            \draw[gray, loosely dotted, thick] (axis cs:0.6296, 0.8996) -- (axis cs:0.6296, -1) node[black, below left, font=\tiny, inner sep=1pt, xshift=1pt] {0.63};

            \draw[gray, loosely dotted, thick] (axis cs:0.5000, 0.3841) -- (axis cs:0, 0.3841) node[black, left, font=\tiny] {0.38};
            \draw[gray, loosely dotted, thick] (axis cs:0.5000, 0.3841) -- (axis cs:0.5000, -1) node[black, below, font=\tiny, inner sep=1pt] {0.50};

            \draw[gray, loosely dotted, thick] (axis cs:0.6396, -0.0603) -- (axis cs:0, -0.0603) node[black, left, font=\tiny, yshift = -2.5pt] {-0.06};
            \draw[gray, loosely dotted, thick] (axis cs:0.6396, -0.0603) -- (axis cs:0.6396, -1) node[black, below right, font=\tiny, inner sep=1pt, xshift=-2pt] {0.64};

            \addplot[only marks, mark=star, mark options={scale=1.8, fill=black}] coordinates {(0.6296, 0.8996)};
            \addplot[only marks, mark=triangle*, mark options={scale=1.5, fill=black, draw=black}] coordinates {(0.5000, 0.3841)};
            \addplot[only marks, mark=square*, mark options={scale=1.0, fill=black, draw=black }] coordinates {(0.6396, -0.0603)};

            \draw[gray, densely dotted, thick] (axis cs:0.7, 0) -- (axis cs:0.7, 0.536);
            \draw[gray, densely dotted, thick] (axis cs:0.7, 0) -- (axis cs:0.7, 0.3500);
            \draw[gray, densely dotted, thick] (axis cs:0.7, -0.7000) -- (axis cs:0.7, -0.3117);

            \addplot[only marks, mark=*, mark options={fill=black, scale=0.5}] coordinates {(0.7, 0.536)};
            \addplot[only marks, mark=*, mark options={fill=white, draw=black, scale=0.5}] coordinates {(0.7, 0)}; 
            
            \addplot[only marks, mark=*, mark options={fill=black, draw=black, scale=0.5}] coordinates {(0.7, 0.3500)};
            \addplot[only marks, mark=*, mark options={fill=white, draw=black, scale=0.5}] coordinates {(0.7, 0.000)};
            
            \addplot[only marks, mark=*, mark options={fill=black, draw = black, scale=0.5}] coordinates {(0.7, -0.3117)};
            \addplot[only marks, mark=*, mark options={fill=white, draw=black, scale=0.5}] coordinates {(0.7, -0.7000)};

            \end{axis}
        \end{tikzpicture}
        \caption{}
        \label{fig:robust_criteria_b}
    \end{subfigure}
    \caption{{Worst-case ratio (solid), worst-case payoff (dashed), and worst-case regret (dash-dotted; see \Cref{def:worst-case_ratio_payoff_regret} for the sign convention) of the censorship with pooled mass $\pooledmass$.
The tick mark $\nonpositivemass$ denotes the total prior mass of states with non-positive values, and the tick mark $\fullyrationalmass$ denotes the pooled mass of the optimal censorship for a fully rational receiver.
The star, triangle, and square mark an optimal pooled mass under the maximin ratio, maximin payoff, and minimax regret criteria, respectively.
Left panel: three-state instance with $\prior = (0.3,0.2,0.5)$ and $\receiverU = (-2,-1,1.6)$, for which $\nonpositivemass = 0.5$ and $\fullyrationalmass = 1$; the worst-case payoff is constant for $\pooledmass \geq 0.5$, so the optimal pooled mass under the maximin payoff criterion is not unique.
Right panel: five-state instance with $\prior = (0.1,0.2,0.2,0.3,0.2)$ and $\receiverU = (-2,-1,0.5,1.5,3)$, for which $\nonpositivemass = 0.3$ and $\fullyrationalmass = 0.7$; the curves are discontinuous at $\fullyrationalmass$, where solid circles mark the values attained and hollow circles mark the one-sided limits.}}
\label{fig:robust_criteria}
\end{figure}
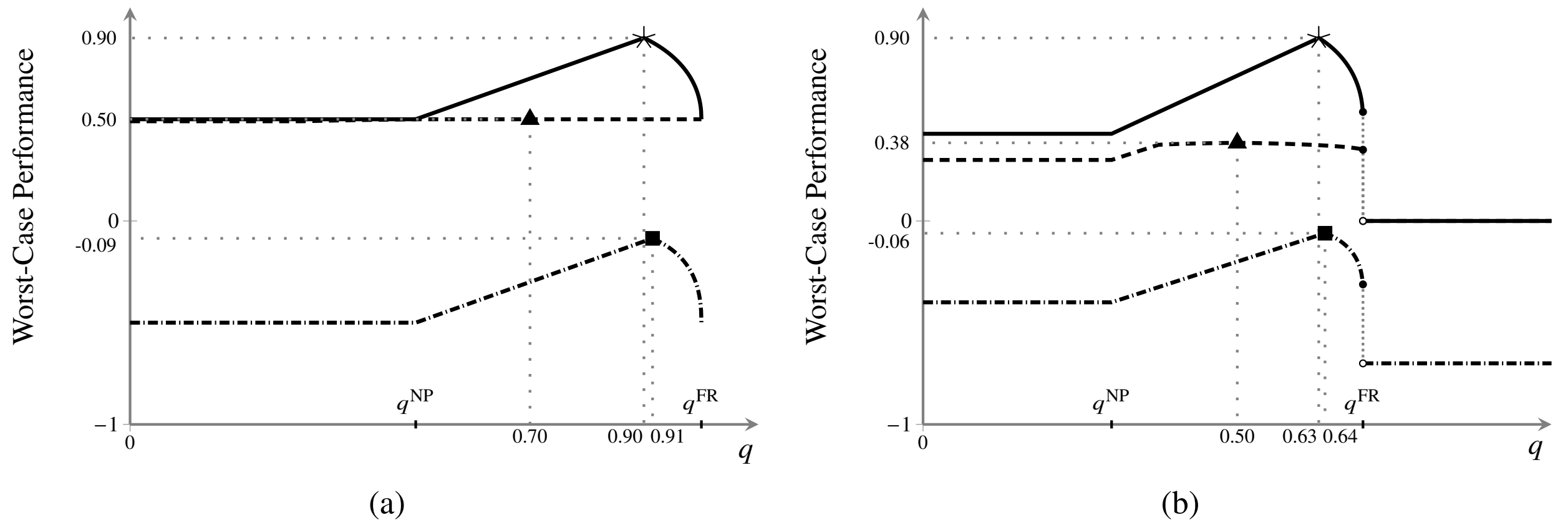

\xhdr{Same structure, varying information disclosure}
\Cref{thm:robust_optimal_censorship_three_criteria} shows that the three robustness criteria share the same optimal disclosure structure, although they may call for different amounts of information disclosure.
For a platform, the disclosure format therefore does not depend on the robustness criterion: the platform can still use a common badge or recommendation for all sufficiently favorable realizations and disclose the remaining realizations individually.
The robustness criterion determines only how broadly this common message applies.
\Cref{fig:robust_criteria} illustrates how the optimal pooled mass varies across criteria and shows that it need not be unique: in the left panel, every pooled mass in an interval attains the maximin optimal payoff.

\xhdr{Computational tractability}
The structural characterization also makes robust policy design computationally tractable.
Because \Cref{thm:robust_optimal_censorship_three_criteria} reduces the design problem to the choice of a single pooled mass, we obtain a fully polynomial-time approximation scheme (FPTAS) that computes an approximately rationality-robust optimal censorship under each of the three criteria: a $(1-\epsilon)$-approximation for the maximin ratio and maximin payoff criteria, and an additive $\epsilon$-approximation for the minimax regret criterion (see \Cref{prop:computation_rationality_robust_scheme} and \Cref{algo:ptas_rationality_robust}).

\xhdr{Greater transparency benefits both parties}
\Cref{prop:SISU_threshold_state_monotonicity} shows that the sender optimally reveals more information to a less rational receiver.
A similar message holds when the receiver's rationality level is unknown.
For each robustness criterion, a rationality-robust optimal censorship can be chosen with a pooled mass $\pooledmass$ satisfying $\nonpositivemass \leq \pooledmass \leq \fullyrationalmass$, where the lower bound $\nonpositivemass$ is the total prior mass of states with non-positive values and the upper bound $\fullyrationalmass$ is the pooled mass of the optimal censorship for a fully rational receiver (see \Cref{fig:robust_criteria} for an illustration and \Cref{prop:same_pooled_mass_interval} for details).
Hence, a sender who does not know how rational the receiver is never needs to be less transparent than a sender who assumes full rationality: she still pools every state in which the receiver weakly prefers action $1$, but she may reveal some states in which the receiver prefers action $0$.
Relative to the scheme designed for a fully rational receiver, this additional transparency benefits both parties: the sender's worst-case performance weakly improves under the chosen robustness criterion, and the receiver's expected utility weakly increases at every rationality level (\Cref{prop:receiver_utility_information_revealing}).
Thus, greater transparency protects the platform against uncertainty about user rationality and serves  users at the same time.

\xhdr{Proof overview for the maximin ratio criterion}
We view $\sup\limits_{\signalscheme \in \schemespace}\inf\limits_{\noiseScale \in \fullsetrationalitylevel}\frac{\Payoff[\noiseScale]{\signalscheme}}{\OPT(\noiseScale)}$ as a zero-sum game between the sender, who chooses a signaling scheme, and an adversary, who chooses the receiver's rationality level.
A natural first idea is to invoke a minimax theorem \citep{V-28,S-58,W-12}.
However, such theorems do not apply directly: for a fixed signaling scheme $\signalscheme$, the ratio $\Payoff[\noiseScale]{\signalscheme}/\OPT(\noiseScale)$ need not be quasi-convex in $\noiseScale$, as these theorems require.
To address this issue, we allow the adversary to randomize over rationality levels.
This relaxation does not change the value of the problem: for any fixed signaling scheme, the expected ratio is linear in the adversary's distribution, so its infimum over distributions equals its infimum over individual rationality levels.
Once the adversary can randomize, the objective is affine in both the signaling scheme and the adversary's distribution, and a standard minimax argument in the spirit of Yao's principle \citep{Y-77} allows us to interchange the order of optimization.
It therefore suffices to study the sender's best response to an arbitrary distribution over rationality levels.

Fix any distribution $\noiseD$ over rationality levels.
Against distribution $\noiseD$, the sender's objective is a weighted average of her quantal response payoffs across rationality levels, with each level $\noiseScale$ weighted by $\noiseD(\noiseScale)/\OPT(\noiseScale)$.
This objective can be rewritten as a linear persuasion problem whose indirect utility function is the corresponding weighted average of the quantal response functions $\Quant\betaed(\cdot)$, and we show that this function is S-shaped.
By the characterization of \citet{KMZ-22}, formally stated in \Cref{lem:sshaped optimality} below, this auxiliary problem therefore admits an optimal censorship.
Thus, against each fixed distribution over rationality levels, the sender can restrict attention to censorships without loss of optimality.
This best-response property alone, however, does not imply that a single \emph{pure} censorship is robustly optimal, because the space of censorships is nonconvex: a mixture of two censorships with different cutoffs is not itself a censorship.
As a result, a minimax theorem cannot be applied directly to interchange the infimum over the adversary's distributions and the supremum over censorships.

To overcome this difficulty, we use the fact that every censorship is uniquely determined by its pooled mass $\pooledmass \in [0,1]$ (\Cref{def:censorship}).
This parameterization transforms the supremum over the nonconvex space of censorships into a supremum over the compact and convex interval $[0,1]$.
We then prove that, against any fixed distribution over rationality levels, the sender's objective is single-peaked, and hence quasi-concave, in the pooled mass $\pooledmass$.
Because this objective is also continuous in $\pooledmass$ and linear in the adversary's distribution, Sion's minimax theorem \citep{S-58} allows us to interchange the infimum over the adversary's distributions and the supremum over pooled masses.
Consequently, a single pure censorship attains the robust optimum.
The formal proof carries out this argument for truncated sets of rationality levels $[0,k]$ and then lets $k \to \infty$, treating the fully rational boundary case $\noiseScale = \infty$ separately.

\begin{lemma}[Theorem 1 in \citealp{KMZ-22}]
\label{lem:sshaped optimality}
Let $\statespace \subseteq \reals$ be a measurable state space, $\prior\in\Delta(\statespace)$ a probability measure, and $\indirectU: \reals\rightarrow\reals$ an indirect utility function, and consider the optimization problem $\sup\nolimits_{\posterior\in\MPC(\prior)} \int \indirectU(x) \posterior(x) \; \dd x$,
where $\MPC(\prior)$ denotes the set of all mean-preserving contractions of the prior $\prior$, that is, the distributions of posterior means that some signaling scheme can induce.
If $\indirectU$ is S-shaped, that is, for some threshold $\bar{x}$, it is either convex on $(-\infty, \bar{x}]$ and concave on $[\bar{x}, \infty)$, or concave on $(-\infty, \bar{x}]$ and convex on $[\bar{x}, \infty)$, then there exists an optimal censorship.
\end{lemma}

We conclude this section by proving \Cref{thm:robust_optimal_censorship_three_criteria} for the maximin ratio criterion.

\begin{proof}[Proof of \Cref{thm:robust_optimal_censorship_three_criteria} for the maximin ratio criterion]
Fix an arbitrary constant $\noiseScaleUB\in(0,\infty)$. We first consider a truncated rationality level set $\rationalitySubclass \triangleq [0,\noiseScaleUB]$.
Let $\noiseD \in \stochastic(\rationalitySubclass)$ denote a probability distribution over $\rationalitySubclass$.
Slightly abusing the notation,
we define the average ratio as 
$\objectiveF[\rationalitySubclass]{\signalscheme, \noiseD}  \triangleq \int_{\noiseScale\in\rationalitySubclass}\frac{\Payoff[\noiseScale]{\signalscheme}}{\OPT(\noiseScale)} \cdot \noiseD(\noiseScale) 
    \;\dd \noiseScale$.
We note that the additional randomization on the rationality level $\noiseScale$ will not change the worst-case ratio, namely,
$
    \inf\nolimits_{\noiseD \in \stochastic(\rationalitySubclass)}\objectiveF[\rationalitySubclass]{\signalscheme, \noiseD} = \inf\nolimits_{\noiseScale \in \rationalitySubclass}\frac{\Payoff[\noiseScale]{\signalscheme}}{\OPT(\noiseScale)}
$,
where equality follows from the fact that the infimum over probability measures is attained by a Dirac mass at a minimizer of $\arg\min\nolimits_{\noiseScale \in \rationalitySubclass} \frac{\Payoff[\noiseScale]{\signalscheme}}{\OPT(\noiseScale)}$.
Since both the signaling scheme space $\schemespace$ and the space of all probability distributions over the truncated rationality level set $\stochastic (\rationalitySubclass)$ are compact and convex, and the average ratio $\objectiveF[\rationalitySubclass]{\signalscheme, \noiseD}$ is continuous and affine in two variables $\signalscheme$ and $\noiseD$, 
by Yao's principle \citep{Y-77}, we have
\begin{align*}
    \sup\nolimits_{\signalscheme \in \schemespace}\inf\nolimits_{\noiseD \in \stochastic(\rationalitySubclass)}\;
    \objectiveF[\rationalitySubclass]{\signalscheme, \noiseD} 
    = \inf\nolimits_{\noiseD \in \stochastic(\rationalitySubclass)}\sup\nolimits_{\signalscheme \in \schemespace}\objectiveF[\rationalitySubclass]{\signalscheme, \noiseD}\,.
\end{align*}
Fixing an arbitrary distribution $\noiseD \in \stochastic(\rationalitySubclass)$, we define the auxiliary function $\kernelF[\noiseD]{\variableX}$ parameterized by $\noiseD$ as follows: for every $\variableX\in(-\infty,\infty)$,
\begin{align*}
    \kernelF[\noiseD]{\variableX} \triangleq\displaystyle\int_{\noiseScale\in\rationalitySubclass}\frac{\Quant\betaed(\variableX)}{\OPT(\noiseScale)}
    \cdot \noiseD(\noiseScale) \; \dd\noiseScale\,.
\end{align*}
With this auxiliary function, we can rewrite $\objectiveF[\rationalitySubclass]{\signalscheme, \noiseD}$ in the following form:
\begin{align*}
    \objectiveF[\rationalitySubclass]{\signalscheme, \noiseD} 
    &= \sum\nolimits_{\state\in[\stateNum]}\prior_{\state} \int_{-\infty}^{\infty} \signalscheme_i(\scaledNoiseDiff)   
    \kernelF[\noiseD]{\scaledNoiseDiff} \;\dd\scaledNoiseDiff = 
    \int_{-\infty}^{\infty}   
    \kernelF[\noiseD]{\scaledNoiseDiff}
    \cdot \tilde{\signalscheme}(\scaledNoiseDiff) \;\dd\scaledNoiseDiff\,,
\end{align*}
where $\tilde{\signalscheme}(\scaledNoiseDiff) \triangleq \sum\nolimits_{\state\in[\stateNum]}\prior_{\state} \signalscheme_i(\scaledNoiseDiff)$.
Thus, the optimization program $\sup_{\signalscheme\in\schemespace} \objectiveF[\rationalitySubclass]{\signalscheme, \noiseD}$
can be equivalently viewed as a linear persuasion problem with the sender's indirect utility function being $\kernelF[\noiseD]{\cdot}$.
We argue that the auxiliary function $\kernelF[\noiseD]{\cdot}$ is S-shaped: it is monotone non-increasing, concave on $(-\infty,0]$ and convex on $[0,\infty)$ (see \Cref{lem:kernel_function_propety}).
Invoking \Cref{lem:sshaped optimality}, 
it is guaranteed that choosing a censorship is the best response to a fixed rationality level distribution $\noiseD$, which is given by
\begin{align*}
   \inf\nolimits_{\noiseD \in \stochastic(\rationalitySubclass)}\sup\nolimits_{\signalscheme \in \schemespace}\objectiveF[\rationalitySubclass]{\signalscheme, \noiseD}  = \inf\nolimits_{\noiseD \in \stochastic(\rationalitySubclass)}\sup\nolimits_{\signalscheme \in \censorshipspace}\objectiveF[\rationalitySubclass]{\signalscheme, \noiseD} = \inf\nolimits_{\noiseD \in \stochastic(\rationalitySubclass)}\sup\nolimits_{\pooledmass \in [0,1]}\objectiveF[\rationalitySubclass]{\parameterizedscheme, \noiseD} \,,
\end{align*}
where the last equality holds since each pooled mass $\pooledmass\in[0,1]$ determines a censorship, denoted by $\parameterizedscheme$.
By reparameterizing pure censorships through the pooled mass $\pooledmass \in [0,1]$, we transform the supremum over the original non-convex space of pure censorships into a supremum over the compact and convex domain $[0,1]$.
Then, we show that the function $\objectiveF[\rationalitySubclass]{\parameterizedscheme,\noiseD}$ is quasi-concave over pooled mass $\pooledmass \in [0,1]$ (see \Cref{lem:quasi-concave}).
Since the domain $[0,1]$ is compact and convex, by Sion's Minimax Theorem \citep{S-58}, we obtain
\begin{align*}
    \inf\nolimits_{\noiseD \in \stochastic(\rationalitySubclass)}\sup\nolimits_{\pooledmass \in [0,1]}
    \objectiveF[\rationalitySubclass]{\parameterizedscheme, \noiseD}= \sup\nolimits_{\pooledmass \in [0,1]}\inf\nolimits_{\noiseD \in \stochastic(\rationalitySubclass)}
    \objectiveF[\rationalitySubclass]{\parameterizedscheme, \noiseD}\,.
\end{align*}
Combining all the pieces above, we have
\begin{align*}
\sup\nolimits_{\signalscheme \in \schemespace}\inf\nolimits_{\noiseD \in \stochastic(\rationalitySubclass)}\;
    \objectiveF[\rationalitySubclass]{\signalscheme, \noiseD} =     \sup\nolimits_{\pooledmass \in [0,1]}\inf\nolimits_{\noiseD \in \stochastic(\rationalitySubclass)}\;
    \objectiveF[\rationalitySubclass]{\parameterizedscheme, \noiseD}\,,
\end{align*}
which establishes the rationality-robust optimality of the pure censorships.

The extension of this argument from the truncated set $\rationalitySubclass = [0, \noiseScaleUB]$ to the full set of rationality levels $\fullsetrationalitylevel$ follows from a standard limiting argument: we let the truncation level $\noiseScaleUB$ grow, take a convergent subsequence of the corresponding optimal pooled masses, and verify that the limit censorship remains optimal, including against a fully rational receiver. We defer the details to \Cref{apx:limiting-arguments}.
\end{proof}

\section{Universal Robust Performance Guarantee}
\label{sec:robust_ratio}

In this section, we establish tight universal performance guarantees under the maximin ratio criterion, where ``universal'' means that the guarantees hold for every instance and every rationality level.
In \Cref{subsec:baseline_baseline_worst_case_ratio}, we show that the optimal censorship for a fully rational receiver guarantees a worst-case ratio of at least $\sfrac{1}{2}$ (\Cref{thm:rationality_robust}\ref{item:baseline_ratio}).
In \Cref{subsec:universal_guarantee_maximin_optimal_ratio}, we show that the rationality-robust optimal censorship improves this guarantee to at least $\worstcaseratio \approx \ratiovalue$ (\Cref{thm:rationality_robust}\ref{item:optimal_ratio}).
Both bounds are tight, and the same binary-state instance in Eqn.~\eqref{ex:binary_worst_case} attains both of them.
\Cref{thm:rationality_robust} summarizes these results.

\begin{restatable}{theorem}{robustoptimalratiocharacterization}
\label{thm:rationality_robust}
For every instance $\instance$, the following hold:
\begin{enumerate}[label=(\roman*)]
    \item\label{item:baseline_ratio} the optimal censorship for a fully rational receiver has a worst-case ratio of at least $\sfrac{1}{2}$; and
    \item\label{item:optimal_ratio} the rationality-robust optimal censorship $\schemeoptimalratio$ (see \Cref{def:worst-case_ratio_payoff_regret}) has a worst-case ratio of at least $\worstcaseratio$, that is, the maximin optimal ratio is at least $\worstcaseratio$, where $\worstcaseratio \approx \ratiovalue$ is defined in Eqn.~\eqref{eq:universal_guarnatee_maximin_optimal_ratio}.
\end{enumerate}
Moreover, both bounds are tight: for the binary-state instance
\begin{align}
\label{ex:binary_worst_case}
    \prior_1 = {2}/{3},\quad\prior_2 = {1}/{3},\quad \receiverU_1 = -1,\quad \receiverU_2 = 2\,,
\end{align}
the optimal censorship for a fully rational receiver has a worst-case ratio of exactly $\sfrac{1}{2}$, and the maximin optimal ratio is exactly $\worstcaseratio$.
\end{restatable}

\begin{figure}[tb]
    \centering
    \begingroup
    \begin{minipage}[b]{0.60\textwidth}
        \centering
\input{plots/figure-hard-instances}
    \end{minipage}
    \endgroup
    \caption{Worst-case ratio of the censorship with pooled mass $\pooledmass$ for the binary-state instance in Eqn.~\eqref{ex:binary_worst_case}.
    The star marks the maximin optimal ratio $\worstcaseratio \approx \ratiovalue$, achieved at a pooled mass of approximately $0.90$, and the square marks the worst-case ratio $\sfrac{1}{2}$ of the optimal censorship for a fully rational receiver, whose pooled mass is $\fullyrationalmass = 1$.}
    \label{fig:hard_instance}
\end{figure}

Achieving a constant worst-case ratio with a censorship is not straightforward.
Indeed, any censorship that pools strictly more mass than the optimal one for a fully rational receiver has a worst-case ratio of zero: every signal it sends is then positive, so a fully rational receiver never takes the sender's preferred action (see \Cref{fig:robust_criteria_b}).
\Cref{thm:rationality_robust} shows that, in contrast, both the optimal censorship for a fully rational receiver\footnote{The censorship structure is essential for this guarantee. A fully rational receiver also admits optimal signaling schemes that are not censorship schemes, and such schemes can perform poorly against a boundedly rational receiver. See \Cref{prop:necessity_censorship}.} and the rationality-robust optimal one guarantee a constant fraction of the sender's optimal expected utility, uniformly over all instances and rationality levels.
These tight guarantees yield the following insights.

\xhdr{The limited value of personalization}
At every rationality level, a platform using the rationality-robust optimal censorship $\schemeoptimalratio$ obtains at least $\ratiopercent\%$ of the maximum achievable utility, that is, the optimal expected utility that she would obtain if she knew the user's rationality level.
Because this guarantee holds at every rationality level, it also applies to a platform that serves users who share the same prior and utilities but differ in their rationality levels.
For any distribution of rationality levels among these users, a single rationality-robust censorship obtains at least $\ratiopercent\%$ of the expected utility that the platform would obtain by tailoring its signaling scheme to each user's rationality level.
The value of such personalization is therefore at most $\losspercent\%$ of the maximum achievable utility.
A platform thus benefits from estimating each user's rationality level and customizing its disclosure only if the cost of doing so is below this bounded gain; when individual-level choice data are scarce or customization is costly, a single rationality-robust censorship captures most of the attainable payoff.

\xhdr{Quantifying the value of transparency}
A platform that ignores bounded rationality and applies the optimal censorship for a fully rational receiver still guarantees at least $50\%$ of the maximum achievable utility, but this bound is tight, so the platform may lose half of the attainable payoff.
As shown in \Cref{sec:robust_optimal_scheme_unkown_rationality_level}, under each robustness criterion, a rationality-robust optimal censorship can be chosen to pool weakly less mass than this fully rational benchmark while weakly increasing the receiver's expected utility.
\Cref{thm:rationality_robust} quantifies the gain from this additional transparency under the maximin ratio criterion: it raises the universal guarantee on the worst-case ratio from $\sfrac{1}{2}$ to approximately $\ratiovalue$.
This improvement can require only a modest amount of additional disclosure.
For the binary-state instance in Eqn.~\eqref{ex:binary_worst_case}, for example, the rationality-robust optimal censorship pools a mass of approximately $0.90$, whereas the fully rational benchmark pools all states (see \Cref{fig:hard_instance}).
By revealing about $10\%$ of the prior mass, the platform gives up about $10\%$ of its payoff against a fully rational receiver and, in exchange, raises its worst-case ratio by approximately $40$ percentage points, from $0.5$ to $\ratiovalue$.
Transparency therefore acts as insurance against uncertainty about user rationality: a small loss when users are fully rational buys a much higher guarantee across all rationality levels.

\xhdr{Behavioral robustness matters even in simple environments}
The binary-state instance in Eqn.~\eqref{ex:binary_worst_case} attains both tight bounds.
Thus, the worst-case losses already arise with only two states and fixed, nonvanishing prior probabilities; they do not require a large state space or a complicated prior.\footnote{This finding is conceptually similar to classic results in prior-independent mechanism design, where the competitive ratio typically worsens as the number of buyers decreases \citep{BK-96,AB-20}.}
For a platform, a simple environment is therefore no reason to apply a disclosure policy calibrated to fully rational users: even in this binary-state instance, that policy loses $50\%$ of the payoff attainable with knowledge of the user's rationality level in the worst case, whereas the rationality-robust policy limits the loss to $\losspercent\%$.

\xhdr{Proof overview of \Cref{thm:rationality_robust}}
Both guarantees in \Cref{thm:rationality_robust} are obtained by comparing, at every rationality level $\noiseScale$, a \emph{candidate} censorship $\candidatecensor$ with pooled mass $\candidatepooledmass$ against a \emph{benchmark} censorship $\benchmarkcensor$ with pooled mass $\benchmarkpooledmass$.
The candidate is the scheme whose guarantee we seek to bound, and the benchmark is ultimately the optimal censorship for rationality level $\noiseScale$, which attains $\OPT(\noiseScale)$ by \Cref{thm:opt_state_independent}.
To make this comparison quantitative, we reduce a payoff comparison involving arbitrarily many states to one involving binary-state payoffs and collect the remaining contributions into remainder terms that we bound separately.
This reduction leaves only a constant number of variables in the ratio $\Payoff[\noiseScale]{\candidatecensor}/\Payoff[\noiseScale]{\benchmarkcensor}$, which makes the ratio tractable to bound.

The reduction rests on a simple observation about any two censorships.
Each scheme pools the states below its threshold state and reveals the remaining states.
The pooled masses of the two schemes therefore partition the state space into three regions: a region pooled by both schemes; an intermediate region pooled by the less informative scheme, which has the larger pooled mass, but revealed by the more informative one; and a region revealed by both.
This observation motivates the following decomposition.

\xhdr{Three-way decomposition}
Fix two censorships $\candidatecensor$ and $\benchmarkcensor$ on the instance $\instance$, with pooled masses $\candidatepooledmass$ and $\benchmarkpooledmass$, respectively. Define
\begin{align}
    \topstates &\triangleq \left\{ i \in [\stateNum] : \sum\nolimits_{j \in [i-1]}\prior_j < \min\{\candidatepooledmass,\benchmarkpooledmass \}\right\}\,, \tag{top states} \label{eq:top_states}\\
    \middlestates &\triangleq  \left\{ i \in [\stateNum] : \max\left\{\sum\nolimits_{j \in [i-1]}\prior_j,\;\min\{\candidatepooledmass,\benchmarkpooledmass \} \right\}<  \min\left\{\sum\nolimits_{j \in [i]}\prior_j,\;\max\{\candidatepooledmass,\benchmarkpooledmass \} \right\}\right\}\,, \tag{middle states} \label{eq:middle_states}\\
    \bottomstates &\triangleq \left\{ i \in [\stateNum] : \sum\nolimits_{j \in [i]}\prior_j  > \max\{\candidatepooledmass,\benchmarkpooledmass \} \right\} \,. \tag{bottom states} \label{eq:bottom_states}
\end{align}
The top states $\topstates$ are pooled by both schemes, and the bottom states $\bottomstates$ are revealed by both.
The middle states $\middlestates$ are pooled by the scheme with the larger pooled mass and revealed by the other.
\Cref{fig:baseline_fully_rational,fig:more_pooled_competitor,fig:less-pooled-t} illustrate these three sets, and we use this decomposition throughout \Cref{subsec:baseline_baseline_worst_case_ratio,subsec:universal_guarantee_maximin_optimal_ratio}.
These sets cover the state space but need not be disjoint; any two of them intersect in at most one state, namely a threshold state that is only partially pooled.
For ease of presentation, we assume that $\topstates$, $\middlestates$, and $\bottomstates$ are pairwise disjoint.\footnote{We can split any shared state into adjacent states with the same value without changing the sender's expected utility under either scheme (see \Cref{apx:state_refinement}), so this assumption only simplifies the comparison between the two schemes.}

Although both guarantees rely on a reduction to binary states, the reduction differs across cases.
For \Cref{thm:rationality_robust}\ref{item:baseline_ratio}, and for the case $\candidatepooledmass \le \benchmarkpooledmass$ in \Cref{thm:rationality_robust}\ref{item:optimal_ratio}, we treat the commonly revealed bottom states as a single ``hypothetical revealed state'' that contributes the same expected utility under both schemes.
We then reduce the top and middle states $\topstates \cup \middlestates$ to a single state under each scheme, which, together with the hypothetical revealed state, yields a binary-state comparison.
For the case $\candidatepooledmass > \benchmarkpooledmass$ in \Cref{thm:rationality_robust}\ref{item:optimal_ratio}, we instead reduce the bottom states $\bottomstates$ explicitly to a single state with their total mass and conditional mean value.
Under the candidate scheme $\candidatecensor$, which pools $\topstates \cup \middlestates$, these states reduce to a single state, which, together with the reduced bottom state, again yields a binary-state space.
Under the benchmark scheme $\benchmarkcensor$, which pools only $\topstates$ and reveals $\middlestates$, we use the convexity of the quantal response function on $[0,\infty)$ to upper bound the contribution of the middle states by that of two states, whose values are $0$ and the conditional mean value of the bottom states.
Together with the pooled top states and the reduced bottom state, this yields three distinct state values, and the resulting payoff equals a convex combination of two binary-state payoffs.

We conclude \Cref{subsec:baseline_baseline_worst_case_ratio} by proving \Cref{thm:rationality_robust}\ref{item:baseline_ratio} and  \Cref{subsec:universal_guarantee_maximin_optimal_ratio} by proving \Cref{thm:rationality_robust}\ref{item:optimal_ratio}.

\definecolor{myblue}{RGB}{79, 118, 164}
\definecolor{myorange}{RGB}{242, 142, 43}
\definecolor{mylightgreen}{RGB}{198, 224, 212}
\definecolor{myhatchedbg}{RGB}{245, 225, 205}
\definecolor{mydarkgreen}{RGB}{46, 117, 86}
\definecolor{mybrown}{RGB}{150, 100, 30}
\definecolor{mygray}{RGB}{234, 234, 234}

\begin{figure}[t]
    \centering
    \resizebox{\textwidth}{!}{%
    \begin{tikzpicture}[font=\sffamily]

        \draw[decorate, decoration={brace, amplitude=5pt},  black, thick] 
            (0, 3.2) -- (6.0, 3.2) node[midway, above=6pt, font=\small] {top states ($\topstates$)};
            
        \draw[decorate, decoration={brace, amplitude=5pt}, black, thick] 
            (6.0, 3.2) -- (9.5, 3.2) node[midway, above=6pt, font=\small] {middle states ($\middlestates$)};
            
        \draw[decorate, decoration={brace, amplitude=5pt}, black, thick] 
            (9.5, 3.2) -- (13.0, 3.2) node[midway, above=6pt, font=\small] {bottom states ($\bottomstates$)};

        \node[font=\normalsize, anchor=east] at (-0.3, 2.0) {$\candidatecensor$};

        \fill[white] (0,1.5) rectangle (9.5,2.5);
        \fill[pattern=dots, pattern color=black] (0,1.5) rectangle (9.5,2.5);
        
        \fill[white] (9.5,1.5) rectangle (13.0,2.5);

        \draw[black, thick] (0,1.5) rectangle (13.0,2.5);
        \foreach \x in {1.5, 3.0, 4.5, 6.0, 7.5, 9.5, 11.5} {
            \draw[black] (\x, 1.5) -- (\x, 2.5);
        }

        \node at (0.75, 2.8) {$v_1$};
        \node at (2.25, 2.8) {$v_2$};
        \node at (3.75, 2.8) {$v_3$};
        \node at (5.25, 2.8) {$v_4$};
        \node at (6.75, 2.8) {$v_5$};
        \node at (8.5,  2.8) {$v_6$};
        \node at (10.5, 2.8) {$v_7$};
        \node at (12.25,2.8) {$v_8$};

        \draw[decorate, decoration={brace, amplitude=5pt, mirror}, thick] 
            (0, 1.35) -- (9.5, 1.35) node[midway, below=8pt] {$\candidatepooledmass \Quant\betaed(\candidatepoolsignal) \ge \sfrac{\candidatepooledmass}{2}$};
            
        \draw[decorate, decoration={brace, amplitude=5pt, mirror}, thick] 
            (9.5, 1.35) -- (13.0, 1.35) node[midway, below=8pt, align=center, font=\normalsize] {$\sum\nolimits_{i \in \bottomstates} \prior_i  \Quant\betaed(\receiverU_i)$};

        \node[anchor=south, align=center, font=\small] (extratext) at (7.75, -0.5) {extra revealed states};
        \draw[->, thick, >=stealth] (7.75, -0.5) -- (7.75, -1.4);

        \node[font=\normalsize, anchor=east] at (-0.3, -2.0) {$\benchmarkcensor$};

        \fill[white] (0,-2.5) rectangle (6.0,-1.5);
        \fill[pattern=vertical lines, pattern color=black] (0,-2.5) rectangle (6.0,-1.5);
        
        \fill[white] (6.0,-2.5) rectangle (9.5,-1.5);
        \fill[pattern=north east lines, pattern color=black] (6.0,-2.5) rectangle (9.5,-1.5);
        
        \fill[white] (9.5,-2.5) rectangle (13.0,-1.5);

        \draw[black, thick] (0,-2.5) rectangle (13.0,-1.5);
        \foreach \x in {1.5, 3.0, 4.5, 6.0, 7.5, 9.5, 11.5} {
            \draw[black] (\x, -2.5) -- (\x, -1.5);
        }

        \node at (0.75, -1.2) {$v_1$};
        \node at (2.25, -1.2) {$v_2$};
        \node at (3.75, -1.2) {$v_3$};
        \node at (5.25, -1.2) {$v_4$};
        \node at (6.75, -1.2) {$v_5$};
        \node at (8.5,  -1.2) {$v_6$};
        \node at (10.5, -1.2) {$v_7$};
        \node at (12.25,-1.2) {$v_8$};

            

        \draw[decorate, decoration={brace, amplitude=5pt, mirror}, thick] 
            (9.5, -2.65) -- (13.0, -2.65) node[midway, below=8pt, font=\normalsize, align=center] {$\sum\nolimits_{i \in \bottomstates} \prior_i  \Quant\betaed(\receiverU_i)$};

        LAYER 2 Bracket (PUSHED LOW to Y = -3.8)
        \draw[decorate, decoration={brace, amplitude=5pt, mirror}, thick] 
            (0, -2.65) -- (9.5, -2.65) node[midway, below=8pt, align=center] {$\benchmarkpooledmass \Quant\betaed(\benchmarkpoolsignal)+\sum\nolimits_{i \in \middlestates} \prior_i  \Quant\betaed(\receiverU_i) \le \candidatepooledmass$};


        \filldraw[fill=white, pattern=dots, draw=black] (13.5, 0.75) rectangle (14.3, 1.05);
        \node[right, font=\small] at (14.4, 0.9) {pooled under $\candidatecensor$};

        \filldraw[fill=white, pattern=vertical lines, draw=black] (13.5, 0.25) rectangle (14.3, 0.55);
        \node[right, font=\small] at (14.4, 0.4) {pooled under $\benchmarkcensor$};
        
        \filldraw[fill=white, draw=black] (13.5, -0.25) rectangle (14.3, 0.05);
        \node[right, font=\small] at (14.4, -0.1) {commonly revealed states};

        \filldraw[fill=white, pattern=north east lines, draw=black] (13.5, -0.75) rectangle (14.3, -0.45);
        \node[right, font=\small] at (14.4, -0.6) {revealed only under $\benchmarkcensor$};

    \end{tikzpicture}%
        }

        \caption{Comparison of the structures and utility decompositions for the optimal censorship 
        for a fully rational receiver $\candidatecensor$ and the benchmark censorship $\benchmarkcensor$ analyzed in \Cref{thm:rationality_robust}\ref{item:baseline_ratio}. 
        Specifically, $\candidatepooledmass\Quant\betaed(\candidatepoolsignal)$ (resp.~$\benchmarkpooledmass\Quant\betaed(\benchmarkpoolsignal)$) represents the expected utility from the pooled states under the signaling scheme $\candidatecensor$ (resp.~$\benchmarkcensor$). The sum $\sum\nolimits_{i \in \middlestates}\prior_i  \Quant\betaed(\receiverU_i)$ represents the utility contribution of the middle states $\middlestates$ under the censorship  $\benchmarkcensor$. The sum $\sum\nolimits_{i \in \bottomstates} \prior_i  \Quant\betaed(\receiverU_i)$ represents the utility contribution of the bottom states $\bottomstates$ under both censorships. }
    \label{fig:baseline_fully_rational}
\end{figure}

\subsection{Robustness of the Fully Rational Censorship Signaling Scheme: \texorpdfstring{\sfrac{1}{2}}{1/2}-Worst-Case Ratio}
\label{subsec:baseline_baseline_worst_case_ratio}

We first evaluate a natural policy when the rationality level is unknown: the optimal censorship signaling scheme designed for a fully rational receiver, which a platform would adopt under the classical model.
By \Cref{thm:rationality_robust}\ref{item:baseline_ratio}, this scheme obtains at least half of the optimal expected utility attainable with knowledge of the receiver's rationality level, for every instance and every rationality level.

We now provide the intuition behind \Cref{thm:rationality_robust}\ref{item:baseline_ratio} (see \Cref{fig:baseline_fully_rational} for an illustration).
Fix a finite rationality level $\noiseScale$ (at $\noiseScale = \infty$, the candidate scheme is optimal by construction).
The candidate censorship signaling scheme $\candidatecensor$ is the optimal one for a fully rational receiver, and the benchmark censorship signaling scheme $\benchmarkcensor$ is the optimal one for rationality level $\noiseScale$.
By \Cref{prop:SISU_threshold_state_monotonicity}, the candidate pools weakly more mass than the benchmark.
Hence, the benchmark additionally reveals the middle states $\middlestates$, while both schemes reveal the bottom states $\bottomstates$, which we can treat as a single ``hypothetical revealed state'' that contributes the same expected utility under both schemes.
Under the candidate, the top and middle states $\topstates \cup \middlestates$ are pooled into a single signal with a non-positive value, so the receiver takes action $1$ after this signal with probability at least $\Quant\betaed(0) = \sfrac{1}{2}$.
Under the benchmark, we upper bound the sender's expected utility from $\topstates \cup \middlestates$ by their total mass, since the receiver takes action $1$ with probability at most $1$.
Both schemes thus reduce to two states: a merged state for $\topstates \cup \middlestates$, whose contribution is at least half as large under the candidate as under the benchmark, and the hypothetical revealed state, whose contribution is identical under both.
Adding the same nonnegative contribution to both payoffs cannot lower their ratio below $\sfrac{1}{2}$, which yields the guarantee.

\begin{proof}[Proof of \Cref{thm:rationality_robust}\ref{item:baseline_ratio}]
We evaluate the worst-case ratio of the optimal censorship $\optschemeInfty$ for a fully rational receiver over all possible rationality levels $\noiseScale \in [0, \infty) \cup \{\infty\}$. There are two cases.

\xhdr{The receiver is not fully rational ($\noiseScale \in [0, \infty)$)}
Fix any finite rationality level $\noiseScale\in[0,\infty)$.
By \Cref{thm:SISU_opt}, it is optimal to choose a censorship signaling scheme for the known rationality level $\noiseScale$.
Let $\optschemebeta$ denote the optimal censorship under the rationality level $\noiseScale$.
Recall that \Cref{prop:SISU_threshold_state_monotonicity}
implies that $\pooledmassInfty \ge \pooledmassbeta$, where $\pooledmassbeta$ is the pooled mass of the censorship signaling scheme $\optschemebeta$.

Then, for the optimal censorship $\optschemeInfty$ designed for a fully rational receiver, we have
\begin{align*}
    \Payoff[\noiseScale]{\optschemeInfty} &= 
     \pooledmassInfty \Quant\betaed(\censorshipsignalInfty) + \sum\nolimits_{i \in \bottomstates} \prior_i \Quant\betaed(\receiverU_i)\,,
\end{align*}
and the sender's optimal expected utility
under the rationality level $\noiseScale$ is 
\begin{align*}
    \OPT(\noiseScale) &=  \pooledmassbeta \Quant\betaed(\censorshipsignalbeta) + \sum\nolimits_{i \in \middlestates} \prior_i \Quant\betaed(\receiverU_i) + \sum\nolimits_{i \in \bottomstates} \prior_i \Quant\betaed(\receiverU_i)
     \le \pooledmassbeta  + \sum\nolimits_{i \in \middlestates} \prior_i  + \sum\nolimits_{i \in \bottomstates} \prior_i \Quant\betaed(\receiverU_i) =  \pooledmassInfty  + 
    \sum\nolimits_{i \in \bottomstates} \prior_i \Quant\betaed(\receiverU_i)\,,
\end{align*}
where the inequality holds since the quantal response function satisfies $\Quant\betaed(\cdot) \le 1$.
Therefore, we obtain
\begin{align*}
    \frac{\Payoff[\noiseScale]{\optschemeInfty}}{\OPT(\noiseScale)}
    \ge 
     \frac{\pooledmassInfty \Quant\betaed(\censorshipsignalInfty) + \sum\nolimits_{i \in \bottomstates} \prior_i \Quant\betaed(\receiverU_i)}{\pooledmassInfty  + 
    \sum\nolimits_{i \in \bottomstates} \prior_i \Quant\betaed(\receiverU_i)}
    \overset{(a)}{\ge} \Quant\betaed(\censorshipsignalInfty) \overset{(b)}{\ge} \frac{1}{2}\,,
\end{align*}
where inequality~(a) holds due to the algebraic fact $\frac{ \variablecapA + \variableX}{\variablecapB + \variableX} \ge \frac{\variablecapA}{\variablecapB}$
    for all $\variableX\in [0,\infty) \text{ and } 0\le \variablecapA \le \variablecapB$, and inequality~(b) holds since $\Quant\betaed(\censorshipsignalInfty) \ge \Quant\betaed(0) =\frac{1}{2} $ due to the fact that the quantal response function $\Quant\betaed(\cdot)$ is monotone non-increasing, and $\censorshipsignalInfty \le 0$ .
    
\xhdr{The receiver is fully rational ($\noiseScale =\infty$)}
By definition, $\optschemeInfty$ is the optimal censorship designed for a fully rational receiver. Therefore, we have $\Payoff[\infty]{\optschemeInfty} = \OPT(\infty)$, and  the utility ratio is exactly 1.

Combining the two cases finishes the proof. 
\end{proof}

Our binary-state instance in \Cref{thm:rationality_robust} shows that the worst-case ratio of $\sfrac{1}{2}$ is tight.
The intuition is as follows.
Because the prior mean of this instance is zero, the optimal censorship signaling scheme for a fully rational receiver pools both states, which yields a pooling signal of $0$ and thus an expected utility of exactly $\sfrac{1}{2}$ at every finite rationality level.
In contrast, a censorship signaling scheme that pools slightly less mass induces a strictly negative pooling signal.
By letting its pooled mass increase to $1$ at a suitable rate as $\noiseScale$ grows (e.g., $1 - \noiseScale^{-1/2}$), we obtain a censorship signaling scheme whose expected utility approaches $1$ as $\noiseScale \to \infty$.
Hence, $\OPT(\noiseScale) \to 1$, so the ratio of the fully rational benchmark approaches $\sfrac{1}{2}$; together with the lower bound in \Cref{thm:rationality_robust}\ref{item:baseline_ratio}, its worst-case ratio is exactly $\sfrac{1}{2}$.
The formal analysis is deferred to \Cref{subsec:baseline_hard_instance}.

\subsection{Characterization of the Universal Guarantee for Maximin Optimal Ratio}
\label{subsec:universal_guarantee_maximin_optimal_ratio}

We now establish the tight universal guarantee for the rationality-robust censorship under the maximin ratio criterion.
Specifically, we prove \Cref{thm:rationality_robust}\ref{item:optimal_ratio} by showing that the maximin optimal ratio is at least $\worstcaseratio \approx \ratiovalue$ for every instance, with equality for the binary-state instance in Eqn.~\eqref{ex:binary_worst_case}.
For the analysis, we restrict attention, without loss of generality, to instances $\instance = (\stateNum,\{\prior_{\state}\}_{\state \in [\stateNum]},\{\receiverU_{\state}\}_{\state \in [\stateNum]})$ with a non-positive prior mean, i.e., $\sum\nolimits_{\state \in [\stateNum]} \prior_{\state}\receiverU_{\state} \leq 0$, and to censorships that pool every state with a non-positive value.\footnote{The formal proofs establishing that these two restrictions are without loss of generality are deferred to \Cref{prop:reduction} and \Cref{prop:same_pooled_mass_interval}, respectively.}
Let $\censorshipspacepositive \subset \censorshipspace$ denote the set of censorships that pool every state with a non-positive value.

We now outline the structure of the proof.
We compare a candidate censorship $\candidatecensor \in \censorshipspacepositive$ with pooled mass $\candidatepooledmass$ against a benchmark censorship $\benchmarkcensor \in \censorshipspacepositive$ with pooled mass $\benchmarkpooledmass$.
The benchmark is arbitrary within $\censorshipspacepositive$; in the proof of \Cref{thm:rationality_robust}\ref{item:optimal_ratio}, we apply the resulting bounds to the optimal censorship at each rationality level, which belongs to $\censorshipspacepositive$ by \Cref{thm:opt_state_independent} and \Cref{lem:opt_fully_rational}.
For each such pair, we partition the states into the top states $\topstates$, middle states $\middlestates$, and bottom states $\bottomstates$ using the three-way decomposition.
The argument then splits into two cases according to the relative pooled masses of the two schemes.
\Cref{lem:more_pooled_comeptitor} treats the case $\candidatepooledmass \le \benchmarkpooledmass$, in which the benchmark pools weakly more mass than the candidate (see \Cref{fig:more_pooled_competitor}).
\Cref{lem:less-pooled-than-t} treats the complementary case $\candidatepooledmass > \benchmarkpooledmass$, in which the benchmark is more informative than the candidate (see \Cref{fig:less-pooled-t}).
Together, these two lemmas bound the payoff ratio $\Payoff[\noiseScale]{\candidatecensor}/\Payoff[\noiseScale]{\benchmarkcensor}$ for every pair of schemes, which yields \Cref{thm:rationality_robust}\ref{item:optimal_ratio}.

\definecolor{myblue}{RGB}{79, 118, 164}
\definecolor{myorange}{RGB}{242, 142, 43}
\definecolor{mygray}{RGB}{234, 234, 234}
\definecolor{myhatchedbg}{RGB}{245, 225, 205}
\definecolor{mybracketbrown}{RGB}{186, 135, 90}

\begin{figure}[t]
    \centering
    \resizebox{\textwidth}{!}{%
    \begin{tikzpicture}[font=\sffamily]


        \node[font=\normalsize, anchor=east] at (-0.3, 2.5) {$\candidatecensor$};

        \fill[pattern=dots, pattern color=black] (0,2) rectangle (7.5,3);
        \fill[pattern=north east lines, pattern color=black] (7.5,2) rectangle (11.0,3);
        \fill[white] (11.0,2) rectangle (13.0,3); 

        \draw[black, thick] (0,2) rectangle (13.0,3);
        \foreach \x in {1.5, 3.0, 4.5, 6.0, 7.5, 9.5,  11.0} {
            \draw[black] (\x, 2) -- (\x, 3);
        }

        \node at (0.75, 3.3) {$\receiverU_1$};
        \node at (2.25, 3.3) {$\receiverU_2$};
        \node at (3.75, 3.3) {$\receiverU_3$};
        \node at (5.25, 3.3) {$\receiverU_4$};
        \node at (6.75, 3.3) {$\receiverU_5$};
        \node at (8.5,  3.3) {$\receiverU_6$};
        \node at (10.5, 3.3) {$\receiverU_7$};
        \node at (12.25,3.3) {$\receiverU_8$};

        \draw[decorate, decoration={brace, amplitude=5pt},  black, thick] 
            (0, 3.6) -- (7.5, 3.6) node[midway, above=6pt, font=\small] {top states ($\topstates$)};
            
        \draw[decorate, decoration={brace, amplitude=5pt}, black, thick] 
            (7.5, 3.6) -- (11.0, 3.6) node[midway, above=6pt, font=\small] {middle states ($\middlestates$)};
            
        \draw[decorate, decoration={brace, amplitude=5pt}, black, thick] 
            (11.0, 3.6) -- (13.0, 3.6) node[midway, above=6pt, font=\small] {bottom states ($\bottomstates$)};

        \draw[decorate, decoration={brace, amplitude=5pt, mirror}, black, thick] 
            (0, 1.85) -- (7.5, 1.85) node[midway, below=6pt] {$\pooledmass\primed \Quant\betaed(\candidatepoolsignal)$};
            
        \draw[decorate, decoration={brace, amplitude=5pt, mirror}, black, thick] 
            (7.5, 1.85) -- (11.0, 1.85) node[midway, below=6pt] {$\sum_{i \in \middlestates} \prior_i \Quant\betaed(\receiverU_i)$};
            
        \draw[decorate, decoration={brace, amplitude=5pt, mirror}, black, thick] 
            (11.0, 1.85) -- (13.0, 1.85) node[midway, below=6pt] {$\sum_{i \in \bottomstates} \prior_i \Quant\betaed(\receiverU_i)$};

        \node[anchor=north, align=center, font=\small] (extratext) at (9.3, 1.0) {extra revealed states \\[-0.5ex] (discarded in the lower bound)};

        \node[font=\normalsize, anchor=east] at (-0.3, -1.5) {$\benchmarkcensor$};

        \fill[pattern=vertical lines, pattern color=black] (0,-2.0) rectangle (11.0,-1.0);
        \fill[white] (11.0,-2.0) rectangle (13.0,-1.0); 

        \draw[black, thick] (0,-2.0) rectangle (13.0,-1.0);
        \foreach \x in {1.5, 3.0, 4.5, 6.0, 7.5, 9.5, 11.0} {
            \draw[black] (\x, -2.0) -- (\x, -1.0);
        }

        \node at (0.75, -0.7) {$\receiverU_1$};
        \node at (2.25, -0.7) {$\receiverU_2$};
        \node at (3.75, -0.7) {$\receiverU_3$};
        \node at (5.25, -0.7) {$\receiverU_4$};
        \node at (6.75, -0.7) {$\receiverU_5$};
        \node at (8.5,  -0.7) {$\receiverU_6$};
        \node at (10.5, -0.7) {$\receiverU_7$};
        \node at (12.25,-0.7) {$\receiverU_8$};

        \draw[decorate, decoration={brace, amplitude=5pt, mirror}, black, thick] 
            (0, -2.15) -- (11.0, -2.15) node[midway, below=6pt] {$\pooledmass\doubleprimed \Quant\betaed(\benchmarkpoolsignal)$};
            
        \draw[decorate, decoration={brace, amplitude=5pt, mirror}, black, thick] 
            (11.0, -2.15) -- (13.0, -2.15) node[midway, below=6pt] {$\sum\nolimits_{i \in \bottomstates}\prior_i\Quant\betaed(\receiverU_i)$};

        \filldraw[fill=white, pattern=dots, draw=black] (13.5, 1.25) rectangle (14.3, 1.55);
        \node[right, font=\small] at (14.4, 1.4) {pooled under $\candidatecensor$};

        \filldraw[fill=white, pattern=vertical lines, draw=black] (13.5, 0.75) rectangle (14.3, 1.05);
        \node[right, font=\small] at (14.4, 0.9) {pooled under $\benchmarkcensor$};

        \filldraw[fill=white, draw=black] (13.5, 0.25) rectangle (14.3, 0.55);
        \node[right, font=\small] at (14.4, 0.4) {commonly revealed states};

        \filldraw[fill=white, pattern=north east lines, draw=black] (13.5, -0.25) rectangle (14.3, 0.05);
        \node[right, font=\small] at (14.4, -0.1) {revealed only under $\candidatecensor$};

    \end{tikzpicture}%
    }
    \caption{Comparison of the structures and utility decompositions for the candidate censorship $\candidatecensor$ and the benchmark censorship $\benchmarkcensor$ analyzed in \Cref{lem:more_pooled_comeptitor}.
$\candidatepooledmass\Quant\betaed(\candidatepoolsignal)$ (resp.~$\benchmarkpooledmass\Quant\betaed(\benchmarkpoolsignal)$) represents the expected utility from the pooled states under signaling scheme $\candidatecensor$ (resp.~$\benchmarkcensor$). The sum $\sum\nolimits_{i \in \bottomstates}\prior_i\Quant\betaed(\receiverU_i)$ represents the utility contribution from the states that are commonly revealed under both signaling schemes. Finally, the sum $\sum\nolimits_{i \in \middlestates} \prior_i \Quant\betaed(\receiverU_i)$ captures the utility contribution of the states that are revealed only under the candidate censorship $\candidatecensor$, which is discarded to establish the lower bound in the proof.}
    \label{fig:more_pooled_competitor}
\end{figure}

\begin{restatable}{lemma}
{morepooledcompetitor}
\label{lem:more_pooled_comeptitor}
Fix any instance $\instance$. Let $\candidatecensor,\benchmarkcensor\in \censorshipspacepositive$ be two censorships with pooled masses $\candidatepooledmass$ and $\benchmarkpooledmass$, respectively. If $\candidatepooledmass \le \benchmarkpooledmass$, then for every rationality level $\noiseScale\in\fullsetrationalitylevel$, we have
\begin{align*}
    \frac{\Payoff[\noiseScale]{\candidatecensor}}{\Payoff[\noiseScale]{\benchmarkcensor}}
    \;\ge\;
    \candidatepooledmass\,.
\end{align*}
\end{restatable}

We now provide the intuition behind \Cref{lem:more_pooled_comeptitor} (see \Cref{fig:more_pooled_competitor} for an illustration).
In contrast to the proof of \Cref{thm:rationality_robust}\ref{item:baseline_ratio}, where the candidate pools weakly more mass than the benchmark, here the middle states $\middlestates$ are revealed by the candidate $\candidatecensor$ but pooled by the benchmark $\benchmarkcensor$.
As before, we treat the commonly revealed bottom states $\bottomstates$ as a single ``hypothetical revealed state'' that contributes the same expected utility under both schemes.
The benchmark pools the top and middle states $\topstates \cup \middlestates$ into a single signal $\benchmarkpoolsignal$, which equals their conditional mean value.
Under the candidate, we lower bound the contribution of these states by that of a single state whose value equals this same conditional mean but whose mass equals only that of the top states $\topstates$.
This lower bound is valid for two reasons: the candidate's pooling signal is weakly lower than the benchmark's, because the benchmark additionally pools the middle states, whose values are weakly higher; and the middle states that the candidate reveals contribute nonnegatively to its expected utility.
Both schemes thus reduce to two states, and the merged state's contribution under the candidate is at least a $\candidatepooledmass/\benchmarkpooledmass$ fraction of that under the benchmark.
Since the hypothetical revealed state contributes equally under both schemes, the payoff ratio is at least $\candidatepooledmass/\benchmarkpooledmass \geq \candidatepooledmass$.


\begin{proof}[Proof of \Cref{lem:more_pooled_comeptitor}]
Fix any rationality level $\noiseScale\in\fullsetrationalitylevel$.
Since $\candidatepooledmass \le \benchmarkpooledmass$, the benchmark censorship $\benchmarkcensor$ pools weakly more mass than the candidate censorship $\candidatecensor$. 
Therefore, we have $\candidatecensorstate \le \benchmarkcensorstate$.


We first show that the respective pooling signals satisfy $\candidatepoolsignal \le \benchmarkpoolsignal$.
By the definition of the pooling signal, we have
$
    \benchmarkpooledmass \benchmarkpoolsignal
    =
    \sum\nolimits_{i\in \topstates} \prior_i \receiverU_i
    +
    \sum\nolimits_{i\in \middlestates} \prior_i \receiverU_i =
    \candidatepooledmass \candidatepoolsignal
    +
    \sum\nolimits_{i\in \middlestates} \prior_i \receiverU_i$.
    For every state $i \in \middlestates$, it holds that state $i$ must lie weakly above the threshold state $\candidatecensorstate$ of candidate censorship. Thus, we have $\receiverU_i \ge \receiverU_{\candidatecensorstate} \ge \candidatepoolsignal$.
Substituting this into the previous equations yields
$
    \benchmarkpooledmass \benchmarkpoolsignal
    \ge
    \candidatepooledmass \candidatepoolsignal
    +
    \left(\benchmarkpooledmass-\candidatepooledmass\right) \candidatepoolsignal
    =
    \benchmarkpooledmass \candidatepoolsignal$,
which proves $\candidatepoolsignal \le \benchmarkpoolsignal$.

Next, using the three-way decomposition, we can lower bound the sender's expected utility $\Payoff[\noiseScale]{\candidatecensor}$ under the candidate censorship $\candidatecensor$ as 
\begin{align*}
    \Payoff[\noiseScale]{\candidatecensor}
     &=
    \candidatepooledmass\,\Quant\betaed(\candidatepoolsignal)
    +
    \sum\limits_{i\in \middlestates} \prior_i\Quant\betaed(\receiverU_i)
    +
    \sum\limits_{i\in \bottomstates} \prior_i\Quant\betaed(\receiverU_i)
    \ge
    \candidatepooledmass\,\Quant\betaed(\candidatepoolsignal)
    +
    \sum\limits_{i\in \bottomstates} \prior_i\Quant\betaed(\receiverU_i) \ge
    \candidatepooledmass\,\Quant\betaed(\benchmarkpoolsignal)
    +
    \sum\limits_{i\in \bottomstates} \prior_i\Quant\betaed(\receiverU_i)\,,
\end{align*}
where the last inequality follows from $\candidatepoolsignal \le \benchmarkpoolsignal$ and the fact that the quantal response function $\Quant\betaed(\cdot)$ is monotone non-increasing.
Note that the sender's expected utility $\Payoff[\noiseScale]{\benchmarkcensor}$ under the benchmark censorship $\benchmarkcensor$ expands exactly to $ \benchmarkpooledmass\,\Quant\betaed(\benchmarkpoolsignal) + \sum\nolimits_{i\in \bottomstates} \prior_i\Quant\betaed(\receiverU_i)$. The decompositions of $\Payoff[\noiseScale]{\candidatecensor}$ and $\Payoff[\noiseScale]{\benchmarkcensor}$ are illustrated in \Cref{fig:more_pooled_competitor}.
Finally, we can bound the utility ratio as 
\begin{align*}
\frac{\Payoff[\noiseScale]{\candidatecensor}}{\Payoff[\noiseScale]{\benchmarkcensor}}
    \ge
    \frac{
    \candidatepooledmass\,\Quant\betaed(\benchmarkpoolsignal)
    +
    \sum\nolimits_{i\in \bottomstates} \prior_i\Quant\betaed(\receiverU_i)}{
    \benchmarkpooledmass\,\Quant\betaed(\benchmarkpoolsignal)
    +
    \sum\nolimits_{i\in \bottomstates} \prior_i\Quant\betaed(\receiverU_i)}
    \ge
    \frac{\candidatepooledmass}{\benchmarkpooledmass}
    \ge
    \candidatepooledmass\,,
    \end{align*}
where the second inequality holds because $\candidatepooledmass \le \benchmarkpooledmass$ and $\sum\nolimits_{i\in \bottomstates} \prior_i\Quant\betaed(\receiverU_i)\ge 0$, allowing us to use the algebraic fact that $\frac{ \variablecapA\, \Quant\betaed(\benchmarkpoolsignal) + \variableX}{\variablecapB\, \Quant\betaed(\benchmarkpoolsignal) + \variableX} \;\ge\; \frac{\variablecapA}{\variablecapB}$
    for all $\variableX\in [0,\infty) \text{ and } 0\le \variablecapA \le \variablecapB$.
This completes the proof of \Cref{lem:more_pooled_comeptitor}.
\end{proof}

We now turn to the complementary case $\candidatepooledmass > \benchmarkpooledmass$, in which the benchmark $\benchmarkcensor$ reveals more states than the candidate $\candidatecensor$.
Unlike in \Cref{lem:more_pooled_comeptitor}, the comparison now involves a trade-off.
Relative to the candidate, the benchmark additionally reveals the middle states $\middlestates$, which weakly lowers the expected utility from these states because their values are positive.
At the same time, by pooling only the top states $\topstates$, the benchmark sends a weakly lower pooling signal, which weakly raises the expected utility from the top states.
We therefore need a quantitative bound on this trade-off.
The same configuration arose in \Cref{subsec:baseline_baseline_worst_case_ratio}, where we simply upper bounded the sender's conditional expected utility from the top and middle states under the benchmark by $1$.
That relaxation, however, yields only the guarantee of $\sfrac{1}{2}$ and is too loose for a sharper bound.
Instead, the following lemma bounds the payoff ratio in terms of three quantities: the candidate's pooled mass $\candidatepooledmass$, the conditional mean value of the bottom states $\bottomstates$ scaled by the rationality level $\noiseScale$, and the ratio $\negativeratio$ between the prior masses of states with positive and non-positive values.

\begin{figure}[t]
\centering
\resizebox{\textwidth}{!}{%
    \begin{tikzpicture}[font=\sffamily]
        \tikzset{
          arr/.style={-{Stealth[length=2mm,width=1.4mm]},line width=.65pt,shorten >=.5pt},
          poolpoint/.style={rectangle,draw=black,fill=black,minimum size=3mm,inner sep=0pt},
          benchpoint/.style={regular polygon,regular polygon sides=3,draw=black,fill=black,thick,minimum size=4.2mm,inner sep=0pt},
          bottompoint/.style={circle,draw=black,fill=black,thick,minimum size=3.4mm,inner sep=0pt},
          zeropoint/.style={star,star points=5,star point ratio=2.2,draw=black,fill=black,minimum size=4.3mm,inner sep=0pt}
        }

        \draw[decorate, decoration={brace, amplitude=5pt},  black, thick] 
            (0, 3.2) -- (6.0, 3.2) node[midway, above=6pt, font=\small] {top states ($\topstates$)};
            
        \draw[decorate, decoration={brace, amplitude=5pt}, black, thick] 
            (6.0, 3.2) -- (9.5, 3.2) node[midway, above=6pt, font=\small] {middle states ($\middlestates$)};
            
        \draw[decorate, decoration={brace, amplitude=5pt}, black, thick] 
            (9.5, 3.2) -- (13.0, 3.2) node[midway, above=6pt, font=\small] {bottom states ($\bottomstates$)};

        \node[font=\normalsize, anchor=east] at (-0.3, 2.0) {$\candidatecensor$};

        \fill[white] (0,1.5) rectangle (9.5,2.5);
        \fill[pattern=dots, pattern color=black] (0,1.5) rectangle (9.5,2.5);
        
        \fill[white] (9.5,1.5) rectangle (13.0,2.5);

        \draw[black, thick] (0,1.5) rectangle (13.0,2.5);
        \foreach \x in {1.5, 3.0, 4.5, 6.0, 7.5, 9.5, 11.5} {
            \draw[black] (\x, 1.5) -- (\x, 2.5);
        }

        \node at (0.75, 2.8) {$v_1$};
        \node at (2.25, 2.8) {$v_2$};
        \node at (3.75, 2.8) {$v_3$};
        \node at (5.25, 2.8) {$v_4$};
        \node at (6.75, 2.8) {$v_5$};
        \node at (8.5,  2.8) {$v_6$};
        \node at (10.5, 2.8) {$v_7$};
        \node at (12.25,2.8) {$v_8$};

        \draw[decorate,decoration={brace,amplitude=5pt,mirror},thick]
            (0,1.35)--(9.5,1.35);
        \draw[decorate,decoration={brace,amplitude=5pt,mirror},thick]
            (9.5,1.35)--(13,1.35);

        \node[anchor=east,align=right,font=\small] at (.65,0) {reduced\\states};
        \draw[->,>=stealth,black!55] (.95,0)--(12.05,0);
        \node[benchpoint] (benchpool) at (3,0) {};
        \node[poolpoint] (candpool) at (4.75,0) {};
        \node[zeropoint] (benchzero) at (6.5,0) {};
        \node[bottompoint] (commonbottom) at (11.25,0) {};

        \draw[arr] ([yshift=-5pt]4.75,1.35)--(candpool.north);
        \draw[arr] ([yshift=-5pt]11.25,1.35)--(commonbottom.north);

        \node[font=\normalsize, anchor=east] at (-0.3, -2.0) {$\benchmarkcensor$};

        \fill[white] (0,-2.5) rectangle (6.0,-1.5);
        \fill[pattern=vertical lines, pattern color=black] (0,-2.5) rectangle (6.0,-1.5);
        
        \fill[white] (6.0,-2.5) rectangle (9.5,-1.5);
        \fill[pattern=north east lines, pattern color=black] (6.0,-2.5) rectangle (9.5,-1.5);
        
        \fill[white] (9.5,-2.5) rectangle (13.0,-1.5);

        \draw[black, thick] (0,-2.5) rectangle (13.0,-1.5);
        \foreach \x in {1.5, 3.0, 4.5, 6.0, 7.5, 9.5, 11.5} {
            \draw[black] (\x, -2.5) -- (\x, -1.5);
        }

        \node at (0.75, -2.8) {$v_1$};
        \node at (2.25, -2.8) {$v_2$};
        \node at (3.75, -2.8) {$v_3$};
        \node at (5.25, -2.8) {$v_4$};
        \node at (6.75, -2.8) {$v_5$};
        \node at (8.5,  -2.8) {$v_6$};
        \node at (10.5, -2.8) {$v_7$};
        \node at (12.25,-2.8) {$v_8$};

        \draw[decorate,decoration={brace,amplitude=5pt},thick]
            (0,-1.35)--(6,-1.35);
        \draw[decorate,decoration={brace,amplitude=5pt},thick]
            (6,-1.35)--(9.5,-1.35);
        \draw[decorate,decoration={brace,amplitude=5pt},thick]
            (9.5,-1.35)--(13,-1.35);

        \coordinate (benchtopstart) at ([yshift=5pt]3,-1.35);
        \coordinate (benchmiddlestart) at ([yshift=5pt]7.75,-1.35);
        \coordinate (benchbottomstart) at ([yshift=5pt]11.25,-1.35);
        \draw[arr] (benchtopstart)--(benchpool.south);
        \draw[arr] (benchbottomstart)--(commonbottom.south);
        \draw[arr] (benchmiddlestart)--(benchzero.south east);
        \draw[arr] (benchmiddlestart)--(commonbottom.south west);

        \filldraw[fill=white,pattern=dots,draw=black]
            (13.5,2.05) rectangle (14.3,2.35);
        \node[right,font=\small] at (14.4,2.2) {pooled under $\candidatecensor$};
        \filldraw[fill=white,pattern=vertical lines,draw=black]
            (13.5,1.55) rectangle (14.3,1.85);
        \node[right,font=\small] at (14.4,1.7) {pooled under $\benchmarkcensor$};
        \filldraw[fill=white,draw=black]
            (13.5,1.05) rectangle (14.3,1.35);
        \node[right,font=\small] at (14.4,1.2) {commonly revealed states};
        \filldraw[fill=white,pattern=north east lines,draw=black]
            (13.5,.55) rectangle (14.3,.85);
        \node[right,font=\small] at (14.4,.7) {revealed only under $\benchmarkcensor$};

        \node[poolpoint] at (13.9,-.75) {};
        \node[right,font=\small] at (14.4,-.75) {pooled value: $\candidatecensor$};
        \node[benchpoint] at (13.9,-1.25) {};
        \node[right,font=\small] at (14.4,-1.25) {pooled value: $\benchmarkcensor$};
        \node[zeropoint] at (13.9,-1.75) {};
        \node[right,font=\small] at (14.4,-1.75) {zero};
        \node[bottompoint] at (13.9,-2.25) {};
        \node[right,font=\small] at (14.4,-2.25) {conditional mean of $\bottomstates$};

    \end{tikzpicture}%
}%
\caption{
Graphical illustration of state reduction for the candidate censorship  $\candidatecensor$ and the benchmark censorship $\benchmarkcensor$ analyzed in \Cref{lem:less-pooled-than-t}.
As shown, the candidate censorship $\candidatecensor$ pools top and middle states $\topstates \cup \middlestates$, whereas the benchmark censorship $\benchmarkcensor$ pools only top states $\topstates$; both censorships commonly reveal bottom states $\bottomstates$.
The middle axis plots the values of the corresponding reduced states.
The square and triangle markers denote the pooled values used in the comparison after separating the prior-mean terms for candidate censorship $\candidatecensor$ and benchmark censorship $\benchmarkcensor$, respectively.
The circle marker denotes the conditional mean of bottom states $\bottomstates$, and the star marker denotes zero.
By leveraging the convexity of the quantal response function $\Quant\betaed(\cdot)$, middle states $\middlestates$ under the benchmark censorship $\benchmarkcensor$ are reduced to two states whose values correspond to the star and the circle.}
\label{fig:less-pooled-t}
\end{figure}

\begin{restatable}{lemma}
{lesspooled}
\label{lem:less-pooled-than-t}
Fix any instance $\instance$. Let $\candidatecensor,\benchmarkcensor\in \censorshipspacepositive$ be two censorships with pooled masses $\candidatepooledmass$ and $\benchmarkpooledmass$, respectively. If $\candidatepooledmass > \benchmarkpooledmass$, then for every rationality level $\noiseScale\in[0,\infty)$, we have \footnote{The case of $\candidatepooledmass=1$ can be addressed directly. In that case, we have $\Payoff[\noiseScale]{\candidatecensor}=\Quant\betaed(\candidatepoolsignal) \ge \frac{1}{2}$ and $\Payoff[\noiseScale]{\benchmarkcensor} \le 1$, resulting in ${\Payoff[\noiseScale]{\candidatecensor}}/
    {\Payoff[\noiseScale]{\benchmarkcensor}} \ge \frac{1}{2}$.}
\begin{align*}
    \frac{\Payoff[\noiseScale]{\candidatecensor}}
    {\Payoff[\noiseScale]{\benchmarkcensor}}
    \;\ge\;
\frac{\censorshipcontribution[\frac{1-\candidatepooledmass}{\candidatepooledmass}]{\noiseScale \cdot \frac{\sum\nolimits_{i\in \bottomstates} \prior_i  \receiverU_i}{1-\candidatepooledmass}}}
    {\optimalcontribution[\negativeratio]{\noiseScale \cdot \frac{\sum\nolimits_{i\in \bottomstates} \prior_i  \receiverU_i}{1-\candidatepooledmass}}}\,,
\end{align*}
where the auxiliary functions are defined by
\begin{align}
\label{eq:n_q_function}
    \censorshipcontribution[\censorshipparameter]{\tailmeanscaled}
    \;\triangleq\;
    \frac{\Quant\betaone(-\censorshipparameter \tailmeanscaled)+\censorshipparameter \Quant\betaone(\tailmeanscaled)}{1+\censorshipparameter}\,, \qquad
    \optimalcontribution[\negativeratio]{\tailmeanscaled}
    \;\triangleq\;
\sup\nolimits_{\censorshipparameter\in[0,\negativeratio]} \censorshipcontribution[\censorshipparameter]{\tailmeanscaled}\,,
\end{align}
and $ \negativeratio \triangleq \frac{1-\sum\nolimits_{i \in [\stateNum]:\, \receiverU_i\le 0}\prior_i}{\sum\nolimits_{i\in [\stateNum]:\, \receiverU_i\le 0}\prior_i} \in (0,\infty)$ is the prior mass ratio between the positive and non-positive states.
\end{restatable}

We now provide the intuition behind \Cref{lem:less-pooled-than-t} (see \Cref{fig:less-pooled-t} for an illustration).
The key idea is to exploit the S-shaped curvature of the quantal response function $\Quant\betaed(\cdot)$ to bound the sender's expected utility under both schemes by binary-state payoffs.
First, under both schemes, we replace the commonly revealed bottom states $\bottomstates$ by a single state with their total mass and conditional mean value.
By the convexity of $\Quant\betaed(\cdot)$ on $[0,\infty)$, this replacement weakly lowers the contribution of the bottom states by the same amount under both schemes; because this common difference is nonnegative, it suffices to bound the ratio of the remaining terms.
Under the candidate $\candidatecensor$, the top and middle states $\topstates \cup \middlestates$ are pooled into a single signal, so the candidate's payoff reduces to a binary-state payoff over this pooled state and the reduced bottom state.
Under the benchmark $\benchmarkcensor$, only the top states $\topstates$ are pooled, and the middle states $\middlestates$ are revealed.
Rather than simply averaging the top and middle states, we again use the convexity of $\Quant\betaed(\cdot)$ on $[0,\infty)$: since the values of the middle states lie between $0$ and the conditional mean value of the bottom states, we can upper bound their contribution by that of two states with exactly these two values.
Together with the pooled top states and the reduced bottom state, the benchmark's payoff is thus bounded by that of three states, with values equal to the benchmark's pooling signal, $0$, and the conditional mean value of the bottom states.
This three-state payoff is a convex combination of two binary-state payoffs, each of which is at most the largest binary-state payoff attainable over all admissible pooled masses.
Comparing the candidate's binary-state payoff with this largest binary-state payoff yields the bound in \Cref{lem:less-pooled-than-t}.

\begin{proof}[Proof of \Cref{lem:less-pooled-than-t}]
Fix any rationality level $\noiseScale \in [0,\infty)$.
For bottom states $\bottomstates$, we define their conditional mean by
$
    \candidatetailmean
    \triangleq
\frac{\sum\nolimits_{i\in \bottomstates} \prior_i  \receiverU_i}{1-\candidatepooledmass}
$.
For notational simplicity, we write $\tailmeanscaled \triangleq \noiseScale\candidatetailmean\in [0,\infty)$ to denote the scaled conditional mean of bottom states $\bottomstates$, and $\candidatepoolpara \triangleq  \frac{1-\candidatepooledmass}{\candidatepooledmass}$ ($\benchmarkpoolpara \triangleq \frac{1-\benchmarkpooledmass}{\benchmarkpooledmass}$) to denote the ratio of disclosed mass to pooled mass under the candidate censorship $\candidatecensor$ (benchmark censorship $\benchmarkcensor$), respectively. 
By assumption, we have $0 \le \candidatepoolpara \le \benchmarkpoolpara \le \negativeratio$. Let $\priormean \in (-\infty, 0]$ denote the prior mean of instance $\instance$, i.e., $\sum\nolimits_{i \in [\stateNum]}\prior_i\receiverU_i = d$.

For the candidate censorship $\candidatecensor$, the sum of expected values of bottom states $\bottomstates$ and the remaining states $\topstates \cup \middlestates$ equals the prior mean, which is given by
$
    \candidatepooledmass \candidatepoolsignal + (1-\candidatepooledmass)\candidatetailmean = \priormean$.
For the benchmark censorship $\benchmarkcensor$, we have
$
    \benchmarkpooledmass \benchmarkpoolsignal+ (\candidatepooledmass-\benchmarkpooledmass)\rivalrevealedmeanratio \candidatetailmean + (1-\candidatepooledmass)\candidatetailmean = d$,
where $\rivalrevealedmeanratio \triangleq \frac{\sum\nolimits_{i \in \middlestates}\prior_i \receiverU_i}{(\candidatepooledmass - \benchmarkpooledmass)\candidatetailmean} \in (0,1]$ is the ratio of conditional mean of middle states $\middlestates$ to that of bottom states $\bottomstates$.
Rearranging these equations, the corresponding pooling signals can be written as 
$
    \candidatepoolsignal = -\candidatepoolpara \candidatetailmean + \frac{\priormean}{\candidatepooledmass}$ and $\benchmarkpoolsignal = -\benchmarkcoefficient\candidatetailmean + \frac{\priormean}{\benchmarkpooledmass}
$,
where $\benchmarkcoefficient \triangleq \frac{(\candidatepooledmass-\benchmarkpooledmass)\rivalrevealedmeanratio+1-\candidatepooledmass}{\benchmarkpooledmass} \in [\candidatepoolpara,\benchmarkpoolpara]$.

Next, using the three-way decomposition framework, we can rewrite the sender's expected utility under the candidate censorship $\candidatecensor$ as
\begin{align*}
    \Payoff[\noiseScale]{\candidatecensor} &= \candidatepooledmass \Quant\betaed(\candidatepoolsignal) + \sum\nolimits_{i \in \bottomstates}\prior_i\Quant\betaed(\receiverU_i) \\
    &= \candidatepooledmass \Quant\betaed(-\candidatepoolpara\candidatetailmean)+(1-\candidatepooledmass)\Quant\betaed(\candidatetailmean) + \left(\candidatepooledmass \Quant\betaed(\candidatepoolsignal)- \candidatepooledmass \Quant\betaed(-\candidatepoolpara\candidatetailmean)\right) 
    +\left(\sum\nolimits_{i \in \bottomstates}\prior_i \Quant\betaed(\receiverU_i) - (1-\candidatepooledmass)\Quant\betaed(\candidatetailmean)\right)\\
    & =  \censorshipcontribution[\candidatepoolpara]{\tailmeanscaled} + \left(\candidatepooledmass \Quant\betaed(\candidatepoolsignal)- \candidatepooledmass \Quant\betaed(-\candidatepoolpara\candidatetailmean)\right)+\left(\sum\nolimits_{i \in \bottomstates}\prior_i\Quant\betaed(\receiverU_i)- (1-\candidatepooledmass)\Quant\betaed(\candidatetailmean)\right)\,,
\end{align*}
where the last two terms in the above equation represent the utility gain from the negative shift of prior mean and the reduction from bottom states to a single state, respectively.
The analogous two terms also appear in the utility decomposition under the benchmark censorship $\benchmarkcensor$.  
For the benchmark censorship $\benchmarkcensor$, the value of state $i \in \middlestates$ satisfies $\receiverU_i \in (0,\candidatetailmean]$.
We first upper bound the utility contribution from middle states $\middlestates$ by the convexity of the quantal response function $\Quant\betaed(\cdot)$ on $[0,\infty)$ as follows:
\begin{align*}
    \sum\nolimits_{i \in \middlestates}\prior_i\Quant\betaed(\receiverU_i) \le \sum\nolimits_{i \in \middlestates}\prior_i \left( \left(1-\frac{\receiverU_i}{\candidatetailmean} \right) \Quant\betaed(0)+ \frac{\receiverU_i}{\candidatetailmean} \Quant\betaed(\candidatetailmean)\right) = (\candidatepooledmass-\benchmarkpooledmass) \left( (1-\rivalrevealedmeanratio)\Quant\betaed(0)+ \rivalrevealedmeanratio\Quant\betaed(\candidatetailmean) \right)\,.
    \end{align*}
Then, with the above inequality, we obtain the upper bound of the sender's expected utility under the benchmark censorship $\benchmarkcensor$, which is given by
\begin{align*}
    \Payoff[\noiseScale]{\benchmarkcensor} &~= \benchmarkpooledmass \Quant\betaed(\benchmarkpoolsignal) + \sum\nolimits_{i \in \middlestates}\prior_i\Quant\betaed(\receiverU_i)+ \sum\nolimits_{i \in \bottomstates}\prior_i\Quant\betaed(\receiverU_i) \\
    & \overset{(a)}{\le} \benchmarkpooledmass \Quant\betaed(\benchmarkpoolsignal) + (\candidatepooledmass-\benchmarkpooledmass) \left( (1-\rivalrevealedmeanratio)\Quant\betaed(0)+ \rivalrevealedmeanratio\Quant\betaed(\candidatetailmean) \right) + \sum\nolimits_{i \in \bottomstates}\prior_i\Quant\betaed(\receiverU_i)\\
    &~= \benchmarkpooledmass \Quant\betaed(-\benchmarkcoefficient\candidatetailmean) + (\candidatepooledmass-\benchmarkpooledmass) \left( (1-\rivalrevealedmeanratio)\Quant\betaed(0)+ \rivalrevealedmeanratio\Quant\betaed(\candidatetailmean) \right) + (1-\candidatepooledmass)\Quant\betaed(\candidatetailmean) \\
    &\qquad +\left( \benchmarkpooledmass \Quant\betaed(\benchmarkpoolsignal) - \benchmarkpooledmass \Quant\betaed(-\benchmarkcoefficient\candidatetailmean)\right) 
    + \left( \sum\nolimits_{i \in \bottomstates}\prior_i\Quant\betaed(\receiverU_i)- (1-\candidatepooledmass)\Quant\betaed(\candidatetailmean)\right)\\
    &\overset{(b)}{=} \benchmarkpooledmass \left( \Quant\betaed(-\benchmarkcoefficient\candidatetailmean) +  \benchmarkcoefficient\Quant\betaed(\candidatetailmean) \right) + \left(1-\benchmarkpooledmass(1+\benchmarkcoefficient)\right) \Quant\betaed(0) +\left( \benchmarkpooledmass \Quant\betaed(\benchmarkpoolsignal) - \benchmarkpooledmass \Quant\betaed(-\benchmarkcoefficient\candidatetailmean)\right) \\
    & \qquad  
    + \left( \sum\nolimits_{i \in \bottomstates}\prior_i\Quant\betaed(\receiverU_i)- (1-\candidatepooledmass)\Quant\betaed(\candidatetailmean)\right)\\
    &~= \benchmarkpooledmass (1+\benchmarkcoefficient)\censorshipcontribution[\benchmarkcoefficient]{\tailmeanscaled} + \left(1-\benchmarkpooledmass(1+\benchmarkcoefficient)\right) \censorshipcontribution[0]{\tailmeanscaled} + \left( \benchmarkpooledmass \Quant\betaed(\benchmarkpoolsignal) - \benchmarkpooledmass \Quant\betaed(-\benchmarkcoefficient\candidatetailmean)\right) \\
    & \qquad + \left( \sum\nolimits_{i \in \bottomstates}\prior_i\Quant\betaed(\receiverU_i)- (1-\candidatepooledmass)\Quant\betaed(\candidatetailmean)\right)\\
    & \overset{(c)}{\le} \optimalcontribution[\negativeratio]{\tailmeanscaled} + \left( \benchmarkpooledmass \Quant\betaed(\benchmarkpoolsignal) - \benchmarkpooledmass \Quant\betaed(-\benchmarkcoefficient\candidatetailmean)\right) + \left( \sum\nolimits_{i \in \bottomstates}\prior_i\Quant\betaed(\receiverU_i)- (1-\candidatepooledmass)\Quant\betaed(\candidatetailmean)\right)\,,
\end{align*}
where inequality~(a) holds due to the upper bound for the utility contributed from the middle states; equality~(b) holds due to the identities $\benchmarkpooledmass\benchmarkcoefficient = (\candidatepooledmass -\benchmarkpooledmass)\rivalrevealedmeanratio + 1-\candidatepooledmass$ and $1-\benchmarkpooledmass(1+\benchmarkcoefficient) = (\candidatepooledmass -\benchmarkpooledmass)(1-\rivalrevealedmeanratio)$; inequality~(c) holds due to the definition of function $\optimalcontribution[\negativeratio]{\cdot}$ in Eqn.~\eqref{eq:n_q_function} and $\benchmarkpooledmass (1+\benchmarkcoefficient) \in (0,1]$.

Finally, we can bound the utility ratio as 
\begin{align*}
   \frac{\Payoff[\noiseScale]{\candidatecensor}}
    {\Payoff[\noiseScale]{\benchmarkcensor}} &~\ge  \frac{\censorshipcontribution[\candidatepoolpara]{\tailmeanscaled} + \left(\candidatepooledmass \Quant\betaed(\candidatepoolsignal)- \candidatepooledmass \Quant\betaed(-\candidatepoolpara\candidatetailmean)\right)+\left(\sum\nolimits_{i \in \bottomstates}\prior_i\Quant\betaed(\receiverU_i)- (1-\candidatepooledmass)\Quant\betaed(\candidatetailmean)\right)}{\optimalcontribution[\negativeratio]{\tailmeanscaled} + \left( \benchmarkpooledmass \Quant\betaed(\benchmarkpoolsignal) - \benchmarkpooledmass \Quant\betaed(-\benchmarkcoefficient\candidatetailmean)\right) + \left( \sum\nolimits_{i \in \bottomstates}\prior_i\Quant\betaed(\receiverU_i)- (1-\candidatepooledmass)\Quant\betaed(\candidatetailmean)\right)} \\
    & \overset{(a)}{\ge} \frac{\censorshipcontribution[\candidatepoolpara]{\tailmeanscaled} + \left(\benchmarkpooledmass \Quant\betaed(\benchmarkpoolsignal) - \benchmarkpooledmass \Quant\betaed(-\benchmarkcoefficient\candidatetailmean)\right)+\left(\sum\nolimits_{i \in \bottomstates}\prior_i\Quant\betaed(\receiverU_i)- (1-\candidatepooledmass)\Quant\betaed(\candidatetailmean)\right)}{\optimalcontribution[\negativeratio]{\tailmeanscaled} + \left( \benchmarkpooledmass \Quant\betaed(\benchmarkpoolsignal) - \benchmarkpooledmass \Quant\betaed(-\benchmarkcoefficient\candidatetailmean)\right) + \left( \sum\nolimits_{i \in \bottomstates}\prior_i \Quant\betaed(\receiverU_i) - (1-\candidatepooledmass)\Quant\betaed(\candidatetailmean)\right)} \overset{(b)}{\ge} \frac{\censorshipcontribution[\candidatepoolpara]{\tailmeanscaled}}{\optimalcontribution[\negativeratio]{\tailmeanscaled}}\,,
\end{align*}
where inequality~(a) holds since $\candidatepooledmass \Quant\betaed(\candidatepoolsignal)- \candidatepooledmass \Quant\betaed(-\candidatepoolpara\candidatetailmean) = \int\nolimits_{\priormean}^{0} -\Quant\betaed^{\prime}(-\candidatepoolpara\candidatetailmean+\frac{\variableZ}{\candidatepooledmass})\dd\variableZ \ge \int\nolimits_{\priormean}^{0} -\Quant\betaed^{\prime}(-\benchmarkcoefficient\candidatetailmean+\frac{\variableZ}{\benchmarkpooledmass})\dd\variableZ=\benchmarkpooledmass \Quant\betaed(\benchmarkpoolsignal) - \benchmarkpooledmass \Quant\betaed(-\benchmarkcoefficient\candidatetailmean)$.
This integral inequality follows from the fact that for every $\variableZ \in [\priormean,0]$, the conditions $\benchmarkcoefficient \in [\candidatepoolpara,\benchmarkpoolpara]$ and $\candidatepooledmass \ge \benchmarkpooledmass$ imply that $-\candidatepoolpara\candidatetailmean+\frac{\variableZ}{\candidatepooledmass} \ge -\benchmarkcoefficient\candidatetailmean+\frac{\variableZ}{\benchmarkpooledmass}$, where both sides are non-positive. 
Because the quantal response function $\Quant\betaed(\cdot)$ is concave on $(-\infty,0]$,  $-\Quant\betaed^{\prime}(\cdot)$ is non-decreasing on $(-\infty,0]$, preserving the direction of the bound.
Inequality~(b) follows from the algebraic fact that $\frac{ \variablecapA + \variableX}{\variablecapB + \variableX} \ge \frac{\variablecapA}{\variablecapB}$
    for all $\variableX\in [0,\infty) \text{ and } 0\le \variablecapA \le \variablecapB$.
    These conditions are satisfied because the convexity of quantal response function $\Quant\betaed(\cdot)$ on $[0,\infty)$ implies that $\sum\nolimits_{i \in \bottomstates}\prior_i \Quant\betaed(\receiverU_i) \ge (1-\candidatepooledmass)\Quant\betaed(\candidatetailmean)$, and the monotonicity of quantal response function $\Quant\betaed(\cdot)$ ensures $\Quant\betaed(\benchmarkpoolsignal)-\Quant\betaed(-\benchmarkcoefficient\candidatetailmean) \ge 0$ due to $\benchmarkpoolsignal \le -\benchmarkcoefficient\candidatetailmean$.
    This completes the proof of \Cref{lem:less-pooled-than-t}.
\end{proof}

We now combine \Cref{lem:more_pooled_comeptitor,lem:less-pooled-than-t} to prove \Cref{thm:rationality_robust}\ref{item:optimal_ratio}.
\begin{proof}[Proof of  \Cref{thm:rationality_robust}\ref{item:optimal_ratio}]
For an instance with prior mass ratio $\negativeratio$ (see its definition in \Cref{lem:less-pooled-than-t}), we express the candidate scheme's pooled mass $\pooledmass$ through the ratio $\censorshipparameter = \left(1-\pooledmass\right)/\pooledmass \in \left[0,\negativeratio\right]$.
The two lemmas yield
\begin{align*}
    \sup\nolimits_{\signalscheme \in \censorshipspace} \instancerobustRatio{\signalscheme}
    &\geq \sup\nolimits_{\censorshipparameter \in \left[0,\negativeratio\right]} \min\left\{\frac{1}{1+\censorshipparameter},\inf\nolimits_{\tailmeanscaled \in \left[0,\infty\right)} \frac{\censorshipcontribution[\censorshipparameter]{\tailmeanscaled}}{\optimalcontribution[\negativeratio]{\tailmeanscaled}}\right\}~.
\end{align*}
The two terms correspond to benchmark schemes that pool weakly more mass and strictly less mass than the candidate, respectively.\footnote{The case of full rationality ($\noiseScale = \infty$) is covered by \Cref{lem:more_pooled_comeptitor}: in this case, the benchmark censorship is fully pooling, which implies that its pooled mass is no less than that of the candidate censorship, satisfying the pooled mass condition of \Cref{lem:more_pooled_comeptitor}.}
By algebra, the right-hand side of the preceding inequality, viewed as a function of the prior mass ratio is minimized at $\negativeratio = 1/2$ (see \Cref{lem:upbound-negativeratiofn}), which gives the universal guarantee
\begin{align}
    \label{eq:universal_guarnatee_maximin_optimal_ratio}
    \worstcaseratio &\triangleq \sup\nolimits_{\censorshipparameter \in \left[0,\frac{1}{2}\right]} \inf\nolimits_{\tailmeanscaled \in \left[0,\infty\right)} \frac{\censorshipcontribution[\censorshipparameter]{\tailmeanscaled}}{\optimalcontribution[\frac{1}{2}]{\tailmeanscaled}} \approx \ratiovalue~,
\end{align}
as desired. This completes the proof of \Cref{thm:rationality_robust}\ref{item:optimal_ratio}.
\end{proof}

The binary-state instance in Eqn.~\eqref{ex:binary_worst_case} attains the universal guarantee: its maximin optimal ratio is exactly $\worstcaseratio$ (see \Cref{fig:hard_instance}; the formal proof is in \Cref{subsec:binary_core_instance}).
To see why, recall the two sources of slack in the proof of \Cref{lem:less-pooled-than-t}, both of which vanish in this instance.
First, all middle and bottom states share the same value $\receiverU_2$, so the convexity bound for the middle states holds with equality, and replacing the bottom states by a single state with their conditional mean value leaves their contribution unchanged.
Second, the prior mean is zero, so the remainder term associated with each pooling signal vanishes.
Hence, at every finite rationality level, the expressions in Eqn.~\eqref{eq:n_q_function} give exactly the expected utilities of the candidate scheme and of the optimal benchmark.
Finally, this instance has prior mass ratio $\negativeratio = \sfrac{1}{2}$, the value at which the lower bound in Eqn.~\eqref{eq:universal_guarnatee_maximin_optimal_ratio} is minimized, so its maximin optimal ratio equals $\worstcaseratio$.

\section{Conclusions and Future Directions}
\label{sec:conclu}

In this work, we study how a platform should disclose information to boundedly rational users who follow the logit quantal response model and whose rationality level the platform may not know. 
We show that the censorship structure, which is optimal for fully rational users, survives both departures from the classical setting. 
When the rationality level is known, censorship remains optimal, and less rational users optimally receive at least as much information as more rational ones. 
When it is unknown, censorship remains robustly optimal under three robust criteria we consider. 
Moreover, rationality-robust censorship achieves at least 89.96\% of the payoff attainable with knowledge of the user's rationality level, whereas the policy designed for fully rational users has a universal guarantee of 50\%; both bounds are tight.

These results mirror the message in our title: less rational, more transparent. Bounded rationality and uncertainty about it change only how much the platform should pool, and only toward greater transparency. 
A platform using a common badge therefore need not redesign its disclosure format; it only needs to narrow the badge's coverage, which also weakly increases users' expected utility. 
Moreover, personalizing disclosure to each user's rationality level is worth at most 10.04\% of the attainable payoff, whereas ignoring bounded rationality can cost up to half of it. 
Our results also open several directions for future research.

Although the logit quantal response model is well established and widely adopted, as discussed in \Cref{sec:introduction}, a natural next step is to examine whether our insights extend to other behavioral models.
The S-shaped curvature of the quantal response function plays a central role in our analysis, so a first candidate is the class of stochastic choice models that share this curvature, such as the probit model, which arises from normally distributed utility shocks \citep{T-27}, and the exponomial choice model, which arises from exponentially distributed ones \citep{AS-16}.
A more ambitious goal is to design disclosure policies that are robust not only to a behavioral parameter but also to the behavioral model itself (see \citealp{LS-21} for a review of robust operations management under model uncertainty).

Our model considers a binary action and a platform whose payoff depends only on the user's action. 
When a user chooses among more than two actions, such as several recommended products, her multinomial logit response depends on a multidimensional summary of her belief. 
In that case, the one-dimensional structure underlying censorship and our pooled-mass representation is no longer directly available. 
Characterizing optimal and rationality-robust disclosure in this setting, possibly by drawing on assortment optimization under the multinomial logit model \citep{TV-04,RT-12}, is a promising avenue. 
Another extension lets the platform's payoff depend on the state, as when it values match quality or long-run user satisfaction beyond adoption. 

\bibliographystyle{plainnat}
\bibliography{myref} 


\renewcommand{\theHchapter}{A\arabic{chapter}}
\renewcommand{\theHsection}{A\arabic{section}}
\newpage
\ECSwitch
\ECDisclaimer

\newpage
\renewcommand{\theHsection}{appendix.\Alph{section}}


\section{Missing Example in \texorpdfstring{\Cref{sec:introduction}}{Section 1}}
\label{apx:full_bounded_rationality}
In this section, we present the example omitted from \Cref{sec:introduction} comparing the receiver's expected utility under full and bounded rationality.
For a fully rational receiver who uses the best response strategy, revealing more information means that her expected utility cannot decrease, which can be shown by Blackwell's theorem \citep{bla-53}.
However, more information need not benefit a quantal-response receiver. 
Here we construct a three-state instance shown below, where with more information, the quantal response receiver's expected utility decreases.
Under a quantal-response model with rationality parameter $\noiseScale > 0$ (formalized in \Cref{sec:preliminary}), 
equivalently, 
given a posterior expected utility difference $\variableX$ between actions $0$ and $1$, the receiver chooses action $1$ with probability $\Quant\betaed(\variableX) = \sfrac{1}{(1 + \exp(\noiseScale\variableX))}$ and action $0$ with probability $1 - \Quant\betaed(\variableX)$.
\begin{example}
\label{exa:full_bounded_rationality}
Consider a three-state instance  with prior probability $\prior_i$, receiver's utilities $\receiverU_i(0)$ and $\receiverU_i(1)$, and receiver's utility difference $\receiverU_i = \receiverU_i(0) - \receiverU_i(1)$ given in the following table:
\[
    \begin{array}{c|c|c|c|c}
 \text{state}~i & \text{prior mass}~\prior_i & \receiverU_i(0) & \receiverU_i(1)
   & \receiverU_i=\receiverU_i(0)-\receiverU_i(1)\\ \hline
 1 & 1/4 & 0 & 5 & -5\\
 2 & 1/4 & 0 & 3 & -3\\
 3& 1/2&4 & 0 & 4
\end{array}
\]
\end{example}
Fix the rationality level $\noiseScale=\log 2$.
Let us compare two signaling schemes: a censorship signaling scheme (formally defined in \Cref{def:censorship}), which pools state~1 and~2 into a single signal while revealing state~3, and the full revelation signaling scheme, which always reveals the true state to the receiver. 
The full revelation signaling scheme is more
informative than this censorship signaling scheme in the Blackwell order.

\xhdr{Censorship that pools state~1 and~2} The pooling signal yields a posterior utility difference of $-4$, under which the receiver takes action 1 with probability $\Quant_{\log2}(-4)=\sfrac{16}{17}$. For state~3, the receiver takes action 0 with probability $1-\Quant_{\log2}(4)= \sfrac{16}{17}$. Thus, the receiver's expected utility is given by
\begin{align*}
    \receiverexpU_{\text{cens}} = \frac14\times5\times\frac{16}{17}
+\frac14\times3\times\frac{16}{17}
+\frac12\times4\times\frac{16}{17}= \frac{64}{17}\,.
\end{align*}

\xhdr{Full revelation} In this case, for state~1, the receiver takes action 1 with probability $\Quant_{\log 2}(-5)=\sfrac{32}{33}$; for state~2, she takes action 1 with probability $\Quant_{\log 2}(-3)=\sfrac{8}{9}$; and for state~3, she takes action 0 with probability $1-\Quant_{\log2}(4)= \sfrac{16}{17}$.
Thus, the receiver's expected utility is given by
\begin{align*}
    \receiverexpU_{\text{reveal}}= \frac{1}{4}\times 5 \times \frac{32}{33}+ \frac{1}{4}\times 3 \times \frac{8}{9} +\frac{1}{2}\times4\times \frac{16}{17} = \frac{2110}{561}\,.
\end{align*}

Therefore, we have $\receiverexpU_{\text{reveal}}-\receiverexpU_{\text{cens}} = -\sfrac{2}{561}<0$, showing that the full revelation signaling scheme strictly lowers the receiver's expected utility under quantal response.
This loss in the receiver's expected utility occurs because the full revelation signaling scheme individually reveals the first two states, which have negative receiver utility differences. 
Because the function $\Quant\betaed(\cdot)$ is concave on $(-\infty,0]$, revealing these two states separately increases the probability of choosing action~$1$ only slightly in state~$1$ (from $\sfrac{16}{17}$ to $\sfrac{32}{33}$) while decreasing it much more sharply in state~$2$ (from $\sfrac{16}{17}$ to $\sfrac{8}{9}$).
In this example, the resulting utility loss in state~$2$ outweighs the gain in state~$1$.
Note that the utility comparison in this example is outside the scope of \Cref{prop:receiver_utility_information_revealing}, which compares censorships that pool every state with a non-positive receiver utility difference.

\section{Missing Proofs in \texorpdfstring{\Cref{sec:optimal_scheme_known_rationality_level}}{Section 3}}
\label{apx:optimal_scheme_known_rationality_level}

In this section, we first present primal and dual programs for the sender's problem, and then provide the proofs of \Cref{thm:opt_state_independent} and \Cref{prop:SISU_threshold_state_monotonicity}.

\subsection{Primal and Dual Programs for Sender's Problem}
\label{apx:primal_dual_program}
We formulate the sender's problem as the following linear program~\ref{eq:opt_lp} and its corresponding dual program~\ref{eq:opt_lp_dual}, where the primal variables $\{\signalscheme_i(\scaledNoiseDiff)\}_{\scaledNoiseDiff \in \R, i\in[\stateNum]}$ represent the signaling scheme.\footnote{While we allow signaling schemes to have a continuous signal space $\signalspace$---meaning $\signalscheme_i(\cdot)$ can be viewed as a probability density function---all signaling schemes considered in this paper (including the optimal ones) have finite signal support. We therefore abuse notation slightly and use $\signalscheme_i(\cdot)$ as a probability mass function when clear from context.}
\begin{align}
    \label{eq:opt_lp}
    \arraycolsep=5.4pt\def\arraystretch{1}
    \tag{$\mathcal{P}_\texttt{OPT-Primal}$}
    &\begin{array}{llll}
     \max\limits_{\boldsymbol\signalscheme\geq \zerobf} ~ &
     \displaystyle\sum\nolimits_{i\in[\stateNum]} \prior_i  
     \displaystyle\int_{-\infty}^{\infty} \signalscheme_i(\scaledNoiseDiff)
     \Quant\betaed(\scaledNoiseDiff)\;\dd\scaledNoiseDiff 
     \quad&  &
     \vspace{1mm}
     \\
     \;\text{s.t.}  & 
       \displaystyle\sum\nolimits_{i\in[\stateNum]}  \prior_i 
      \left(\receiverU_i - \scaledNoiseDiff \right)
       \signalscheme_i(\scaledNoiseDiff)
       = 0
       &\quad  \forall \scaledNoiseDiff\in(-\infty,\infty) 
       \quad&  
       \qquad \quad \langle\noiseDualVar(\scaledNoiseDiff)\rangle
     \vspace{1mm}
       \\
       &
       \displaystyle\int_{-\infty}^{\infty} \signalscheme_i(\scaledNoiseDiff)\;\dd\scaledNoiseDiff = 1
       &
       \quad \forall i\in[\stateNum] &
       \qquad\quad \langle\distDualVar(i)\rangle
     \vspace{1mm}
     \\
    \end{array}
    \vspace{2mm}
    \\
    \label{eq:opt_lp_dual}
    \tag{$\mathcal{P}_\texttt{OPT-Dual}$}
    &\begin{array}{llll}
     \min\limits_{\boldsymbol\noiseDualVar,
     \boldsymbol\distDualVar} ~ &
     \displaystyle\sum\nolimits_{i\in[\stateNum]} \distDualVar(i)
     & &
     \vspace{1mm}
     \\
      \;\text{s.t.} & 
      \prior_i 
      \left(\receiverU_i - \scaledNoiseDiff \right)
      \noiseDualVar(\scaledNoiseDiff) 
      +
      \distDualVar(i)
      \geq 
      \prior_i \Quant\betaed(\scaledNoiseDiff)
      &
      \quad\, \forall \scaledNoiseDiff\in(-\infty,\infty),\,
      i\in[\stateNum] 
     \quad
      &
      \langle
      \signalscheme_i(\scaledNoiseDiff)
      \rangle
     \\
    \end{array}
\end{align}

\begin{restatable}{proposition}{optimallp}
\label{prop:opt_lp}
For every feasible solution
$\{\signalscheme_i(\scaledNoiseDiff)\}$
in program~\ref{eq:opt_lp},
there exists a signaling scheme
where for each state $i\in[\stateNum]$,
the boundedly rational receiver 
takes action 1 with probability 
$\int_{-\infty}^{\infty}
\signalscheme_i(\scaledNoiseDiff)\Quant\betaed(\scaledNoiseDiff)\,d\scaledNoiseDiff$.
Furthermore, the sender's optimal expected utility 
(in the optimal signaling scheme)
is equal to the optimal objective value of program~\ref{eq:opt_lp}.
\end{restatable}
\Cref{prop:opt_lp} reduces the design of an optimal signaling scheme to the linear program \ref{eq:opt_lp} without loss of optimality.
The first constraint in the program 
\ref{eq:opt_lp} ensures that whenever a signal $\scaledNoiseDiff$
is realized, the probability that the receiver 
takes action $1$ is exactly $\Quant\betaed(\scaledNoiseDiff)$.


\begin{proof}[Proof of \Cref{prop:opt_lp}]
Fix an arbitrary feasible solution 
$\{\signalscheme_i(\scaledNoiseDiff)\}$
in program~\ref{eq:opt_lp}.
We construct a signaling scheme $\signalscheme\primed$ 
as follows.\footnote{Here we use the superscript $\dagger$
to denote the constructed signaling scheme.}
Let the signal space $\signalspace\primed \gets  
\{\scaledNoiseDiff:\exists i\in[\stateNum],
\signalscheme_i(\scaledNoiseDiff) > 0\}$.
For each realized state $i$, let $\signalscheme\primed_i(\scaledNoiseDiff)
\gets \signalscheme_i(\scaledNoiseDiff)$ 
for each $\scaledNoiseDiff\in \signalspace\primed$.
Due to the second constraint 
in program~\ref{eq:opt_lp},
the constructed signaling scheme 
$\signalscheme\primed$ is valid.
When signal $\scaledNoiseDiff\in\signalspace\primed$
is realized, the posterior belief 
$\posterior_i\primed(\scaledNoiseDiff)$
equals 
$\frac{
\prior_i\signalscheme_i(\scaledNoiseDiff)}
{\sum_{j\in[\stateNum]}\prior_j \signalscheme_j(\scaledNoiseDiff)}$.
Due to the first constraint 
in program~\ref{eq:opt_lp},
we know that 
$\sum_{i\in[\stateNum]}\posterior_i\primed(\scaledNoiseDiff)\cdot 
(\receiverU_i - \scaledNoiseDiff) = 0$,
which implies $\scaledNoiseDiff = 
\sum_{i\in[\stateNum]}
\posterior_i\primed(\scaledNoiseDiff)\cdot 
\receiverU_i$.
Thus,
given realized signal $\scaledNoiseDiff$,
the receiver takes action 1 with probability 
$
\Quant\betaed(\scaledNoiseDiff)$.

So far, we have shown that the sender's optimal expected utility
is weakly higher than the optimal objective value of program~\ref{eq:opt_lp}.
We finish our proof by converting 
the optimal signaling scheme $\signalscheme^*$
(with signal space $\signalspace^*$)
into a feasible solution 
$\{\signalscheme_i\doubleprimed(\scaledNoiseDiff)\}$
of program~\ref{eq:opt_lp}
whose objective value equals the optimal sender's expected utility 
in $\signalscheme^*$.\footnote{Here we use 
the superscript $\ddagger$
to denote the constructed feasible solution.}
The construction works as follows.
First, we initialize $\signalscheme\doubleprimed_i(\scaledNoiseDiff)\gets 0$
for all $i\in[\stateNum]$, $\scaledNoiseDiff\in(-\infty,\infty)$.
Next, we enumerate each signal $\signal \in \signalspace^*$,
let $\posterior^*(\signal)$ be the induced posterior belief and 
$\scaledNoiseDiff \triangleq 
\sum_{i\in[\stateNum]}\posterior^*_i(\signal)\receiverU_i$.
By definition, given posterior belief $\posterior^*(\signal)$,
the boundedly rational receiver takes action 1 
with probability $\Quant\betaed(\scaledNoiseDiff)$.
We update $\signalscheme\doubleprimed_i(\scaledNoiseDiff) \gets 
\signalscheme\doubleprimed_i(\scaledNoiseDiff) + \signalscheme^*_i(\signal)
$
for every state $i\in[\stateNum]$.
After enumerating every signal $\signal \in\signalspace^*$,
it can be verified that the constructed solution 
$\signalscheme\doubleprimed$ is feasible 
and its objective value equals the sender's expected utility.
\end{proof}



\subsection{Proof of \texorpdfstring{\Cref{thm:SISU_opt}}{Theorem 1}}
\label{apx:SISU_opt}
In this subsection, we present the omitted proof of \Cref{thm:opt_state_independent}, which establishes the optimality of censorship signaling schemes for any known rationality level.
\knownrationalityoptimality*
\begin{proof}[Proof]
When the receiver is fully rational, i.e., $\noiseScale = \infty$, a censorship signaling scheme is known to be optimal (see, e.g., \citealp{RSV-17}; we restate this result as \Cref{lem:opt_fully_rational} in \Cref{apx:feasibility_for_fully_rational}).
At the rationality level $\noiseScale = 0$, the receiver chooses each action with probability one half regardless of the signal, so every signaling scheme, including any censorship signaling scheme, yields the sender an expected utility of $\sfrac{1}{2}$ and is therefore optimal.

We next focus on finite positive rationality levels $\noiseScale \in (0, \infty)$ in the remainder of this analysis.
We prove the optimality of the signaling scheme $\optscheme$ defined in \Cref{thm:SISU_opt}
by constructing a feasible dual solution to 
the dual program \ref{eq:opt_lp_dual} 
that satisfies complementary slackness. 
Let $(\censorshipsignal,\censorshipsignalInverse,
\censorshipstate,\censorshipprob)$
be the unique feasible solution to program~\ref{eq:SISU_opt_condition}, where the uniqueness and feasibility are argued in \Cref{lem:existence_SISU_opt_condition}.

\xhdr{Assignment construction}
To facilitate the analysis, we explicitly write out the 
optimal signaling scheme $\optscheme$ as follows,
\begin{align*}
     i\in[\censorshipstate - 1]: 
     & \qquad\optscheme_i(\censorshipsignal) = 1; 
    \\ 
    i=\censorshipstate:& \qquad \optscheme_{i}(\scaledNoiseDiff) = \censorshipprob \cdot \indicator {\scaledNoiseDiff = \censorshipsignal}+(1-\censorshipprob)\cdot \indicator {\scaledNoiseDiff=\receiverU_{\censorshipstate}};
    \\ 
    i\in[\censorshipstate + 1:m]: 
     & \qquad\optscheme_i\left(\receiverU_i\right) = 1.
\end{align*}

Due to constraint \pf\ in program~\ref{eq:SISU_opt_condition},
signaling scheme $\optscheme$ is feasible. 
Now, consider the following dual assignment 
$\{\noiseDualVar(\scaledNoiseDiff), \distDualVar(i)\}$
of program~\ref{eq:opt_lp_dual}.
Depending on whether $\censorshipstate<\stateNum$ or $\censorshipstate=\stateNum$, we set $\noiseDualVar(\scaledNoiseDiff)$ as follows:
\[
\begin{aligned}[t]
    &\text{If } \censorshipstate<\stateNum:
    \\[1mm]
    &\noiseDualVar(\scaledNoiseDiff) =
    \begin{cases}
    -\Quant'\betaed(\censorshipsignal),
    & \scaledNoiseDiff \in (-\infty, \censorshipsignalInverse],
    \\[1mm]
    -\dfrac{
    \Quant\betaed(\scaledNoiseDiff)
    -\Quant\betaed(\receiverU_{\censorshipstate+1})
    }{
    \scaledNoiseDiff-\receiverU_{\censorshipstate+1}
    },
    & \scaledNoiseDiff \in (\censorshipsignalInverse, \receiverU_{\censorshipstate+1}),
    \\[3mm]
    -\dfrac{
    \Quant\betaed(\scaledNoiseDiff)
    -\Quant\betaed(\receiverU_{j+1})
    }{
    \scaledNoiseDiff-\receiverU_{j+1}
    },
    & \scaledNoiseDiff \in [\receiverU_j, \receiverU_{j+1}), \; j\in [\censorshipstate+1,\stateNum-1],
    \\[3mm]
    0,
    & \scaledNoiseDiff \in [\receiverU_{\stateNum}, \infty).
    \end{cases}
\end{aligned}
\qquad\qquad
\begin{aligned}[t]
    &\text{If } \censorshipstate=\stateNum:
    \\[1mm]
    &\noiseDualVar(\scaledNoiseDiff) =
    \begin{cases}
    -\Quant'\betaed(\censorshipsignal),
    & \scaledNoiseDiff \in (-\infty, \receiverU_{\stateNum}),
    \\[1mm]
    0,
    & \scaledNoiseDiff \in [\receiverU_{\stateNum}, \infty).
    \end{cases}
\end{aligned}
\]
In both cases, we set $\distDualVar(i)$ as follows:
\[
\distDualVar(i) =
\begin{cases}
    \prior_i\left(
    \Quant\betaed(\censorshipsignal)
    +(\censorshipsignal-\receiverU_i)
    \noiseDualVar(\censorshipsignal)
    \right),
    & i\in[\censorshipstate],
    \\[1mm]
    \prior_i\Quant\betaed(\receiverU_i),
    & i\in[\censorshipstate+1:\stateNum].
\end{cases}
\]

\xhdr{Complementary slackness} 
We now verify the complementary slackness 
of the constructed assignment.
Namely, for each state $i$
and $\scaledNoiseDiff\in(-\infty,\infty)$
such that $\optscheme_i(\scaledNoiseDiff) > 0$,
its corresponding dual constraint holds with equality,
i.e., 
\begin{align}
      \label{eq:SISU_cs}
      \Quant\betaed(\scaledNoiseDiff) + 
      ( \scaledNoiseDiff - \receiverU_i)
      \noiseDualVar(\scaledNoiseDiff) &= 
      \frac{\distDualVar(i)}{\prior_i}\,.
\end{align}
We verify this for each state $i\in[\stateNum]$ separately.
\begin{itemize}[label={-}]
    \item Fix an arbitrary
    state $i\in[\censorshipstate - 1]$,
    note that $\optscheme_i(\scaledNoiseDiff) > 0$
    for $\scaledNoiseDiff = \censorshipsignal$ only.
    Here Eqn.~\eqref{eq:SISU_cs} holds 
    by construction straightforwardly.
    \item Now consider threshold state $\censorshipstate$, 
    the same argument holds for 
    Eqn.~\eqref{eq:SISU_cs}
    with $\optscheme_{\censorshipstate}
    (\censorshipsignal) > 0$.
    It remains to verify 
    Eqn.~\eqref{eq:SISU_cs} 
    associated with
    $\optscheme_{\censorshipstate}
    (\receiverU_{\censorshipstate})
    = 1 - \censorshipprob > 0$.
    When $\censorshipprob < 1$, 
    constraint~\cs\ in program~\ref{eq:SISU_opt_condition}
    ensures that $\censorshipsignalInverse =
    \receiverU_{\censorshipstate}$.
    Thus,
    $
      \Quant\betaed(\receiverU_{\censorshipstate}) + 
      ( \receiverU_{\censorshipstate} -
      \receiverU_{\censorshipstate})
      \noiseDualVar(\receiverU_{\censorshipstate})
      \overset{(a)}{=}
      \Quant\betaed(\censorshipsignalInverse)
      \overset{(b)}{=}
      \Quant\betaed(\censorshipsignal) +
      (\censorshipsignal - \censorshipsignalInverse)
      \noiseDualVar(\censorshipsignal)
      \overset{(c)}{=}
      \Quant\betaed(\censorshipsignal) +
      (\censorshipsignal - \receiverU_{\censorshipstate})
      \noiseDualVar(\censorshipsignal)
      \overset{(d)}{=}
      \frac{\distDualVar(\censorshipstate)}{\prior_{\censorshipstate}}$,
where equalities~(a), (c) hold since $\receiverU_{\censorshipstate} = \censorshipsignalInverse$;
equality~(b) holds 
due to constraint \dfone\ in program~\ref{eq:SISU_opt_condition}
and the construction of $\noiseDualVar(\censorshipsignal)$;
and equality (d) holds by the definition of
$\distDualVar(\censorshipstate)$.
This argument also applies when $\censorshipstate=\stateNum$.
If $\censorshipprob<1$, then
$\censorshipsignalInverse=\receiverU_{\stateNum}$, and setting
$\noiseDualVar(\receiverU_{\stateNum})=0$ preserves the equality since its coefficient is
$0$.
\item Fix an arbitrary state $i\in[\censorshipstate + 1:\stateNum]$,
note that $\optscheme_i(\scaledNoiseDiff) > 0$
for $\scaledNoiseDiff = \receiverU_i$ only.
Here Eqn.~\eqref{eq:SISU_cs} holds 
by construction straightforwardly.
\end{itemize}

\xhdr{Dual feasibility}
To verify whether the dual constraints
associated with $\optscheme_i(\scaledNoiseDiff)$
for state $i \in [\censorshipstate]$ hold,
note that 
$
    \frac{\distDualVar(i)}{\prior_i}
      \overset{(a)}{=}
      \Quant\betaed(\censorshipsignal) +
      (\censorshipsignal - \receiverU_i)
      \noiseDualVar(\censorshipsignal)
      \overset{(b)}{=}
      \Quant\betaed(\censorshipsignal) -
      (\censorshipsignal - \receiverU_i)
      \Quant\betaed'(\censorshipsignal)$,
where equality~(a) follows from the definition of $\distDualVar(i)$;
and equality~(b) holds by the construction of 
$\noiseDualVar(\censorshipsignal)$.
Thus, 
we can rewrite those dual constraints
associated with $\optscheme_i(\scaledNoiseDiff)$
for state $i \in [\censorshipstate]$ 
as
\begin{align}
    \label{eq:SISU_dual_feasibility_pooling_state}
    \Quant\betaed(\scaledNoiseDiff) + 
      ( \scaledNoiseDiff - \receiverU_i)
      \noiseDualVar(\scaledNoiseDiff) \leq 
      \Quant\betaed(\censorshipsignal) -
      (\censorshipsignal - \receiverU_i)
      \Quant\betaed'(\censorshipsignal) \;.
\end{align}
To verify whether the dual constraints
associated with $\optscheme_i(\scaledNoiseDiff)$
for state $i \in [\censorshipstate+1:\stateNum]$ hold,  
for each state $i\in[\censorshipstate+1:\stateNum]$,
the assignment
$\distDualVar(i)=\prior_i\Quant\betaed(\receiverU_i)$
reduces to
\begin{align}
    \label{eq:SISU_dual_feasibility_revealing_state}
    \Quant\betaed(\scaledNoiseDiff) + 
      ( \scaledNoiseDiff - \receiverU_i)
      \noiseDualVar(\scaledNoiseDiff) \leq 
      \Quant\betaed(\receiverU_i)\;.
\end{align}

We first consider $\censorshipstate=\stateNum$.
Constraint \dftwo\ gives
$\receiverU_{\stateNum}\le\censorshipsignalInverse$.
Since $\censorshipsignal\le0<\receiverU_{\stateNum}$,
the assignment gives
$\noiseDualVar(\censorshipsignal)
=-\Quant'\betaed(\censorshipsignal)$.
Fix an arbitrary state $i\in[\stateNum]$.
If $\scaledNoiseDiff<\receiverU_{\stateNum}$, then
$
\Quant\betaed(\scaledNoiseDiff)
+(\scaledNoiseDiff-\receiverU_i)
 \noiseDualVar(\scaledNoiseDiff)
=
\Quant\betaed(\scaledNoiseDiff)
-(\scaledNoiseDiff-\receiverU_i)
 \Quant'\betaed(\censorshipsignal)
\le
\Quant\betaed(\censorshipsignal)
+(\receiverU_i-\censorshipsignal)
 \Quant'\betaed(\censorshipsignal) =
\frac{\distDualVar(i)}{\prior_i}.
$
For $\scaledNoiseDiff\le0$, the inequality follows from
the concavity of the quantal response function $\Quant\betaed(\cdot)$ on $(-\infty,0]$.
For $0<\scaledNoiseDiff<\receiverU_{\stateNum}
\le\censorshipsignalInverse$, it follows from the convexity of the quantal response function $\Quant\betaed(\cdot)$
on $[0,\censorshipsignalInverse]$, using the concavity of the quantal response function $\Quant\betaed(\cdot)$ on
$(-\infty,0]$ for the comparison at $0$ and constraint
\dfone\ for the comparison at $\censorshipsignalInverse$.
If $\scaledNoiseDiff\ge\receiverU_{\stateNum}$, then
$\noiseDualVar(\scaledNoiseDiff)=0$, and
$
\Quant\betaed(\scaledNoiseDiff)
+(\scaledNoiseDiff-\receiverU_i)
 \noiseDualVar(\scaledNoiseDiff)
=
\Quant\betaed(\scaledNoiseDiff)
\overset{(a)}{\le}
\Quant\betaed(\receiverU_{\stateNum})
\overset{(b)}{\le}
\Quant\betaed(\censorshipsignal)
+(\receiverU_{\stateNum}-\censorshipsignal)
 \Quant'\betaed(\censorshipsignal)
\overset{(c)}{\le}
\Quant\betaed(\censorshipsignal)
+(\receiverU_i-\censorshipsignal)
 \Quant'\betaed(\censorshipsignal)
=
\frac{\distDualVar(i)}{\prior_i},
$
where inequality~(a) follows from the monotonicity of the quantal response function
$\Quant\betaed(\cdot)$,
inequality~(b) follows from the same curvature argument and
constraint \dfone, since
$0<\receiverU_{\stateNum}\le\censorshipsignalInverse$; and inequality~(c) follows from the fact that
$\receiverU_i\le\receiverU_{\stateNum}$ and
$\Quant'\betaed(\censorshipsignal)\le0$.
Thus, all dual constraints hold when
$\censorshipstate=\stateNum$.

It remains to consider $\censorshipstate<m$.
We verify both Eqn.~\eqref{eq:SISU_dual_feasibility_pooling_state}
and Eqn.~\eqref{eq:SISU_dual_feasibility_revealing_state}
for different values of $\scaledNoiseDiff$ 
in four cases separately: 
$\scaledNoiseDiff \in 
(-\infty,\censorshipsignalInverse]$;
$\scaledNoiseDiff\in 
(\censorshipsignalInverse, \receiverU_{\censorshipstate + 1})$;
$\scaledNoiseDiff\in[\receiverU_j, \receiverU_{j + 1})$
for some $j\in[\censorshipstate + 1: \stateNum - 1]$;
and $\scaledNoiseDiff\in[\receiverU_\stateNum, \infty)$.
The argument mainly uses the curvature of 
function $\Quant\betaed(\cdot)$ 
and the constraints in feasibility program~\ref{eq:SISU_opt_condition}.
\begin{figure}[tb]
\centering
\subfloat[]{
\begin{tikzpicture}[scale=0.55, transform shape]
\begin{axis}[
axis line style=gray,
axis lines=middle,
xtick style={draw=none},
ytick style={draw=none},
xticklabels=\empty,
yticklabels=\empty,
xmin=-9,xmax=9,ymin=-0.15,ymax=1.3,
width=0.9\textwidth,
height=0.5\textwidth,
samples=50]

\node[anchor=north east, font=\large, inner sep=3pt] at (axis cs:0,0) {0};

\addplot[domain=-9:9, gray!40!white, line width=2.mm] (x, {1/(1+e^(0.7*x))});

\addplot[mark=*,only marks, fill=white] coordinates {(6.34907,0.0116084)} node[above, pos=1]{};
\addplot[gray, thick] coordinates {(6.34907,0.01)
(6.34907,-0.01)};
\addplot[] coordinates {(6.34907,0.)} node[below, pos=1]{\Large$\censorshipsignalInverse$};

\addplot[mark=*,only marks, fill=white] coordinates {(-2.34798,0.838022)} node[above, pos=1]{};
\addplot[gray, thick] coordinates {(-2.34798,0.01)
(-2.34798,-0.01)};

\addplot[gray, dotted] coordinates {(-2.34798,0.838022)
(-2.34798,0.)};
\addplot[] coordinates {(-2.34798,0.)} node[below, pos=1]{\Large$\censorshipsignal$};

\addplot[mark=*,only marks, fill=white] coordinates {(-8,0.99631576)} node[above, pos=1]{};
\addplot[gray, thick] coordinates {(-8,0.01)
(-8,-0.01)};
\addplot[] coordinates {(-8,-0.015)} node[below, pos=1]{\Large$\scaledNoiseDiff$};
\addplot[gray, dotted] coordinates {(-8,0.99631576)
(-8,0.)};

\addplot[mark=*,only marks, fill=white] coordinates {(4,0.0573242)} node[above, pos=1]{};
\addplot[gray, thick] coordinates {(4,0.01)
(4,-0.01)};
\addplot[] coordinates {(4,-0.015)} node[below, pos=1]{\Large$\scaledNoiseDiff$};
\addplot[gray, dotted] coordinates {(4,0.0573242)
(4,0.)};

\addplot[domain=-2.34798:6.34907, black,line width=0.5mm] (x, {-0.09502 * x + 0.614897});

\addplot[domain=-8:-2.34798, dotted, black,line width=0.5mm] (x, {-0.0280066 * x + 0.772263});

\addplot[domain=-2.34798:4, dash pattern={on 8pt off 2pt}, black,line width=0.5mm] (x, {-0.122984 * x + 0.549259});

\end{axis}

\end{tikzpicture}
\label{fig:case a}
}
\quad
\subfloat[]{
\begin{tikzpicture}[scale=0.55, transform shape]
\begin{axis}[
axis line style=gray,
axis lines=middle,
xtick style={draw=none},
ytick style={draw=none},
xticklabels=\empty,
yticklabels=\empty,
xmin=-6,xmax=12,ymin=-0.15,ymax=1.3,
width=0.9\textwidth,
height=0.5\textwidth,
samples=50]

\node[anchor=north east, font=\large, inner sep=3pt] at (axis cs:0,0) {0};

\addplot[domain=-6:12, gray!40!white, line width=2.mm] (x, {1/(1+e^(0.7*x))});

\addplot[mark=*,only marks, fill=white] coordinates {(6.34907,0.0116084)} node[above, pos=1]{};
\addplot[gray, thick] coordinates {(6.34907,0.01)
(6.34907,-0.01)};
\addplot[] coordinates {(6.34907,0.)} node[below, pos=1]{\Large$\censorshipsignalInverse$};

\addplot[mark=*,only marks, fill=white] coordinates {(-2.34798,0.838022)} node[above, pos=1]{};
\addplot[gray, thick] coordinates {(-2.34798,0.01)
(-2.34798,-0.01)};
\addplot[] coordinates {(-2.34798,0.)} node[below, pos=1]{\Large$\censorshipsignal$};

\addplot[gray, dotted] coordinates {(-2.34798,0.838022)
(-2.34798,0.)};

\addplot[mark=*,only marks, fill=white] coordinates {(11,0.000452622)} node[above, pos=1]{};
\addplot[gray, thick] coordinates {(11,0.01)
(11,-0.01)};
\addplot[] coordinates {(11,-0.035)} node[below, pos=1]{\Large$\receiverU_{\censorshipstate+1}$};

\addplot[mark=*,only marks, fill=white] coordinates {(9,0.00183294)} node[above, pos=1]{};
\addplot[gray, thick] coordinates {(9,0.01)
(9,-0.01)};
\addplot[] coordinates {(9,-0.015)} node[below, pos=1]{\Large$\scaledNoiseDiff$};
\addplot[gray, dotted] coordinates {(9,0.00183294)
(9,0.)};

\addplot[mark=*,only marks, fill=white] coordinates {(1.2,0.500873)} node[above, pos=1]{};
\addplot[gray, thick] coordinates {(1.2,0.01)
(1.2,-0.01)};
\addplot[] coordinates {(1.2,-0.035)} node[below, pos=1]{\Large$\receiverU_i$};
\addplot[gray, dotted] coordinates {(1.2,0.500873)
(1.2,0.)};

\addplot[domain=-2.34798:6.34907, black,line width=0.5mm] (x, {-0.09502 * x + 0.614897});

\addplot[domain=6.34907:9, dash pattern={on 8pt off 2pt}, black,line width=0.5mm] (x, {-0.00368756 * x + 0.035021});

\addplot[domain=9:11, dotted, black,line width=0.5mm] (x, {-0.000690159 * x + 0.00804437});

\addplot[domain=1.2:9, dash pattern={on 7pt off 2pt on 1pt off 3pt}, black,line width=0.5mm] (x, {-0.0639795  * x + 0.577648});

\end{axis}

\end{tikzpicture}
\label{fig:case b}
}
\caption{\label{fig:SISU dual feasibility}Graphical illustration of 
the dual assignment feasibility 
for Eqn.~\eqref{eq:SISU_dual_feasibility_pooling_state}
in the proof of \Cref{thm:SISU_opt}.
The gray solid curve is the quantal response function $\Quant\betaed(\cdot)$.
(a):
For $\scaledNoiseDiff\in 
(-\infty,\censorshipsignalInverse\text{]}$, 
the dual assignment $\noiseDualVar(\scaledNoiseDiff)$
is the absolute value of the slope 
of the black solid line.
We prove that Eqn.~\eqref{eq:SISU_dual_feasibility_pooling_state} holds by 
showing that
the slope of the black dotted (resp.\ dashed)
line 
is larger (resp.\ smaller) than 
the slope of the black solid line 
if $\scaledNoiseDiff\in (-\infty,\censorshipsignal\text{]}$
(resp.\ 
$\scaledNoiseDiff\in\text{[}\censorshipsignal,\censorshipsignalInverse\text{]}$).
(b):
For $\scaledNoiseDiff\in 
\text{(}\censorshipsignalInverse,
\receiverU_{\censorshipstate+1}\text{)}$, 
the dual assignment $\noiseDualVar(\scaledNoiseDiff)$
is the absolute value of the slope 
of the black dotted line.
We prove that Eqn.~\eqref{eq:SISU_dual_feasibility_pooling_state} holds by 
showing that
the slope of the black dash-dotted 
line 
is smaller than 
the slope of the black dotted line.
More specifically, 
we rewrite Eqn.~\eqref{eq:SISU_dual_feasibility_pooling_state}
as Eqn.~\eqref{eq:SISU_dual_feasibility_pooling_state_case_2}.
The right-hand side of Eqn.~\eqref{eq:SISU_dual_feasibility_pooling_state_case_2}
is the absolute value of the slope of the black dash-dotted line.
We lower bound this term
by the absolute value of the slope of the black dashed line,
which is due to the convexity of the quantal response function $\Quant\betaed(\cdot)$,
is larger than the absolute value of the slope of the black dotted line,
i.e., the left-hand side of Eqn.~\eqref{eq:SISU_dual_feasibility_pooling_state_case_2}.
}
\end{figure}
We present the analysis for state $i\in[\censorshipstate]$
for the first two cases,
together with a graphical illustration
of our argument (see \Cref{fig:SISU dual feasibility}).
The analysis of the third case is similar to the second case,
and the fourth case is trivial.
\begin{itemize}[label={-}]
    \item Fix an arbitrary $\scaledNoiseDiff \in 
    (-\infty,\censorshipsignalInverse]$.
    We illustrate this case in \Cref{fig:case a}.
    Note that
        $
        \Quant\betaed(\scaledNoiseDiff) + 
      ( \scaledNoiseDiff - \receiverU_i)
      \noiseDualVar(\scaledNoiseDiff)
      \overset{(a)}{=}
      \Quant\betaed(\scaledNoiseDiff) -
      (\scaledNoiseDiff - \receiverU_i)
      \Quant\betaed'(\censorshipsignal) 
      =
       \Quant\betaed(\censorshipsignal) -
      (\censorshipsignal - \receiverU_i)
      \Quant\betaed'(\censorshipsignal) 
      +
      \left(\Quant\betaed(\scaledNoiseDiff) - 
      \Quant\betaed(\censorshipsignal)\right)
      -
      (\scaledNoiseDiff - \censorshipsignal)
      \Quant\betaed'(\censorshipsignal)$,
    where equality~(a) holds due to 
    the construction of $\noiseDualVar(\scaledNoiseDiff)$.
    Hence, to show Eqn.~\eqref{eq:SISU_dual_feasibility_pooling_state} in this case,
    it suffices to argue that 
$(\censorshipsignal - \scaledNoiseDiff)
        \Quant\betaed'(\censorshipsignal) \leq 
        \Quant\betaed(\censorshipsignal) - 
        \Quant\betaed(\scaledNoiseDiff)$
holds. 
    Specifically,
    if $\scaledNoiseDiff\in(-\infty,0]$,
    the above inequality
    holds since the quantal response function $\Quant\betaed(\cdot)$
    is concave in $(-\infty, 0]$;
    if $\scaledNoiseDiff\in[0, \censorshipsignalInverse]$,
    the above inequality follows from the convexity of the quantal response function $\Quant\betaed(\cdot)$ on $[0,\censorshipsignalInverse]$, using the concavity of the quantal response function $\Quant\betaed(\cdot)$ on $(-\infty,0]$ for the comparison at 0 and constraint~\dfone\ for the comparison at $\censorshipsignalInverse$.
    \item Fix an arbitrary $\scaledNoiseDiff\in 
    (\censorshipsignalInverse, \receiverU_{\censorshipstate + 1})$.
    We illustrate this case in \Cref{fig:case b}.
    By construction, $\noiseDualVar(\scaledNoiseDiff) 
    = -
    (\Quant\betaed(\scaledNoiseDiff) - 
    \Quant\betaed(\receiverU_{\censorshipstate + 1}))/(
    \scaledNoiseDiff - 
    \receiverU_{\censorshipstate + 1})$.
    After rearranging the terms,
    Eqn.~\eqref{eq:SISU_dual_feasibility_pooling_state}
    becomes\footnote{Here we use the fact 
    that $\scaledNoiseDiff - \receiverU_i \geq 0$
    for every state $i\in[\censorshipstate]$, 
    since
    $\scaledNoiseDiff \geq 
    \censorshipsignalInverse\geq 
    \receiverU_{\censorshipstate} \geq 
    \receiverU_i$ 
    where the second inequality holds 
    due to 
    constraint~{\small \dftwo}\ 
    in program~\ref{eq:SISU_opt_condition}.}
    \begin{align}
\label{eq:SISU_dual_feasibility_pooling_state_case_2}
        -\frac{
        \Quant\betaed(\scaledNoiseDiff) - \Quant\betaed(\receiverU_{\censorshipstate+1})}
        {\scaledNoiseDiff - 
        \receiverU_{\censorshipstate+1}}
        \leq 
        - \frac{ 
        \Quant\betaed(\scaledNoiseDiff) - 
        \left(\Quant\betaed(\censorshipsignal)
        -\Quant\betaed'(\censorshipsignal) 
        (\censorshipsignal - \receiverU_i) \right)}
        {\scaledNoiseDiff - \receiverU_i}\,.
   \end{align}
    Here we argue that it is sufficient to show 
    the above inequality holds when we replace $\receiverU_i$
    with $\censorshipsignalInverse \geq \receiverU_i$.
    To see this, note that
    the right-hand side of the above inequality
    is monotone decreasing as a function of 
    $\receiverU_i$.
    In particular, consider function
    $\tangentfunc(\variableX) \triangleq
    - ({ 
    \Quant\betaed(\scaledNoiseDiff) - 
    (\Quant\betaed(\censorshipsignal)
    -\Quant\betaed'(\censorshipsignal) 
    (\censorshipsignal - \variableX) )})/(
    {\scaledNoiseDiff - \variableX})$,
    and compute its derivative
    $
    \tangentfunc'(\variableX) =
    -
    \frac{
    \Quant\betaed(\scaledNoiseDiff) - 
    \Quant\betaed(\censorshipsignal)
    -
    \Quant\betaed'(\censorshipsignal)
    (\scaledNoiseDiff - \censorshipsignal)
    }
    {(\scaledNoiseDiff - \variableX)^2}
    \leq 0
    $
    where the last inequality holds since 
    $\Quant\betaed'(\censorshipsignal)
    (\scaledNoiseDiff -
    \censorshipsignal)
    \leq \Quant\betaed(\scaledNoiseDiff) - 
    \Quant\betaed(\censorshipsignal)$
    if
    $\scaledNoiseDiff \geq \censorshipsignalInverse$,
    which is implied by 
    constraint~\dfone\ 
    and the convexity of quantal response function $\Quant\betaed(\cdot)$
    on $[0,\infty)$.
    Hence, we have
    $
    - \frac{ 
    \Quant\betaed(\scaledNoiseDiff) - 
    \left(\Quant\betaed(\censorshipsignal)
    -\Quant\betaed'(\censorshipsignal) 
    (\censorshipsignal - \receiverU_i) \right)}
    {\scaledNoiseDiff - \receiverU_i}
    \overset{(a)}{\geq} 
    f(\censorshipsignalInverse)
    \overset{}{=}
        - \frac{ 
    \Quant\betaed(\scaledNoiseDiff) - 
    \left(\Quant\betaed(\censorshipsignal)
    -\Quant\betaed'(\censorshipsignal) 
    (\censorshipsignal - \censorshipsignalInverse) \right)}
    {\scaledNoiseDiff - \censorshipsignalInverse}
    \overset{(b)}{=}
     -\frac{
    \Quant\betaed(\scaledNoiseDiff) - \Quant\betaed(\censorshipsignalInverse)}
    {\scaledNoiseDiff - 
    \censorshipsignalInverse}
    \overset{(c)}{\geq }
    -\frac{
    \Quant\betaed(\scaledNoiseDiff) - \Quant\betaed(\receiverU_{\censorshipstate+1})}
    {\scaledNoiseDiff - 
    \receiverU_{\censorshipstate+1}}$,
    where inequality~(a) holds due to the 
    monotonicity of function $f(\cdot)$,
    equality~(b) holds 
    due to constraint~\dfone\ in program~\ref{eq:SISU_opt_condition},
    and inequality~(c)
    holds
    due to the convexity of the quantal response function $\Quant\betaed(\cdot)$
    on $[0,\infty)$. 
\end{itemize}

\begin{itemize}[label={-}]
        \item Fix an arbitrary $j\in
        [\censorshipstate + 1:
        \stateNum - 1]$
        and an arbitrary $\scaledNoiseDiff\in[\receiverU_j,
        \receiverU_{j + 1})$.
        Then, we have 
        $\noiseDualVar(\scaledNoiseDiff) = -\frac{\Quant\betaed(\scaledNoiseDiff)-\Quant\betaed(\receiverU_{j+1})}{\scaledNoiseDiff-\receiverU_{j+1}} \overset{(a)}{\le} -\frac{\Quant\betaed(\scaledNoiseDiff)-\Quant\betaed(\censorshipsignalInverse)}{\scaledNoiseDiff-\censorshipsignalInverse} = f(\censorshipsignalInverse) \overset{(b)}{\le} f(\receiverU_i) $, where inequality~(a) follows from the convexity of quantal response function $\Quant\betaed(\cdot)$ on $[0,\infty)$, since $\censorshipsignalInverse < \scaledNoiseDiff < \receiverU_{j+1}$, and inequality~(b) follows from the monotonicity of $\tangentfunc(\variableX)$ established in the preceding case and $\receiverU_i \le \censorshipsignalInverse$.
        Thus, Eqn.~\eqref{eq:SISU_dual_feasibility_pooling_state} holds.
        
        \item Fix an arbitrary 
        $\scaledNoiseDiff\in[\receiverU_m, \infty)$.
        By construction $\noiseDualVar(\scaledNoiseDiff) 
        = 0$.
        Eqn.~\eqref{eq:SISU_dual_feasibility_pooling_state}
        becomes 
        $
            \Quant\betaed(\scaledNoiseDiff) \leq 
      \Quant\betaed(\censorshipsignal) -
      (\censorshipsignal - \receiverU_i)
      \Quant\betaed'(\censorshipsignal)
        $.
        If $\receiverU_i \leq \censorshipsignal$,
        the above inequality holds since the quantal response function $\Quant\betaed(\cdot)$
        is monotone decreasing.
        Otherwise, i.e., 
        if $\censorshipsignal\leq \receiverU_i \leq
        \censorshipsignalInverse$,
        the above inequality holds since
        $
            \Quant\betaed(\censorshipsignal) -
      (\censorshipsignal - \receiverU_i)
      \Quant\betaed'(\censorshipsignal) 
      \geq 
            \Quant\betaed(\censorshipsignal) -
      (\censorshipsignal - \censorshipsignalInverse)
      \Quant\betaed'(\censorshipsignal) 
      \overset{(a)}{=}\Quant\betaed(\censorshipsignalInverse)
      \overset{(b)}{\geq} 
      \Quant\betaed(\scaledNoiseDiff)$,
        where equality~(a) holds due to 
        constraint~\dfone\ in program~\ref{eq:SISU_opt_condition};
        and inequality~(b) holds due to the monotonicity
        of the quantal response function $\Quant\betaed(\cdot)$.
\end{itemize}
Finally, we verify dual constraints
associated with $\optscheme_i(\scaledNoiseDiff)$
for state $i \in [\censorshipstate+1:\stateNum]$. 
\begin{itemize}[label={-}]
\item Fix an arbitrary $\scaledNoiseDiff
\in(-\infty,\censorshipsignalInverse]$.
By construction $\noiseDualVar(\scaledNoiseDiff) 
= - \Quant\betaed^{\prime}(\censorshipsignal)$.
Then, we have 
$\Quant\betaed(\scaledNoiseDiff)+(\scaledNoiseDiff-\receiverU_i)\noiseDualVar(\scaledNoiseDiff) = \Quant\betaed(\scaledNoiseDiff)+(\receiverU_i-\scaledNoiseDiff)\Quant\betaed^{\prime}(\censorshipsignal) \overset{(a)}{\le} \Quant\betaed(\censorshipsignal)+(\receiverU_i-\censorshipsignal)\Quant\betaed^{\prime}(\censorshipsignal) \overset{(b)}{\le}\Quant\betaed(\receiverU_i)$, where inequality~(a) follows from the first pooled-state case, since $\scaledNoiseDiff \le \censorshipsignalInverse$, and inequality~(b) follows from the
    constraint~\dfone\ 
    and the convexity of quantal response function $\Quant\betaed(\cdot)$
    on $[0,\infty)$, since $\receiverU_i>\censorshipsignalInverse$. Thus, Eqn.~\eqref{eq:SISU_dual_feasibility_revealing_state} holds.

\item Fix an arbitrary $\scaledNoiseDiff\in
(\censorshipsignalInverse, \receiverU_{\censorshipstate+1})$.
By construction, $\noiseDualVar(\scaledNoiseDiff) 
= - (\Quant\betaed(\scaledNoiseDiff) - \Quant\betaed(\receiverU_{\censorshipstate + 1}))
/
(\scaledNoiseDiff - \receiverU_{\censorshipstate + 1})$.
By rearranging the terms, Eqn.~\eqref{eq:SISU_dual_feasibility_revealing_state} becomes 
$
    \frac{
    \Quant\betaed(\receiverU_{\censorshipstate + 1}) - \Quant\betaed(\scaledNoiseDiff)}{
    \receiverU_{\censorshipstate + 1} - \scaledNoiseDiff}
     \leq 
    \frac{
    \Quant\betaed(\receiverU_i) - \Quant\betaed(\scaledNoiseDiff)}{
    \receiverU_i - \scaledNoiseDiff}$,
which holds due to the convexity of function $\Quant\betaed(\cdot)$
on $[0,\infty)$ and 
$\scaledNoiseDiff \leq\receiverU_{\censorshipstate + 1}
\leq \receiverU_i$.

\item Fix an arbitrary $j\in
        [\censorshipstate + 1:
        \stateNum - 1]$
        and an arbitrary $\scaledNoiseDiff\in[\receiverU_j,
        \receiverU_{j + 1})$.
By construction, $\noiseDualVar(\scaledNoiseDiff) 
= - (\Quant\betaed(\scaledNoiseDiff) - \Quant\betaed(\receiverU_{j + 1}))
/
(\scaledNoiseDiff - \receiverU_{j + 1})$.
For state $i\in[\censorshipstate + 1:j]$,
by rearranging the terms, 
Eqn.~\eqref{eq:SISU_dual_feasibility_revealing_state} becomes 
$
    \frac{
    \Quant\betaed(\receiverU_{j + 1}) - \Quant\betaed(\scaledNoiseDiff)}{
    \receiverU_{j + 1} - \scaledNoiseDiff}
     \geq 
    \frac{
    \Quant\betaed(\receiverU_i) - \Quant\betaed(\scaledNoiseDiff)}{
    \receiverU_i - \scaledNoiseDiff}$,
which holds due to the convexity of the quantal response function $\Quant\betaed(\cdot)$
on $[0,\infty)$ and 
$\receiverU_i \leq \scaledNoiseDiff 
\leq \receiverU_{j+1}$.
Similarly, for state $i\in[j+1:\stateNum]$,
by rearranging the terms, 
Eqn.~\eqref{eq:SISU_dual_feasibility_revealing_state} becomes 
$
    \frac{
    \Quant\betaed(\receiverU_{j + 1}) - \Quant\betaed(\scaledNoiseDiff)}{
    \receiverU_{j + 1} - \scaledNoiseDiff}
     \leq 
    \frac{
    \Quant\betaed(\receiverU_i) - \Quant\betaed(\scaledNoiseDiff)}{
    \receiverU_i - \scaledNoiseDiff}$,
which holds due to the convexity of the quantal response function $\Quant\betaed(\cdot)$
on $[0,\infty)$ and 
$  \scaledNoiseDiff 
\leq \receiverU_{j+1}\leq \receiverU_i$.

\item Fix an arbitrary $\scaledNoiseDiff\in[\receiverU_m,\infty)$.
By construction, $\noiseDualVar(\scaledNoiseDiff) = 0$.
Here Eqn.~\eqref{eq:SISU_dual_feasibility_revealing_state} follows from the monotonicity of the quantal response function $\Quant\betaed(\cdot)$.
\end{itemize}
This completes the proof of \Cref{thm:opt_state_independent}.
\end{proof}

\subsection{Interpreting 
Program in \texorpdfstring{\Cref{thm:opt_state_independent}}{Theorem 1}
for Optimal Censorship of a Fully Rational Receiver.}
\label{apx:feasibility_for_fully_rational}
In this subsection, we first state the established optimality result for censorship signaling schemes under a fully rational receiver, and then explain how it relates to our program~\ref{eq:SISU_opt_condition}.
\begin{restatable}[See for example \citealp{RSV-17}]{lemma}{optfullyrational}
\label{lem:opt_fully_rational}
Fix any instance $\instance$.
For a fully rational receiver 
(i.e., 
with rationality level $\noiseScale = \infty$), it is optimal for the sender to adopt a censorship
 signaling scheme such that the pooled mass $\pooledmass = \sum\nolimits_{i \in [\censorshipstate-1]}\prior_i+\censorshipprob\prior_{\censorshipstate}$, where threshold state $\censorshipstate = \arg\max\nolimits_{i\in[\stateNum]} \left\{\receiverU_i: \sum_{j \in [i-1]} \prior_j\receiverU_j \leq 0 \right\}$ and threshold-state fraction $\censorshipprob = \max\left\{p\in[0, 1]: p\prior_{\censorshipstate}\receiverU_{\censorshipstate} + \sum_{i\in[\censorshipstate-1]} \prior_i\receiverU_i \leq 0 \right\}$.
\end{restatable}
The feasibility program~\ref{eq:SISU_opt_condition}
recovers the structure of the optimal censorship signaling scheme
for a fully rational receiver.
For a fully rational receiver 
(whose rationality level $\noiseScale=\infty$),
the quantal response function $\Quant\betaed(\cdot)$ becomes $\Quant\betaed(\variableX) = \indicator{\variableX \leq 0}$.
In this case, there is no longer a bijection between 
$\censorshipsignalInverse\in[0,\infty)$
and $\censorshipsignal\in(-\infty,0]$
satisfying constraint \dfone.
Instead, the feasible solutions of constraint
\dfone\ admit one of the two forms:
either (i) $\{\censorshipsignalInverse \in [0,\infty), \censorshipsignal = 0\}$;
or (ii) $\{\censorshipsignalInverse = \infty, 
\censorshipsignal\in(-\infty, 0]\}$.
Note that (i) $\{\censorshipsignalInverse \in [0,\infty), \censorshipsignal = 0\}$
corresponds to instances where $\sum_{i\in[\stateNum]}\prior_i\receiverU_i > 0$ and 
thus the optimal censorship signaling scheme 
in \Cref{lem:opt_fully_rational}
selects the threshold state and 
the threshold-state fraction such that 
$\censorshipsignal=0$, i.e.,
the fully rational receiver is indifferent between
action 0 and action 1 
when the pooling
signal $\censorshipsignal$ is realized.
On the other side, 
(ii) $\{\censorshipsignalInverse = \infty, 
\censorshipsignal\in(-\infty, 0]\}$
corresponds to instances where $\sum_{i\in[\stateNum]}\prior_i\receiverU_i \le 0$ and 
thus the optimal censorship signaling scheme
in \Cref{lem:opt_fully_rational}
sets the threshold state $\censorshipstate = \argmax_{i \in [\stateNum]}\{
\receiverU_i:\receiverU_i\leq \censorshipsignalInverse\} = \stateNum$,
i.e.,
pools all states together
and reveals no information.

\subsection{Proof of 
\texorpdfstring{\Cref{prop:SISU_threshold_state_monotonicity}}{}}
\label{apx:SISU_proof_SISU_threshold_state monotonicity}
Throughout this subsection, we focus on rationality level $\noiseScale \in (0,\infty)$.
To simplify the analysis, we first introduce the following definition.
For any $\scaledNoiseDiff\in[0, \infty)$, we define its tangent contact point
$\tangentIntersect(\scaledNoiseDiff)\in(- \infty, 0]$ such that 
\begin{align}
    \label{eq:tagent_line}
    \Quant\betaed'(\tangentIntersect(\scaledNoiseDiff)) 
    = \frac{ \Quant\betaed(\scaledNoiseDiff)  - \Quant\betaed(\tangentIntersect(\scaledNoiseDiff))}{\scaledNoiseDiff - \tangentIntersect(\scaledNoiseDiff)}~.
\end{align}
Clearly, we have $\censorshipsignal = \tangentIntersect(\censorshipsignalInverse)$
where $\censorshipsignal$ and $\censorshipsignalInverse$ are defined
in \Cref{thm:opt_state_independent}.
By the curvature of the quantal response function $\Quant\betaed(\cdot)$, namely, $\Quant\betaed(\cdot)$ is 
concave over $(-\infty, 0]$ and convex over $[0, \infty)$, 
we have the following property about tangent contact point function
$\tangentIntersect(\cdot)$:
\begin{lemma}
\label{lem:kappa_property}
Tangent contact point function $\tangentIntersect(\cdot)$ is a bijection from $[0, \infty)$ to $ (- \infty, 0]$,
i.e., for any $\scaledNoiseDiff \in[0, \infty)$, 
there exists a unique $\tangentIntersect(\scaledNoiseDiff) \in (- \infty, 0]$  such that Eqn.~\eqref{eq:tagent_line} holds.
Moreover, tangent contact point function $\tangentIntersect(\cdot)$ is decreasing 
as $\scaledNoiseDiff \in[0, \infty)$ increases. 
\end{lemma}
\begin{proof}
Recall that $\Quant\betaed(\variableX) = \frac{1}{1+\exp(\noiseScale \variableX)}$,
and $\frac{\partial \Quant\betaed(\variableX)}{\partial \variableX}
= -\noiseScale \cdot \frac{\exp(\noiseScale \variableX)}{(1+\exp(\noiseScale \variableX))^2}$,
from Eqn.~\eqref{eq:tagent_line}, for a fixed $\scaledNoiseDiff \ge 0$,
its tangent contact point $\tangentIntersect(\scaledNoiseDiff)$
is the root of the  function
$
    \tangentfunc(\variableX, \scaledNoiseDiff) \triangleq 
    -\noiseScale \cdot \frac{\exp(\noiseScale \variableX)}{(1+\exp(\noiseScale \variableX))^2} \cdot(x - \scaledNoiseDiff) - \frac{1}{1+\exp(\noiseScale \variableX)} + \frac{1}{1+\exp(\noiseScale \scaledNoiseDiff)}$.
Inspecting its first-order partial derivatives, we can see that
$\frac{\partial \tangentfunc(\variableX, \scaledNoiseDiff)}{\partial \variableX}
> 0, \forall \variableX < 0; 
\left.\frac{\partial \tangentfunc(\variableX, \scaledNoiseDiff)}{\partial \scaledNoiseDiff}\right|_{\variableX = \tangentIntersect(\scaledNoiseDiff)} > 0$.
Note that $
\lim\nolimits_{\variableX\to-\infty}\tangentfunc(\variableX,\scaledNoiseDiff)<0, \tangentfunc(0,\scaledNoiseDiff)>0$ for $\scaledNoiseDiff>0$.
Hence there is a unique negative root
$\tangentIntersect(\scaledNoiseDiff)$.
Then, differentiating $\tangentfunc(\tangentIntersect(\scaledNoiseDiff),\scaledNoiseDiff)=0$ with respect to $\scaledNoiseDiff$, we obtain $\frac{\partial \tangentIntersect(\scaledNoiseDiff)}{\partial \scaledNoiseDiff} = - \left(\left.\frac{\partial \tangentfunc(\variableX, \scaledNoiseDiff)}{\partial \scaledNoiseDiff}\right|_{\variableX = \tangentIntersect(\scaledNoiseDiff)}\right)\Big/\frac{\partial \tangentfunc(\tangentIntersect(\scaledNoiseDiff),\scaledNoiseDiff)}{\partial \tangentIntersect(\scaledNoiseDiff)} <0$. 
As a consequence, given $\scaledNoiseDiff_2 > \scaledNoiseDiff_1 > 0$, we have $\tangentIntersect(\scaledNoiseDiff_2) < \tangentIntersect(\scaledNoiseDiff_1)$.
Finally, we consider the endpoint limits.
Note that $\tangentfunc(-\scaledNoiseDiff,\scaledNoiseDiff)<0$.
Since $\tangentfunc(\cdot,\scaledNoiseDiff)$ is strictly increasing on
$(-\infty,0)$, this implies
$-\scaledNoiseDiff<\tangentIntersect(\scaledNoiseDiff)<0$.
Hence $\tangentIntersect(\scaledNoiseDiff)\to0$ when $\scaledNoiseDiff\to 0$.
If $\tangentIntersect(\scaledNoiseDiff)$ remained bounded as
$\scaledNoiseDiff\to\infty$, then
$\Quant\betaed'(\tangentIntersect(\scaledNoiseDiff))(\tangentIntersect(\scaledNoiseDiff)-\scaledNoiseDiff)$
would tend to positive infinity, contradicting
$\tangentfunc(\tangentIntersect(\scaledNoiseDiff),\scaledNoiseDiff)=0$.
Thus $\tangentIntersect(\scaledNoiseDiff)\to-\infty$ as $\scaledNoiseDiff\to\infty$.
This completes the proof of \Cref{lem:kappa_property}.
\end{proof}
With the above definition in Eqn.~\eqref{eq:tagent_line},
we also define the candidate threshold fraction $\thresholdfrac_i$ for each state $i \in [\stateNum]$ with $\receiverU_i \ge 0$ as
\begin{align}
    \label{eq:pooled_prob_defn}
    \thresholdfrac_i \triangleq
    -\frac{\sum_{j\in [i-1]} \prior_j  \cdot (\receiverU_j - \tangentIntersect(\receiverU_i) )}{\prior_i 
    \cdot( \receiverU_i  - \tangentIntersect(\receiverU_i))}~.
\end{align}

\begin{lemma}
\label{lem:unique_valid_p_i}
For any state $i\in[\stateNum]$ with $\receiverU_i \ge  0$, 
if the corresponding candidate threshold fraction $\thresholdfrac_i < 0$, then it holds that $\thresholdfrac_j < 0, \forall j \in [i+1,\stateNum]$.
\end{lemma}
\begin{proof}
Consider two states $i_1, i_2 \in [\stateNum]$ where 
$i_1 < i_2$ and $ \receiverU_{i_1} \ge 0$.
Then, we have candidate threshold fractions $
    \thresholdfrac_{i_1} = -\frac{\sum_{j \in [i_1-1]} \prior_j  (\receiverU_j - \tangentIntersect(\receiverU_{i_1}))}{\prior_{i_1} \cdot(\receiverU_{i_1} - \tangentIntersect(\receiverU_{i_1}))}~\text{and}~
    \thresholdfrac_{i_2} = -\frac{\sum_{j \in [i_2-1]} \prior_j (\receiverU_j - \tangentIntersect(\receiverU_{i_2}))}{\prior_{i_2} \cdot(\receiverU_{i_2} - \tangentIntersect(\receiverU_{i_2}))}$.
Suppose candidate threshold fraction $\thresholdfrac_{i_1} < 0$. 
Since $\receiverU_{i_1} - \tangentIntersect(\receiverU_{i_1}) > 0$, it must imply 
that $\sum_{j\in [i_1-1]} \prior_j  (\receiverU_j - \tangentIntersect(\receiverU_{i_1})) > 0$.
Observe that 
$
    \sum_{j \in [i_2-1]} \prior_j (\receiverU_j - \tangentIntersect(\receiverU_{i_2}))
    = \sum_{j \in [i_1-1]} \prior_j (\receiverU_j - \tangentIntersect(\receiverU_{i_2}))
    + \sum_{j \in [i_1,i_2-1]} \prior_j (\receiverU_j - \tangentIntersect(\receiverU_{i_2}))
    > \sum_{j \in [i_1,i_2-1]} \prior_j (\receiverU_j - \tangentIntersect(\receiverU_{i_2})) > 0$,
where the last inequality follows from 
the fact that $\receiverU_j > 0, \forall i_1 \le j < i_2$
and $\tangentIntersect(\receiverU_{i_2}) < 0$.
Hence, with the fact that $\receiverU_{i_2} - \tangentIntersect(\receiverU_{i_2}) > 0$, 
one must have $\thresholdfrac_{i_2} < 0$. This completes the proof of \Cref{lem:unique_valid_p_i}.
\end{proof}

We are ready to prove \Cref{prop:SISU_threshold_state_monotonicity}. 
\sisustatemonotone*
\begin{proof}[Proof]
We begin the analysis by showing the following observation: 
Fix a $\scaledNoiseDiff \in [0, \infty)$, 
let $\tangentIntersect(\scaledNoiseDiff)$ (resp. $\hat{\tangentIntersect}(\scaledNoiseDiff)$)
be the tangent contact point that satisfies Eqn.~\eqref{eq:tagent_line}
for the rationality level $\noiseScale$ 
(resp. $\hat{\noiseScale}$).
Then we have 
\begin{align*}
    \tangentIntersect(\scaledNoiseDiff)
    \le 
    \hat{\tangentIntersect}(\scaledNoiseDiff)
    < 0, 
    \quad
    \text{ if }\noiseScale \le \hat{\noiseScale}~.
\end{align*}
To see this, 
recall that $\Quant\betaed(\variableX) = \frac{1}{1+\exp(\noiseScale \variableX)}$,
and $\frac{\partial \Quant\betaed(\variableX)}{\partial \variableX}
= -\noiseScale \cdot \frac{\exp(\noiseScale \variableX)}{(1+\exp(\noiseScale \variableX))^2}$,
from Eqn.~\eqref{eq:tagent_line}, 
the tangent contact point $\tangentIntersect(\scaledNoiseDiff)$
is the root of the function
$
    \tangentfunc(\variableX, \noiseScale) \triangleq 
    -\noiseScale \cdot \frac{\exp(\noiseScale \variableX)}{(1+\exp(\noiseScale \variableX))^2} \cdot(\variableX - \scaledNoiseDiff) - \frac{1}{1+\exp(\noiseScale \variableX)} + \frac{1}{1+\exp(\noiseScale \scaledNoiseDiff)}$.
Inspecting its first-order partial derivatives, we can see that
$\frac{\partial \tangentfunc(\variableX, \noiseScale)}{\partial \variableX}
\ge 0$, for all $\variableX\le 0$, and $
\left.\frac{\partial \tangentfunc(\variableX, \noiseScale)}{\partial \noiseScale}\right|_{\variableX = \tangentIntersect(\scaledNoiseDiff)} \le 0$.
Then, differentiating $f(\tangentIntersect(\scaledNoiseDiff),\noiseScale)= 0 $ with respect to $\noiseScale$, we obtain
$\frac{\partial \tangentIntersect(\scaledNoiseDiff)}{\partial \noiseScale} = -\frac{\partial \tangentfunc( \tangentIntersect(\scaledNoiseDiff), \noiseScale)}{\partial \noiseScale}\big/\frac{\partial \tangentfunc(\tangentIntersect(\scaledNoiseDiff), \noiseScale)}{\partial \tangentIntersect(\scaledNoiseDiff)} \ge 0 $. Thus,
given $\hat{\noiseScale} \ge \noiseScale$, we have $\hat{\tangentIntersect}(\scaledNoiseDiff) \ge \tangentIntersect(\scaledNoiseDiff)$.


Now given a state $i$ where $\receiverU_i \ge 0$,
consider the rationality levels $\noiseScale, \hat{\noiseScale}$ ($\noiseScale\le \hat{\noiseScale}$),
from the above inequality related to the comparison of tangent contact points,
we have
$\tangentIntersect(\receiverU_i)
\le 
\hat{\tangentIntersect}(\receiverU_i)$, 
implying
\begin{align*}
    \thresholdfracHat_i-\thresholdfrac_i = \frac{(\hat{\tangentIntersect}(\receiverU_i)-\tangentIntersect(\receiverU_i))\sum\nolimits_{j \in [i-1]}\prior_j(\receiverU_i-\receiverU_j)}{\prior_i(\receiverU_i-\tangentIntersect(\receiverU_i))(\receiverU_i-\hat{\tangentIntersect}(\receiverU_i))} \ge 0\,.
\end{align*}
The above inequality ensures that 
the threshold state $\censorshipstateHat$ for a larger
rationality level $\hat{\noiseScale}$
is no smaller than the threshold state $\censorshipstate$ for a smaller
rationality level $\noiseScale$. 
If $\censorshipstateHat = \censorshipstate$, one still has
$\censorshipprob = \min\{\thresholdfrac_{\censorshipstate}, 1\}
\le \censorshipprobHat = \min\{\thresholdfracHat_{\censorshipstateHat}, 1\}$.

Let \((\censorshipstate,\censorshipprob)\) and \((\censorshipstateHat,\censorshipprobHat)\) be the optimal threshold state and optimal threshold-state fraction selected at rationality levels \(\noiseScale \le \hat{\noiseScale}\). The preceding argument gives \(\censorshipstate\le\censorshipstateHat\) and, when the thresholds coincide, \(\censorshipprob\le\censorshipprobHat\).
If $\censorshipstate = \censorshipstateHat$, we have $\pooledmassHat - \pooledmass = \prior_{\censorshipstate} (\censorshipprobHat - \censorshipprob)\ge0$.
If $\censorshipstate < \censorshipstateHat$, we have $\pooledmass = \sum\nolimits_{i \in [\censorshipstate-1]}\prior_i+ \prior_{\censorshipstate}\censorshipprob \le \sum\nolimits_{i \in [\censorshipstate]}\prior_i \le \sum\nolimits_{i \in [\censorshipstateHat-1]}\prior_i \le \sum\nolimits_{i \in [\censorshipstateHat-1]}\prior_i+ \prior_{\censorshipstateHat}\censorshipprobHat = \pooledmassHat$.
Thus, the optimal pooled mass is monotone non-decreasing with respect to the rationality level.
This completes the proof of \Cref{prop:SISU_threshold_state_monotonicity}.
\end{proof}

\subsection{Uniqueness of the Feasible Solution for Program in \texorpdfstring{\Cref{thm:opt_state_independent}}{Theorem 1}}
\label{apx:SISU_proof_opt_condition_existence}

\begin{lemma}
\label{lem:existence_SISU_opt_condition}
Fix any instance $\instance$ and any rationality level
$\noiseScale\in(0,\infty)$.
There exists a unique feasible solution 
in program~\ref{eq:SISU_opt_condition}.
\end{lemma}

\Cref{lem:existence_SISU_opt_condition} guarantees that the constraints in program~\ref{eq:SISU_opt_condition} always admit a feasible solution, establishing the existence of an optimal censorship for every $\noiseScale \in (0,\infty)$, and that the resulting tuple $(\censorshipstate, \censorshipprob, \censorshipsignal, \censorshipsignalInverse)$ is uniquely determined by the program.
With the two properties (see \Cref{lem:kappa_property} and \Cref{lem:unique_valid_p_i}) 
for tangent contact point function $\tangentIntersect(\cdot)$ and candidate threshold fraction $\thresholdfrac_i$
we established in \Cref{apx:SISU_proof_SISU_threshold_state monotonicity}, we 
prove \Cref{lem:existence_SISU_opt_condition} as follows:

\begin{proof}[Proof of \Cref{lem:existence_SISU_opt_condition}]
Let
$\censorshipstate=\max\{i\in[\stateNum]:
\receiverU_i>0,\ \thresholdprob_i\ge0\}$
and
$\censorshipprob=\min\{\thresholdprob_{\censorshipstate},1\}$.
If this set is empty, let
$\censorshipstate=\max\{i\in[\stateNum]:\receiverU_i\le0\}$
and $\censorshipprob=1$.
Let $\censorshipsignal$ be the corresponding pooling signal.

We first verify feasibility.
When $\censorshipprob<1$, Eqn.~\eqref{eq:pooled_prob_defn}
gives
$\censorshipsignal=\tangentIntersect(\receiverU_{\censorshipstate})$.
Hence, setting
$\censorshipsignalInverse=\receiverU_{\censorshipstate}$
satisfies all constraints.
When $\censorshipprob=1$, we have $\censorshipsignal<0$:
if $\receiverU_{\censorshipstate}>0$, this follows from
$\thresholdprob_{\censorshipstate}\ge1$ and
Eqn.~\eqref{eq:pooled_prob_defn}; otherwise, all non-positive
states are pooled.
By \Cref{lem:kappa_property}, there is a unique $\censorshipsignalInverse>0$
such that
$\censorshipsignal=\tangentIntersect(\censorshipsignalInverse)$.
Moreover, we have $\censorshipsignalInverse\ge\receiverU_{\censorshipstate}$ since
$\censorshipsignal\le
\tangentIntersect(\receiverU_{\censorshipstate})$
when $\receiverU_{\censorshipstate}>0$, and the inequality is
immediate otherwise.
If $\censorshipstate<\stateNum$, then
$\thresholdprob_{\censorshipstate+1}<0$, so
Eqn.~\eqref{eq:pooled_prob_defn} gives
$
\censorshipsignal>
\tangentIntersect(\receiverU_{\censorshipstate+1})$ and $
\censorshipsignalInverse<\receiverU_{\censorshipstate+1}$.
If $\censorshipstate=\stateNum$,
$\censorshipsignalInverse\ge\receiverU_{\stateNum}$ suffices.
Thus, constraint \dftwo\ holds, and the other constraints follow from
the construction.

Conversely, every feasible solution with
$\receiverU_{\censorshipstate}>0$ must have
$\censorshipprob=\min\{\thresholdprob_{\censorshipstate},1\}$
by constraints \dftwo, \cs, and Eqn.~\eqref{eq:pooled_prob_defn}.
If $\censorshipstate<\stateNum$, then the pooling signal
$\censorshipsignal
=\tangentIntersect(\censorshipsignalInverse)
>\tangentIntersect(\receiverU_{\censorshipstate+1})$.
Pooling the entire threshold state weakly increases this signal,
so Eqn.~\eqref{eq:pooled_prob_defn} implies the candidate threshold fraction
$\thresholdprob_{\censorshipstate+1}<0$.
By \Cref{lem:unique_valid_p_i}, the threshold state must therefore
be the one specified above.
If $\receiverU_{\censorshipstate}\le0$, the constraints require
$\censorshipstate=\max\{i\in[\stateNum]:\receiverU_i\le0\}$
and $\censorshipprob=1$.
Hence the threshold state and threshold-state fraction are unique.
They determine the pooling signal $\censorshipsignal$, and \Cref{lem:kappa_property} uniquely
determines the auxiliary variable $\censorshipsignalInverse$.
\end{proof}

\section{Missing Proofs in \texorpdfstring{\Cref{sec:robust_optimal_scheme_unkown_rationality_level}}{Section 4}}
\label{apx:robust_optimal_scheme_unkown_rationality_level}
In this section, we establish the admissible range for the rationality-robust optimal pooled mass, explain why strategically revealing information benefits the receiver, and complete the proof of \Cref{thm:robust_optimal_censorship_three_criteria} by providing the remaining steps for the maximin ratio criterion along with the proofs for the maximin payoff and minimax regret criteria.

\subsection{Bounds on Rationality-Robust Optimal Pooled Mass}
\label{apx:same_pooled_mass_interval}

\begin{proposition}
\label{prop:same_pooled_mass_interval}
 Fix any instance $\instance$. For each of the three robustness criteria (maximin ratio, maximin payoff, and minimax regret), a rationality-robust optimal censorship can be chosen
with pooled mass in $[\nonpositivemass,\fullyrationalmass]$ where $\nonpositivemass$ is the total mass of states with non-positive values and $\fullyrationalmass$ is the pooled mass of optimal censorship for a fully rational receiver. 
\end{proposition}
Note that the proof of \Cref{prop:same_pooled_mass_interval} establishes a stronger pointwise dominance property: any censorship with pooled mass outside $[\nonpositivemass,\fullyrationalmass]$ yields a weakly lower expected utility for the sender, at every rationality level $\noiseScale$, than one with pooled mass $\nonpositivemass$ or $\fullyrationalmass$.
Restricting attention to $[\nonpositivemass,\fullyrationalmass]$ provides the basis for both computing rationality-robust optimal signaling schemes in \Cref{prop:computation_rationality_robust_scheme} and establishing the universal robust performance guarantee in \Cref{subsec:universal_guarantee_maximin_optimal_ratio}.
\begin{proof}[Proof of \Cref{prop:same_pooled_mass_interval}]
We argue that for any rationality level $\noiseScale$, the sender's expected utility under the censorship $\signalscheme^{(\nonpositivemass)}$ with pooled mass $\nonpositivemass$ is weakly larger than that under the censorship $\signalscheme^{(\pooledmass)}$ with pooled mass $\pooledmass \in [0,\nonpositivemass)$, and the sender's expected utility under the censorship $\signalscheme^{(\fullyrationalmass)}$ with pooled mass $\fullyrationalmass$ is weakly larger than that under the censorship $\signalscheme^{(\pooledmass)}$ with pooled mass $\pooledmass \in (\fullyrationalmass,1]$.

\xhdr{Case 1: $\pooledmass \in [0,\nonpositivemass)$}
Fix a finite rationality level $\noiseScale \in [0,\infty)$.
The sender's expected utility under the censorship $\signalscheme^{(\pooledmass)}$ is given by
\begin{align*}
    \Payoff[\noiseScale]{\signalscheme^{(\pooledmass)}} &= \pooledmass \Quant\betaed(\censorshipsignal(\pooledmass)) + (1-\censorshipprob)\prior_{\censorshipstate(\pooledmass)}\Quant\betaed(\receiverU_{\censorshipstate(\pooledmass)}) + \sum\nolimits_{i\in [\censorshipstate(\pooledmass)+1:\censorshipstate(\nonpositivemass)]}\prior_i\Quant\betaed(\receiverU_{i})+ \sum\nolimits_{i\in (\censorshipstate(\nonpositivemass):\stateNum]}\prior_i\Quant\betaed(\receiverU_{i})\\
    & \le \nonpositivemass \Quant\betaed\left(\frac{\pooledmass\censorshipsignal(\pooledmass)+ (1-\censorshipprob(\pooledmass))\prior_{\censorshipstate(\pooledmass)}\receiverU_{\censorshipstate(\pooledmass)}+\sum\nolimits_{i \in [\censorshipstate(\pooledmass)+1:\censorshipstate(\nonpositivemass)]}\prior_i\receiverU_i}{\nonpositivemass}\right) + \sum\nolimits_{i\in (\censorshipstate(\nonpositivemass):\stateNum]}\prior_i\Quant\betaed(\receiverU_{i})\\
    &= \nonpositivemass \Quant\betaed(\censorshipsignal(\nonpositivemass))+\sum\nolimits_{i\in (\censorshipstate(\nonpositivemass):\stateNum]}\prior_i\Quant\betaed(\receiverU_{i}) = \Payoff[\noiseScale]{\signalscheme^{(\nonpositivemass)}}\,,
\end{align*}
where the inequality holds due to the concavity of quantal response function $\Quant\betaed(\cdot)$ on $(-\infty,0]$.
For the fully rational case $\noiseScale = \infty$, every non-positive signal induces action \(1\), while every positive revealing signal induces action \(0\). Therefore, pooling all non-positive states leaves their total utility contribution unchanged, i.e., $\Payoff[\infty]{\signalscheme^{(\pooledmass)}} = \nonpositivemass = \Payoff[\infty]{\signalscheme^{(\nonpositivemass)}} $.

\xhdr{Case 2. $\pooledmass \in (\fullyrationalmass,1]$}
In that case, we have $\censorshipsignal(\fullyrationalmass)=0$ (see \Cref{lem:opt_fully_rational}).
Fix a finite rationality level $\noiseScale \in [0,\infty)$.
The sender's expected utility under the censorship $\signalscheme^{(\pooledmass)}$ is given by
\begin{align*}
    \Payoff[\noiseScale]{\signalscheme^{(\pooledmass)}} &= \pooledmass \Quant\betaed(\censorshipsignal(\pooledmass)) + (1-\censorshipprob)\prior_{\censorshipstate(\pooledmass)}\Quant\betaed(\receiverU_{\censorshipstate(\pooledmass)}) + \sum\limits_{i\in [\censorshipstate(\pooledmass)+1:\stateNum]}\prior_i\Quant\betaed(\receiverU_{i})\\
    & \le \fullyrationalmass\Quant\betaed(\censorshipsignal(\fullyrationalmass)) + (1-\censorshipprob(\fullyrationalmass))\prior_{\censorshipstate(\fullyrationalmass)}\Quant\betaed(\receiverU_{\censorshipstate(\fullyrationalmass)}) +\sum\limits_{i \in(\censorshipstate(\fullyrationalmass),\censorshipstate(\pooledmass)] }\prior_i\Quant\betaed(\receiverU_i) 
    + \sum\limits_{i\in [\censorshipstate(\pooledmass)+1:\stateNum]}\prior_i\Quant\betaed(\receiverU_{i})\\
    &= \Payoff[\noiseScale]{\signalscheme^{(\fullyrationalmass)}} \,,
\end{align*}
where the inequality holds due to the convexity of quantal response function $\Quant\betaed(\cdot)$ on $[0,\infty)$.
For the fully rational case $\noiseScale=\infty$, the pooling signal $\censorshipsignal(\pooledmass)$ is strictly positive, and all remaining revealing signals are also strictly positive. Hence, we have $\Payoff[\infty]{\signalscheme^{(\pooledmass})}=0$.
For the censorship $\signalscheme^{(\fullyrationalmass)}$, the zero pooling signal $\censorshipsignal(\fullyrationalmass)$ induces action 1 by sender-favorable tie-breaking.
Consequently, we have $\Payoff[\infty]{\signalscheme^{(\fullyrationalmass)}}=\fullyrationalmass>0$.

Note that dividing by $\OPT(\noiseScale)$, leaving the payoff unchanged,
or subtracting $\OPT(\noiseScale)$ preserves the pointwise comparison.
Therefore, combining the above cases, we can conclude that the pooled mass of rationality-robust optimal censorship can be chosen from $[\nonpositivemass,\fullyrationalmass]$.
This completes the proof of \Cref{prop:same_pooled_mass_interval}.
\end{proof}

\subsection{Strategically Revealing More Information Benefits the Receiver}
\label{apx:receiver_utility_information_revealing}

\begin{proposition}
\label{prop:receiver_utility_information_revealing}
Fix any instance $\instance$. For any two censorships $\signalscheme$ and $\rivalsignalscheme$ with pooled mass $\pooledmass$ and $\rivalpooledmass$, if $\nonpositivemass\le \pooledmass \le \rivalpooledmass\le \fullyrationalmass$, where $\nonpositivemass$ is the total mass of non-positive states and  $\fullyrationalmass$ is the pooled mass for the optimal censorship for a fully rational receiver,
the receiver's expected utility  under the censorship $\signalscheme$ is weakly larger than that under the censorship $\rivalsignalscheme$ for any rationality level.
\end{proposition}

The condition $\nonpositivemass \le \pooledmass \le \rivalpooledmass\le \fullyrationalmass$ in
\Cref{prop:receiver_utility_information_revealing}
is essential for this utility comparison.
In particular, \Cref{exa:full_bounded_rationality} shows that disclosing additional information (by reducing the pooled mass below $\nonpositivemass$) can strictly decrease a quantal-response receiver's expected utility when this condition fails.

\begin{proof}[Proof of \Cref{prop:receiver_utility_information_revealing}]
Fix any rationality level $\noiseScale\in\fullsetrationalitylevel$.
Let $\censorshipsignal$ be the pooling signal of censorship $\signalscheme$ and 
$\censorshipsignalHat$ be the pooling signal of censorship $\rivalsignalscheme$.
Since the state values are ordered increasingly, it means that if a censorship has a larger pooled mass, it also has a larger pooling signal, i.e., $\censorshipsignal\le\censorshipsignalHat\le0$.

For the censorship $\signalscheme$, the receiver's expected utility is given by
\begin{align*}
    \receiverexpU &= \sum\nolimits_{i\in [\stateNum]}\prior_i \int_{-\infty}^{\infty} \signalscheme_{i}(\scaledNoiseDiff)\left( (1-\Quant\betaed(\scaledNoiseDiff))\receiverU_i(0)+\Quant\betaed(\scaledNoiseDiff)\receiverU_i(1)\right)\dd \scaledNoiseDiff \\
    &= \sum\nolimits_{i \in [\stateNum]}\prior_i\receiverU_i(0)-\sum\nolimits_{i \in [\stateNum]}\prior_i\receiverU_i \int_{-\infty}^{\infty} \signalscheme_{i}(\scaledNoiseDiff)\Quant\betaed(\scaledNoiseDiff) \dd \scaledNoiseDiff\\
    & = \sum\nolimits_{i\in[\stateNum]}\prior_i\receiverU_i(0)
-\pooledmass\censorshipsignal
 \Quant_{\noiseScale}(\censorshipsignal)-\sum\nolimits_{i\in [\stateNum]}
 \prior_i(1-\signalscheme_{i}(\censorshipsignal))\receiverU_i\Quant_{\noiseScale}(\receiverU_i)\,,
\end{align*}
where the last equality holds since $\pooledmass\censorshipsignal = \sum\nolimits_{i \in [\censorshipstate-1]}\prior_i\receiverU_i+\signalscheme_{\censorshipstate}(\censorshipsignal)\prior_{\censorshipstate}\receiverU_{\censorshipstate}$.

The analogous expression holds for receiver's expected utility $\rivalreceiverexpU$ under the censorship $\rivalsignalscheme$ with
$\rivalpooledmass,\censorshipsignalHat$.
Next, let us compare them by subtraction.
\begin{align*}
   \receiverexpU-\rivalreceiverexpU & = \rivalpooledmass\censorshipsignalHat\Quant\betaed(\censorshipsignalHat)- \pooledmass\censorshipsignal\Quant\betaed(\censorshipsignal)-\sum\nolimits_{i \in [\stateNum]}\prior_i(\rivalsignalscheme_{i}(\censorshipsignalHat)-\signalscheme_{i}(\censorshipsignal))\receiverU_i\Quant\betaed(\receiverU_i) \\
   & = \pooledmass\censorshipsignal(\Quant\betaed(\censorshipsignalHat)-\Quant\betaed(\censorshipsignal)) + \sum\nolimits_{i \in [\stateNum]}\prior_i(\rivalsignalscheme_{i}(\censorshipsignalHat)-\signalscheme_{i}(\censorshipsignal))\receiverU_i(\Quant\betaed(\censorshipsignalHat)-\Quant\betaed(\receiverU_i))\,,
\end{align*}
where the equality holds due to identities $\rivalpooledmass\censorshipsignalHat-\pooledmass\censorshipsignal=\sum\nolimits_{i \in [\stateNum]}\prior_i(\rivalsignalscheme_{i}(\censorshipsignalHat)-\signalscheme_{i}(\censorshipsignal))\receiverU_i$.
Note that the first term is non-negative: both
$\pooledmass\censorshipsignal$ and
$\Quant_{\noiseScale}(\censorshipsignalHat)
 -\Quant_{\noiseScale}(\censorshipsignal)$ are non-positive, where $\censorshipsignal \le \censorshipsignalHat$ and the quantal response function $\Quant_{\noiseScale}(\cdot)$ is non-increasing.
The second term with positive probability mass has
$\receiverU_i>0$, and its first bracket is non-negative since censorship $\rivalsignalscheme$ pools weakly more states than censorship $\signalscheme$ and second bracket is non-negative because
$\censorshipsignalHat\le0<\receiverU_i$ and
the quantal response function $\Quant_{\noiseScale}(\cdot)$ is non-increasing.
Therefore, we can conclude that $\receiverexpU\ge\rivalreceiverexpU$.
This completes the proof of \Cref{prop:receiver_utility_information_revealing}.
\end{proof}

\subsection{Completing the Proof of \texorpdfstring{\Cref{thm:robust_optimal_censorship_three_criteria}}{Theorem 2} for the Maximin Ratio Criterion}
\label{apx:limiting-arguments}
We extend the argument from compact domains to the full set of rationality levels $\fullsetrationalitylevel$. For each integer $\integer\in \mathbb{N}$, let $\rationalityIntclass \triangleq [0, \integer]$.
By the previous argument on compact domain  $\rationalityIntclass$, there exists a censorship $\signalscheme_\integer$ such that
    $\ratiointeger 
    \triangleq 
    \inf\nolimits_{\noiseScale \in \rationalityIntclass} 
    \frac{\Payoff[\noiseScale]{\signalscheme_\integer}}{\OPT(\noiseScale)}=  
    \sup\nolimits_{\signalscheme \in \schemespace}\inf\nolimits_{\noiseScale \in \rationalityIntclass} \frac{\Payoff[\noiseScale]{\signalscheme}}{\OPT(\noiseScale)}$.
Since $\rationalityIntclass \subset \rationalityClass_{\integer+1}$, the sequence $\{\ratiointeger\}_{\integer\in \mathbb{N}}$ is monotone non-increasing. We define its limit as
$
    \ratiolimit \triangleq \lim\nolimits_{\integer \rightarrow \infty} \ratiointeger$.
Because the space of censorships $\censorshipspace$ is compact, 
there exists a convergent subsequence of censorships $\{\signalscheme_{\integer_{\integernew}}\}_{\integernew \in \mathbb{N}}$ approaching a limit censorship $\schemeoptimal \in \censorshipspace$.
For any finite rationality level $\noiseScale \in [0,\infty)$, the receiver's quantal response function $\Quant\betaed(\cdot)$ is smooth, and thus the sender's expected utility varies continuously with the chosen signaling scheme. Taking the limit as $\integernew \to \infty$ where $\signalscheme_{\integer_\integernew} \to \schemeoptimal$, we obtain
\begin{align*} 
\frac{\Payoff[\noiseScale]{\schemeoptimal}}{\OPT(\noiseScale)} = \lim\nolimits_{\integernew \to \infty} \frac{\Payoff[\noiseScale]{\signalscheme_{\integer_\integernew}}}{\OPT(\noiseScale)} \ge \lim\nolimits_{\integernew \to \infty} \ratio_{\integer_\integernew} 
    = \ratiolimit\,.
\end{align*}
We next analyze the boundary case of a fully rational receiver, i.e., $\noiseScale = \infty$. 
Because $\schemeoptimal$ is a fixed limit censorship, it induces a fixed marginal distribution over the signals $\scaledNoiseDiff$.
For finite rationality level \(\noiseScale\), the quantal response function $\Quant\betaed(\variableX)$ 's pointwise limit as \(\noiseScale\to\infty\) is
$\lim\nolimits_{\noiseScale\to\infty}\Quant\betaed(\variableX)=1$ for $\variableX<0$, $\lim\nolimits_{\noiseScale\to\infty}\Quant\betaed(\variableX)=\sfrac{1}{2}$ for $\variableX=0$, or $\lim\nolimits_{\noiseScale\to\infty}\Quant\betaed(\variableX)=0$ for $\variableX>0$.
By our sender-favorable tie-breaking convention for the fully rational receiver, we have
$
\lim\nolimits_{\noiseScale\to\infty}\Quant\betaed(\variableX)\le \Quant_\infty(\variableX)$.
Because the limit censorship \(\schemeoptimal\) is fixed and the sequence of quantal response functions $\{\Quant\betaed(\cdot)\}_{\noiseScale \ge 0}$ is uniformly bounded, Lebesgue's dominated convergence theorem \citep{L-09} guarantees that we can interchange the limit and the integral. 
Therefore, the sender's expected utility under the limit censorship $\schemeoptimal$ converges to the
expected utility evaluated under the pointwise limit of the quantal response functions.
Recall that this pointwise limit is no larger than the sender-favorable
fully rational response $\Quant_\infty(\cdot)$. 
Thus, we have
$
    \lim\nolimits_{\noiseScale \to \infty} \Payoff[\noiseScale]{\schemeoptimal} 
    = \lim\nolimits_{\noiseScale \to \infty}  \sum\nolimits_{i\in[\stateNum]}\prior_{i}  \int\nolimits_{-\infty}^{\infty} \schemeoptimal_i(\scaledNoiseDiff) \Quant\betaed(\scaledNoiseDiff) \dd\scaledNoiseDiff = \sum\nolimits_{i\in[\stateNum]}\prior_{i}  \int_{-\infty}^{\infty} \schemeoptimal_i(\scaledNoiseDiff)   \left( \lim\nolimits_{\noiseScale \to \infty}  \Quant\betaed(\scaledNoiseDiff) \right) \dd\scaledNoiseDiff
    \le \sum\nolimits_{i\in[\stateNum]}\prior_{i}  \int_{-\infty}^{\infty} \schemeoptimal_i(\scaledNoiseDiff)      \Quant_\infty(\scaledNoiseDiff)  \,\dd\scaledNoiseDiff  = \Payoff[\infty]{\schemeoptimal}.
$
Furthermore, we have $\lim_{\noiseScale \to \infty} \OPT(\noiseScale) = \OPT(\infty)$.
Let $\fullyrationalmass \triangleq \OPT(\infty)$.
To see this, first note that it holds that $\OPT(\noiseScale) \le \fullyrationalmass + \sum_{i \in [\stateNum]:\, \receiverU_i > 0} \prior_i \Quant\betaed(\receiverU_i)$.
Taking the limit supremum yields $\limsup_{\noiseScale\to\infty}\OPT(\noiseScale) \le \fullyrationalmass$. Conversely, consider a censorship $\signalscheme^{(\pooledmass)}$ with a pooled mass $\pooledmass \in [\sum\nolimits_{i \in [\stateNum]:\receiverU_i\le0}\prior_i, \fullyrationalmass)$.
Then, we have $\lim_{\noiseScale\to\infty} \Payoff[\noiseScale]{\signalscheme^{(\pooledmass)}} = \pooledmass$. Since $\OPT(\noiseScale) \ge \Payoff[\noiseScale]{\signalscheme^{(\pooledmass)}}$, letting $\pooledmass \to \fullyrationalmass$ gives the reverse bound $\liminf_{\noiseScale \to \infty} \OPT(\noiseScale) \ge \fullyrationalmass$.
Because both limits exist and the sender's optimal expected utility at the full rationality boundary, $\OPT(\infty)$, is strictly positive, taking the limit as $\noiseScale \to \infty$ on $\frac{\Payoff[\noiseScale]{\schemeoptimal}}{\OPT(\noiseScale)} \ge \ratiolimit$ yields
$
    \frac{\Payoff[\infty]{\schemeoptimal}}{\OPT(\infty)} 
    \ge \lim\nolimits_{\noiseScale \to \infty} \frac{\Payoff[\noiseScale]{\schemeoptimal}}{\OPT(\noiseScale)} 
    \ge \ratiolimit$.
Because the inequality $\frac{\Payoff[\noiseScale]{\schemeoptimal}}{\OPT(\noiseScale)} \ge \ratiolimit$ independently holds for every finite rationality level $\noiseScale \in [0, \infty)$ and also for the full rational boundary $\noiseScale = \infty$, we can take the infimum over the extended domain $\noiseScale \in \fullsetrationalitylevel$ to conclude that
$\inf\nolimits_{\noiseScale \in \fullsetrationalitylevel} \frac{\Payoff[\noiseScale]{\schemeoptimal}}{\OPT(\noiseScale)}
    \ge 
    \ratiolimit$.
Conversely, for every integer $\integer \in \mathbb{N}$, we have
$
\sup\nolimits_{\signalscheme \in \schemespace}\inf\nolimits_{\noiseScale \in \fullsetrationalitylevel}  
    \frac{\Payoff[\noiseScale]{\signalscheme}}{\OPT(\noiseScale)} \le
    \sup\nolimits_{\signalscheme \in \schemespace}\inf\nolimits_{\noiseScale \in \rationalityIntclass}  
     \frac{\Payoff[\noiseScale]{\signalscheme}}{\OPT(\noiseScale)}
    =  \ratiointeger$.
Letting $\integer \to \infty$ yields
$\sup\nolimits_{\signalscheme \in \schemespace}\inf\nolimits_{\noiseScale \in \fullsetrationalitylevel}  
    \frac{\Payoff[\noiseScale]{\signalscheme}}{\OPT(\noiseScale)} \le 
    \ratiolimit$.
Therefore, the censorship $\schemeoptimal \in \censorshipspace$ attains the robust optimum on the extended domain $\noiseScale \in \fullsetrationalitylevel$. This completes the proof.

\subsection{Auxiliary Lemmas in the Proof of \texorpdfstring{\Cref{thm:robust_optimal_censorship_three_criteria}}{Theorem 2} for the Maximin Ratio Criterion}
\label{apx:missing_lemma_robust_optimal_ratio}

In this subsection, we present the auxiliary lemmas used in the proof of \Cref{thm:robust_optimal_censorship_three_criteria} for the maximin ratio criterion: the S-shaped property of weighted combinations of quantal-response functions (\Cref{lem:kernel_function_propety}), and the single-peakedness and quasi-concavity of the average ratio function with respect to the pooled mass (\Cref{lem:quasi-concave,lem:derivative_gap}).

\begin{lemma}
    \label{lem:kernel_function_propety}
    Fix an arbitrary distribution $\noiseD$ over the truncated rationality level set $\rationalitySubclass \triangleq [0,\noiseScaleUB]$, where $\noiseScaleUB$ is an arbitrary positive constant.
Consider a function
    \begin{align*}
       \generalkernelF[\noiseD]{\variableX} \triangleq \displaystyle\int_{\noiseScale\in\rationalitySubclass}\generalkernelFweight\betaed \Quant\betaed(\variableX)
    \cdot \noiseD(\noiseScale) \; \dd\noiseScale\,,
    \end{align*}
where $\noiseScale \mapsto\generalkernelFweight\betaed$ is positive, bounded,
and measurable. Then, the function $\generalkernelF[\noiseD]{\variableX}$ is S-shaped, where
\begin{enumerate}[label=(\roman*)]
    \item\label{item:kernel_bullet1} $\generalkernelF[\noiseD]{\variableX}$ is monotone non-increasing on $(-\infty,\infty)$.
    \item\label{item:kernel_bullet2} $\generalkernelF[\noiseD]{\variableX}$ is concave on $(-\infty,0]$ and convex on $[0,\infty)$.
    \item\label{item:kernel_bullet3} $\generalkernelF[\noiseD]{\variableX}$ is centrally symmetric with respect to the point $(0,\generalkernelF[\noiseD]{0})$, i.e., $\generalkernelFderivative[\noiseD]{\variableX} = \generalkernelFderivative[\noiseD]{-\variableX}$.
\end{enumerate}
\end{lemma}
\begin{proof}
 Taking the first and second derivatives of $\generalkernelF[\noiseD]{\variableX}$ with respect to $\variableX$, we have
$
    \frac{\partial \generalkernelF[\noiseD]{\variableX}}{\partial \variableX} = \int_{\noiseScale\in\rationalitySubclass}\generalkernelFweight\betaed \cdot \frac{\partial \Quant\betaed(\variableX)}{\partial \variableX}\,\noiseD(\noiseScale) \,\dd\noiseScale\,
$
and 
$
    \frac{\partial^2 \generalkernelF[\noiseD]{\variableX}}{\partial \variableX^2} = \int_{\noiseScale\in\rationalitySubclass}\generalkernelFweight\betaed \cdot \frac{\partial^2 \Quant\betaed(\variableX)}{\partial \variableX^2}\,\noiseD(\noiseScale) \,\dd\noiseScale\,.
$
Recall that for every rationality level $\noiseScale\in[0,\infty)$, the quantal response function $\Quant\betaed(\cdot)$ is monotone non-increasing,
and concave on $(-\infty,0]$ and convex on $[0,\infty)$. 
Because integration is a linear operator and the weight $\generalkernelFweight\betaed$ is strictly positive, the function $\generalkernelF[\noiseD]{\variableX}$ preserves these derivative bounds.
Therefore, it satisfies $\frac{\partial \generalkernelF[\noiseD]{\variableX}}{\partial \variableX}\leq0$ for all $\variableX\in(-\infty,\infty)$, $\frac{\partial^2 \generalkernelF[\noiseD]{\variableX}}{\partial \variableX^2} \leq 0$ for all $\variableX\in(-\infty,0]$, $\frac{\partial^2 \generalkernelF[\noiseD]{\variableX}}{\partial \variableX^2} \geq 0$ for all $\variableX\in[0, \infty)$, establishing that it is S-shaped.
Additionally, for the quantal response function $\Quant\betaed(\cdot)$, it is centrally symmetric with respect to the point $(0,\frac{1}{2})$.
In other words, we have the symmetry identity $
\Quant\betaed(\variableX)+\Quant\betaed(-\variableX) = 2\Quant\betaed(0)=1$.
Thus, the function $\generalkernelF[\noiseD]{\variableX}$ also has the following symmetry identity:
$
 \generalkernelF[\noiseD]{\variableX} +\generalkernelF[\noiseD]{-\variableX} = 2\generalkernelF[\noiseD]{0} 
$.
Hence, $\generalkernelF[\noiseD]{\variableX}$ is centrally symmetric with respect to the point $(0,\generalkernelF[\noiseD]{0})$ and $\generalkernelFderivative[\noiseD]{\variableX} = \generalkernelFderivative[\noiseD]{-\variableX}$.
This completes the proof of \Cref{lem:kernel_function_propety}.
\end{proof}

\begin{lemma}
    \label{lem:quasi-concave}
    Fix an arbitrary distribution $\noiseD$ over the truncated rationality level set $\rationalitySubclass \triangleq [0,\noiseScaleUB]$, where $\noiseScaleUB$ is an arbitrary positive constant. Define
$\objectiveF[\rationalitySubclass]{\parameterizedscheme, \noiseD} \triangleq \sum\nolimits_{i\in[\stateNum]}\prior_{i} \int_{-\infty}^{\infty} \parameterizedscheme_i(\scaledNoiseDiff)   
    \kernelF[\noiseD]{\scaledNoiseDiff} \,\dd\scaledNoiseDiff$,
where $\parameterizedscheme$ is the censorship parameterized by the pooled mass $\pooledmass \in [0,1]$ and the function $\kernelF[\noiseD]{\variableX} \triangleq\int_{\noiseScale\in\rationalitySubclass}\frac{\Quant\betaed(\variableX)}{\OPT(\noiseScale)}
    \cdot \noiseD(\noiseScale) \; \dd\noiseScale$.
Then, the function $\objectiveF[\rationalitySubclass]{\parameterizedscheme, \noiseD}$ is single-peaked (and thus quasi-concave) with respect to the pooled mass $\pooledmass\in[0,1]$.
\end{lemma}

\begin{proof}
For ease of presentation, we write 
$\objectiveF[\noiseD]{\pooledmass} \triangleq \objectiveF[\rationalitySubclass]{\parameterizedscheme,\noiseD}$
as the average ratio under the censorship $\parameterizedscheme$ when the distribution over the truncated rationality level set $\rationalitySubclass$ is $\noiseD \in \stochastic(\rationalitySubclass)$.
For every pooled mass $\pooledmass \in [0,1]$, by the definitions of the pooling signal $\censorshipsignal(\pooledmass)$ and threshold state $\censorshipstate(\pooledmass)$ in \Cref{def:censorship}, the function $\objectiveF[\noiseD]{\pooledmass}$ can be written as $
    \objectiveF[\noiseD]{\pooledmass} = \pooledmass\kernelF[\noiseD]{\censorshipsignal(\pooledmass)}+ \left(\sum\nolimits_{i \in [\censorshipstate(\pooledmass)]}\prior_i - \pooledmass\right)\kernelF[\noiseD]{\receiverU_{\censorshipstate(\pooledmass)}}+\sum\nolimits_{i \in [\censorshipstate(\pooledmass)+1\,:\,m]}\prior_i \kernelF[\noiseD]{\receiverU_i}$.
By the above expression and the fact that $\kernelF[\noiseD]{\cdot}$ is continuously differentiable, it can be verified that the function $\objectiveF[\noiseD]{\pooledmass}$ is continuous and continuously differentiable everywhere except at a finite set of breakpoints given by 
$
 \left\{\sum\nolimits_{i \in [j]}\prior_i\right\}_{j \in [0\,:\,m]}   
$.
Accordingly, we divide the cumulative mass interval $(0,1)$ into $m$ open subintervals based on these breakpoints, namely $(0,\prior_1),\ldots,(\sum\nolimits_{i \in [m-1]}\prior_i,1)$. 

Consider an arbitrary open cumulative mass subinterval $(\sum\nolimits_{i \in [\censorshipstate\primed-1]}\prior_i\,,\,\sum\nolimits_{i \in [\censorshipstate\primed]}\prior_i)$ where $\censorshipstate\primed \in [\stateNum]$.
Because $\censorshipstate(\pooledmass)=\censorshipstate\primed$
throughout this open cumulative mass subinterval $(\sum\nolimits_{i \in [\censorshipstate\primed-1]}\prior_i\,,\,\sum\nolimits_{i \in [\censorshipstate\primed]}\prior_i)$,
the corresponding threshold state value $\receiverU_{\censorshipstate(\pooledmass)}$ remains constant.
Let $\kernelFderivative[\noiseD]{\cdot}$ and  $\censorshipsignalderivative(\cdot)$  denote the derivatives of the functions $\kernelF[\noiseD]{\cdot}$ and $\censorshipsignal(\cdot)$, respectively.
Taking the derivative of the function $\objectiveF[\noiseD]{\pooledmass}$ on this subinterval yields
$
    \frac{\partial\, \objectiveF[\noiseD]{\pooledmass}}{\partial \pooledmass} = \kernelF[\noiseD]{\censorshipsignal(\pooledmass)}+\pooledmass\kernelFderivative[\noiseD]{\censorshipsignal(\pooledmass)}\cdot\censorshipsignalderivative(\pooledmass) - \kernelF[\noiseD]{\receiverU_{\censorshipstate(\pooledmass)}}= \kernelF[\noiseD]{\censorshipsignal(\pooledmass)}+\kernelFderivative[\noiseD]{\censorshipsignal(\pooledmass)}\cdot\left(\receiverU_{\censorshipstate(\pooledmass)}-\censorshipsignal(\pooledmass)\right) - \kernelF[\noiseD]{\receiverU_{\censorshipstate(\pooledmass)}}$,
where the last equality holds since we have
$\frac{\partial\,\censorshipsignal(\pooledmass)}{\partial \pooledmass} = \frac{\receiverU_{\censorshipstate(\pooledmass)}-\censorshipsignal(\pooledmass)}{\pooledmass}$.
From the above derivation, it can be seen that the sign of the derivative
$\frac{\partial\, \objectiveF[\noiseD]{\pooledmass}}{\partial \pooledmass}$
is the same as that of 
$
    \kernelF[\noiseD]{\censorshipsignal(\pooledmass)}+\kernelFderivative[\noiseD]{\censorshipsignal(\pooledmass)}\cdot\left(\receiverU_{\censorshipstate(\pooledmass)}-\censorshipsignal(\pooledmass)\right) - \kernelF[\noiseD]{\receiverU_{\censorshipstate(\pooledmass)}}$.
 Then, for the sake of analysis, we define the following auxiliary function $\derivativegap(\variableX,\variableY)$:
 \begin{align*}
\derivativegap(\variableX,\variableY) \triangleq \kernelF[\noiseD]{\variableX}+\kernelFderivative[\noiseD]{\variableX}\cdot\left(\variableY-\variableX\right) - \kernelF[\noiseD]{\variableY}\,. 
 \end{align*}
Geometrically, $\derivativegap(\variableX,\variableY)$ represents the difference between the tangent line to $\kernelF[\noiseD]{\cdot}$ at $\variableX$, evaluated at $\variableY$, and the actual value $\kernelF[\noiseD]{\variableY}$.
To establish the single-peaked nature of the function $\objectiveF[\noiseD]{\pooledmass}$, we analyze the sign structure of $\frac{\partial\, \objectiveF[\noiseD]{\pooledmass}}{\partial \pooledmass}$ by considering the following two cases for the threshold state value $\receiverU_{\censorshipstate(\pooledmass)}$. We also illustrate the subsequent analysis in \Cref{fig:tangent_derivative}.

\input{plots/tangent_derivative}

\xhdr{Case 1: threshold state value $\receiverU_{\censorshipstate(\pooledmass)}\le 0$}
In that case, we have $\censorshipsignal(\pooledmass)\le \receiverU_{\censorshipstate(\pooledmass)} \le 0$ by the definition of pooling signal in \Cref{def:censorship}.
Then, we obtain $\derivativegap(\censorshipsignal(\pooledmass),\receiverU_{\censorshipstate(\pooledmass)})\ge 0$  by \Cref{lem:derivative_gap}\ref{item:tangentgap_bullet1}, implying that $\frac{\partial\, \objectiveF[\noiseD]{\pooledmass}}{\partial \pooledmass} \ge 0$ (see \Cref{fig:tangent_case_a} for an illustration).
Thus, we can conclude that for any open cumulative mass subinterval whose associated threshold state value is non-positive, the derivative of the function $\objectiveF[\noiseD]{\pooledmass}$ is non-negative, meaning that the function $\objectiveF[\noiseD]{\pooledmass}$ is non-decreasing.  

\xhdr{Case 2: threshold state value $\receiverU_{\censorshipstate(\pooledmass)}> 0$} 
We analyze this under two possible cases for the pooling signal $\censorshipsignal(\pooledmass)$.
First, suppose that $\censorshipsignal(\pooledmass) > 0$.
We obtain $\derivativegap(\censorshipsignal(\pooledmass),\receiverU_{\censorshipstate(\pooledmass)})\le 0$ by \Cref{lem:derivative_gap}\ref{item:tangentgap_bullet2}, implying that $\frac{\partial\, \objectiveF[\noiseD]{\pooledmass}}{\partial \pooledmass} \le 0$ (see \Cref{fig:tangent_case_b} for an illustration).
In other words, for any open cumulative mass subinterval in which both the threshold state value and the pooling signal are positive, the derivative of the function $\objectiveF[\noiseD]{\pooledmass}$ is non-positive, implying that the function $\objectiveF[\noiseD]{\pooledmass}$ is non-increasing.
Second, suppose that $\censorshipsignal(\pooledmass) \le 0$.
In that case, although there is no fixed global sign of $\frac{\partial\, \objectiveF[\noiseD]{\pooledmass}}{\partial \pooledmass}$, we argue that one of the following three subcases occurs:
\begin{enumerate}[label=(\arabic*)]
    \item $\frac{\partial\, \objectiveF[\noiseD]{\pooledmass}}{\partial \pooledmass} \ge 0$ for all pooled mass $\pooledmass \in (\sum\nolimits_{i \in [\censorshipstate\primed-1]}\prior_i\,,\,\sum\nolimits_{i \in [\censorshipstate\primed]}\prior_i)$ (a geometric realization of this subcase is illustrated in \Cref{fig:tangent_case_c}).
    \item $\frac{\partial\, \objectiveF[\noiseD]{\pooledmass}}{\partial \pooledmass} \le 0$ for all pooled mass $\pooledmass \in (\sum\nolimits_{i \in [\censorshipstate\primed-1]}\prior_i\,,\,\sum\nolimits_{i \in [\censorshipstate\primed]}\prior_i)$.
    \item\label{subcase:sign_change} The sign of $\frac{\partial\, \objectiveF[\noiseD]{\pooledmass}}{\partial \pooledmass}$ transitions from non-negative to non-positive exactly once for all pooled mass $\pooledmass \in (\sum\nolimits_{i \in [\censorshipstate\primed-1]}\prior_i\,,\,\sum\nolimits_{i \in [\censorshipstate\primed]}\prior_i)$.
\end{enumerate}
To see this, we first claim that a necessary and sufficient condition for $\derivativegap(\censorshipsignal(\pooledmass),\receiverU_{\censorshipstate(\pooledmass)})\ge 0$ is given by
\begin{align*}
\derivativegap(\censorshipsignal(\pooledmass),\receiverU_{\censorshipstate(\pooledmass)})\ge 0 \iff \censorshipsignal(\pooledmass) \le \tangentpoint(\receiverU_{\censorshipstate(\pooledmass)})\,,
\end{align*}
where $\tangentpoint(\receiverU_{\censorshipstate(\pooledmass)}) \triangleq \sup\{\variableX \in (-\infty, 0]\,:\,\derivativegap(\variableX,\receiverU_{\censorshipstate(\pooledmass)})\ge 0\}$.
Since this set $\{\variableX \in (-\infty, 0]\,:\,\derivativegap(\variableX,\receiverU_{\censorshipstate(\pooledmass)})\ge 0\}$ is contained in $(-\infty,0]$, it is bounded above by 0. 
Thus, whenever the set is non-empty, the supremum $\tangentpoint(\receiverU_{\censorshipstate(\pooledmass)}) $ is well-defined.
Notice that we have
\begin{align*}
    \derivativegap(-\receiverU_{\censorshipstate(\pooledmass)},\receiverU_{\censorshipstate(\pooledmass)})
    &=  \kernelF[\noiseD]{-\receiverU_{\censorshipstate(\pooledmass)}}+2\receiverU_{\censorshipstate(\pooledmass)}\kernelFderivative[\noiseD]{-\receiverU_{\censorshipstate(\pooledmass)}} - \kernelF[\noiseD]{\receiverU_{\censorshipstate(\pooledmass)}} \nonumber
    \overset{(a)}{=}2\left(\kernelF[\noiseD]{0}-\kernelF[\noiseD]{\receiverU_{\censorshipstate(\pooledmass)}}+\receiverU_{\censorshipstate(\pooledmass)}\kernelFderivative[\noiseD]{\receiverU_{\censorshipstate(\pooledmass)}}\right)  \overset{(b)}{\ge}0\,\\
    \derivativegap(0,\receiverU_{\censorshipstate(\pooledmass)})
    & =  \kernelF[\noiseD]{0}+\receiverU_{\censorshipstate(\pooledmass)}\kernelFderivative[\noiseD]{0} - \kernelF[\noiseD]{\receiverU_{\censorshipstate(\pooledmass)}} \overset{(c)}{\le} 0\,,
\end{align*}
where equality~(a) holds since we have the symmetry equality $\kernelF[\noiseD]{\receiverU_{\censorshipstate(\pooledmass)}}+\kernelF[\noiseD]{-\receiverU_{\censorshipstate(\pooledmass)}} = 2\kernelF[\noiseD]{0}$ (see \Cref{lem:kernel_function_propety}\ref{item:kernel_bullet3}), and inequalities~(b) and (c) hold due to the convexity of the function $\kernelF[\noiseD]{\cdot}$ on $[0,\infty)$ by \Cref{lem:kernel_function_propety}\ref{item:kernel_bullet2}, specifically, $\kernelFderivative[\noiseD]{\receiverU_{\censorshipstate(\pooledmass)}} \ge \frac{\kernelF[\noiseD]{\receiverU_{\censorshipstate(\pooledmass)}}-\kernelF[\noiseD]{0}}{\receiverU_{\censorshipstate(\pooledmass)}-0}$.
Since the function $\derivativegap(\variableX,\receiverU_{\censorshipstate(\pooledmass)})$ is monotone non-increasing with respect to $\variableX$ on $(-\infty, 0]$ given that $\receiverU_{\censorshipstate(\pooledmass)}\ge0$ (see \Cref{lem:derivative_gap}\ref{item:tangentgap_bullet3}), combining above two inequalities together, it follows that the set $\{\variableX \le 0 \,:\,\derivativegap(\variableX,\receiverU_{\censorshipstate(\pooledmass)})\ge 0\}$ is indeed non-empty.
Then, because the pooling signal $\censorshipsignal(\pooledmass)$ is monotone non-decreasing with respect to $\pooledmass$ and $\tangentpoint(\receiverU_{\censorshipstate(\pooledmass)})$ is a constant in the open cumulative mass subinterval $(\sum\nolimits_{i \in [\censorshipstate\primed-1]}\prior_i\,,\,\sum\nolimits_{i \in [\censorshipstate\primed]}\prior_i)$, 
the quantity
$
    \censorshipsignal(\pooledmass)
-
\tangentpoint(\receiverU_{\censorshipstate(\pooledmass)})
$
is monotone non-decreasing on the open cumulative mass subinterval $(\sum\nolimits_{i \in [\censorshipstate\primed-1]}\prior_i\,,\,\sum\nolimits_{i \in [\censorshipstate\primed]}\prior_i)$.
Hence, one of the following three situations occurs:
\begin{enumerate}[label=(\arabic*)]
    \item
    $\censorshipsignal(\pooledmass)
    \le
    \tangentpoint(\receiverU_{\censorshipstate(\pooledmass)})$
    for all pooled mass $\pooledmass \in (\sum\nolimits_{i \in [\censorshipstate\primed-1]}\prior_i\,,\,\sum\nolimits_{i \in [\censorshipstate\primed]}\prior_i)$.

    \item
    $\censorshipsignal(\pooledmass)
    \ge
    \tangentpoint(\receiverU_{\censorshipstate(\pooledmass)})$
    for all pooled mass $\pooledmass \in (\sum\nolimits_{i \in [\censorshipstate\primed-1]}\prior_i\,,\,\sum\nolimits_{i \in [\censorshipstate\primed]}\prior_i)$.

    \item
    $\censorshipsignal(\pooledmass)$ crosses
    $\tangentpoint(\receiverU_{\censorshipstate(\pooledmass)})$
    exactly once from below for all pooled mass $\pooledmass \in (\sum\nolimits_{i \in [\censorshipstate\primed-1]}\prior_i\,,\,\sum\nolimits_{i \in [\censorshipstate\primed]}\prior_i)$.
\end{enumerate}
By the equivalence established in the necessary and sufficient condition, these three situations correspond respectively to the three proposed subcases.
To complete the argument, it remains to show that once the sign of $\frac{\partial\, \objectiveF[\noiseD]{\pooledmass}}{\partial \pooledmass}$ transitions from positive to negative, this negative property is preserved across all subsequent open cumulative mass subintervals.
Suppose that the derivative is negative at a point that is not a cumulative-mass breakpoint, i.e., there exists a pooled mass $\pooledmass\primed \in (\sum\nolimits_{i \in [\censorshipstate\primed-1]}\prior_i\,,\,\sum\nolimits_{i \in [\censorshipstate\primed]}\prior_i)$ such that $\derivativegap(\censorshipsignal(\pooledmass\primed),\receiverU_{\censorshipstate(\pooledmass\primed)})< 0$, where $\censorshipsignal(\pooledmass\primed) \le 0$.
We now claim that
\begin{align*}
\receiverU_{\censorshipstate(\pooledmass\primed)} \ge -\censorshipsignal(\pooledmass\primed)\,.
\end{align*}
To see this, note that since $\censorshipsignal(\pooledmass\primed) \le 0$, we have $\derivativegap(\censorshipsignal(\pooledmass\primed), \variableY) \ge 0$ for all $\variableY \in [\censorshipsignal(\pooledmass\primed), 0]$ (see \Cref{lem:derivative_gap}\ref{item:tangentgap_bullet1}). Furthermore, $\derivativegap(\censorshipsignal(\pooledmass\primed), \variableY)$ is monotone non-decreasing with respect to $\variableY \in [0, -\censorshipsignal(\pooledmass\primed)]$ (see \Cref{lem:derivative_gap}\ref{item:tangentgap_bullet4}). 
Combining these two properties yields $\derivativegap(\censorshipsignal(\pooledmass\primed), \variableY) \ge 0$ for all $\variableY \in [\censorshipsignal(\pooledmass\primed), -\censorshipsignal(\pooledmass\primed)]$. 
Therefore, if  $\receiverU_{\censorshipstate(\pooledmass\primed)} < -\censorshipsignal(\pooledmass\primed)$, then $\derivativegap(\censorshipsignal(\pooledmass\primed),\receiverU_{\censorshipstate(\pooledmass\primed)})\ge 0$, contradicting the assumption that $\derivativegap(\censorshipsignal(\pooledmass\primed),\receiverU_{\censorshipstate(\pooledmass\primed)})<0$. 
Thus, $\receiverU_{\censorshipstate(\pooledmass\primed)} \ge -\censorshipsignal(\pooledmass\primed)$ must hold.
Now, consider any open cumulative mass subinterval $(\sum\nolimits_{i \in [\censorshipstate\doubleprimed-1]}\prior_i\,,\,\sum\nolimits_{i \in [\censorshipstate\doubleprimed]}\prior_i)$, where $\censorshipstate\doubleprimed \in [\censorshipstate\primed+1\,:\,m]$. Choose an arbitrary pooled mass $\pooledmass\doubleprimed \in (\sum\nolimits_{i \in [\censorshipstate\doubleprimed-1]}\prior_i\,,\,\sum\nolimits_{i \in [\censorshipstate\doubleprimed]}\prior_i)$. 
If $\censorshipsignal(\pooledmass\doubleprimed) \le 0$, we obtain
$
\derivativegap(\censorshipsignal(\pooledmass\doubleprimed),\receiverU_{\censorshipstate(\pooledmass\doubleprimed)})
\overset{(a)}{\le}
\derivativegap(\censorshipsignal(\pooledmass\primed),\receiverU_{\censorshipstate(\pooledmass\doubleprimed)}) 
\overset{(b)}{\le}
\derivativegap(\censorshipsignal(\pooledmass\primed),\receiverU_{\censorshipstate(\pooledmass\primed)}) < 0$,
where inequality~(a) holds since $0\ge \censorshipsignal(\pooledmass\doubleprimed) \ge \censorshipsignal(\pooledmass\primed)$ and the function $\derivativegap(\variableX,\variableY)$ is monotone non-increasing with respect to $\variableX$ on $ (-\infty,0]$ whenever $\variableY \ge 0$ (see \Cref{lem:derivative_gap}\ref{item:tangentgap_bullet3}).
Inequality~(b) holds since $\censorshipsignal(\pooledmass\primed) \le 0$ and $\receiverU_{\censorshipstate(\pooledmass\doubleprimed)} \ge \receiverU_{\censorshipstate(\pooledmass\primed)} \ge -\censorshipsignal(\pooledmass\primed)$, where the first inequality holds since $\pooledmass\doubleprimed \ge \pooledmass\primed$, and the second inequality follows from the above inequality. Moreover, the function $\derivativegap(\variableX,\variableY)$ is monotone non-increasing with respect to $\variableY$ on $[-\variableX,\infty)$ whenever $\variableX \le 0$ (see \Cref{lem:derivative_gap}\ref{item:tangentgap_bullet4}).
If $\censorshipsignal(\pooledmass\doubleprimed) > 0$, we directly have $\derivativegap(\censorshipsignal(\pooledmass\doubleprimed),\receiverU_{\censorshipstate(\pooledmass\doubleprimed)}) \le 0$ due to \Cref{lem:derivative_gap}\ref{item:tangentgap_bullet2}.  

Next, we prove that the function $\objectiveF[\noiseD]{\pooledmass}$ is absolutely continuous on $[0,1]$, which allows us to invoke the fundamental theorem of calculus to establish the global monotonicity of $\objectiveF[\noiseD]{\pooledmass}$.
We first show that the function $\objectiveF[\noiseD]{\pooledmass}$ is Lipschitz continuous on $[0,1]$.
Recall that for an arbitrary open cumulative mass subinterval $(\sum\nolimits_{i \in [\censorshipstate\primed-1]}\prior_i\,,\,\sum\nolimits_{i \in [\censorshipstate\primed]}\prior_i)$, where $\censorshipstate\primed \in [\stateNum]$, the derivative of the function $\objectiveF[\noiseD]{\pooledmass}$ is given by
$\derivativegap(\censorshipsignal(\pooledmass),\receiverU_{\censorshipstate(\pooledmass)})$.
Note that $\censorshipsignal(\pooledmass) \in [\receiverU_1,\receiverU_m]$ and $\receiverU_{\censorshipstate(\pooledmass)} \in \{\receiverU_i\}_{i \in [\stateNum]}$.
Because the domain of these pairs is the Cartesian product of a compact interval and a finite set, it is compact. 
Since the function $\derivativegap(\variableX,\variableY)$ is continuous on this compact domain, it is bounded. 
Thus, there exists a positive constant $\maxderivative > 0$ such that
$
\left|\derivativegap(\censorshipsignal(\pooledmass),\receiverU_{\censorshipstate(\pooledmass)})\right| \le \maxderivative 
\Rightarrow
\left|\frac{\partial\,\objectiveF[\noiseD]{\pooledmass}}{\partial \pooledmass}\right|\le \maxderivative$.
Let $\objectiveFderivative[\noiseD]{\cdot}$ denote the derivative of the function $\objectiveF[\noiseD]{\cdot}$.
For any two points $\variableX,\variableY \in (\sum\nolimits_{i \in [\censorshipstate\primed-1]}\prior_i\,,\,\sum\nolimits_{i \in [\censorshipstate\primed]}\prior_i)$, where $\variableX <\variableY$, by the fundamental theorem of calculus and the above inequality, we have
\begin{align*}
   \left|\objectiveF[\noiseD]{\variableX}- \objectiveF[\noiseD]{\variableY}\right| = 
   \left| \int_{\variableX}^{\variableY} \objectiveFderivative[\noiseD]{\variableZ}\dd\variableZ\right| \le  \int_{\variableX}^{\variableY} \left|\objectiveFderivative[\noiseD]{\variableZ}\right|\dd\variableZ  \le \maxderivative\left|\variableX-\variableY\right|\,.
\end{align*}
Now, fix two points $\variableX\primed,\variableY\primed$ in the closed cumulative mass subinterval $[\sum\nolimits_{i \in [\censorshipstate\primed-1]}\prior_i\,,\,\sum\nolimits_{i \in [\censorshipstate\primed]}\prior_i]$.
We can construct sequences $\{\variableX_{\integer}\}_{\integer \in \mathbb{N}}$ and $\{\variableY_{\integer}\}_{\integer \in \mathbb{N}}$ contained in the open cumulative mass subinterval $(\sum\nolimits_{i \in [\censorshipstate\primed-1]}\prior_i\,,\,\sum\nolimits_{i \in [\censorshipstate\primed]}\prior_i)$, such that 
$\variableX\primed = \lim\nolimits_{\integer \rightarrow \infty}\variableX_{\integer}$ and  $ \variableY\primed = \lim\nolimits_{\integer \rightarrow \infty}\variableY_{\integer}$.
By the above inequality, we obtain
$
   \left|\objectiveF[\noiseD]{\variableX_{\integer}}- \objectiveF[\noiseD]{\variableY_{\integer}}\right|  \le \maxderivative\left|\variableX_{\integer}-\variableY_{\integer}\right|$.
By the continuity of the function $\objectiveF[\noiseD]{\pooledmass}$, taking the limit as $\integer \rightarrow \infty$ yields
\begin{align*}
 \left|\objectiveF[\noiseD]{\variableX\primed}- \objectiveF[\noiseD]{\variableY\primed}\right|  \le \maxderivative\left|\variableX\primed-\variableY\primed\right|\,,   
\end{align*}
implying that the function $\objectiveF[\noiseD]{\pooledmass}$ is Lipschitz continuous on the closed cumulative mass subinterval $[\sum\nolimits_{i \in [\censorshipstate\primed-1]}\prior_i\,,\,\sum\nolimits_{i \in [\censorshipstate\primed]}\prior_i]$.
Then, take two points $\variableA, \variableB \in [0,1]$, where $\variableA<\variableB$.
Suppose that, for some $\integer \in \mathbb
{N}$, there are $\integer$ breakpoints $\{\breakpoint_{i}\}_{i \in [\integer]} \subseteq \{\sum\nolimits_{i \in [j]}\prior_i\}_{j \in [0\,:\,m]}$ lying in $[\variableA,\variableB]$, ordered such that $\breakpoint_1 < 
\cdots < \breakpoint_{\integer}$.
Applying the triangle inequality together with the above inequality on each closed cumulative mass subinterval yields
$
\left|\objectiveF[\noiseD]{\variableB}- \objectiveF[\noiseD]{\variableA}\right|  \le \left|\objectiveF[\noiseD]{\variableB} - \objectiveF[\noiseD]{\breakpoint_{\integer}}\right|+\cdots+\left|\objectiveF[\noiseD]{\breakpoint_{1}} - \objectiveF[\noiseD]{\variableA}\right| \le \maxderivative\left|\variableB-\breakpoint_{\integer}\right|+ \cdots+ 
    \maxderivative\left|\breakpoint_{1}-\variableA\right| = \maxderivative\left(\variableB-\breakpoint_{\integer}\right)+ \cdots+ 
    \maxderivative\left(\breakpoint_{1}-\variableA\right)
    = \maxderivative\left(\variableB-\variableA\right) 
    = \maxderivative\left|\variableB-\variableA\right|$.
Thus, the function $\objectiveF[\noiseD]{\pooledmass}$ is globally Lipschitz continuous on $[0,1]$, implying that it is absolutely continuous on $[0,1]$.

Because $\objectiveF[\noiseD]{\pooledmass}$ is differentiable on every open cumulative mass subinterval and the set of breakpoints is finite, which has Lebesgue measure zero, $\objectiveFderivative[\noiseD]{\pooledmass}$ exists almost everywhere.\footnote{Because a finite set of breakpoints has Lebesgue measure zero, the exact location of the peak---whether it falls inside an open subinterval or exactly on a breakpoint---does not affect the almost everywhere properties of the derivative.}
By the previous analysis of the sign structure of $\frac{\partial\,\objectiveF[\noiseD]{\pooledmass}}{\partial\pooledmass}$, we conclude that there exists $\pooledmass^* \in [0,1]$ such that $\frac{\partial\,\objectiveF[\noiseD]{\pooledmass}}{\partial\pooledmass} \ge 0$ holds almost everywhere for $\pooledmass \in [0,\pooledmass^*]$ and $\frac{\partial\,\objectiveF[\noiseD]{\pooledmass}}{\partial\pooledmass} \le 0$ holds almost everywhere for $\pooledmass \in [\pooledmass^*,1]$.
Consider two points $\variableA,\variableB \in [0,1]$ where $\variableA <\variableB$.
Because the function $\objectiveF[\noiseD]{\pooledmass}$ is absolutely continuous, by the fundamental theorem of calculus, we obtain
$
    \objectiveF[\noiseD]{\variableB}-\objectiveF[\noiseD]{\variableA} = \int_{\variableA}^{\variableB} \objectiveFderivative[\noiseD]{\pooledmass}\dd\pooledmass$.
If $0\le \variableA<\variableB\le \pooledmass^*$, then $\objectiveFderivative[\noiseD]{\pooledmass} \ge 0$ holds almost everywhere, ensuring the integral is non-negative.
This implies that the function $\objectiveF[\noiseD]{\pooledmass}$ is weakly increasing on $[0,\pooledmass^*]$.
Conversely, if $\pooledmass^*\le \variableA<\variableB\le 1$, then $\objectiveFderivative[\noiseD]{\pooledmass} \le 0$ holds almost everywhere, ensuring the integral is non-positive. 
This implies that the function $\objectiveF[\noiseD]{\pooledmass}$ is weakly decreasing on $[\pooledmass^*,1]$.
Therefore, the function $\objectiveF[\noiseD]{\pooledmass}$ is single-peaked, and hence quasi-concave. 
This completes the proof of \Cref{lem:quasi-concave}.
\end{proof}

\begin{lemma}
    \label{lem:derivative_gap}
    Fix an arbitrary distribution $\noiseD$ over the truncated rationality level set $\rationalitySubclass \triangleq [0,\noiseScaleUB]$, where $\noiseScaleUB$ is an arbitrary positive constant.
    Define function
$\derivativegap(\variableX,\variableY) \triangleq \kernelF[\noiseD]{\variableX}+\kernelFderivative[\noiseD]{\variableX}\cdot\left(\variableY-\variableX\right) - \kernelF[\noiseD]{\variableY}$,
where the function $\kernelF[\noiseD]{\variableX} \triangleq\int_{\noiseScale\in\rationalitySubclass}\frac{\Quant\betaed(\variableX)}{\OPT(\noiseScale)}
    \cdot \noiseD(\noiseScale) \; \dd\noiseScale$.
    Then, we have
    \begin{enumerate}[label=(\roman*)]
        \item\label{item:tangentgap_bullet1} If $\variableX \le \variableY \le 0$, then, $\derivativegap(\variableX,\variableY) \ge 0$.
        \item\label{item:tangentgap_bullet2} If $0 \le \variableX \le \variableY$, then, $\derivativegap(\variableX,\variableY) \le 0$.
        \item\label{item:tangentgap_bullet3} If $\variableY \ge 0$, then, $\derivativegap(\variableX,\variableY)$ is non-increasing on $\variableX \in (-\infty,0]$ .
        \item\label{item:tangentgap_bullet4} If $\variableX \le 0$, then, $\derivativegap(\variableX,\variableY)$ is non-decreasing on $\variableY \in [0,-\variableX]$ and non-increasing on $\variableY \in [-\variableX,\infty)$ .
    \end{enumerate}
\end{lemma}
\begin{proof}
    We show these properties individually.
    For part~(i), the function $\kernelF[\noiseD]{\cdot}$ is concave on $(-\infty,0]$ (see \Cref{lem:kernel_function_propety}\ref{item:kernel_bullet2}).
    For any $\variableX \le \variableY \le 0$, it can be verified that
   $
        \kernelFderivative[\noiseD]{\variableX} \ge \frac{\kernelF[\noiseD]{\variableY}-\kernelF[\noiseD]{\variableX}}{\variableY-\variableX}$,
implying that $\derivativegap(\variableX,\variableY)\ge 0$.
For part~(ii), the function $\kernelF[\noiseD]{\cdot}$ is convex on $[0,\infty)$ (see \Cref{lem:kernel_function_propety}\ref{item:kernel_bullet2}).
    For any $0 \le \variableX \le \variableY $, it can be verified that
   $
        \kernelFderivative[\noiseD]{\variableX} \le \frac{\kernelF[\noiseD]{\variableY}-\kernelF[\noiseD]{\variableX}}{\variableY-\variableX}$,
implying that $\derivativegap(\variableX,\variableY)\le 0$.
For part~(iii), let $\kernelFsecondderivative[\noiseD]{\cdot}$ denote the second order derivative of $\kernelF[\noiseD]{\cdot}$.
Then, for any $\variableX \in (-\infty,0]$, taking the derivative of $\derivativegap(\variableX,\variableY)$ with respect to $\variableX$ yields
$
\frac{\partial\derivativegap(\variableX,\variableY)}{\partial \variableX} = \kernelFsecondderivative[\noiseD]{\variableX}(\variableY-\variableX) \le 0$,
where the inequality holds since $\variableY-\variableX >0$ and $\kernelFsecondderivative[\noiseD]{\variableX} \le 0$, which follows from the concavity of $\kernelF[\noiseD]{\cdot}$ on $(-\infty,0]$.
Thus, the function $\derivativegap(\variableX,\variableY)$ is monotone non-increasing with respect to $\variableX$ on $(-\infty,0]$.
For part~(iv), let $\kernelFderivative[\noiseD]{\cdot}$ denote the derivative of $\kernelF[\noiseD]{\cdot}$. 
Then, taking the derivative of $\derivativegap(\variableX,\variableY)$ with respect to $\variableY$ yields
$
\frac{\partial\derivativegap(\variableX,\variableY)}{\partial \variableY} = \kernelFderivative[\noiseD]{\variableX}- \kernelFderivative[\noiseD]{\variableY}$.
Note that we have the symmetry identity $\kernelFderivative[\noiseD]{\variableX} = \kernelFderivative[\noiseD]{-\variableX}$ (see \Cref{lem:kernel_function_propety}\ref{item:kernel_bullet3}). 
Furthermore, because $\variableX \le 0$, we know $-\variableX \ge 0$. 
For any $\variableY \in [0,-\variableX]$, since both $\variableY$ and $-\variableX$ reside on the domain $[0, \infty)$ where the function $\kernelF[\noiseD]{\cdot}$ is convex, its derivative $\kernelFderivative[\noiseD]{\cdot}$ is monotone non-decreasing on this domain (see \Cref{lem:kernel_function_propety}\ref{item:kernel_bullet2}). Thus, $\kernelFderivative[\noiseD]{\variableY} \le \kernelFderivative[\noiseD]{-\variableX} = \kernelFderivative[\noiseD]{\variableX}$, yielding $\frac{\partial\derivativegap(\variableX,\variableY)}{\partial \variableY} \ge 0$.
Conversely, for any $\variableY \in [-\variableX,\infty)$, the non-decreasing property yields $\kernelFderivative[\noiseD]{\variableY} \ge \kernelFderivative[\noiseD]{-\variableX} = \kernelFderivative[\noiseD]{\variableX}$, yielding $\frac{\partial\derivativegap(\variableX,\variableY)}{\partial \variableY} \le 0$.
This completes the proof of \Cref{lem:derivative_gap}.
\end{proof}

\subsection{Proof of \texorpdfstring{\Cref{thm:robust_optimal_censorship_three_criteria}}{Theorem 2} for the Maximin Payoff and Minimax Regret Criteria}
\label{apx:missing_proof_robust_optimal_payoff_regret}
In this subsection, we present the omitted proofs of \Cref{thm:robust_optimal_censorship_three_criteria} for the maximin payoff and minimax regret criteria.
\robustoptimalcensorship*

\begin{proof}[Proof of \Cref{thm:robust_optimal_censorship_three_criteria} for the Maximin Payoff Criterion]
We establish the result by relating the average payoff to the average ratio studied in \Cref{sec:robust_optimal_scheme_unkown_rationality_level} and then applying the limiting argument in \Cref{apx:limiting-arguments}.

We first consider the truncated rationality level set $\rationalitySubclass=\left[0,\noiseScaleUB\right]$, where $\noiseScaleUB>0$.
For a distribution $\noiseD\in\stochastic\left(\rationalitySubclass\right)$, we define the average payoff $\PayoffF[\rationalitySubclass]{\signalscheme,\noiseD}\triangleq\int\nolimits_{\rationalitySubclass}\Payoff[\noiseScale]{\signalscheme}\,\noiseD\left(\noiseScale\right)\; \dd\noiseScale$.
To relate this objective to the average ratio, we introduce the positive normalization constant $\regretconstant\triangleq\int\nolimits_{\rationalitySubclass}\OPT\left(\noiseScale\right)\,\noiseD\left(\noiseScale\right)\; \dd\noiseScale$ and the reweighted probability distribution 
$\noiseDscaled\left(\noiseScale\right)
\triangleq
\frac{\OPT\left(\noiseScale\right)\noiseD\left(\noiseScale\right)}
{\regretconstant}$.
These definitions give, for every signaling scheme $\signalscheme\in\schemespace$,
\begin{align*}
    \PayoffF[\rationalitySubclass]{\signalscheme,\noiseD}
    &= \regretconstant\objectiveF[\rationalitySubclass]{\signalscheme,\noiseDscaled}~.
\end{align*}
Because the factor $\regretconstant$ is positive and independent of the signaling scheme, the censorship best-response property established in the maximin ratio proof in \Cref{sec:robust_optimal_scheme_unkown_rationality_level} also holds for the average payoff.
Moreover, \Cref{lem:quasi-concave} and the continuity established in its proof imply that the average payoff $\PayoffF[\rationalitySubclass]{\parameterizedscheme,\noiseD}$ is continuous and quasi-concave in the pooled mass $\pooledmass\in\left[0,1\right]$.

Randomization over rationality levels does not change the worst-case payoff, so $\inf\nolimits_{\noiseD\in\stochastic\left(\rationalitySubclass\right)}\PayoffF[\rationalitySubclass]{\signalscheme,\noiseD}=\inf\nolimits_{\noiseScale\in\rationalitySubclass}\Payoff[\noiseScale]{\signalscheme}$.
Since the average payoff is also continuous and affine in the distribution $\noiseD$, we obtain
\begin{align*}
    \sup\nolimits_{\signalscheme\in\schemespace}
    \inf\nolimits_{\noiseD\in\stochastic\left(\rationalitySubclass\right)}
    \PayoffF[\rationalitySubclass]{\signalscheme,\noiseD}
    &\le
    \inf\nolimits_{\noiseD\in\stochastic\left(\rationalitySubclass\right)}
    \sup\nolimits_{\signalscheme\in\schemespace}
    \PayoffF[\rationalitySubclass]{\signalscheme,\noiseD}\\
    &=
    \inf\nolimits_{\noiseD\in\stochastic\left(\rationalitySubclass\right)}
    \max\nolimits_{\pooledmass\in\left[0,1\right]}
    \PayoffF[\rationalitySubclass]{\parameterizedscheme,\noiseD}
    =
    \max\nolimits_{\pooledmass\in\left[0,1\right]}
    \inf\nolimits_{\noiseD\in\stochastic\left(\rationalitySubclass\right)}
    \PayoffF[\rationalitySubclass]{\parameterizedscheme,\noiseD}~.
\end{align*}
The inequality is the weak minimax inequality, the first equality follows from the censorship best-response property, and the final equality follows from Sion's minimax theorem \citep{S-58}.
The reverse inequality follows from $\censorshipspace\subseteq\schemespace$.
Thus, a censorship attains the maximin payoff on every truncated set $\rationalitySubclass$.

We extend this conclusion to the full rationality level set using the argument in \Cref{apx:limiting-arguments}.
For each integer $\integer\ge1$, we choose an optimal pooled mass $\pooledmass_{\integer}$ on $\left[0,\integer\right]$ and denote the corresponding optimal value by $\payoffinteger\triangleq\sup\nolimits_{\signalscheme\in\schemespace}\inf\nolimits_{\noiseScale\in\left[0,\integer\right]}\Payoff[\noiseScale]{\signalscheme}$.
The values $\payoffinteger$ decrease to a limit $\payoffinfty$, and compactness of $\left[0,1\right]$ yields a subsequence $\pooledmass_{\integer_{\integernew}}\to\pooledmass^*$.
For every finite rationality level $\noiseScale$, the inequality $\Payoff[\noiseScale]{\signalscheme^{\left(\pooledmass_{\integer_{\integernew}}\right)}}\ge\payoff_{\integer_{\integernew}}$ holds for all sufficiently large $\integernew$.
Continuity in the pooled mass therefore gives $\Payoff[\noiseScale]{\signalscheme^{\left(\pooledmass^*\right)}}\ge\payoffinfty$.
The boundary inequality established in \Cref{apx:limiting-arguments} further implies
\begin{align*}
    \Payoff[\infty]{\signalscheme^{\left(\pooledmass^*\right)}}
    &\ge
    \lim\nolimits_{\noiseScale\to\infty}
    \Payoff[\noiseScale]{\signalscheme^{\left(\pooledmass^*\right)}}
    \ge \payoffinfty~.
\end{align*}
Since the full-domain optimum is bounded above by every truncated optimum $\payoffinteger$, it is at most $\payoffinfty$.
The censorship $\signalscheme^{\left(\pooledmass^*\right)}$ therefore attains the maximin payoff over $\fullsetrationalitylevel$.
\end{proof}

\begin{proof}[Proof of \Cref{thm:robust_optimal_censorship_three_criteria} for the Minimax Regret Criterion]
We use the average payoff $\PayoffF[\rationalitySubclass]{\cdot, \cdot}$ and normalization constant $\regretconstant$ defined in the preceding proof for maximin payoff criterion.
For a truncated rationality level set $\rationalitySubclass=\left[0,\noiseScaleUB\right]$ and a distribution $\noiseD\in\stochastic\left(\rationalitySubclass\right)$, the average regret satisfies
\begin{align*}
    \RegretF[\rationalitySubclass]{\signalscheme,\noiseD}
    &\triangleq
    \int\nolimits_{\rationalitySubclass}
    \left(\Payoff[\noiseScale]{\signalscheme}-\OPT\left(\noiseScale\right)\right)
    \,\noiseD\left(\noiseScale\right)\; \dd\noiseScale
    =
    \PayoffF[\rationalitySubclass]{\signalscheme,\noiseD}-\regretconstant~.
\end{align*}
Because the subtracted term $\regretconstant$ is independent of the signaling scheme, the results established in the preceding proof for maximin payoff criterion imply that a censorship maximizes the average regret against each fixed distribution $\noiseD$, and that the average regret $\RegretF[\rationalitySubclass]{\parameterizedscheme,\noiseD}$ is continuous and quasi-concave in the pooled mass.
The average regret is also continuous and affine in the distribution $\noiseD$.
Consequently, the same weak minimax and Sion argument as in the preceding proof for maximin payoff criterion establishes that a censorship attains the minimax regret on every truncated set $\rationalitySubclass$.

For each integer $\integer\ge1$, we choose an optimal pooled mass $\pooledmass_{\integer}$ on $\left[0,\integer\right]$ and denote the corresponding optimal value by $\regretinteger\triangleq\sup\nolimits_{\signalscheme\in\schemespace}\inf\nolimits_{\noiseScale\in\left[0,\integer\right]}\left(\Payoff[\noiseScale]{\signalscheme}-\OPT\left(\noiseScale\right)\right)$.
The values $\regretinteger$ decrease to a limit $\regretinfty$, and a subsequence of the pooled masses converges to some $\pooledmass^*\in\left[0,1\right]$.
As in the preceding proof for maximin payoff criterion, continuity implies $\Payoff[\noiseScale]{\signalscheme^{\left(\pooledmass^*\right)}}-\OPT\left(\noiseScale\right)\ge\regretinfty$ for every finite rationality level $\noiseScale$.
\Cref{apx:limiting-arguments} establishes both the boundary payoff inequality and the convergence $\OPT\left(\noiseScale\right)\to\OPT\left(\infty\right)$.
Hence,
\begin{align*}
    \Payoff[\infty]{\signalscheme^{\left(\pooledmass^*\right)}}-\OPT\left(\infty\right)
    &\ge
    \lim\nolimits_{\noiseScale\to\infty}
    \left(\Payoff[\noiseScale]{\signalscheme^{\left(\pooledmass^*\right)}}-\OPT\left(\noiseScale\right)\right)
    \ge\regretinfty~.
\end{align*}
The full-domain optimum is bounded above by every truncated optimum $\regretinteger$, and therefore by $\regretinfty$.
The censorship $\signalscheme^{\left(\pooledmass^*\right)}$ consequently attains the minimax regret over $\fullsetrationalitylevel$.
\end{proof}




\section{Missing Proofs in \texorpdfstring{\Cref{sec:robust_ratio}}{Section 5}}
\label{apx:robust_ratio}

In this section, we present our results on the necessity of the censorship structure for a fully rational receiver, the non-positive prior mean assumption, and the tightness of \Cref{thm:rationality_robust}.

\subsection{Fully Rational Optimality Alone Does Not Guarantee Robustness}
\label{apx:necessity_censorship}
\begin{proposition}
\label{prop:necessity_censorship}
    For every $\varepsilon>0$, there exists an instance such that the optimal non-censorship designed for the fully rational receiver has a worst-case ratio less than $\varepsilon$.
\end{proposition}
\Cref{prop:necessity_censorship} shows that full-rationality optimality alone is insufficient for rationality robustness: a simple binary-signal signaling scheme that is optimal at full rationality level $\noiseScale = \infty$ may have an arbitrarily small worst-case ratio.
Because pooling positive states weakly reduces the sender's expected utility under the convexity of quantal response function $\Quant\betaed(\cdot)$ on $[0,\infty)$, the censorship structure which reveals some positive states is critical for $\sfrac{1}{2}$-worst case ratio guarantee in \Cref{subsec:baseline_baseline_worst_case_ratio}.

\begin{proof}[Proof of \Cref{prop:necessity_censorship}]
Given a sufficiently small $\varepsilon > 0$,
consider following instance 
with three states:
\begin{align*}
    \prior_1 = \varepsilon, \qquad
    \prior_2 = \prior_3 = \frac{1 - \varepsilon}{2}, \qquad
    \receiverU_1 = -0.01, \qquad
    \receiverU_2 = 0.01, \qquad
    \receiverU_3 = 3\,.
\end{align*}
For a fully rational receiver, suppose that we restrict attention to a signaling scheme that maps each state to a probability distribution over a signal space containing exactly two signals. Specifically, this signaling scheme pools the lowest states into one signal and the remaining states into another, splitting the threshold state between the two groups.
Then, the optimal signaling scheme $\optdirectInfy$ satisfying this requirement for a fully rational receiver is characterized as follows:
\begin{align*}
    \optdirectInfy_1(\scaledNoiseDiff) 
    & = \indicator{\scaledNoiseDiff = 0}\,, \\
    \optdirectInfy_2(\scaledNoiseDiff)
    & = \frac{2\varepsilon}{1-\varepsilon}
    \indicator{\scaledNoiseDiff = 0} + \left(1 - \frac{2\varepsilon}{1-\varepsilon}\right)
    \indicator{\scaledNoiseDiff = \frac{ \frac{0.01(1-3\varepsilon)}{2} + 
    \frac{3(1-\varepsilon)}{2}}{1-2\varepsilon}}\,, \\
    \optdirectInfy_3(\scaledNoiseDiff)
    & = \indicator{\scaledNoiseDiff = \frac{ \frac{0.01(1-3\varepsilon)}{2} + 
    \frac{3(1-\varepsilon)}{2}}{1-2\varepsilon}}\,.
\end{align*}
On the other hand, 
for a sufficiently small $\varepsilon$,
it can be shown that from \Cref{thm:SISU_opt},
the optimal signaling scheme for any finite rationality level $\noiseScale < \infty$ is full revelation, 
yielding 
$\OPT(\noiseScale)
= \varepsilon \Quant\betaed(-0.01) + \sfrac{\Quant\betaed(0.01)(1-\varepsilon)}{2}  + \sfrac{\Quant\betaed(3)(1-\varepsilon)}{2}$. 
Numerically, 
it can be verified that 
$\lim_{\noiseScale \rightarrow \infty} \lim_{\varepsilon \rightarrow 0}
\frac{\Payoff[\noiseScale]{\optdirectInfy}}{\OPT(\noiseScale)}
= 0$,
which completes the proof.
\end{proof}

\subsection{The Equivalence of State Splitting}
\label{apx:state_refinement}

Fix any instance $\instance=(\stateNum,\{\prior_i\}_{i\in[\stateNum]},\{\receiverU_i\}_{i\in[\stateNum]})$. We can construct a refined instance $\refinedinstance$ by replacing any state $i$ with multiple adjacent states with the same value $\receiverU_i$. 
Then every censorship in $\instance$ induces a censorship in $\refinedinstance$ with the same
pooled mass, the same pooling signal, and the same sender's expected utility for every rationality level $\noiseScale\in\fullsetrationalitylevel$,
and any censorship in $\refinedinstance$ maps back to an equivalent scheme in $\instance$.


By the state splitting procedure, fix an instance $\instance$ with candidate censorship $\candidatecensor$ (threshold state $\candidatecensorstate$ and threshold-state fraction $\candidatecensorprob$) and benchmark censorship $\benchmarkcensor$ (threshold state $\benchmarkcensorstate$ and threshold-state fraction $\benchmarkcensorprob$).
If threshold states are different, i.e., $\candidatecensorstate \neq \benchmarkcensorstate$, for the candidate threshold state $\candidatecensorstate$, we can split it into the two adjacent states with probabilities $\candidatecensorprob \prior_{\candidatecensorstate}$ and $(1-\candidatecensorprob)\prior_{\candidatecensorstate}$ by the candidate threshold-state fraction $\candidatecensorprob$, and we apply the same split for the benchmark threshold $\benchmarkcensorstate$.
If threshold states are identical, i.e., $\candidatecensorstate = \benchmarkcensorstate$, we can split the threshold state into three adjacent states with probabilities $ \prior_{\candidatecensorstate}\min\{\candidatecensorprob,\benchmarkcensorprob\}$, $ \prior_{\candidatecensorstate}\mid\candidatecensorprob-\benchmarkcensorprob\mid$, and $ \prior_{\candidatecensorstate}\left(1-\max\{\candidatecensorprob,\benchmarkcensorprob\}\right)$.

\subsection{Tightness of  \texorpdfstring{\Cref{thm:rationality_robust}\ref{item:baseline_ratio}}{Theorem 3(i)}}
\label{subsec:baseline_hard_instance}

Below, we provide the omitted proof of the tightness of the bound in \Cref{thm:rationality_robust}\ref{item:baseline_ratio}.

\begin{proof}[Proof of Tightness of \Cref{thm:rationality_robust}\ref{item:baseline_ratio}]
Consider the binary-state instance in Eqn.~\eqref{ex:binary_worst_case}. 
The optimal censorship signaling
scheme $\candidatecensor$ for a fully rational receiver pools
both states, with $\candidatepooledmass=1$ and
$\candidatepoolsignal=0$. For every finite rationality
level $\noiseScale$, the sender's expected utility is
$
 \Payoff[\noiseScale]{\candidatecensor}
 =\Quant\betaed(0)=\frac12.
$
To prove tightness, for every finite rationality level $\noiseScale>9$, consider
the censorship with pooled mass
$
 \pooledmass_{\noiseScale}
 =1-\noiseScale^{-1/2}
$.
Because $\pooledmass_{\noiseScale}>\prior_1=2/3$, this censorship
pools all of state 1 and part of state 2. Its pooling signal is
$\censorshipsignal(\pooledmass_{\noiseScale})
 =
 \frac{
   \prior_1\receiverU_1+
   (\pooledmass_{\noiseScale}-\prior_1)\receiverU_2
 }{\pooledmass_{\noiseScale}}
 =
 -\frac{2\noiseScale^{-1/2}}
        {\pooledmass_{\noiseScale}}
$.
Therefore, we have $\pooledmass_{\noiseScale}
 \Quant\betaed\!\left(
   \censorshipsignal(\pooledmass_{\noiseScale})
 \right)
\to 1$ for $\noiseScale \to\infty$ since 
$
 \noiseScale
 \censorshipsignal(\pooledmass_{\noiseScale})
 =
 -\frac{2\sqrt{\noiseScale}}
        {\pooledmass_{\noiseScale}}
 \to-\infty
$.
It follows that
$
\OPT(\noiseScale)
 \ge
 \Payoff[\noiseScale]{\signalscheme^{(\pooledmass\betaed)}}
 \longrightarrow1
$ for $\noiseScale \to \infty$.
Consequently, we obtain
$
 \instancerobustRatio{\candidatecensor}
 =
 \inf\nolimits_{\noiseScale\in\fullsetrationalitylevel}
 \frac{\Payoff[\noiseScale]{\candidatecensor}}
      {\OPT(\noiseScale)}
 \le
 \lim_{\noiseScale\to\infty}
 \frac{1}{2\OPT(\noiseScale)}
 =
 \frac12.
$
This completes the proof.
\end{proof}



\subsection{The Non-Positive Prior Mean Instance Assumption in \texorpdfstring{\Cref{subsec:universal_guarantee_maximin_optimal_ratio}}{Section 5.2}}
\label{apx:reduction}

\begin{restatable}{lemma}
{corereduction}
\label{prop:reduction}
For any instance $\instance$ whose prior mean is positive, i.e., $\sum\nolimits_{i \in [\stateNum]}\prior_i\receiverU_i>0$, there exists a corresponding reduced instance $\coreinstance$ such that its prior mean is zero and 
$\sup\nolimits_{\signalscheme \in \schemespace}\instancerobustRatio{\signalscheme, \instance} \geq \sup\nolimits_{\signalscheme \in \censorshipspace}\instancerobustRatio{\signalscheme, \coreinstance}$.
\end{restatable}

\Cref{prop:reduction} allows us to restrict attention to instances with zero prior mean when deriving a universal lower bound on the worst-case ratio for instances with positive prior mean.
The key intuition is that any state above those pooled states under the optimal censorship for a fully rational receiver is never pooled at any rationality level (by \Cref{prop:SISU_threshold_state_monotonicity}), so it contributes the same utility to both the candidate and benchmark censorship signaling schemes.
Removing these revealed states and renormalizing the remaining states to a zero-mean instance $\coreinstance$ can therefore only lower the worst-case ratio.

\xhdr{Reduction}
We first clarify what we mean by a zero-mean truncation point. 
A lower-tail truncation point is represented by a pair \((\splitstate,\splitprob)\), where \(\splitstate\in[\stateNum]\) 
and \(\splitprob \in(0,1]\). It consists of all states $i \in [\splitstate-1]$, together with an 
\(\splitprob\)-fraction of state \(\splitstate\), and its unnormalized mean is
$\lowmean(\splitstate,\splitprob)
\triangleq
\sum\nolimits_{i\in[\splitstate-1]}\prior_i \receiverU_i+\alpha\prior_\splitstate \receiverU_\splitstate
$.
We call such a truncation point zero-mean if the corresponding unnormalized mean is zero, 
i.e., \(\lowmean(\splitstate,\splitprob)=0\) for a lower-tail truncation. Since the selected states have positive mass, this is equivalent 
to saying that its conditional mean is zero.

By the state splitting operation in \Cref{apx:state_refinement}, we may, without loss of generality, refine the instance so that any zero-mean truncation point appearing in the construction is a whole
state. Similarly, whenever we compare two censorship signaling schemes, we may split their
threshold states so that both thresholds are whole states. Since these refinements preserve the sender's expected utility at every rationality level and hence leave $\OPT(\noiseScale)$ and $\sup\nolimits_{\signalscheme \in \censorshipspace}\instancerobustRatio{\signalscheme}$ unchanged, we
henceforth work with such refined instances.


Next, by the definition of lower-tail zero-mean truncation above, and because the lower-tail
unnormalized mean starts negative and the total prior mean is positive,   
    there exists a lower-tail
zero-mean truncation point \((\positivestate,\positivefraction)\) such that
    $  \sum\nolimits_{i\in[\positivestate-1]}\prior_i \receiverU_i + \positivefraction\prior_{\positivestate}\receiverU_{\positivestate}=0
    $.
    After splitting the state $\positivestate$ into two adjacent fragments of masses
    $\positivefraction\prior_{\positivestate}$ and $(1-\positivefraction)\prior_{\positivestate}$, both with value $\receiverU_{\positivestate}$, we may
    assume that there is a whole lower block $\lowerblock\subseteq[\stateNum]$ such that $\sum\nolimits_{i\in \lowerblock}\prior_i \receiverU_i = 0$,
    and every state outside $\lowerblock$ lies weakly above every state in $\lowerblock$.
    Let $\lowerblockmass \triangleq \sum\nolimits_{i\in \lowerblock}\prior_i$,
    and let $\coreinstance$ be the normalized restriction of $\instance$ to the block $\lowerblock$, namely,
    \begin{align*}
        \corestate \gets |\lowerblock|\,;
        \quad 
        \corereceiverU_i \gets \receiverU_i\,, \quad 
        \coreprior_i \;\gets\; \frac{\prior_i}{\lowerblockmass}\,,
        \qquad i\in \lowerblock\,.
    \end{align*}

With the above constructions, we are ready to prove \Cref{prop:reduction}.
When necessary, we write $\Payoff[\noiseScale]{\signalscheme}, \OPT(\noiseScale), \instancerobustRatio{\signalscheme}$ as $\Payoff[\noiseScale]{\signalscheme, \instance}, \OPT(\noiseScale, \instance), \instancerobustRatio{\signalscheme, \instance}$ to indicate their dependence on the considered instance $\instance$.


\begin{proof}[Proof of \Cref{prop:reduction}]
Fix any instance $\instance$, and let the reduced instance $\coreinstance$ be constructed as described above.
Fix a rationality level $\noiseScale\in\fullsetrationalitylevel$.

Now fix any pooled mass $\corecandidatepooledmass\in[\frac{\sum\nolimits_{i \in [\stateNum]:\, \receiverU_i\le 0}\prior_i}{\lowerblockmass}, 1]$ in the admissible range of the reduced instance $\coreinstance$, and let
$\corecandidatesignalscheme$ be the corresponding base candidate censorship with pooled mass $\corecandidatepooledmass$ in
$\coreinstance$. We extend it to a censorship signaling scheme in $\censorshipspace$ in the original
instance $\instance$ by applying exactly the same pooled fractions on the states in $\lowerblock$, and fully
revealing every state outside $\lowerblock$. 
We denote such an extended candidate censorship by $\candidatecensor$.
Let $\revealedstates =[\stateNum]\setminus \lowerblock$.
Then, we have
\begin{align*}
    \Payoff[\noiseScale]{\candidatecensor, \instance}
    =
    \lowerblockmass \Payoff[\noiseScale]{\corecandidatesignalscheme, \coreinstance}
    +
    \sum\nolimits_{i\in \revealedstates}\prior_i \Quant\betaed(\receiverU_i)\,.
\end{align*}
The optimal censorship signaling scheme for a fully rational receiver in the original instance $\instance$ pools exactly the block $\lowerblock$.
By \Cref{prop:SISU_threshold_state_monotonicity}, the optimal pooled mass weakly increases as the rationality level
$\noiseScale$ increases; equivalently, as the rationality level $\noiseScale$ decreases from $\infty$, the optimal pooled mass can only
weakly decrease. Hence, for any rationality level $\noiseScale$, the benchmark censorship in the original instance $\instance$ fully reveals the
entire block $\revealedstates$. Therefore, its restriction to $\lowerblock$ induces a censorship signaling scheme in the reduced instance
$\coreinstance$.
This restricted censorship signaling scheme is optimal for the reduced instance.
Indeed, if another reduced signaling scheme gave a strictly
larger payoff, we could apply that signaling scheme on $\lowerblock$ and
continue to fully reveal $\revealedstates$. The resulting feasible
signaling scheme would strictly improve the original benchmark,
contradicting its optimality. Thus, we have
\begin{align*}
\OPT(\noiseScale, \instance)
=
\lowerblockmass \OPT(\noiseScale, \coreinstance) + \sum\nolimits_{i\in \revealedstates}\prior_i \Quant\betaed(\receiverU_i)\,.
\end{align*}
Consequently, we can bound the utility ratio as follows:
\begin{align*}
\frac{\Payoff[\noiseScale]{\candidatecensor, \instance}}{\OPT(\noiseScale, \instance)}
=
\frac{
\lowerblockmass \Payoff[\noiseScale]{\corecandidatesignalscheme, \coreinstance} + \sum\nolimits_{i\in \revealedstates}\prior_i \Quant\betaed(\receiverU_i)
}{
\lowerblockmass \OPT(\noiseScale, \coreinstance) + \sum\nolimits_{i\in \revealedstates}\prior_i \Quant\betaed(\receiverU_i)
}
\;\ge\;
\frac{\Payoff[\noiseScale]{\corecandidatesignalscheme, \coreinstance}}{\OPT(\noiseScale, \coreinstance)}\,,
\end{align*}
where the inequality follows because
$\Payoff[\noiseScale]{\corecandidatesignalscheme, \coreinstance}\le \OPT(\noiseScale, \coreinstance)$, $\sum\nolimits_{i\in \revealedstates}\prior_i \Quant\betaed(\receiverU_i) \ge 0$ (which naturally follows from  $\Quant\betaed(\cdot) \ge 0$), and the function
$\variableX \;\mapsto\; \frac{\lowerblockmass \variablecapA + \variableX}{\lowerblockmass \variablecapB + \variableX}$
is monotone non-decreasing for $\variableX \in [0,\infty)$, provided that $0\le \variablecapA \le \variablecapB$.

Taking the infimum over $\noiseScale \in \fullsetrationalitylevel$, we obtain
$
\instancerobustRatio{\candidatecensor, \instance} \;\ge\; \instancerobustRatio{\corecandidatesignalscheme, \coreinstance}
$.
Finally, taking the supremum over $\corecandidatepooledmass \in [\frac{\sum\nolimits_{i \in [\stateNum]:\, \receiverU_i\le 0}\prior_i}{\lowerblockmass}, 1]$ gives us
$
\sup\nolimits_{\signalscheme \in \censorshipspace}\instancerobustRatio{\signalscheme, \instance} \ge \sup\nolimits_{\signalscheme \in \censorshipspace}\instancerobustRatio{\signalscheme, \coreinstance}$, implying that $\sup\nolimits_{\signalscheme \in \schemespace}\instancerobustRatio{\signalscheme, \instance} \ge \sup\nolimits_{\signalscheme \in \censorshipspace}\instancerobustRatio{\signalscheme, \coreinstance}$.
This completes the proof of \Cref{prop:reduction}.
\end{proof}

\subsection{Tightness of \texorpdfstring{\Cref{thm:rationality_robust}\ref{item:optimal_ratio}}{Theorem 3(ii)}}
\label{subsec:binary_core_instance}

Below, we provide the omitted proof of the tightness of the bound in \Cref{thm:rationality_robust}\ref{item:optimal_ratio}.
\begin{proof}[Proof of Tightness of \Cref{thm:rationality_robust}\ref{item:optimal_ratio}]
Consider the binary-state instance in Eqn.~\eqref{ex:binary_worst_case}. 
Invoking \Cref{prop:same_pooled_mass_interval}, it suffices to consider censorship signaling schemes that pool at least all non-positive states.
Because the instance has only two states, every such censorship $\signalscheme$ is uniquely determined
by its pooled mass $\pooledmass\in[\prior_1,1]=[\sfrac{2}{3},1]$, with $\censorshipsignal$ as its pooling signal. 
Let $\censorshipparameter \triangleq \sfrac{(1-\pooledmass)}{\pooledmass}$ denote the ratio of disclosed mass to pooled mass. 
Now fix any rationality level $\noiseScale\in[0,\infty)$.
Because state 1 is fully pooled, the only state that is partially revealed is state 2. 
Let $\tailmeanscaled \triangleq  2\noiseScale$.
Then, we have
$
     \Payoff[\noiseScale]{\signalscheme} 
     =\pooledmass \Quant\betaed(\censorshipsignal)+ (1-\pooledmass)\Quant\betaed(2)
=\censorshipcontribution[\censorshipparameter]{\tailmeanscaled}
$
and $
     \OPT(\noiseScale) 
     = \sup\nolimits_{c \in [0,\frac{1}{2}]}\censorshipcontribution[c]{\tailmeanscaled}
     =\optimalcontribution[\sfrac{1}{2}]{\tailmeanscaled}$.
For the rationality level $\noiseScale = \infty$, we have
$
    \Payoff[\infty]{\signalscheme} = \frac{1}{1+\censorshipparameter}$ and $
    \OPT(\infty)  = 1$.
Thus, for any censorship signaling scheme $\signalscheme$ with pooled mass $\pooledmass$, we have 
$
    \instancerobustRatio{\signalscheme}
    =
    \inf\nolimits_{\noiseScale\in\fullsetrationalitylevel}
    \frac{\Payoff[\noiseScale]{\signalscheme}}{\OPT(\noiseScale)}
    =
    \min\left\{
    \frac{1}{1+\censorshipparameter},\;
    \inf\nolimits_{\tailmeanscaled\in[0,\infty)}\frac{\censorshipcontribution[\censorshipparameter]{\tailmeanscaled}}{\optimalcontribution[\sfrac{1}{2}]{\tailmeanscaled}}
    \right\}$.
Therefore, we obtain
$
    \sup\nolimits_{\signalscheme \in \censorshipspace}\instancerobustRatio{\signalscheme}
    =
    \sup\nolimits_{\censorshipparameter\in[0,\frac{1}{2}]}
    \min\left\{
    \frac{1}{1+\censorshipparameter},\;
    \inf\nolimits_{\tailmeanscaled \in[0,\infty)}\frac{\censorshipcontribution[\censorshipparameter]{\tailmeanscaled}}{\optimalcontribution[\sfrac{1}{2}]{\tailmeanscaled}}
    \right\}$.
Notice that the right-hand side of this expression is exactly the function $\negativeratiofn(\negativeratio)$ defined in \Cref{lem:upbound-negativeratiofn}, evaluated at $\negativeratio = \sfrac{1}{2}$.
Then, we show that the minimum operator is redundant, allowing the expression to simplify to a pure supremum-infimum program (see \Cref{lem:upbound-negativeratiofn}\ref{item:lemma_simplification}). Furthermore, we prove that the infimum of $\negativeratiofn(\negativeratio)$ over all $\negativeratio \in (0,\infty)$ is achieved at $\negativeratio = \sfrac{1}{2}$ (see \Cref{lem:upbound-negativeratiofn}\ref{item:lemma_infimum}), yielding the universal guarantee for the maximin optimal ratio $\worstcaseratio$. 
Thus, we conclude that 
$
    \sup\nolimits_{\signalscheme \in \censorshipspace}\instancerobustRatio{\signalscheme}
    = 
    \sup\nolimits_{\censorshipparameter\in[0,\frac{1}{2}]}
    \inf\nolimits_{\tailmeanscaled \in[0,\infty)}\frac{\censorshipcontribution[\censorshipparameter]{\tailmeanscaled}}{\optimalcontribution[\sfrac{1}{2}]{\tailmeanscaled}}
    = \worstcaseratio$.
This completes the proof.
\end{proof}

\subsection{Auxiliary Lemma for \texorpdfstring{\Cref{thm:rationality_robust}\ref{item:optimal_ratio}}{Theorem 3(ii)}}
\label{apx:missing_lemma_maximin_optimal_ratio}
\begin{lemma}
\label{lem:upbound-negativeratiofn}
Suppose that $\negativeratio \in (0,\infty)$. 
Define function
$
    \negativeratiofn(\negativeratio) \triangleq \sup\nolimits_{\variableX\in [0,\negativeratio]}\min \left\{\frac{1}{1+\variableX},\; \inf\nolimits_{\tailmeanscaledf \in[0,\infty)}\frac{\censorshipcontribution[\variableX]{\tailmeanscaledf}}{\optimalcontribution[\negativeratio]{\tailmeanscaledf}}\right\}$, where functions $\censorshipcontribution[\variableX]{\tailmeanscaledf}$ and $\optimalcontribution[\negativeratio]{\tailmeanscaledf}$ are defined in Eqn.~\eqref{eq:n_q_function}. 
Then the following hold:
\begin{enumerate}[label=(\roman*)]
    \item\label{item:lemma_simplification}$\negativeratiofn(\negativeratio) = \sup\nolimits_{\variableX\in [0,\negativeratio]} \inf\nolimits_{\tailmeanscaledf \in[0,\infty)}\frac{\censorshipcontribution[\variableX]{\tailmeanscaledf}}{\optimalcontribution[\negativeratio]{\tailmeanscaledf}}$.

    \item\label{item:lemma_infimum}$\inf_{\negativeratio \in (0,\infty)} \negativeratiofn(\negativeratio) 
= 
\negativeratiofn(\sfrac{1}{2})
=
\worstcaseratio \approx \ratiovalue$, where $\worstcaseratio$ is defined as in Eqn.~\eqref{eq:universal_guarnatee_maximin_optimal_ratio}.
\end{enumerate}

\end{lemma}
\begin{proof}
 We first argue that for any $\negativeratio \in (0,\infty)$ and $\variableX \in [0,\negativeratio]$, it holds that 
 $
   \inf\nolimits_{\variableZ \in [0, \infty)}\frac{\censorshipcontribution[\variableX]{\variableZ}}{\optimalcontribution[\negativeratio]{\variableZ}} \leq \frac{1}{1+\variableX}$.
Let us consider $\lim\nolimits_{\variableZ \rightarrow\infty}\frac{\censorshipcontribution[\variableX]{\variableZ}}{\optimalcontribution[\negativeratio]{\variableZ}}$.
First, we evaluate $\lim\nolimits_{\variableZ\rightarrow \infty}\optimalcontribution[\negativeratio]{\variableZ}$.
As $\variableZ \rightarrow \infty$, we have $\variableZnegativesqrt \rightarrow 0$.
Thus, for sufficiently large $\variableZ$, the inequality $\variableZnegativesqrt \leq \negativeratio$ holds. 
Then, when $\variableX  = \variableZnegativesqrt$, for the function $\censorshipcontribution[\variableX]{\variableZ}$, we have 
$
    \lim\nolimits_{\variableZ \rightarrow \infty}\censorshipcontribution[\variableZnegativesqrt]{\variableZ} = \lim\nolimits_{\variableZ \rightarrow \infty} \frac{\Quant\betaone(-\variableZpositivesqrt)+\variableZnegativesqrt\Quant\betaone(\variableZ)}{1+\variableZnegativesqrt} = 1$,
where the last equality holds due to $\Quant\betaone(-\infty) = 1$ and $\Quant\betaone(\infty) = 0$. 
By the definition of the function $\optimalcontribution[\negativeratio]{\cdot}$ in Eqn.~\eqref{eq:n_q_function}, we have $\optimalcontribution[\negativeratio]{\variableZ} \geq \censorshipcontribution[\variableZnegativesqrt]{\variableZ}$ for sufficiently large $\variableZ$, implying that $\liminf\nolimits_{\variableZ \rightarrow \infty}\optimalcontribution[\negativeratio]{\variableZ} \in [1,\infty)$.
By the definition of the function $\censorshipcontribution[\variableX]{\cdot}$ in Eqn.~\eqref{eq:n_q_function}, we have $
\censorshipcontribution[\variableX]{\variableZ} = \frac{\Quant\betaone(-\variableX\variableZ)+\variableX\Quant\betaone(\variableZ)}{1+\variableX} \leq 1$,
where the inequality holds due to $\Quant\betaone(\cdot)\le 1$.
Thus, we have $\optimalcontribution[\negativeratio]{\variableZ} \le 1$, implying that $\lim\nolimits_{\variableZ \rightarrow \infty}\optimalcontribution[\negativeratio]{\variableZ} =1$.
Next, we evaluate $\lim\nolimits_{\variableZ \rightarrow \infty}\frac{\censorshipcontribution[\variableX]{\variableZ}}{\optimalcontribution[\negativeratio]{\variableZ}}$.
If $\variableX = 0$, we have $\censorshipcontribution[0]{\variableZ}= \Quant\betaone(0) = \frac{1}{2}$, implying that
$  \lim\nolimits_{\variableZ \rightarrow \infty}\frac{\censorshipcontribution[0]{\variableZ}}{\optimalcontribution[\negativeratio]{\variableZ}} = \frac{1}{2} < \left. \frac{1}{1+\variableX}\, \right|_{\variableX = 0}$.
If $\variableX\in (0,\infty)$, we have $\lim\nolimits_{\variableZ \rightarrow \infty}\censorshipcontribution[\variableX]{\variableZ} = \frac{\Quant\betaone(-\infty)+\variableX\Quant\betaone(\infty)}{1+\variableX} = \frac{1}{1+\variableX}$, implying that 
$ \lim\nolimits_{\variableZ \rightarrow \infty}\frac{\censorshipcontribution[\variableX]{\variableZ}}{\optimalcontribution[\negativeratio]{\variableZ}} = \frac{1}{1+\variableX}$.
Considering the above two cases, we have 
\begin{align*}
    \inf\nolimits_{\variableZ \in [0, \infty)}\frac{\censorshipcontribution[\variableX]{\variableZ}}{\optimalcontribution[\negativeratio]{\variableZ}}\leq\lim\nolimits_{\variableZ \rightarrow \infty}\frac{\censorshipcontribution[\variableX]{\variableZ}}{\optimalcontribution[\negativeratio]{\variableZ}} \le \frac{1}{1+\variableX}\,.
    \end{align*}
Therefore, for the function $\negativeratiofn(\negativeratio)$, the outer $\min$ operator is mathematically redundant.
In other words, we can simplify the function $\negativeratiofn(\negativeratio)$ as 
$\negativeratiofn(\negativeratio) = \sup\nolimits_{\variableX \in [0, \negativeratio]}\inf\nolimits_{\variableZ \in [0, \infty)} \frac{\censorshipcontribution[\variableX]{\variableZ}}{\optimalcontribution[\negativeratio]{\variableZ}}$.

Next, we claim that for any $\candidatenegativeratio, \benchmarknegativeratio \in (0, \infty)$ such that $\candidatenegativeratio < \benchmarknegativeratio$, it holds that $\negativeratiofn(\benchmarknegativeratio) \le \negativeratiofn(\candidatenegativeratio)$.
Because $[0, \candidatenegativeratio] \subset [0,\benchmarknegativeratio]$, by the definition of the function $\optimalcontribution[\negativeratio]{\cdot}$ in Eqn.~\eqref{eq:n_q_function}, we have $\optimalcontribution[\candidatenegativeratio]{\variableZ} \le \optimalcontribution[\benchmarknegativeratio]{\variableZ}$.
Because the function $\censorshipcontribution[\variableX]{\variableZ}$ is independent of $\negativeratio$, for every $\variableX \in [0, \candidatenegativeratio]$, we have 
$
    \inf\nolimits_{\variableZ \in [0, \infty)} \frac{\censorshipcontribution[\variableX]{\variableZ}}{\optimalcontribution[\benchmarknegativeratio]{\variableZ}} \le \inf\nolimits_{\variableZ \in [0, \infty)} \frac{\censorshipcontribution[\variableX]{\variableZ}}{\optimalcontribution[\candidatenegativeratio]{\variableZ}}
$.
Then, we partition the domain $[0, \benchmarknegativeratio]$ into two intervals: the base interval $[0, \candidatenegativeratio]$ and the extended interval $(\candidatenegativeratio,\benchmarknegativeratio]$, and then we can rewrite $\negativeratiofn(\benchmarknegativeratio)$ into the  equivalent form $\negativeratiofn(\benchmarknegativeratio) = \max\left\{\sup\nolimits_{\variableX \in [0, \candidatenegativeratio]}\inf\nolimits_{\variableZ \in [0, \infty)} \frac{\censorshipcontribution[\variableX]{\variableZ}}{\optimalcontribution[\benchmarknegativeratio]{\variableZ}},~ \sup\nolimits_{\variableX \in (\candidatenegativeratio, \benchmarknegativeratio]}\inf\nolimits_{\variableZ \in [0, \infty)} \frac{\censorshipcontribution[\variableX]{\variableZ}}{\optimalcontribution[\benchmarknegativeratio]{\variableZ}}\right\}$.
For the first term of $\negativeratiofn(\benchmarknegativeratio)$, by the above inequality, we have
\begin{align*}
    \sup\nolimits_{\variableX \in [0, \candidatenegativeratio]}\inf\nolimits_{\variableZ \in [0, \infty)} \frac{\censorshipcontribution[\variableX]{\variableZ}}{\optimalcontribution[\benchmarknegativeratio]{\variableZ}} \le \sup\nolimits_{\variableX \in [0, \candidatenegativeratio]}\inf\nolimits_{\variableZ \in [0, \infty)} \frac{\censorshipcontribution[\variableX]{\variableZ}}{\optimalcontribution[\candidatenegativeratio]{\variableZ}} = \negativeratiofn(\candidatenegativeratio)\,.
\end{align*}
For the second term of $\negativeratiofn(\benchmarknegativeratio)$, we first establish an auxiliary lower bound of $\negativeratiofn(\candidatenegativeratio)$. 
Let us consider the function $\censorshipcontribution[\variableX]{\variableZ}$.
For $\variableX \in [0, \candidatenegativeratio]$, we have 
$
\censorshipcontribution[\variableX]{\variableZ} = \frac{\Quant\betaone(-\variableX\variableZ)+\variableX\Quant\betaone(\variableZ)}{1+\variableX} \le \frac{\Quant\betaone(-\candidatenegativeratio\variableZ)+\candidatenegativeratio\Quant\betaone(\variableZ)}{1} = (1+\candidatenegativeratio)\censorshipcontribution[\candidatenegativeratio]{\variableZ}$,
where the inequality holds because the function $\variableX\, \mapsto\, \Quant\betaone(-\variableX\variableZ)+\variableX\Quant\betaone(\variableZ)$ is monotone non-decreasing on $[0,\candidatenegativeratio]$.
Then, taking the supremum over $\variableX \in [0,\candidatenegativeratio]$ yields 
$
\optimalcontribution[\candidatenegativeratio]{\variableZ} \le (1+\candidatenegativeratio)\censorshipcontribution[\candidatenegativeratio]{\variableZ}.
$
Rearranging the above inequality gives 
$\frac{\censorshipcontribution[\candidatenegativeratio]{\variableZ}}{\optimalcontribution[\candidatenegativeratio]{\variableZ}} \ge \frac{1}{1+\candidatenegativeratio}$.
Taking the infimum over $\variableZ \in [0, \infty)$, we have 
$
\inf\nolimits_{\variableZ \in [0, \infty)} \frac{\censorshipcontribution[\candidatenegativeratio]{\variableZ}}{\optimalcontribution[\candidatenegativeratio]{\variableZ}} \ge \frac{1}{1+\candidatenegativeratio}    
$,
implying that 
$
\negativeratiofn(\candidatenegativeratio) = \sup\nolimits_{\variableX \in [0, \candidatenegativeratio]}\inf\nolimits_{\variableZ \in [0, \infty)} \frac{\censorshipcontribution[\variableX]{\variableZ}}{\optimalcontribution[\candidatenegativeratio]{\variableZ}}  \geq \frac{1}{1+\candidatenegativeratio} $.
Next, for any $\variableX \in (\candidatenegativeratio,\benchmarknegativeratio]$, it holds that 
\begin{align*}
    \inf\nolimits_{\variableZ \in [0, \infty)}\frac{\censorshipcontribution[\variableX]{\variableZ}}{\optimalcontribution[\benchmarknegativeratio]{\variableZ}} \overset{(a)}{\le} \frac{1}{1+\variableX} \overset{(b)}{<} \frac{1}{1+\candidatenegativeratio} \overset{(c)}{\le} \negativeratiofn(\candidatenegativeratio)\,,
\end{align*}
where inequality~(a) follows from the inequality built for the two inner terms comparison in $\negativeratiofn(\negativeratio)$, inequality~(b) holds since $\variableX>\candidatenegativeratio$, and inequality~(c) holds due to the above inequality.
Therefore, combining the above two inequalities of $\negativeratiofn(\candidatenegativeratio)$, we obtain
$\negativeratiofn(\benchmarknegativeratio) = \max\left\{\sup\nolimits_{\variableX \in [0, \candidatenegativeratio]}\inf\nolimits_{\variableZ \in [0, \infty)} \frac{\censorshipcontribution[\variableX]{\variableZ}}{\optimalcontribution[\benchmarknegativeratio]{\variableZ}},~ \sup\nolimits_{\variableX \in (\candidatenegativeratio, \benchmarknegativeratio]}\inf\nolimits_{\variableZ \in [0, \infty)} \frac{\censorshipcontribution[\variableX]{\variableZ}}{\optimalcontribution[\benchmarknegativeratio]{\variableZ}}\right\}  \le \max \left\{\negativeratiofn(\candidatenegativeratio),~ \negativeratiofn(\candidatenegativeratio)\right\}= \negativeratiofn(\candidatenegativeratio)$.

Then, we show that when $\variableX \in [\frac{1}{2},\infty)$, the function $\censorshipcontribution[\variableX]{\variableZ}$ is monotone non-increasing. Taking the first derivative of $\censorshipcontribution[\variableX]{\variableZ}$ with respect to $\variableX$, and denoting $\Quant\betaone'(\cdot)$ as the first derivative of the quantal response function, we have
$
    \frac{\partial\censorshipcontribution[\variableX]{\variableZ}}{\partial \variableX} = \frac{(1+\variableX)\left(-\variableZ\Quant\betaone'(-\variableX\variableZ) + \Quant\betaone(\variableZ)\right) - \left(\Quant\betaone(-\variableX\variableZ) + \variableX\Quant\betaone(\variableZ)\right)}{(1+\variableX)^2} = \frac{-\variableZ(1+\variableX)\Quant\betaone'(-\variableX\variableZ)+\Quant\betaone(\variableZ)-\Quant\betaone(-\variableX\variableZ)}{(1+\variableX)^2}$.
Since the denominator is always positive, the sign of this derivative is dictated by the numerator. Let us define the auxiliary function $\judgefunctionf(\variableX)$ as $
    \judgefunctionf(\variableX) \triangleq -\variableZ(1+\variableX)\Quant\betaone'(-\variableX\variableZ)+\Quant\betaone(\variableZ)-\Quant\betaone(-\variableX\variableZ)$.
Taking the derivative of $\judgefunctionf(\variableX)$ with respect to $\variableX$, we obtain
$\frac{\partial\judgefunctionf(\variableX)}{\partial\variableX} = \variableZ^2(1+\variableX) \Quant\betaone''(-\variableX\variableZ) \leq 0$,
where the inequality holds because $-\variableX\variableZ \leq 0$, and the quantal response function $\Quant\betaone(\cdot)$ is concave on the non-positive domain $(-\infty, 0]$, meaning its second derivative $\Quant\betaone''(\cdot)$ is non-positive.
Because $\judgefunctionf(\variableX)$ is monotone non-increasing, if we can show that $\judgefunctionf(\frac{1}{2}) \leq 0$, it implies that $\censorshipcontribution[\variableX]{\variableZ}$ is monotone non-increasing for all $\variableX \in [\frac{1}{2},\infty)$. This would guarantee that the maximum of $\censorshipcontribution[\variableX]{\variableZ}$ can be restricted to $\variableX \in [0,\frac{1}{2}]$.
Evaluating $\judgefunctionf(\frac{1}{2})$ with the substitution $\variableH \triangleq \exp(\frac{1}{2}\variableZ) \geq 1$, we obtain $
    \judgefunctionf\left(\frac{1}{2}\right) = -\frac{3}{2}\variableZ \Quant\betaone'\left(-\frac{1}{2}\variableZ\right)+\Quant\betaone(\variableZ)-\Quant\betaone\left(-\frac{1}{2}\variableZ\right)= \frac{3\variableH\ln \variableH}{(1+\variableH)^2} + \frac{1}{1+\variableH^2} - \frac{\variableH}{1+\variableH}$.
We wish to show that the evaluated expression is non-positive for all $\variableH \in [1,\infty)$. Let us define this expression as 
$
    \judgefunction(\variableH) \triangleq \frac{3\variableH\ln \variableH}{(1+\variableH)^2} + \frac{1}{1+\variableH^2} - \frac{\variableH}{1+\variableH}$.
To bound the logarithmic term compactly, we use the substitution $\variableR = \frac{\variableH-1}{\variableH+1} \in [0, 1)$ and the Taylor series expansion for $\ln\left(\frac{1+\variableR}{1-\variableR}\right)$. By replacing the denominators of all higher-order terms with $3$, we establish a rational upper bound as
$
    \ln \variableH = 2 \sum\nolimits_{k=0}^\infty \frac{\variableR^{2k+1}}{2k+1} 
    \le 2\variableR + \frac{2\variableR^3}{3} \sum\nolimits_{k=0}^\infty \variableR^{2k}= 2\variableR + \frac{2\variableR^3}{3(1-\variableR^2)} = \frac{2(\variableH-1)}{\variableH+1} + \frac{2(\variableH-1)^3}{12\variableH(\variableH+1)} 
    = \frac{(\variableH-1)(\variableH^2+10\variableH+1)}{6\variableH(\variableH+1)}$.
Substituting this bound back into the definition of $\judgefunction(\variableH)$ and factoring out the common term $\frac{\variableH-1}{1+\variableH}$, we obtain $
    \judgefunction(\variableH) 
    \le \frac{3\variableH}{(1+\variableH)^2} \left( \frac{(\variableH-1)(\variableH^2+10\variableH+1)}{6\variableH(\variableH+1)} \right) + \frac{1}{1+\variableH^2} - \frac{\variableH}{1+\variableH} 
    = \frac{\variableH-1}{1+\variableH} \left( \frac{\variableH^2+10\variableH+1}{2(1+\variableH)^2} - \frac{\variableH^2+\variableH+1}{1+\variableH^2} \right) 
    = -\frac{(\variableH-1)^5}{2(1+\variableH)^3(1+\variableH^2)} \leq 0$,
where the last inequality holds since for all $\variableH \in [1,\infty)$, the term $(\variableH-1)^5$ is non-negative, and the denominator is strictly positive. 

Finally, we characterize the global minimum of $\negativeratiofn(\negativeratio)$ for $\negativeratio \in (0,\infty)$.
Fix an arbitrary $\negativeratio \in [\frac{1}{2},\infty)$.
Note that $\censorshipcontribution[\variableX]{\variableZ} \le \censorshipcontribution[\frac{1}{2}]{\variableZ}$ holds for $\variableX\in [\frac{1}{2},\negativeratio]$.  
Thus, extending the supremum's domain of $\optimalcontribution[\negativeratio]{\variableZ}$ beyond the threshold of $\frac{1}{2}$ does not increase the supremum.
In other words, for every $\variableZ \in [0,\infty)$, it holds that
$
\optimalcontribution[\negativeratio]{\variableZ} = \optimalcontribution[\frac{1}{2}]{\variableZ}  
$.
Dividing $\censorshipcontribution[\variableX]{\variableZ} \le \censorshipcontribution[\frac{1}{2}]{\variableZ}$ by the strictly positive denominator $\optimalcontribution[\frac{1}{2}]{\variableZ}$, where $\optimalcontribution[\frac{1}{2}]{\variableZ} = \sup\nolimits_{\variableX \in [0, \frac{1}{2}]}\censorshipcontribution[\variableX]{\variableZ}  \ge \censorshipcontribution[0]{\variableZ} = \frac{1}{2}$, for every $\variableZ \in [0,\infty)$, we get 
$
\frac{\censorshipcontribution[\variableX]{\variableZ}}{\optimalcontribution[\frac{1}{2}]{\variableZ}} \leq \frac{\censorshipcontribution[\frac{1}{2}]{\variableZ}}{\optimalcontribution[\frac{1}{2}]{\variableZ}}$.
Taking the infimum over $\variableZ \in [0,\infty)$ of both sides of the above inequality, for any $\variableX \in (\frac{1}{2}, \negativeratio]$, we have
\begin{align*}
\inf\nolimits_{\variableZ \in [0, \infty)} \frac{\censorshipcontribution[\variableX]{\variableZ}}{\optimalcontribution[\negativeratio]{\variableZ}} = \inf\nolimits_{\variableZ \in [0, \infty)} \frac{\censorshipcontribution[\variableX]{\variableZ}}{\optimalcontribution[\frac{1}{2}]{\variableZ}} \le \inf\nolimits_{\variableZ \in [0, \infty)} \frac{\censorshipcontribution[\frac{1}{2}]{\variableZ}}{\optimalcontribution[\frac{1}{2}]{\variableZ}}\,,
\end{align*}
where the first equality holds due to 
$\optimalcontribution[\negativeratio]{\variableZ} = \optimalcontribution[\frac{1}{2}]{\variableZ}$.
Then, we partition the supremum domain of $\negativeratiofn(\negativeratio)$, given by $[0,\negativeratio]$, at the constant $\frac{1}{2}$ and then rewrite $\negativeratiofn(\negativeratio)$ into $
\negativeratiofn(\negativeratio) = \max\left\{\sup\nolimits_{\variableX \in [0, \frac{1}{2}]}\inf\nolimits_{\variableZ \in [0, \infty)} \frac{\censorshipcontribution[\variableX]{\variableZ}}{\optimalcontribution[\negativeratio]{\variableZ}},~ \sup\nolimits_{\variableX \in (\frac{1}{2}, \negativeratio]}\inf\nolimits_{\variableZ \in [0, \infty)} \frac{\censorshipcontribution[\variableX]{\variableZ}}{\optimalcontribution[\negativeratio]{\variableZ}}\right\}$.
For the second term, we have
\begin{align*}
    \sup\nolimits_{\variableX \in (\frac{1}{2}, \negativeratio]}\inf\nolimits_{\variableZ \in [0, \infty)} \frac{\censorshipcontribution[\variableX]{\variableZ}}{\optimalcontribution[\negativeratio]{\variableZ}} \le \inf\nolimits_{\variableZ \in [0, \infty)} \frac{\censorshipcontribution[\frac{1}{2}]{\variableZ}}{\optimalcontribution[\frac{1}{2}]{\variableZ}} \le \sup\nolimits_{\variableX \in [0, \frac{1}{2}]}\inf\nolimits_{\variableZ \in [0, \infty)} \frac{\censorshipcontribution[\variableX]{\variableZ}}{\optimalcontribution[\frac{1}{2}]{\variableZ}}=  \negativeratiofn(\frac{1}{2})\,,
\end{align*}
where the first inequality holds due to the above inequality. 
For the first term of $\negativeratiofn(\negativeratio)$, we have
\begin{align*}
   \sup\nolimits_{\variableX \in [0, \frac{1}{2}]}\inf\nolimits_{\variableZ \in [0, \infty)} \frac{\censorshipcontribution[\variableX]{\variableZ}}{\optimalcontribution[\negativeratio]{\variableZ}} = \sup\nolimits_{\variableX \in [0, \frac{1}{2}]}\inf\nolimits_{\variableZ \in [0, \infty)} \frac{\censorshipcontribution[\variableX]{\variableZ}}{\optimalcontribution[\frac{1}{2}]{\variableZ}} = \negativeratiofn(\frac{1}{2})\,, 
\end{align*}
where the first equality holds due to $\optimalcontribution[\negativeratio]{\variableZ} = \optimalcontribution[\frac{1}{2}]{\variableZ}$.
Considering the above two inequalities of $\negativeratiofn(\frac{1}{2})$, for any $\negativeratio \geq \frac{1}{2}$, we have
$
\negativeratiofn(\negativeratio) = \max\left\{\sup\nolimits_{\variableX \in [0, \frac{1}{2}]}\inf\nolimits_{\variableZ \in [0, \infty)} \frac{\censorshipcontribution[\variableX]{\variableZ}}{\optimalcontribution[\negativeratio]{\variableZ}},~ \sup\nolimits_{\variableX \in (\frac{1}{2}, \negativeratio]}\inf\nolimits_{\variableZ \in [0, \infty)} \frac{\censorshipcontribution[\variableX]{\variableZ}}{\optimalcontribution[\negativeratio]{\variableZ}}\right\}\leq \max\left\{\negativeratiofn(\frac{1}{2}),~ \negativeratiofn(\frac{1}{2})\right\} = \negativeratiofn(\frac{1}{2})$.
Conversely, because the interval $[0, \frac{1}{2}]$ is a subset of the full domain $[0, \negativeratio]$, the supremum over the full domain $[0, \negativeratio]$  satisfies 
$
\negativeratiofn(\negativeratio)  = \sup\nolimits_{\variableX \in [0,\negativeratio]}\inf\nolimits_{\variableZ \in [0, \infty)} \frac{\censorshipcontribution[\variableX]{\variableZ}}{\optimalcontribution[\negativeratio]{\variableZ}} = \sup\nolimits_{\variableX \in [0,\negativeratio]}\inf\nolimits_{\variableZ \in [0, \infty)} \frac{\censorshipcontribution[\variableX]{\variableZ}}{\optimalcontribution[\frac{1}{2}]{\variableZ}}\ge \sup\nolimits_{\variableX \in [0,\frac{1}{2}]}\inf\nolimits_{\variableZ \in [0, \infty)} \frac{\censorshipcontribution[\variableX]{\variableZ}}{\optimalcontribution[\frac{1}{2}]{\variableZ}} = \negativeratiofn(\frac{1}{2})$,
where the second equality holds due to $\optimalcontribution[\negativeratio]{\variableZ} = \optimalcontribution[\frac{1}{2}]{\variableZ}$.
Combining these bounds yields $\negativeratiofn(\negativeratio) = \negativeratiofn(\frac{1}{2})$ for all  $\negativeratio \in [\frac{1}{2},\infty)$.

In summary, because the function $\negativeratiofn(\negativeratio)$ is monotone non-increasing on $\negativeratio \in (0, \frac{1}{2}]$ and becomes constant for every $\negativeratio \in [\frac{1}{2},\infty)$, one global minimum of $\negativeratiofn(\negativeratio)$ over the entire domain of $\negativeratio \in (0,\infty)$ is attained at $\negativeratio=\frac{1}{2}$.
By numerical computation of $\negativeratiofn(\frac{1}{2})$, we obtain that the universal guarantee for the maximin optimal ratio is $\worstcaseratio \approx \ratiovalue$.
This finishes the proof of \Cref{lem:upbound-negativeratiofn}.
\end{proof}

\section{Computational Aspects}
\label{sec:computation}
In this section, we present the computational results for computing the optimal signaling scheme for a known rationality level and 
the approximately rationality-robust optimal signaling scheme.

\subsection{Computing the Optimal Censorship for Known Rationality Level}
\label{apx:complexity_SISU_opt}
\begin{proposition}
\label{prop:complexity_SISU_opt}
Fix any instance $\instance = (\stateNum,\{\prior_i\}_{i\in[\stateNum]},\{\receiverU_i\}_{i\in[\stateNum]})$.
Given any known rationality level $\noiseScale \in \fullsetrationalitylevel$, there exists a $\poly(m)$ time algorithm that computes the optimal signaling scheme.
\end{proposition}
\begin{proof}
When the receiver is fully rational, by \Cref{lem:opt_fully_rational}, the optimal signaling scheme can be computed in polynomial time.
When the receiver is not fully rational, recall that by \Cref{thm:opt_state_independent}, the optimal 
signaling scheme is a censorship
signaling scheme. 
Thus, to find the optimal signaling scheme, it suffices to find the 
threshold state $\censorshipstate$ and the threshold-state fraction 
$\censorshipprob$.
To identify $\censorshipstate, \censorshipprob$, consider the following procedure:
for every state $i\in[\stateNum]$ where $\receiverU_i \ge 0$, compute the corresponding candidate threshold fraction 
$p_i$, where $p_i$ is defined as in Eqn.~\eqref{eq:pooled_prob_defn}. 
If the set $\{i \in [\stateNum]: \receiverU_i \ge 0, p_i \ge 0 \}$ is non-empty, we set the threshold state to $\censorshipstate = \max\{i \in [\stateNum]: \receiverU_i \ge 0, p_i \ge 0 \}$ and the threshold-state fraction to $\censorshipprob = \min\{p_{\censorshipstate}, 1\}$. 
Otherwise, if no such state exists, we set $\censorshipstate = \max\{i \in [\stateNum]: \receiverU_i \le 0\}$ and $\censorshipprob=1$.
It is easy to see that the above procedure has 
the complexity at most $O(m)$. 
This completes the proof of \Cref{prop:complexity_SISU_opt}.
\end{proof}

\subsection{Computing the Approximately Rationality-Robust Optimal Censorship}
\label{apx:computation_rationality_robust_scheme}

\begin{proposition}
\label{prop:computation_rationality_robust_scheme}
  Fix any instance $\instance = (\stateNum,\{\prior_i\}_{i\in[\stateNum]},\{\receiverU_i\}_{i\in[\stateNum]})$. Given any $\accuracy \in (0,1)$, for each robustness criterion, there exists an FPTAS (\Cref{algo:ptas_rationality_robust}) that has the running time of $\poly(\stateNum, \sfrac{1}{\epsilon})$ and computes a corresponding censorship $\fptascensorratio$~/~$\fptascensorpayoff$~/~$\fptascensorregret$ such that
  \begin{align*}
      \instancerobustRatio{\fptascensorratio} \ge (1-\accuracy)\instancerobustRatio{\schemeoptimalratio}\,,\quad \instancerobustPayoff{\fptascensorpayoff} \ge (1-\accuracy)\instancerobustPayoff{\schemeoptimalpayoff}\,,\quad \instancerobustRegret{\fptascensorregret} \ge \instancerobustRegret{\schemeoptimalregret} -\accuracy\,.
  \end{align*}
\end{proposition}
The intuition behind \Cref{prop:computation_rationality_robust_scheme} is threefold.  
First, it suffices to restrict attention to censorships, which reduces the problem to a one-dimensional search over pooled mass $\pooledmass \in [0,1]$ (see \Cref{thm:robust_optimal_censorship_three_criteria}). This allows us to uniformly discretize the relevant pooled mass domain.
Second, recall from \Cref{prop:complexity_SISU_opt} that for any known rationality level $\noiseScale$, the optimal censorship can be computed in $\poly(m)$ time.
Third, for the discretization of rationality level set, rather than discretizing it evenly, we place non-uniform grid points so that for each candidate signal, the receiver's quantal response probability $\Quant\betaed(\cdot)$ changes by at most $O(\epsilon)$ between consecutive points, which can further control the change of sender's expected utility across grid.

\begin{proof}[Proof of \Cref{prop:computation_rationality_robust_scheme}]
Choose a small positive number $\smallnumber > 0 $.
Let $\nonpositivemass \triangleq \sum\nolimits_{i\in [\stateNum]:\receiverU_i \le 0}\prior_i$ denote the total mass for non-positive states, $\fullyrationalmass$ denote the pooled mass of optimal censorship for a fully rational receiver and $\signalscheme^{(\pooledmass)}$ denote the censorship with the pooled mass $\pooledmass$.

We first introduce the discretization of the pooled mass domain.
We choose a pooled mass grid $\pooledmassgrid \subseteq [\nonpositivemass, \fullyrationalmass]$ that contains both endpoints, with a grid spacing of at most $\gridspace \le \frac{\smallnumber \fullyrationalmass}{2}$.\footnote{We restrict our search to this range due to \Cref{prop:same_pooled_mass_interval}.}
Thus, there are $O(\sfrac{1}{\smallnumber})$ pooled mass grid points.
Next, we focus on the discretizations of rationality level set $\fullsetrationalitylevel$.
Rather than simply partitioning it evenly, we would like to achieve a goal such that between two neighboring rationality level grid points, for a non-zero signal, the value of quantal response $\Quant\betaed(\scaledNoiseDiff)$ changes by at most $\smallnumber$.
Note that we have $O(\stateNum+\sfrac{1}{\smallnumber})$ candidate signals in total: $\scaledNoiseDiff \in \{\censorshipsignal(\pooledmass):\pooledmass \in \pooledmassgrid\} \cup \{\receiverU_i: \receiverU_i\ge 0, i \in [\stateNum]\}$.
For each non-zero candidate signal $\scaledNoiseDiff$, we generate $O(\sfrac{1}{\smallnumber})$ finite rationality level grid points using the explicit formula $\noiseScale(\scaledNoiseDiff,\variableK) = \frac{1}{\scaledNoiseDiff} \log\left(\frac{\lceil\sfrac{1}{\smallnumber}\rceil}{\variableK}-1\right)$, where $\variableK \in [\lceil \sfrac{1}{\smallnumber} \rceil-1]$. We also unconditionally include the two boundary rationality levels $\{0, \infty\}$.
For a zero signal ($\scaledNoiseDiff = 0$), we do not need to generate specific grid points since $\Quant\betaed(0)=\frac{1}{2}$ for any finite rationality level and $\Quant_{\infty}(0)=1$.
Applying this procedure across all candidate signals yields a rationality level grid $\rationalitylevelgrid$ containing $O(\sfrac{m}{\smallnumber}+\sfrac{1}{\smallnumber^2})$ points.
By this construction, it can be seen that between consecutive rationality levels, the quantal response $\Quant\betaed(\scaledNoiseDiff)$ for each candidate signal $\scaledNoiseDiff$ changes by at most $\smallnumber$.

Fix a pooled mass $\pooledmass \in \pooledmassgrid$.
For any finite rationality level $\noiseScale$, let $\noiseScaleHat \in \rationalitylevelgrid$ be the smallest grid point strictly greater than $\noiseScale$.
Then, we can decompose sender's expected utility $\Payoff[\noiseScale]{\signalscheme^{(\pooledmass)}}$ under the censorship $\signalscheme^{(\pooledmass)}$ for the rationality level $\noiseScale$ into two parts: one is the utility contribution from the accumulated mass $\fullyrationalmass$ and the other is from utility contribution from the remaining mass $(1-\fullyrationalmass)$.
For ease of presentation, let $\payofffullyrational[\noiseScale]{\pooledmass}$ denote the utility contribution from the accumulated mass $\fullyrationalmass$ under censorship $\signalscheme^{(\pooledmass)}$, where if $\censorshipstate(\pooledmass)<\censorshipstate(\fullyrationalmass)$, $\payofffullyrational[\noiseScale]{\pooledmass} \triangleq \pooledmass \Quant\betaed(\censorshipsignal(\pooledmass)) + (1-\censorshipprob(\pooledmass))\prior_{\censorshipstate(\pooledmass)}\Quant\betaed\left(\receiverU_{\censorshipstate(\pooledmass)}\right) + \sum\nolimits_{i \in [\censorshipstate(\pooledmass)+1,\censorshipstate(\fullyrationalmass)-1]}\prior_i \Quant\betaed(\receiverU_i) + \censorshipprob(\fullyrationalmass)\prior_{\censorshipstate(\fullyrationalmass)}\Quant\betaed\left(\receiverU_{\censorshipstate(\fullyrationalmass)}\right)$ and if $\censorshipstate(\pooledmass) = \censorshipstate(\fullyrationalmass)$, $\payofffullyrational[\noiseScale]{\pooledmass} \triangleq \pooledmass \Quant\betaed(\censorshipsignal(\pooledmass))+\left(\censorshipprob(\fullyrationalmass)-\censorshipprob(\pooledmass)\right)\prior_{\censorshipstate(\fullyrationalmass)}\Quant\betaed\left(\receiverU_{\censorshipstate(\fullyrationalmass)}\right)$, and $\payoffcommon[\noiseScale] \triangleq (1-\censorshipprob(\fullyrationalmass))\prior_{\censorshipstate(\fullyrationalmass)}\Quant\betaed\left(\receiverU_{\censorshipstate(\fullyrationalmass)}\right) + \sum\nolimits_{i \in [\censorshipstate(\fullyrationalmass)+1:\stateNum]}\prior_i \Quant\betaed(\receiverU_i) $ denote the utility contribution from the remaining mass $(1-\fullyrationalmass)$, which only depends on $\fullyrationalmass$ and $\noiseScale$.
The analogous expressions hold for $\payofffullyrational[\noiseScaleHat]{\pooledmass}$ and $\payoffcommon[\noiseScaleHat]$ for the rationality level $\noiseScaleHat$.
The candidate with a zero-mean pooling signal requires separate treatment.
If $\censorshipsignal(\fullyrationalmass)=0$, we directly obtain its worst-case performance: $\instancerobustRatio{\signalscheme^{(\fullyrationalmass)}} = \sfrac{1}{2}$, $\instancerobustPayoff{\signalscheme^{(\fullyrationalmass)}} = \sfrac{\fullyrationalmass}{2}$, and $\instancerobustRegret{\signalscheme^{(\fullyrationalmass)}} = -\sfrac{\fullyrationalmass}{2}$.
For all remaining candidates, we use the rationality grid to estimate their objective values.

For the candidate comparisons below, fix $\pooledmass \in \pooledmassgrid$
with $\censorshipsignal(\pooledmass)<0$. 
We have the following two observations about these two utility contributions:
(1) for the utility contribution from the accumulated mass $\fullyrationalmass$, 
\begin{align}
\label{eq:payoff_fully_rational_mass_gap}
    \left\lvert \payofffullyrational[\noiseScale]{\pooledmass} -\payofffullyrational[\noiseScaleHat]{\pooledmass} \right\rvert \le \smallnumber \fullyrationalmass\,,
\end{align}
since $\lvert \Quant\betaed(\scaledNoiseDiff) -\Quant\betahat(\scaledNoiseDiff) \rvert \le \smallnumber$ for any non-zero signal $\scaledNoiseDiff$, and the coefficients in $\payofffullyrational[\noiseScale]{\pooledmass}$ or $\payofffullyrational[\noiseScaleHat]{\pooledmass}$ sum to $\fullyrationalmass$; and
(2) for the utility contribution from the remaining mass $(1-\fullyrationalmass)$, we have $\payoffcommon[\noiseScale] \ge \payoffcommon[\noiseScaleHat]$ since $\receiverU_i \ge 0$ for any state $ i \ge \censorshipstate(\fullyrationalmass)$ and the function $\noiseScale \mapsto \frac{1}{1+\exp(\noiseScale\scaledNoiseDiff)}$ is monotone non-increasing for a fixed signal $\scaledNoiseDiff>0$.

Next, fix a rationality level $\noiseScale$ and compare the sender's expected utility under censorships with two pooled masses $\candidatepooledmass<\benchmarkpooledmass$. Then, we have
\begin{align}
\label{eq:additive_loss}
    \Payoff[\noiseScale]{\signalscheme^{(\candidatepooledmass)}}-\Payoff[\noiseScale]{\signalscheme^{(\benchmarkpooledmass)}} 
    \ge \candidatepooledmass\Quant\betaed(\candidatepoolsignal) - \benchmarkpooledmass\Quant\betaed(\benchmarkpoolsignal) \overset{(a)}{\ge} (\candidatepooledmass - \benchmarkpooledmass)\Quant\betaed(\benchmarkpoolsignal) \overset{(b)}{\ge} \candidatepooledmass - \benchmarkpooledmass\,,
\end{align}
where inequality~(a) holds since these two pooling signals satisfy $\candidatepoolsignal \le \benchmarkpoolsignal$ and the quantal response function $\Quant\betaed(\cdot)$ is monotone non-increasing, and inequality~(b) holds due to $\Quant\betaed(\cdot)\le 1$ and $ \candidatepooledmass-\benchmarkpooledmass<0$.
By the above inequality, we can directly have
$
 \OPT(\noiseScale) \le \max\nolimits_{\pooledmass \in \pooledmassgrid}\Payoff[\noiseScale]{\signalscheme^{(\pooledmass)}}+\gridspace\,$.
Subtracting the utility contribution $\payoffcommon[\noiseScale]$ from the remaining mass $(1-\fullyrationalmass)$, the above inequality becomes
$  \max\nolimits_{\pooledmass \in [\nonpositivemass,\fullyrationalmass]}\payofffullyrational[\noiseScale]{\pooledmass} \le \max\nolimits_{\pooledmass \in \pooledmassgrid}\payofffullyrational[\noiseScale]{\pooledmass}+\gridspace$.
Combining with Eqn.~\eqref{eq:payoff_fully_rational_mass_gap}, we obtain 
\begin{align}
    \label{eq:opt_payoff_fully_rational_mass_gap}
    \left\lvert \max\nolimits_{\pooledmass \in [\nonpositivemass,\fullyrationalmass]}\payofffullyrational[\noiseScale]{\pooledmass}-\max\nolimits_{\pooledmass \in [\nonpositivemass,\fullyrationalmass]}\payofffullyrational[\noiseScaleHat]{\pooledmass} \right\rvert \le  \smallnumber \fullyrationalmass +\gridspace \le \frac{3}{2}\smallnumber\fullyrationalmass\,,
\end{align}
where $\noiseScaleHat \in \rationalitylevelgrid$ is the smallest grid point strictly greater than $\noiseScale$.
\footnote{For the pooled mass $\fullyrationalmass$ with $\censorshipsignal(\fullyrationalmass)=0$, Eqn.~\eqref{eq:payoff_fully_rational_mass_gap} fails for the final rationality level grid interval with the right endpoint of $\noiseScaleHat = \infty$. However, Eqn.~\eqref{eq:opt_payoff_fully_rational_mass_gap} still holds for this interval.
That is because $\max\nolimits_{\pooledmass \in [\nonpositivemass,\fullyrationalmass]}\payofffullyrational[\infty]{\pooledmass} = \fullyrationalmass$ and $\max\nolimits_{\pooledmass \in [\nonpositivemass,\fullyrationalmass]}\payofffullyrational[\noiseScale]{\pooledmass} \ge \payofffullyrational[\noiseScale]{\pooledmass^{\prime}} \ge \pooledmass^{\prime}(1-\smallnumber)\ge \fullyrationalmass-\gridspace-\smallnumber\fullyrationalmass$, where $\pooledmass^{\prime}$ is the largest pooled mass grid point below $\fullyrationalmass$.}

Then, we would like to bound ratio, payoff and regret gap under the censorship $\signalscheme^{(\pooledmass)}$ for any rationality level points $\noiseScale$ and the smallest grid point $\noiseScaleHat \in \rationalitylevelgrid$ strictly greater than $\noiseScale$.

\xhdr{Ratio} In this case, we have
\begin{align*}
\frac{\Payoff[\noiseScaleHat]{\signalscheme^{(\pooledmass)}}}{\OPT(\noiseScaleHat)} &= \frac{\payofffullyrational[\noiseScaleHat]{\pooledmass}+\payoffcommon[\noiseScaleHat]}{\max\nolimits_{\pooledmass^{\prime}\in [\nonpositivemass,\fullyrationalmass]}\payofffullyrational[\noiseScaleHat]{\pooledmass^{\prime}}+\payoffcommon[\noiseScaleHat]} \overset{(a)}{\le} \frac{\payofffullyrational[\noiseScaleHat]{\pooledmass}+\payoffcommon[\noiseScale]}{\max\nolimits_{\pooledmass^{\prime}\in [\nonpositivemass,\fullyrationalmass]}\payofffullyrational[\noiseScaleHat]{\pooledmass^{\prime}}+\payoffcommon[\noiseScale]} \nonumber \\
& \overset{(b)}{\le} \frac{\payofffullyrational[\noiseScale]{\pooledmass}+\payoffcommon[\noiseScale]}{\max\nolimits_{\pooledmass^{\prime}\in [\nonpositivemass,\fullyrationalmass]}\payofffullyrational[\noiseScale]{\pooledmass^{\prime}}+\payoffcommon[\noiseScale]} + 5\smallnumber = \frac{\Payoff[\noiseScale]{\signalscheme^{(\pooledmass)}}}{\OPT(\noiseScale)} + 5\smallnumber\,,
\end{align*}
where inequality~(a) holds since the function $\variableX \mapsto \frac{\variablecapA+\variableX}{\variablecapB+\variableX}$ is monotone non-decreasing when $0\le \variablecapA \le \variablecapB$ and $\payoffcommon[\noiseScaleHat]\le \payoffcommon[\noiseScale]$, and inequality~(b) holds since 
$
    \left\lvert \frac{\payofffullyrational[\noiseScaleHat]{\pooledmass}+\payoffcommon[\noiseScale]}{\max\nolimits_{\pooledmass^{\prime}\in [\nonpositivemass,\fullyrationalmass]}\payofffullyrational[\noiseScaleHat]{\pooledmass^{\prime}}+\payoffcommon[\noiseScale]} - \frac{\payofffullyrational[\noiseScale]{\pooledmass}+\payoffcommon[\noiseScale]}{\max\nolimits_{\pooledmass^{\prime}\in [\nonpositivemass,\fullyrationalmass]}\payofffullyrational[\noiseScale]{\pooledmass^{\prime}}+\payoffcommon[\noiseScale]} \right\rvert
    \le \frac{\left\vert \payofffullyrational[\noiseScaleHat]{\pooledmass}-\payofffullyrational[\noiseScale]{\pooledmass}\right\rvert+\left\vert \max\nolimits_{\pooledmass \in [\nonpositivemass,\fullyrationalmass]}\payofffullyrational[\noiseScale]{\pooledmass}-\max\nolimits_{\pooledmass \in [\nonpositivemass,\fullyrationalmass]}\payofffullyrational[\noiseScaleHat]{\pooledmass}\right\rvert }{\max\nolimits_{\pooledmass^{\prime}\in [\nonpositivemass,\fullyrationalmass]}\payofffullyrational[\noiseScaleHat]{\pooledmass^{\prime}}+\payoffcommon[\noiseScale]} \le \frac{\smallnumber\fullyrationalmass+\frac{3}{2}\smallnumber\fullyrationalmass}{\frac{\fullyrationalmass}{2}} = 5\smallnumber$,
where the last inequality holds due to Eqns.~\eqref{eq:payoff_fully_rational_mass_gap} and \eqref{eq:opt_payoff_fully_rational_mass_gap}, and $\max\nolimits_{\pooledmass^{\prime}\in [\nonpositivemass,\fullyrationalmass]}\payofffullyrational[\noiseScaleHat]{\pooledmass^{\prime}} \ge \payofffullyrational[\noiseScaleHat]{\fullyrationalmass}\ge \frac{\fullyrationalmass}{2}$.

\xhdr{Payoff} In this case, we have
\begin{align*}
  \Payoff[\noiseScaleHat]{\signalscheme^{(\pooledmass)}}= \payofffullyrational[\noiseScaleHat]{\pooledmass}+\payoffcommon[\noiseScaleHat] \le \payofffullyrational[\noiseScale]{\pooledmass} +\smallnumber\fullyrationalmass+\payoffcommon[\noiseScale] = \Payoff[\noiseScale]{\signalscheme^{(\pooledmass)}}+\smallnumber\fullyrationalmass\,,
\end{align*}
where the inequality holds due to Eqn.~\eqref{eq:payoff_fully_rational_mass_gap} and $\payoffcommon[\noiseScaleHat]\le \payoffcommon[\noiseScale]$.

\xhdr{Regret} In this case, we have
\begin{align*}
&\left\lvert\left(\Payoff[\noiseScaleHat]{\signalscheme^{(\pooledmass)}} - \OPT(\noiseScaleHat)\right) - \left(\Payoff[\noiseScale]{\signalscheme^{(\pooledmass)}} - \OPT(\noiseScale)\right)\right\rvert \nonumber\\
    =& \left\lvert \left(\payofffullyrational[\noiseScaleHat]{\pooledmass} -\payofffullyrational[\noiseScale]{\pooledmass}\right) - \left( \max\nolimits_{\pooledmass \in [\nonpositivemass,\fullyrationalmass]}\payofffullyrational[\noiseScaleHat]{\pooledmass}-\max\nolimits_{\pooledmass \in [\nonpositivemass,\fullyrationalmass]}\payofffullyrational[\noiseScale]{\pooledmass} \right) \right\rvert \le \frac{5}{2}\smallnumber\fullyrationalmass\,,
\end{align*}
where the inequality holds due to Eqns.~\eqref{eq:payoff_fully_rational_mass_gap} and \eqref{eq:opt_payoff_fully_rational_mass_gap}.


By the above results, compared to the rationality-robust optimal censorship $\schemeoptimalratio$~/~$\schemeoptimalpayoff$~/~$\schemeoptimalregret$ with the pooled mass $\pooledmassratio$~/~$\pooledmasspayoff$~/~$\pooledmassregret$, respectively, we can bound the additive loss for the estimated rationality-robust signaling scheme $\fptascensorratio$~/~$\fptascensorpayoff$~/~$\fptascensorregret$ with the pooled mass $\pooledmassratioestimation$~/~$\pooledmasspayoffestimation$~/~$\pooledmassregretestimation$, respectively.
Let $\instancerobustRatioestimation{\signalscheme} \triangleq \inf\nolimits_{\noiseScale \in \rationalitylevelgrid} \frac{\Payoff[\noiseScale]{\signalscheme}}{\OPT(\noiseScale)}$, $\instancerobustPayoffestimation{\signalscheme} \triangleq \inf\nolimits_{\noiseScale \in \rationalitylevelgrid} \Payoff[\noiseScale]{\signalscheme}$ and $\instancerobustRegretestimation{\signalscheme} \triangleq \inf\nolimits_{\noiseScale \in \rationalitylevelgrid} \Payoff[\noiseScale]{\signalscheme}-\OPT(\noiseScale)$. Thus, we have $\pooledmassratioestimation \in \arg\max\nolimits_{\pooledmass \in \pooledmassgrid } \instancerobustRatioestimation{\signalscheme^{(\pooledmass)}}$, $\pooledmasspayoffestimation \in \arg\max\nolimits_{\pooledmass \in \pooledmassgrid } \instancerobustPayoffestimation{\signalscheme^{(\pooledmass)}}$, and $\pooledmassregretestimation \in \arg\max\nolimits_{\pooledmass \in \pooledmassgrid } \instancerobustRegretestimation{\signalscheme^{(\pooledmass)}}$.

\xhdr{Ratio} In this case, we have
\begin{align*}
    \instancerobustRatio{\fptascensorratio} \overset{(a)}{\ge} \instancerobustRatioestimation{\fptascensorratio} - 5\smallnumber \overset{(b)}{\ge} \instancerobustRatioestimation{\signalscheme^{(\pooledmass\primed_{\text{ratio}})}}-5\smallnumber \overset{(c)}{\ge} \instancerobustRatio{\signalscheme^{(\pooledmass\primed_{\text{ratio}})}} - 5\smallnumber \overset{(d)}{\ge} \instancerobustRatio{\schemeoptimalratio} - 6\smallnumber\,,
\end{align*}
where $\pooledmass\primed_{\text{ratio}} \in \pooledmassgrid$ is the largest pooled mass grid point no greater than $\pooledmassratio$, inequality~(a) holds since $\instancerobustRatioestimation{\fptascensorratio} \le \frac{\Payoff[\noiseScaleHat]{\fptascensorratio}}{\OPT(\noiseScaleHat)} \le \instancerobustRatio{\fptascensorratio} +5\smallnumber$, where $\noiseScaleHat \in \rationalitylevelgrid$ is the smallest rationality level grid point no less than $\noiseScale \in \arg\min\nolimits_{\noiseScale^{\prime} \in \fullsetrationalitylevel} \frac{\Payoff[\noiseScale^{\prime}]{\fptascensorratio}}{\OPT(\noiseScale^{\prime})}$, inequality~(b) holds since the selected pooled mass maximizes the estimated objective,
inequality~(c) holds due to the chosen rationality level set $\rationalitylevelgrid$ is the subset of the full rationality level set, and inequality~(d) holds since by Eqn.~\eqref{eq:additive_loss} and $\OPT(\noiseScale) \ge \Payoff[\noiseScale]{\signalscheme^{(\fullyrationalmass)}} \ge \fullyrationalmass\Quant\betaed\left(\censorshipsignal(\fullyrationalmass)\right) \ge  \frac{\fullyrationalmass}{2}$, where the last inequality follows from the fact that the quantal response function $\Quant\betaed(\cdot)$ is monotone non-increasing, and $\censorshipsignal(\fullyrationalmass) \le 0$, for any finite rationality level $\noiseScale$, rounding a maximin optimal pooled mass down to $\pooledmassgrid$ loses at most $\frac{2\gridspace}{\fullyrationalmass} \le \smallnumber$.

\xhdr{Payoff} In this case, we have
\begin{align*}
    \instancerobustPayoff{\fptascensorpayoff}  \overset{(a)}{\ge} \instancerobustPayoffestimation{\fptascensorpayoff} -\smallnumber\fullyrationalmass  \overset{(b)}{\ge} \instancerobustPayoffestimation{\signalscheme^{(\pooledmass\primed_{\text{payoff}} )}}-\smallnumber\fullyrationalmass \overset{(c)}{\ge} \instancerobustPayoff{\signalscheme^{(\pooledmass\primed_{\text{payoff})} }} - \smallnumber\fullyrationalmass \overset{(d)}{\ge} \instancerobustPayoff{\schemeoptimalpayoff} -\frac{3}{2}\smallnumber\fullyrationalmass\,,
\end{align*}
 where $\pooledmass\primed_{\text{payoff}} \in \pooledmassgrid$ is the largest pooled mass grid point no greater than $\pooledmasspayoff$, inequality~(a) holds since $\instancerobustPayoffestimation{\fptascensorpayoff} \le \Payoff[\noiseScale\primed]{\fptascensorpayoff} \le \instancerobustPayoff{\fptascensorpayoff} +\smallnumber\fullyrationalmass$ where $\noiseScale\primed \in \rationalitylevelgrid$ is the smallest rationality level grid point no less than $\noiseScale \in \arg\min\nolimits_{\noiseScale^{\prime} \in \fullsetrationalitylevel} \Payoff[\noiseScale^{\prime}]{\fptascensorpayoff}$, inequality~(b) holds since the selected pooled mass maximizes the estimated objective,
 inequality~(c) holds due to the chosen rationality level set $\rationalitylevelgrid$ is the subset of the full rationality level set, and inequality~(d) holds since by Eqn.~\eqref{eq:additive_loss}, rounding a maximin optimal pooled mass down to $\pooledmassgrid$ loses at most $\gridspace \le \frac{\smallnumber \fullyrationalmass}{2}$.

\xhdr{Regret} In this case, we have
\begin{align*}
      \instancerobustRegret{\fptascensorregret}  \overset{(a)}{\ge} \instancerobustRegretestimation{\fptascensorregret} -\frac{5}{2}\smallnumber\fullyrationalmass \overset{(b)}{\ge} \instancerobustRegretestimation{\signalscheme^{(\pooledmass\primed_{\text{regret}})}}-\frac{5}{2}\smallnumber\fullyrationalmass \overset{(c)}{\ge} \instancerobustRegret{\signalscheme^{(\pooledmass\primed_{\text{regret}})}} - \frac{5}{2}\smallnumber\fullyrationalmass \overset{(d)}{\ge}\instancerobustRegret{\schemeoptimalregret} -3\smallnumber\fullyrationalmass\,,
\end{align*}
where $\pooledmass\primed_{\text{regret}} \in \pooledmassgrid$ is the largest pooled mass grid point no greater than $\pooledmassregret$, inequality~(a) holds since $\instancerobustRegretestimation{\fptascensorregret} \le \Payoff[\noiseScale\doubleprimed]{\fptascensorregret} -\OPT(\noiseScale\doubleprimed) \le \instancerobustRegret{\fptascensorregret} +\frac{5}{2}\smallnumber\fullyrationalmass$ where $\noiseScale\doubleprimed \in \rationalitylevelgrid$ is the smallest rationality level grid point no less than $\noiseScale \in \arg\min\nolimits_{\noiseScale^{\prime} \in \fullsetrationalitylevel} \Payoff[\noiseScale^{\prime}]{\fptascensorregret}-\OPT(\noiseScale^{\prime})$,
inequality~(b) holds since the selected pooled mass maximizes the estimated objective,
inequality~(c) holds due to the chosen rationality level set $\rationalitylevelgrid$ is the subset of the full rationality level set, and inequality~(d) holds since by Eqn.~\eqref{eq:additive_loss}, rounding a maximin optimal pooled mass down to $\pooledmassgrid$ loses at most $\gridspace \le \frac{\smallnumber \fullyrationalmass}{2}$.

Notice that $\instancerobustRatio{\schemeoptimalratio} \ge \instancerobustRatio{\signalscheme^{(\fullyrationalmass)}} \ge \frac{1}{2}$ and $\instancerobustPayoff{\schemeoptimalpayoff} \ge \instancerobustPayoff{\signalscheme^{(\fullyrationalmass)}} \ge \frac{\fullyrationalmass}{2}$. Taking $\smallnumber = \sfrac{\accuracy}{12} $, we can get the stated approximation guarantees.
Finally, we analyze the running time. By our construction, there are $O(\sfrac{1}{\accuracy})$ pooled mass grid points and $O(\sfrac{\stateNum}{\accuracy}+\sfrac{1}{\accuracy^2})$ rationality level grid points.
At each rationality level, computing the benchmark censorship needs $O(\stateNum)$ by \Cref{prop:complexity_SISU_opt} and evaluating all possible candidate censorships needs $O(\sfrac{1}{\accuracy})$.
Therefore, the total running time is at most $O(\sfrac{(\stateNum+\sfrac{1}{\accuracy})^2}{\accuracy})$, which is polynomial in $\stateNum$ and $\sfrac{1}{\accuracy}$.
This completes the proof of \Cref{prop:computation_rationality_robust_scheme}.
\end{proof}

\clearpage

\subsection{FPTAS for Computing Approximately Rationality-Robust Censorships}
\label{apx:ptas_rationality_robust}

We present the formal description of the FPTAS for computing approximately rationality-robust censorhships in \Cref{algo:ptas_rationality_robust}.

\begin{algorithm}[ht]
    \caption{FPTAS for computing approximately rationality-robust censorships.}
    \label{algo:ptas_rationality_robust}
    \small
    \begin{algorithmic}[1]
    \State \textbf{Input:} instance
    $\instance = (\stateNum,\{\prior_i\}_{i\in[\stateNum]},\{\receiverU_i\}_{i\in[\stateNum]})$,
    rationality level set
    $\fullsetrationalitylevel$,
    and $\accuracy\in(0,1)$.

    \State \textbf{Output:} approximately rationality-robust censorships
    $\fptascensorratio$, $\fptascensorpayoff$, and
    $\fptascensorregret$.



    \State Compute the total mass for non-positive states
$\nonpositivemass=\sum_{i\in [\stateNum]:\receiverU_i\le0}\prior_i$
    and the optimal censorship for a fully rational receiver's pooled mass $\fullyrationalmass$.

    \State Set
    $\smallnumber=\accuracy/12$,
    $N=\lceil2/\smallnumber\rceil$, and
    $K=\lceil1/\smallnumber\rceil$.

    \State Construct the pooled mass grid
    \[
    \pooledmassgrid=
    \left\{
    \nonpositivemass+
    \frac{j}{N}(\fullyrationalmass-\nonpositivemass):
    j=0,\ldots,N
    \right\}\,.
    \]


    \State Initialize the rationality level grid $\rationalitylevelgrid=\{0,\infty\}$.
    For each nonzero signal
    \[
    \scaledNoiseDiff \in
    \{\censorshipsignal(\pooledmass):
       \pooledmass\in\pooledmassgrid\}
    \cup\{\receiverU_i:\receiverU_i>0\},
    \]
    add every non-negative value
    $\noiseScale=\scaledNoiseDiff^{-1}\log((K-k)/k)$,
    $k=1,\ldots,K-1$, to $\rationalitylevelgrid$.

    \State For each $\noiseScale\in\rationalitylevelgrid $,
    compute $\OPT(\noiseScale)$ using \Cref{prop:complexity_SISU_opt}.
For each $\pooledmass\in\pooledmassgrid$, compute \footnotemark 
    \[
    \begin{aligned}
    \instancerobustRatioestimation{\parameterizedscheme}
    =
    \inf\nolimits_{\noiseScale\in\rationalitylevelgrid}
    \frac{\Payoff[\noiseScale]{\parameterizedscheme}}
         {\OPT(\noiseScale)}\,,\,
    \instancerobustPayoffestimation{\parameterizedscheme}
    =
    \inf\nolimits_{\noiseScale\in\rationalitylevelgrid}
    \Payoff[\noiseScale]{\parameterizedscheme}\,,\,
    \instancerobustRegretestimation{\parameterizedscheme}
    =
    \inf\nolimits_{\noiseScale\in\rationalitylevelgrid}
    \Payoff[\noiseScale]{\parameterizedscheme}
    -\OPT(\noiseScale)\,.
    \end{aligned}
    \]

    \State Choose $\pooledmassratioestimation, \pooledmasspayoffestimation \in \pooledmassgrid$ to maximize the estimated worst-case ratio and payoff, respectively, and choose $\pooledmassregretestimation \in \pooledmassgrid$ to 
    maximize the estimated negative-regret objective.

    \State \Return
    $\fptascensorratio=
        \signalscheme^{(\pooledmassratioestimation)}$,
    $\fptascensorpayoff=
        \signalscheme^{(\pooledmasspayoffestimation)}$, and
    $\fptascensorregret=
        \signalscheme^{(\pooledmassregretestimation)}$.
    \end{algorithmic}
\end{algorithm}
\footnotetext{For the pooled mass $\fullyrationalmass$, if $\censorshipsignal(\fullyrationalmass)=0$, we directly obtain its worst-case performance $\instancerobustRatio{\signalscheme^{(\fullyrationalmass)}} = {1}/{2}$, $\instancerobustPayoff{\signalscheme^{(\fullyrationalmass)}} = {\fullyrationalmass}/{2}$ and $\instancerobustRegret{\signalscheme^{(\fullyrationalmass)}} = -{\fullyrationalmass}/{2}$, without performing a grid search. }


\end{document}